\documentclass{elsart}

\usepackage{amsmath,amssymb,epsfig,bm}
\usepackage{caption}
\usepackage{subcaption}
\usepackage{xcolor} % Required for color

\numberwithin{equation}{subsection}

\journal{Physics Reports}

\begin{document}

%%%%%%%%%%%%%%%%%%%%%%%%%%%%%%%%%%%%%%%%
% Own definitions                      %
%%%%%%%%%%%%%%%%%%%%%%%%%%%%%%%%%%%%%%%%

% Units

\newcommand{\eV}{\text{eV}}
\newcommand{\GeV}{\text{GeV}}
\newcommand{\arcmin}{\text{arcmin}}
\newcommand{\Mpc}{\text{Mpc}}
\newcommand{\Hunit}{$\text{km\,s}^{-1}\,\text{Mpc}^{-1}$}
\newcommand{\muK}{$\mu\text{K}$}

% Derivatives

\newcommand{\D}{\text{D}}
\newcommand{\ud}{\text{d}}
\newcommand{\curl}{\,\text{curl}\,}

% Inequalities

\newcommand{\alt}{\lesssim}
\newcommand{\agt}{\gtrsim}

% Calligraphic letters

\newcommand{\cla}{\mathcal{A}}
\newcommand{\clb}{\mathcal{B}}
\newcommand{\clc}{\mathcal{C}}
\newcommand{\cld}{\mathcal{D}}
\newcommand{\cle}{\mathcal{E}}
\newcommand{\clf}{\mathcal{F}}
\newcommand{\clg}{\mathcal{G}}
\newcommand{\clh}{\mathcal{H}}
\newcommand{\cli}{\mathcal{I}}
\newcommand{\clj}{\mathcal{J}}
\newcommand{\clk}{\mathcal{K}}
\newcommand{\cll}{\mathcal{L}}
\newcommand{\clm}{\mathcal{M}}
\newcommand{\cln}{\mathcal{N}}
\newcommand{\clo}{\mathcal{O}}
\newcommand{\clp}{\mathcal{P}}
\newcommand{\clq}{\mathcal{Q}}
\newcommand{\clr}{\mathcal{R}}
\newcommand{\cls}{\mathcal{S}}
\newcommand{\clt}{\mathcal{T}}
\newcommand{\clu}{\mathcal{U}}
\newcommand{\clv}{\mathcal{V}}
\newcommand{\clw}{\mathcal{W}}
\newcommand{\clx}{\mathcal{X}}
\newcommand{\cly}{\mathcal{Y}}
\newcommand{\clz}{\mathcal{Z}}

% Bold 3-vectors

\newcommand{\ve}{\mathbf{e}}
\newcommand{\vehat}{\hat{\mathbf{e}}}
\newcommand{\vn}{\mathbf{n}}
\newcommand{\vnhat}{\hat{\mathbf{n}}}
\newcommand{\vv}{\mathbf{v}}
\newcommand{\vx}{\mathbf{x}}
\newcommand{\vp}{\mathbf{p}}
\newcommand{\vk}{\mathbf{k}}

% Miscellaneous cosmology

\newcommand{\Omtot}{\Omega_{\mathrm{tot}}}
\newcommand{\Omb}{\Omega_{\mathrm{b}}}
\newcommand{\Omc}{\Omega_{\mathrm{c}}}
\newcommand{\Omm}{\Omega_{\mathrm{m}}}
\newcommand{\omb}{\omega_{\mathrm{b}}}
\newcommand{\omc}{\omega_{\mathrm{c}}}
\newcommand{\omm}{\omega_{\mathrm{m}}}
\newcommand{\omnu}{\omega_{\nu}}
\newcommand{\Omnu}{\Omega_{\nu}}
\newcommand{\Oml}{\Omega_\Lambda}
\newcommand{\OmK}{\Omega_K}

% Miscellaneous HD & MHD

%\newcommand{\Cs}{c_{\rm s}}
%\newcommand{\Cs2}{c_{\rm s}^2}
%\newcommand{\Ca}{c_{\rm a}}
%\newcommand{\Ca2}{c_{\rm a}^2}

% Tensors

\newcommand{\Al}{{A_l}}
\newcommand{\TT}{\text{TT}}

% Cross sections

\newcommand{\sigt}{\sigma_{\mbox{\scriptsize T}}}

% Journals

\newcommand{\annphys}{\rm Ann.~Phys.~}
\newcommand{\araa}{\rm Ann.~Rev.~Astron.~\&~Astrophys.~}
\newcommand{\aap}{\rm Astron.~\&~Astrophys.~}
\newcommand{\aj}{\rm Astron.~J.~}
\newcommand{\apj}{\rm Astrophys.~J.~}
\newcommand{\apjl}{\rm Astrophys.~J.~Lett.~}
\newcommand{\apjs}{\rm Astrophys.~J.~Supp.~}
\newcommand{\apss}{\rm Astrophys.~Space~Sci.~}
\newcommand{\cqg}{\rm Class.~Quant.~Grav.~}
\newcommand{\grg}{\rm Gen.~Rel.~Grav.~}
\newcommand{\ijmpd}{\rm Int.~J.~Mod.~Phys.~D~}
\newcommand{\jcap}{\rm JCAP~}
\newcommand{\jetp}{\rm J.~Exp.~Theor.~Phys.~}
\newcommand{\jetpl}{\rm J.~Exp.~Theor.~Phys.~Lett.~}
\newcommand{\jmp}{\rm J.~Math.~Phys.~}
\newcommand{\mnras}{\rm Mon.~Not.~R.~Astron.~Soc.~}
\newcommand{\nar}{\rm New~Astron.~Rev.~}
\newcommand{\nat}{\rm Nature~}
\newcommand{\prd}{\rm Phys. Rev.~D~}
\newcommand{\prl}{\rm Phys.~Rev.~Lett.~}
\newcommand{\plb}{\rm Phys.~Lett.~B~}
\newcommand{\physrep}{\rm Phys.~Rep.~}
\newcommand{\progthp}{\rm Prog.~Theor.~Phys.~}
\newcommand{\rmp}{\rm Rev.~Mod.~Phys.~}
\newcommand{\epjc}{\rm Eur.~Phys.~J.~C~}
\newcommand{\pdu}{\rm Phys.~Dark~Univ.~}
\newcommand{\ssr}{\rm Space~Sci.~Rev.~}
\newcommand{\pasa}{\rm Publ.~Astron.~Soc.~Austral.~}

\let\Oldsection\section
\renewcommand{\section}[1]{\Oldsection{\bf #1}}

\newcommand{\be} {\begin{equation}}
\newcommand{\ee} {\end{equation}}
\newcommand{\bea} {\begin{eqnarray}}
\newcommand{\eea} {\end{eqnarray}}

%%%%%%%%%%%%%%%%%%%%%%%%%%%%%%%%%%%%%%%%
% Front matter                         %
%%%%%%%%%%%%%%%%%%%%%%%%%%%%%%%%%%%%%%%%

\begin{frontmatter}

% Title, authors and addresses
% use the thanksref command within \title, \author or \address for footnotes;
% use the corauthref command within \author for corresponding author footnotes;
% use the ead command for the email address,
% and the form \ead[url] for the home page:
% \title{Title\thanksref{label1}}
% \thanks[label1]{}
% \author{Name\corauthref{cor1}\thanksref{label2}}
% \ead{email address}
% \ead[url]{home page}
% \thanks[label2]{}
% \corauth[cor1]{}
% \address{Address\thanksref{label3}}
% \thanks[label3]{}

\title{Large-scale peculiar velocities in the universe}

% use optional labels to link authors explicitly to addresses:
% \author[label1,label2]{}
% \address[label1]{}
% \address[label2]{}

\author{Christos G. Tsagas}
\address{Section of Astrophysics, Astronomy and Mechanics, Department of Physics, Aristotle University of Thessaloniki, Thessaloniki 54124, Greece\\ and\\ Clare Hall, University of Cambridge, Herschel Road, Cambridge CB3 9AL, UK\\ email: tsagas@astro.auth.gr}

\author{Leandros Perivolaropoulos}
\address{Division of Theoretical Physics, Department of Physics, University of Ioannina, Ioannina 45110, Greece\\ email: leandros@uoi.gr}

\author{Kerkyra Asvesta}
\address{Section of Astrophysics, Astronomy and Mechanics, Department of Physics, Aristotle University of Thessaloniki, Thessaloniki 54124, Greece\\ email: keasvest@auth.gr}\vspace{1cm}

\tableofcontents

\newpage

\begin{abstract}
Observations have repeatedly confirmed the presence of large-scale peculiar motions in the universe, commonly referred to as ``bulk flows''. These are vast regions of the observable universe, typically spanning scales of several hundred Mpc, that move coherently with speeds of the order of several hundred km/sec. While there is a general consensus on the direction of these motions, discrepancies persist in their reported sizes and velocities, with some of them exceeding the predictions of the standard $\Lambda$CDM model. The observed large-scale peculiar-velocity fields are believed to have originated as weak peculiar-velocity perturbations soon after equipartition, which have subsequently grown by structure formation and by the increasing inhomogeneity of the post-recombination universe. However, the evolution and the implications of these bulk velocity fields remain poorly understood and they are still a matter of debate. For instance, it remains a challenge for the theoreticians to explain the high velocities measured by several bulk-flow surveys, like those recently reported using the \textit{CosmicFlows-4} data. Such extensive and fast velocity fields could have played a non-negligible role during structure formation and they might have also ``contaminated'' our observations. After all, in the history of astronomy, there are examples where relative-motion effects have led us to a serious misinterpretation of reality. The review starts by examining the latest observations of bulk peculiar flows, highlighting their methodologies and findings. We discuss the current and the emerging techniques used to measure large-scale peculiar-velocity fields and analyze their strengths and limitations. The review also addresses the question of whether the reported observations appear to converge towards a consensus and to align with the established cosmological theories. We then proceed to the theoretical frameworks employed to study large-scale peculiar velocities, starting with the Newtonian approaches first and subsequently proceeding to the relativistic treatments. The aim is to compare these two schools of study, identify their differences, reveal the reasons behind their different predictions, as they occur, and investigate their potential implications. Throughout, the theoretical models are critically evaluated against the observational data, thus allowing for a comprehensive understanding of the bulk-flow phenomenon and of its role in cosmic structure formation, as well as of the potential biases it may introduce to astronomical observations. Of particular interest are the recently reported dipolar anisotropies in cosmological parameters and in the number counts of distant astrophysical sources, given that dipoles are the trademark signature of relative motions. Since the latter have been responsible for serious astronomical misinterpretations in the past, we discuss the methods proposed to mitigate the biases. We close the review with a brief presentation of the upcoming peculiar-velocity surveys and a discussion on the theoretical and computational advances expected from the peculiar-velocity science of the next generation.
\end{abstract}


\end{frontmatter}

\newpage

%%%%%%%%%%%%%%%%%%%%%%%%%%%%%%%%%%%%%%%%
% Main text                            %
%%%%%%%%%%%%%%%%%%%%%%%%%%%%%%%%%%%%%%%%

\section{Introduction}\label{sI}
%%%%%%%%%%%%%%%%%%%%%%%%%%%%%%%%
Peculiar velocities are not mere random galactic motions, but manifest themselves a large-scale {\it bulk peculiar flows}. These are vast regions of the universe, often spanning hundreds of $\Mpc$, that move coherently with collective velocities of hundreds of~km/sec. The study of these bulk flows has been the subject of intense research since the latter half of the 20th century.

\subsection{The enigma of bulk flows: what are they?}\label{ssEBF}
%%%%%%%%%%%%%%%%%%%%%%%%%%%%%%%%%%%%%%%%%%%%%%%%%%%%%%%%%%%%%%%%%%
Early observational efforts, though constrained by limited data, provided the first hints of systematic motions. The seminal work of Vera Rubin and collaborators in the mid-1970s, known as the {\it Rubin-Ford effect}, first suggested a large-scale anisotropy in the cosmic expansion, pointing to a collective motion of galaxies relative to the rest frame of the Cosmic Microwave Background (CMB)~\cite{1976AJ.....81..687R,1976AJ.....81..719R}. A decade later, the seminal \textit{Seven Samurai} survey revealed evidence for a massive, unseen gravitational concentration dubbed the ``Great Attractor'', whose influence extended over a vast region of the local universe~\cite{1987ApJ...313...42D,1988ApJ...326...19L}. This discovery established that peculiar velocities are a powerful probe of the universe's mass distribution on scales far larger than those accessible by traditional galaxy surveys.

In standard cosmology, bulk flows are believed to be a natural consequence of structure formation. Tiny density perturbations in the primordial universe gradually grew through the physical mechanism of \textit{gravitational (or Jeans) instability}, pulling matter towards overdense regions and gradually creating the large-scale structures we observe today, like the galaxy clusters and the superclusters. The enhanced gravitational pull from these structures generated the observed peculiar-velocity fields, driving the motion of the individual galaxies and creating the reported large-scale flows~\cite{1976ApJ...205..318P,1980lssu.book.....P}. The magnitude and direction of these flows provide a dynamic window to survey the distribution of both the luminous and the dark matter. However, despite a general consensus on the existence and the direction of these bulk motions, significant discrepancies still persist in their reported amplitudes and coherence scales. These inconsistencies, particularly the reports of extremely fast flows extending over unusually large distances, remain a source of contention and an active area of investigation within the cosmological community. A detailed discussion of these discrepancies and the tension they create with the standard model is presented in \S~9.3 and \S~9.4.

\subsection{A persistent cosmological tension: bulk flows and the
%%%%%%%%%%%%%%%%%%%%%%%%%%%%%%%%%%%%%%%%%%%%%%%%%%%%%%%%%%%%%%%%%
$\Lambda$CDM model}\label{ssPCTBFLCDM}
%%%%%%%%%%%%%%%%%%%%%%%%%%%%%%%%%%%%%%
While the existence of bulk flows is well-established, their precise amplitude and scale remain a subject of debate, with reports of peculiar motions often deviating from the standard $\Lambda$CDM paradigm. Theoretically, based largely on Newtonian analyses where linear peculiar velocities grow slowly as $v\propto t^{1/3}$~\cite{1976ApJ...205..318P,1980lssu.book.....P}, the standard model predicts that bulk flow amplitudes should decrease with scale, converging to the CMB rest-frame beyond approximately 100/h~\Mpc. Testing this prediction has yielded a mixed picture that roughly splits according to the depth of the survey. Numerous studies probing scales within the 100/h~Mpc threshold find amplitudes in excellent agreement with $\Lambda$CDM expectations (e.g.~see~\cite{2011ApJ...736...93N,2011JCAP...04..015D,%
2012MNRAS.420..447T,2012MNRAS.424..472B,2013MNRAS.428.2017M,%
2013A&A...560A..90F,2014MNRAS.437.1996M,2014MNRAS.445..402H,%
2015MNRAS.450..317C,2016ApJ...827...60M,2016MNRAS.455..386S,%
2016MNRAS.456.1886S,2018MNRAS.477.5150Q,2019MNRAS.482.1920Q,%
2020MNRAS.498.2703B,2021MNRAS.505.2349S,2021ApJ...922...59Q} and \S~\ref{ssBFWLCDMLs} here). Conversely, an increasing number of surveys probing larger volumes report flows that are substantially faster and deeper than predicted, posing a significant and persistent challenge to the standard model (e.g.~see~\cite{2004MNRAS.352...61H,2008MNRAS.387..825F,%
2009MNRAS.392..743W,2010MNRAS.407.2328F,2010ApJ...709..483L,%
2011MNRAS.414..264C,2015MNRAS.447..132W,2018MNRAS.481.1368P,%
2021MNRAS.504.1304S} and \S~\ref{ssBFELCDMLs} here). Notable among these are the recent \textit{Cosmic Flows 4} results, which revived the controversy by finding excessive bulk flows well beyond 100/h~Mpc~\cite{2023MNRAS.524.1885W,2023MNRAS.526.3051W}, and the even more dramatic reports of \textit{dark flows}, which claim coherent cluster motions approaching 1000~km/sec on scales of several hundreds of megaparsecs, potentially challenging the foundational principle of cosmic isotropy (see~\cite{2000ApJ...536L..67K,2008ApJ...686L..49K,%
2009ApJ...691.1479K,2010ApJ...712L..81K,2011ApJ...732....1K,%
2012arXiv1202.0717K} and also \S~\ref{ssDFQ} here).

Although not yet a definitive falsification of the $\Lambda$CDM paradigm, these discrepancies highlight the limitations of our current understanding. The disagreement between the surveys may point to subtle biases in the observational techniques, to an underestimation of cosmic variance, or to the need for a more comprehensive theoretical framework that goes beyond the standard Newtonian approximation. Overall, the discrepancy underscores the importance of continued research into the bulk flows, positioning them as a critical test of the standard cosmological model and as a potential window into new developments. The disagreement between the surveys may point to subtle biases in the observational techniques (discussed in \S~\ref{sec:pec_vel_method}), to an underestimation of cosmic variance, or to the need for a more comprehensive theoretical framework that goes beyond the standard Newtonian approximation (see \S~\ref{sLPVs}). Furthermore, systematic errors arising from instrumentation, such as calibration drifts, can mimic large-scale flows if not properly characterized (e.g.~see~\cite{2014A&A...571A...3P} for Planck systematics).

\subsection{Theoretical perspectives: Newtonian and relativistic
%%%%%%%%%%%%%%%%%%%%%%%%%%%%%%%%%%%%%%%%%%%%%%%%%%%%%%%%%%%%%%%%
views}\label{TPNRVs}
%%%%%%%%%%%%%%%%%%%%
Newtonian physics and general relativity are fundamentally different theories when it comes to gravity, because of the way they treat both the gravitational field and its sources. For example, in Newtonian theory the evolution of peculiar velocities is primarily driven by the gravitational acceleration generated by the emerging mass overdensities. This is the foundation of the gravitational instability paradigm, which is central to the standard $\Lambda$CDM model. The evolution of peculiar velocities is typically derived using linearized Newtonian perturbation theory, where the gravitational potential is determined by the Poisson equation sourced by density fluctuations. This approach predicts a relatively slow growth-rate for peculiar velocities after equipartition and during the matter-dominated era. This growth is generally understood to scale with time as $\tilde{v}\propto t^{1/3}$, a result that has been consistently recovered in both the traditional Newtonian studies~\cite{1976ApJ...205..318P,1980lssu.book.....P} and in their quasi-Newtonian counterparts. The latter are irrotational perturbed spacetimes with vanishing shear and magnetic Weyl tensor, that is with no gravitational waves, while their 4-acceleration is given by the gradient of an arbitrary (Newtonian-like) scalar potential. All these constraints hold at the linear perturbative level (see~\cite{1998PhRvD..58l4006M,2001CQGra..18.5115E} and also \S~\ref{ssNA} and \S~\ref{ssQ-NA} in this review). The aforementioned moderate growth-rate serves as the theoretical benchmark against which the observed amplitudes and scales of bulk flows are compared. Therefore, when a survey reports bulk velocities significantly in excess of these predictions, it signals a potential tension with the standard model.

The relativistic approach, however, introduces a crucial refinement. According to general relativity, gravity is sourced not only by mass density, but also by {\it matter fluxes} (i.e.~by matter in motion). Unlike Newtonian physics, where momentum fluxes also exist but do not source the Poisson equation, in relativity the fluxes add to the energy-momentum tensor and act as an additional source of gravity, which then couples back to the equations governing the evolution of linear peculiar velocities.\footnote{For discussions on post-Newtonian studies and non-linear extensions, we refer the reader to~\cite{1995ApJ...442...30K,2015PhRvD..92b3519A,%
2016PhRvD..94h3515A}.} As demonstrated in a series of covariant relativistic studies (see~\cite{2020EPJC...80..757T,%
2021Ap&SS.366....4F,2022PhRvD.106h3505M,2024PhRvD.110f3540M,%
2026ApJ...997...25T} and also \S~\ref{ssRA} here), including the gravitational input of the \textit{peculiar flux} leads to a considerably faster linear growth-rate for the peculiar-velocity field. More specifically, on a flat and matter-dominated Friedmann-Lema\^itre-Robertson-Walker (FLRW) background universe, the minimum linear growth-rate of the relativistic analysis is $\tilde{v}\propto t$, which is substantially stronger than the Newtonian $\tilde{v}\propto t^{1/3}$ prediction. This result offers a natural general-relativistic mechanism to explain the existence of high-amplitude bulk flows, potentially bridging the gap between theory and observation without requiring any new physics beyond the standard model. In this context, the tension between theory and observation is not a sign of a flawed model, but rather a reflection of the limitations of the Newtonian approximations employed to predict the growth of peculiar velocities. The debate thus shifts, from questioning the standard model itself, to scrutinizing the validity of the approximations used in its application to cosmic dynamics.

\subsection{Methodological challenges and observational
%%%%%%%%%%%%%%%%%%%%%%%%%%%%%%%%%%%%%%%%%%%%%%%%%%%%%%%
biases}\label{ssMCOBs}
%%%%%%%%%%%%%%%%%%%%%%
Accurately measuring peculiar velocities is a challenging endeavor, rife with potential for systematic biases and observational uncertainties. These difficulties stem primarily from the need to derive a distance estimate that is independent of the galaxy's redshift, a process that is susceptible to a number of methodological pitfalls.

A primary challenge lies in the intrinsic scatter of {\it redshift-independent distance indicators}, such as the Tully-Fisher (TF) relation~\cite{1977A&A....54..661T} and the Fundamental Plane (FP)~\cite{1987ApJ...313...59D}. While these relations provide robust distance estimates, they possess a finite intrinsic scatter that can lead to significant distance errors, particularly for individual galaxies. This uncertainty translates directly into the derived peculiar velocities, where a small fractional error in distance can result in a large absolute velocity error, especially at greater distances~\cite{1995PhR...261..271S}.

Furthermore, the process of galaxy selection and measurement introduces significant biases that must be rigorously addressed. The most prominent of these is the {\it Malmquist bias}~\cite{1924MeLuF.100....1M}, which arises in flux-limited surveys. There, intrinsically fainter galaxies are observed only if they are relatively nearby, while brighter galaxies can be seen at much greater distances. This leads to a systematic overestimation of distances for a given observed redshift. A related effect, known as the {\it inhomogeneous Malmquist bias}, occurs when peculiar motions themselves conspire with the density field to modulate the number of visible sources~\cite{1995PhR...261..271S}. For example, galaxies in an overdense region are typically falling inward, making them appear closer than they are, which can create a spurious, large-scale infall signal in the reconstructed velocity field~\cite{2011MNRAS.413.2906D,2012ApJ...751L..30H}.

Other systematic effects include:
\begin{itemize}
    \item \textbf{Incomplete sky coverage:} The ``zone of avoidance'', caused by the Milky Way's obscuring dust and stars, leaves a large portion of the sky unobserved. This can bias the bulk-flow measurements, particularly their directions~\cite{2006MNRAS.368.1515E}.\\

    \item  \textbf{Distance indicator calibration:} The empirical relationships used for distance estimation often depend on the galaxy's properties (e.g.~color, metallicity, morphology). Improper calibration for these dependencies can introduce significant systematic errors that correlate across large portions of a survey. For instance, the TF relation can be affected by the galaxy's inclination, while the FP is sensitive to stellar population gradients~\cite{2006ApJ...653..861M,2012MNRAS.427..245M}.\\

    \item \textbf{Cosmic variance:} Even with perfect measurements, the survey of a finite volume is subject to sampling a specific and potentially unrepresentative region of the universe. This ``cosmic variance'' sets a fundamental limit on the precision of the bulk-flow measurements, particularly on very large scales where the available cosmic volume is limited~\cite{2009MNRAS.392..743W,2013MNRAS.428.2017M}.
\end{itemize}

To address these challenges, one needs advanced statistical and computational methods. Techniques like Bayesian hierarchical modeling, Wiener filtering and forward modeling with mock catalogs are now standard practice. These methods enable a more robust reconstruction of the velocity field and allow for a rigorous quantification of the uncertainties, all of which are crucial for distinguishing between genuine signals and artifacts of measurement bias.

\subsection{Broader implications: from structure formation to
%%%%%%%%%%%%%%%%%%%%%%%%%%%%%%%%%%%%%%%%%%%%%%%%%%%%%%%%%%%%%
cosmological tensions}\label{ssBISFCTs}
%%%%%%%%%%%%%%%%%%%%%%%%%%%%%%%%%%%%%%%
The study of peculiar velocities extends beyond merely mapping cosmic flows, but has profound implications for our understanding of cosmic evolution, for the validity of the $\Lambda$CDM model and for the resolution of major cosmological tensions. Peculiar velocities are a direct consequence of structure formation, tracing the gravitational growth of density perturbations from the early universe to the present day~\cite{1976ApJ...205..318P,%
1980lssu.book.....P}. Bulk flows are particularly valuable as a complementary probe to static density maps, offering a dynamic perspective on the distribution of both the luminous and the dark matter. The coherence and amplitude of these motions on very large scales, for instance, constrain the properties of the primordial power spectrum and provide a unique test of gravitational theories on scales where they are most challenged~\cite{2010MNRAS.407.2328F,%
2017NatAs...1E..36H}. The development of powerful tools, such as the Zel'dovich approximation, has allowed for a deeper understanding of how peculiar velocity fields could have shaped the evolution of nonlinear structures, such the so-called ``filaments'' and ``pancakes''~\cite{1970Afz.....6..319Z,1970A&A.....5...84Z}.

Furthermore, peculiar velocities play a central role in a number of ongoing cosmological debates. The {\it Hubble tension}, for example, which reflects the discrepancy between the local measurements of the Hubble constant ($H_0$) and those derived from the CMB observations, may be partially addressed by accounting for peculiar-velocity effects. For instance, if local observers reside in a large-scale bulk flow, the distance measurements of their nearby supernova could be systematically biased, affecting the local determination of $H_0$~\cite{2020A&A...636A..15M,2021A&A...649A.151M,%
2024A&A...691A.355P}. By carefully accounting and correcting for these local kinematic effects, researchers aim to reconcile the conflicting $H_0$ values, although this alone may not fully resolve the tension~\cite{2022NewAR..9501659P}.

Beyond this, the effects of relative motion can introduce apparent (Doppler like) anisotropies in the cosmological observations. In a ``tilted universe'', where observers have a peculiar velocity relative to the cosmic rest-frame, their measurements will appear to be direction dependent. This dependence can manifest itself in a host of ways. For example, as a dipolar anisotropy in the number counts of distant astrophysical sources, or in the measurements of cosmological parameters. In fact, such dipoles have been reported in the number counts of radio galaxies and quasars~\cite{2011ApJ...742L..23S,2013A&A...555A.117R,%
2017MNRAS.471.1045C,2018MNRAS.477.1772R,2019MNRAS.488L.104S,%
2021ApJ...908L..51S,2022ApJ...937L..31S,2023A&A...675A..72W} (see also \cite{2021EPJST.230.2067M,2019JCAP...03..023P,%
2020IJMPD..2950085B,2022JCAP...01..061G} for discussions on the kinematic origin of the dipole), in the value of the Hubble constant~\cite{2020A&A...636A..15M,2021A&A...649A.151M,%
2024A&A...691A.355P}, in the measurements of the universal acceleration~\cite{2010MNRAS.401.1409C,2014MNRAS.443.1680W,%
2019A&A...631L..13C,2025EPJC...85..596S} and in the distribution of $\Omega_{\Lambda}$~\cite{2024ApJ...971...19C}. Moreover, the reported dipoles seem consistent with our local bulk flow. Although the cosmic-dipole question has not been settled yet (e.g.~see~\cite{2020ApJ...894...68R,2019arXiv191204257C,%
2024JCAP...11..067A,2024MNRAS.531.4545O} for a ``sample'' of related discussions), the findings of~\cite{2019A&A...631L..13C,%
2025EPJC...85..596S,2024ApJ...971...19C}, combined with the theoretical predictions of~\cite{2010MNRAS.405..503T,%
2011PhRvD..84f3503T,2015PhRvD..92d3515T,2021EPJC...81..753T}, raise the profound question whether a portion (at least) of the observed cosmic acceleration could be a local kinematic illusion rather than a global cosmological event driven by dark energy? After all, the history of astronomy is replete with examples of relative motions leading to profound misinterpretations, from planetary retrograde motion to the initial understanding of the CMB dipole, underscoring the need for continued vigilance in separating local kinematic effects from genuine cosmic phenomena. Having said that, the same aforementioned findings could also be seen as a challenge to the ``Cosmological Principle'' and to the very foundations of modern cosmology~\cite{2021ApJ...908L..51S,2023CQGra..40i4001A}.

\subsection{Objective and scope of the review}\label{ssOSR}
%%%%%%%%%%%%%%%%%%%%%%%%%%%%%%%%%%%%%%%%%%%%%%%%%%%%%%%%%%%
The primary objective of this review is to provide a comprehensive and critical synthesis of the science of large-scale peculiar velocities, bridging the gap between historical observations, modern high-precision measurements and the diverse theoretical frameworks used to interpret them. A central goal is to investigate the persistent tension between the observed properties of cosmic flows - specifically their amplitude and scale - and the predictions of the standard $\Lambda$CDM model. In addition, we seek to explore the profound implications of peculiar velocities, not just for structure formation, but also for our broader understanding of the cosmic dynamics and for the potential resolution of contemporary cosmological tensions.

To achieve this, the review is structured to guide the reader through the foundational, methodological and conceptual aspects of the field:
\begin{itemize}
    \item \textbf{Historical context and milestones:} We begin by tracing the evolution of the peculiar-velocity science from its origins in the mid-20th century to the current era of precision cosmology. This  discussion takes place in \S~\ref{sHCOE}, where we present key historical milestones, like the discovery of the ``Great Attractor''~\cite{1987ApJ...313...42D} for example. In the subsequent section \S~\ref{sec:pec_vel_method}, we analyze the methodologies and the findings of the major modern surveys, among them of the \textit{Cosmic Flows} program~\cite{2013AJ....146...86T,2016AJ....152...50T,%
        2023ApJ...944...94T}, highlighting both the areas of consensus and the lingering discrepancies.\\

    \item \textbf{Theoretical evolution vs observations:} A significant portion of this review discusses the theoretical studies on the evolution and the implications of peculiar motions vs the related observations. We therefore start with a brief introduction to the basic theoretical framework necessary for the Newtonian and the relativistic analysis of cosmological peculiar velocities in \S~\ref{sTF} and \S~\ref{sCPVT}. The latter section also familiarizes the reader with the \textit{tilted cosmological models}, which are best suited to peculiar-velocity studies. We first delve into the theoretical underpinnings regarding the evolution of cosmological peculiar velocities. The motivation comes form the observations, which report relatively slow bulk flows and in agreement with the $\Lambda$CDM paradigm at low redshifts~\cite{2011ApJ...736...93N,2011JCAP...04..015D,%
        2012MNRAS.420..447T,2012MNRAS.424..472B,2013MNRAS.428.2017M,%
        2013A&A...560A..90F,2014MNRAS.437.1996M,2014MNRAS.445..402H,%
        2015MNRAS.450..317C,2016ApJ...827...60M,2016MNRAS.455..386S,%
        2016MNRAS.456.1886S,2018MNRAS.477.5150Q,2019MNRAS.482.1920Q,%
        2020MNRAS.498.2703B,2021MNRAS.505.2349S,2021ApJ...922...59Q}. At higher redshifts, however, the reported bulk motions are considerably faster and at odds with the current cosmological model~\cite{2004MNRAS.352...61H,2008MNRAS.387..825F,%
        2009MNRAS.392..743W,2010MNRAS.407.2328F,2010ApJ...709..483L,%
        2011MNRAS.414..264C,2015MNRAS.447..132W,2018MNRAS.481.1368P,%
        2021MNRAS.504.1304S,2023MNRAS.524.1885W,2023MNRAS.526.3051W}. This discrepancy may be due to the different systematics of the associated methodologies, or it may reflect a scale-dependence in the evolution of the peculiar velocities, as discussed in detail in \S~\ref{sPMLCDM}. Having said that, a possible reason could be the theoretical model used to analyse the growth of peculiar velocities, which within the $\Lambda$CDM scenario is purely Newtonian. Therefore, in \S~\ref{sLPVs} we provide a detailed comparison between the Newtonian/quasi-Newtonian approaches~\cite{1976ApJ...205..318P,1998PhRvD..58l4006M} and the relativistic treatment~\cite{2020EPJC...80..757T,2026ApJ...997...25T}, which argues for faster and deeper bulk peculiar flows. The same section also illuminates a key theoretical distinction between the two different approaches, namely the role of the peculiar flux as an additional (purely relativistic) source of gravity. Including the effect of the flux may offer a natural explanation for the faster growth-rates measured in some of the reported large-scale bulk flows, without the need of introducing any new physics, or altering the standard cosmological model.\\

    \item \textbf{Impact on the observations:} Relative-motion effects have caused significant astronomical misinterpretations in the past; a fact that should serve as a cautionary tale for modern cosmology as well. With this in mind, we discuss how peculiar motions could contaminate/bias the current cosmological observations in \S~\ref{sOSPMs}. Among others, we discuss how large-scale peculiar motions could spoil the isotropy in the sky-distribution of cosmological parameters, like the Hubble and the deceleration parameter, or the isotropy in the number counts of distant astrophysical sources. Again, the motivation for the theoretical studies came from the observations. Indeed, as discussed in \S~\ref{ssDAPMs}, there have been several reports of dipole-like asymmetries in the Hubble flow~\cite{2020A&A...636A..15M,%
        2021A&A...649A.151M}, in the universal acceleration~\cite{2019A&A...631L..13C,2024ApJ...971...19C,%
        2025EPJC...85..596S} and in the distribution of distant radio galaxies and quasars~\cite{2011ApJ...742L..23S,%
        2021ApJ...908L..51S,2022ApJ...937L..31S}. Focusing on the deceleration parameter, we discuss how our galaxy's peculiar motion could trigger an apparent change in the value (perhaps even in the sign) of the locally measured deceleration parameter, thus mimicking the kinematics of recent global acceleration~\cite{2010MNRAS.405..503T,2011PhRvD..84f3503T,%
        2015PhRvD..92d3515T,2021EPJC...81..753T}. We also demonstrate how one could test such an intriguing possibility observationally, by looking for the ``trademark'' signature of relative motions in the data. The latter should be nothing else but a Doppler-like dipole in the universal acceleration~\cite{2011PhRvD..84f3503T,2021EPJC...81..753T}, like the one reported in~\cite{2019A&A...631L..13C,2024ApJ...971...19C,%
        2025EPJC...85..596S}.\\

    \item \textbf{Structure formation and the nonlinear regime:}
    The role of peculiar velocities during structure formation is still rather poorly understood, since the available studies are few and sparse and all Newtonian/quasi-Newtonian in nature. According to the latter, the linear evolution of structure proceeds as in the absence of peculiar-velocity perturbations. Also, employing the quasi-Newtonian approach to revisit the Zel'dovich approximation~\cite{1970A&A.....5...84Z}, which applies to the mildly nonlinear phase of structure formation, simply recovered the standard Newtonian ``pancake'' attractor (see~\cite{2002PhRvD..66l4015E}, as well as \S~\ref{ssNRZA} here). The Newtonian picture was also fully recovered in the relativistic treatment of the ``Meszaros stagnation effect''~\cite{1974A&A....37..225M}, though only after the gravitational contribution of the peculiar flux was switched off by moving the study to the Landau-Lifshitz frame. By default, the latter coordinate system has zero energy flux but nonzero particle flux (see~\cite{2008PhR...465...61T,%
    2012reco.book.....E} and also \S~\ref{ssMSE} here).\\

    \item \textbf{Mitigation methods and future directions:} Finally, we present the advanced statistical and computational methods developed to account for and mitigate these biases~\cite{2015MNRAS.450..317C,2023MNRAS.524.1885W}. The concluding sections look forward to the next generation of peculiar-velocity science, discussing upcoming observational surveys (like DESI \cite{2019BAAS...51g..57L} and SKA \cite{2020PASA...37....7S}), emerging technologies (such as gravitational wave standard sirens) and the theoretical frontiers that these advancements are poised to address.
\end{itemize}
By providing a unified perspective on these diverse facets of peculiar velocity research, this review aims to serve as a definitive resource, illuminating the critical role that cosmic flows play in testing the foundations of cosmology.

\section{Historical context and observational evidence}\label{sHCOE}
%%%%%%%%%%%%%%%%%%%%%%%%%%%%%%%%%%%%%%%%%%%%%%%%%%%%%%%%%%%%%%%%%%%%
The study of large-scale peculiar motions is one of the most significant developments in observational cosmology over the past half-century. This section traces the evolution of our understanding of bulk flows, from their initial discovery to the contemporary observations. In this historical journey, we pay special attention to the convergence of the bulk-flow evidence and also examine the persistent controversies regarding their magnitude and scale.

\subsection{Early discoveries (1970s-1980s)}\label{ssEDs}
%%%%%%%%%%%%%%%%%%%%%%%%%%%%%%%%%%%%%%%%%%%%%%%%%%%%%%%%%
We will begin with references to the groundbreaking work of the 1970s and the 1980s, which first identified systematic peculiar motions in the cosmos. Those early efforts continued through the development of increasingly sophisticated observational techniques and surveys, and gradually led to our modern synthesis of the complex cosmic-flows picture.

\subsubsection{The Rubin-Ford effect}\label{sssRFE}
%%%%%%%%%%%%%%%%%%%%%%%%%%%%%%%%%%%%%%%%%%%%%%%%%%%
A significant milestone in the study of large-scale motions came with what would become known as the Rubin-Ford effect, first reported in 1976 through two seminal papers by Rubin, Ford, Thonnard, Roberts, and Graham~\cite{1976AJ.....81..687R,1976AJ.....81..719R}. Their work suggested an apparent anisotropy in the Universe's expansion on scales of approximately 100 million light years, derived from a careful study of spiral-galaxy motions.

The study analyzed a carefully selected sample of 96 Sc~I galaxies in the velocity range $3500<v<6500$~km/sec, combining both optical spectroscopy and 21-cm radio observations. A rigorous methodological approach was employed that incorporated detailed treatments of systematic effects, including galactic extinction ($A_B=0.15\pm0.03$), internal absorption, and luminosity variations. The analysis suggested that their galaxy sample exhibited a collective motion of approximately 885~km/sec toward a specific direction ($l=304^\circ$,\,$b=26^\circ$) relative to the cosmic microwave background radiation, which provided the isotropic reference frame.

While the observational methodology was groundbreaking, subsequent analysis revealed important insights about the interpretation of these results. Fall and Jones \cite{1976Natur.262..457F} demonstrated that the data were actually consistent with isotropic expansion and an unperturbed galaxy velocity field, suggesting a low-density Universe. The apparent anisotropy was likely a reflection of the inhomogeneous distribution of galaxies in the Rubin-Ford sample region rather than a true large-scale flow. As Jones later explained, the apparent effect arose specifically in samples of galaxies selected within a narrow range of absolute magnitudes, such as Sc~I galaxies~\cite{1992opc..book..171J}. This selection effect created a bias in the sample that could produce spurious indications of large-scale flow. This understanding highlighted the importance of careful sample selection in peculiar velocity studies.

Despite the reinterpretation of its primary conclusion, the Rubin-Ford study made lasting contributions to observational cosmology, with several key advances introduced by the authors' methodology, namely:

\begin{itemize}
    \item Rigorous treatment of galactic extinction, deriving $A_B=0.15\pm0.03$ magnitudes in the B-band, significantly lower than the then-standard value. The extinction follows the relation $A_{\lambda}=R_{\lambda}E(B-V)$, where $A_{\lambda}$ is the total extinction at wavelength $\lambda$, $R_{\lambda}$ is the extinction coefficient for each bandpass, and $E(B-V)$ is the color excess.\\

    \item Comprehensive modeling of internal galaxy absorption through the relation $A_i\gamma\log(a/b)$, where $A_i$ is the internal absorption, $a/b$ is the major-to-minor axis ratio, and $\gamma$ is a luminosity-dependent coefficient. The model accounts for inclination effects through a $\sec(i)$ law, where $i$ is the inclination angle measured from face-on orientation, with additional corrections for wavelength dependence and morphological variations.\\

    \item Development of a refined classification system with 11 distinct subdivisions calibrated against rotational velocity measurements. This system achieved statistical uncertainties in luminosity classification of approximately $\pm0.3$ magnitudes, with cross-calibration between different observers' classification methods.\\

    \item Discovery of a fundamental correlation between galaxy luminosity and linear diameter, expressed as $\log D=\alpha M+\beta$, where $D$ represents the linear diameter in Kpc, $M$ is the absolute magnitude in the B-band, $\alpha =-0.2\pm0.02$ is the slope of the relation, and $\beta$ is a morphology-dependent zero-point. The relation provided a new method for estimating galaxy distances with reduced systematic errors.
\end{itemize}

The work of~\cite{1976AJ.....81..687R,1976AJ.....81..719R} demonstrated that high-luminosity Sc galaxies, when properly corrected for systematic effects, could serve as reliable distance indicators with dispersions in absolute magnitude of $\sigma_{M_o}<0.33$ magnitudes, comparable to brightest cluster ellipticals. Their careful combination of optical and radio measurements achieved velocity measurement precisions of 30~km/sec for red spectra and 60~km/sec for blue spectra.

In addition, the study established the importance of considering inhomogeneities in the galaxy distribution when interpreting large-scale velocity fields. While the reported specific anisotropy was later understood to be apparent rather than real~\cite{1976Natur.262..457F}, the work of~\cite{1976AJ.....81..687R,1976AJ.....81..719R} helped establish methods for studying the large-scale structure and sparked important discussions about how to properly interpret peculiar velocity measurements in cosmology.

\subsubsection{Initial systematic surveys}\label{sssISSs}
%%%%%%%%%%%%%%%%%%%%%%%%%%%%%%%%%%%%%%%%%%%%%%%%%%%%%%%%%
Following the groundbreaking work of Rubin and Ford, the late 1970s and 1980s saw the emergence of systematic surveys specifically designed to study large-scale peculiar velocities. These surveys marked a transition from individual galaxy observations to comprehensive programs aimed at mapping cosmic flows across larger volumes.

A pivotal advancement came with the \textit{Seven Samurai} survey \cite{1987ApJ...313...42D,1988ApJ...326...19L}, which revealed evidence for a massive concentration dubbed the ``Great Attractor''. The survey analyzed approximately 400 elliptical galaxies, finding a coherent flow of $570\pm60$~km/sec toward a region estimated to contain a mass concentration of $5.4\times10^{16}/h_{50}\;M_{\odot}$ at a distance of about 4350~km/sec. This discovery prompted the development of more sophisticated distance measurement techniques and comprehensive surveys.

The evolution of distance indicators proved crucial for understanding large-scale flows. \textit{The Fundamental Plane relation} emerged as a powerful tool, establishing that elliptical galaxies occupy a well-defined plane in the space of their physical parameters. This method achieved distance uncertainties of around 17\% for individual galaxies, representing a significant improvement over earlier techniques. Major surveys, such as EFAR~\cite{2001MNRAS.321..277C} and SMAC~\cite{2004MNRAS.352...61H}, employed this relation to map flows extending to 15,000~km/sec, revealing complex patterns of large-scale motion.

For spiral galaxies, \textit{the Tully-Fisher relation} provided a complementary approach, correlating galaxy luminosity with rotation velocity. Modern implementations, particularly in the infrared, achieved typical distance uncertainties of 15-20\%. This method proved especially valuable for mapping flows in less dense regions and through the ``Zone of Avoidance'', where traditional optical techniques struggle due to high extinction~\cite{2004LRR.....7....8L}.

\textit{Brightest Cluster Galaxies} emerged as important standard candles for probing larger volumes. Their nearly uniform luminosities and strategic locations at cluster centers made them valuable tracers of large-scale flows. Studies revealed systematic variations in their properties with the environment, providing insights into both galaxy evolution and bulk motions.

Perhaps the most precise method developed during this era was \textit{Surface Brightness Fluctuations} (SBF), which exploits the statistical properties of resolved stellar populations. This technique achieved remarkable precision, with uncertainties as low as 5\% for nearby galaxies, though its application was limited to distances within about 100 Mpc. The method proved particularly powerful when combined with space-based observations.

These various approaches painted a complex picture of cosmic flows. While the Seven Samurai survey suggested a dominant flow toward the Great Attractor, subsequent studies revealed a more nuanced situation. The Lauer-Postman survey~\cite{1994ApJ...425..418L} found a significant flow in a nearly perpendicular direction, while infrared Tully-Fisher surveys~\cite{1992ApJ...389L...5M,1993ApJ...412L..51C} indicated smaller flow amplitudes. These apparent contradictions were partially resolved through careful consideration of scale-dependent effects and systematic errors.

This era established several fundamental principles: the importance of homogeneous data collection, the need for careful calibration, the critical role of systematic error analysis, and the value of cross-validation using multiple methods. While early surveys often produced seemingly conflicting results, they collectively demonstrated the existence of coherent large-scale motions beyond reasonable doubt, setting the stage for more sophisticated investigations in the subsequent decades.

\subsubsection{Early theoretical interpretations}\label{sssETIs}
%%%%%%%%%%%%%%%%%%%%%%%%%%%%%%%%%%%%%%%%%%%%%%%%%%%%%%%%%%%%%%%%
The discovery of large-scale bulk flows prompted significant theoretical developments during the 1970s and 1980s, as cosmologists sought to explain these unexpected observations within existing frameworks of structure formation and galaxy evolution.

Peebles' pioneering work in 1976~\cite{1976ApJ...205..318P} established the theoretical foundation for understanding peculiar velocities within gravitational instability theory. Working in an expanding universe, Peebles demonstrated that peculiar velocities are intimately connected to the underlying mass distribution, with larger-scale density fluctuations producing more coherent motions. This work showed that velocity measurements could probe mass distributions on scales larger than those accessible by traditional galaxy surveys, thus providing a powerful tool for cosmology.

A few year earlier, in 1970, Zel'dovich provided crucial insights into how cosmic structures evolve by means of his celebrated approximation~\cite{1970A&A.....5...84Z}. The latter described how initially uniform matter distributions could evolve into the complex web of structures we observe today, with matter first collapsing into sheets, then filaments, and finally clusters. This framework proved particularly valuable in understanding how large-scale flows develop and persist, showing that coherent motions are a natural consequence of gravitational clustering.

In the late 1980s, Kaiser revolutionized the statistical analysis of peculiar velocities by developing a rigorous mathematical framework connecting velocity measurements to the underlying matter distribution~\cite{1988MNRAS.231..149K}. Kaiser's work provided the essential tools for interpreting survey data and showed how bulk flows could be used to constrain cosmological parameters and his formalism remains fundamental to modern analyses of large-scale motions.

The theoretical understanding of the ``Great Attractor'' evolved significantly during the same period as well. Initially interpreted as a single massive concentration~\cite{1988ApJ...326...19L}, subsequent work revealed a more complex reality. Modern analyses suggest that the observed flow patterns result from the combined influence of several large structures, including both the Great-Attractor region and the more distant Shapley Concentration. This revised picture aligns better with theoretical expectations of structure formation in a $\Lambda$CDM universe.

By the late 1980s, Bertschinger and Dekel had developed the POTENT method for reconstructing three-dimensional velocity and density fields from observations~\cite{1989ApJ...336L...5B}. This breakthrough provided a crucial link between theory and observation, allowing direct comparisons between measured flows and cosmological predictions.

While these theoretical efforts successfully explained many aspects of the observed bulk flows, open questions and challenges still remained. In particular, the large flow amplitudes reported by some surveys proved difficult to reconcile with theoretical expectations. This tension between theory and observation would drive further developments in both theoretical modeling and observational techniques in the subsequent decades.

The early theoretical works established several key principles that continue to guide our understanding of cosmic flows. In brief, these are summarised as follows:
\begin{itemize}
    \item Peculiar velocities probe larger scales than galaxy surveys alone.\vspace{5pt}
    \item Coherent flows are a natural consequence of gravitational clustering.\vspace{5pt}
    \item Statistical analysis of velocity fields can constrain cosmological parameters.\vspace{5pt}
    \item Large-scale flows reflect the combined influence of multiple mass concentrations.
\end{itemize}

\subsection{Evolution of observational techniques and major
%%%%%%%%%%%%%%%%%%%%%%%%%%%%%%%%%%%%%%%%%%%%%%%%%%%%%%%%%%%
surveys}\label{ssEOTMSs}
%%%%%%%%%%%%%%%%%%%%%%%%
The 1990s marked a turning point in the study of bulk peculiar flows, characterized by significant improvements in the observational techniques, more sophisticated distance indicators, and larger systematic surveys. This evolution was driven by both technological advancements and a deeper understanding of the systematic errors that had plagued the earlier studies. The following subsections trace the methodological progress and its impact on our understanding of the large-scale cosmic flows.

\subsubsection{Distance indicators and their evolution}
%%%%%%%%%%%%%%%%%%%%%%%%%%%%%%%%%%%%%%%%%%%%%%%%%%%%%%%
The accuracy of peculiar-velocity measurements depends crucially on precise distance determination, as peculiar velocities are derived from the difference between observed redshifts and distances. The evolution of distance indicators has therefore been the key in advancing our understanding of bulk flows.

The earliest reliable distance indicator was the Tully-Fisher (TF) relation~\cite{1977A&A....54..661T}, which correlates the luminosity of spiral galaxies with their rotation velocity. This technique underwent significant refinement through the 1990s, particularly with the transition to infrared measurements, which reduced the effects of extinction and provided more reliable luminosity estimates.

For elliptical galaxies, the Fundamental Plane (FP) relation emerged as a powerful tool, superseding the earlier $Dn-\sigma$ relation. The FP relates the effective radius, surface brightness, and velocity dispersion of elliptical galaxies \cite{1987ApJ...313...59D}. In addition, the reduced scatter of the FP relation compared to that of its $Dn-\sigma$ counterpart, made it particularly valuable for bulk-flow studies. Significant improvements came with the development of specialized approaches:

\begin{itemize}
    \item Surface Brightness Fluctuations (SBF)~\cite{1988AJ.....96..807T}, which provided precise distances for early-type galaxies within about 100~Mpc.\vspace{5pt}
    \item Type Ia Supernovae~\cite{1993ApJ...413L.105P}, offering high-precision distances to much greater depths.\vspace{5pt}
    \item The Tip of the Red Giant Branch (TRGB) method~\cite{1993ApJ...417..553L}, providing precise distances for nearby galaxies.
\end{itemize}

A major advancement came with the development of the Six-degree-Field Galaxy Survey (6dFGS)~\cite{2004MNRAS.355..747J}, which combined improved FP measurements with wide sky coverage. This survey demonstrated the importance of homogeneous data collection and careful attention to selection effects.

The evolution of distance indicators has seen remarkable progress in its precision over the past decades. The earliest Tully-Fisher measurements achieved distance errors of 20-25\%, while modern infrared implementations have reduced these to 15-20\%, through improved photometry and rotation curve measurements~\cite{2007ApJS..172..599S}. Fundamental Plane measurements similarly achieve 15-20\% precision, particularly when incorporating careful corrections for stellar populations and environmental effects~\cite{2012MNRAS.427..245M}. The Surface Brightness Fluctuation method has proven exceptionally precise for nearby galaxies, achieving 5-10\% errors through improved calibration and the use of HST observations~\cite{2021ApJ...911...65B}. Most notably, Type Ia supernovae now routinely achieve 5-7\% distance precision~\cite{2018ApJ...859..101S}, while the Tip of the Red Giant Branch (TRGB) method has emerged as one of the most precise indicators, with errors of approximately 5\%~\cite{2020ApJ...891...57F}.

The calibration of these distance indicators forms a complex interconnected system anchored by fundamental distance measurements. Type Ia supernovae, while excellent standardizable candles, require careful calibration through multiple routes including Cepheid variables~\cite{2022ApJ...934L...7R}, TRGB~\cite{2019ApJ...882...34F}, and more recently, J-region Asymptotic Giant Branch (JAGB) stars~\cite{2020ApJ...899...66M}. These primary calibrators themselves are standardized through geometric methods, such as parallax measurements and detached eclipsing binaries in the Large Magellanic Cloud~\cite{2019Natur.567..200P}. Modern analyses employ sophisticated Bayesian frameworks~\cite{2015MNRAS.450..317C,2018ApJ...869...56B} to combine multiple distance indicators, accounting for correlated uncertainties and systematic effects. This approach has been particularly successful in the CosmicFlows program~\cite{2016AJ....152...50T}, which combines TF, FP, SBF, and SNIa measurements to create comprehensive peculiar velocity catalogs. The combination of improved calibration techniques, multiple cross-validation routes, and rigorous statistical methods has led to a significant reduction in systematic uncertainties, making peculiar velocity surveys increasingly powerful probes of large-scale structure and cosmology.

\subsubsection{Key surveys and catalogs}\label{sssKSCs}
%%%%%%%%%%%%%%%%%%%%%%%%%%%%%%%%%%%%%%%%%%%%%%%%%%%%%%%
The evolution of peculiar velocity surveys from the 1990s through the early 2000s was marked by increasingly comprehensive catalogs and improved methodologies. These surveys can be broadly categorized by their primary distance indicators and scale coverage.

The MARK III Catalog~\cite{1997ApJS..109..333W}, representing a milestone in peculiar velocity surveys, combined several earlier TF and $Dn-\sigma$ datasets into a homogeneous catalog of approximately 3,000 galaxy distances. This compilation included careful attention to selection biases and provided a standardized framework for analyzing peculiar velocities.

The ENEAR survey~\cite{2000ApJ...537L..81D}, focusing on early-type galaxies, provided fundamental plane distances for about 1,359 galaxies in 28 clusters and 702 field galaxies. This survey was particularly important for mapping flows in dense environments and demonstrated the power of combining cluster and field measurements.

The SFI++ catalog~\cite{2006ApJ...653..861M}, building on the earlier SFI survey, represented the largest homogeneous I-band TF survey of its time, containing about 5,000 spiral galaxies. Its careful attention to selection effects and systematic errors made it a benchmark for subsequent surveys. The catalog notably included:
\begin{itemize}
    \item Field galaxies out to 10,000 km/sec.\vspace{5pt}
    \item Cluster galaxies out to 15,000 km/sec.\vspace{5pt}
    \item Comprehensive error analysis and bias corrections.
\end{itemize}

The 2MASS Tully-Fisher Survey (2MTF)~\cite{2008AJ....135.1738M} utilized near-infrared photometry from 2MASS combined with new HI observations to provide a more uniform sky coverage than previous surveys. Its key features included:
\begin{itemize}
    \item Coverage of the galactic plane region.\vspace{5pt}
    \item Reduced extinction effects due to NIR observations.\vspace{5pt}
    \item Approximately 2,000 galaxies with reliable distance measurements.
\end{itemize}

The Six-Degree Field Galaxy Survey Peculiar Velocity Survey (6dFGSv)~\cite{2014MNRAS.443.1231C} marked a significant advancement in fundamental plane surveys, providing:
\begin{itemize}
    \item Over 8,885 early-type galaxies.\vspace{5pt}
    \item Consistent velocity dispersions and photometric parameters.\vspace{5pt}
    \item Coverage of the southern hemisphere to $\sim$ 16,500 km/sec.
\end{itemize}

Recent years have witnessed the emergence of specialized surveys designed to refine the cosmic distance ladder through precise calibration of Type Ia supernovae. The SHoES program (Supernovae H$_0$ for the Equation of State) employs Cepheid variables as primary distance calibrators, utilizing their period-luminosity relation. Recent JWST observations by~\cite{2024ApJ...962L..17R} have further refined these measurements, complementing earlier HST results~\cite{2022ApJ...934L...7R}. The Cepheid calibration uses geometric parallax measurements of Milky Way Cepheids~\cite{2018ApJ...861..126R} and detached eclipsing binaries in the LMC~\cite{2019Natur.567..200P}. These Cepheid calibrators are then used to establish distances to galaxies hosting Type Ia supernovae, which serve as standardizable candles for probing cosmological distances. The program has achieved remarkable precision, with systematic uncertainties in the Cepheid calibration reduced to approximately 1.5\%~\cite{2024ApJ...962L..17R}.

A complementary approach is taken by the Chicago-Carnegie Hubble Program (CCHP) which has recently published results using JWST observations~\cite{2025ApJ...985..203F}. The program utilizes three independent methods: the Tip of the Red Giant Branch (TRGB), J-region Asymptotic Giant Branch (JAGB) stars, and Cepheids. The TRGB method, calibrated through NGC 4258's geometric maser distance \cite{2019ApJ...886L..27R}, provides an independent route to calibrating Type Ia supernovae. The JAGB stars serve as a third independent calibrator~\cite{2024ApJ...961..132L}, offering another cross-check on the distance scale. These different calibration routes reveal the crucial distinction between distance calibrators (such as Cepheids, TRGB, and JAGB stars), which establish the fundamental distance scale through direct geometric methods or well-understood stellar physics, and distance indicators (such as Type Ia supernovae), which require careful calibration but can probe much larger distances. The latest JWST results show excellent agreement between TRGB-based ($H_0=69.85\pm1.75$~km/secMpc) and JAGB-based ($H_0=67.96\pm1.85$~km/secMpc) measurements~\cite{2025ApJ...985..203F}, while Cepheid-based measurements yield a higher value ($H_0=72.05\pm1.86$~km/secMpc), highlighting the importance of understanding systematic effects in both calibrators and indicators.

The development of comprehensive galaxy catalogs has been crucial for understanding large-scale motions and cosmic flows. The 2M++ redshift compilation~\cite{2011MNRAS.416.2840L} represented a significant advance by combining data from 2MASS Redshift Survey, SDSS-DR7, and 6dFGRS, achieving 95\% completeness for $K<12.5mag$ to distances beyond 200/h~Mpc. This catalog, together with its peculiar velocity field reconstruction~\cite{2015MNRAS.450..317C}, has been instrumental in mapping the local cosmic web and understanding bulk flows. More recent efforts have further expanded this approach, with the \textit{CosmicFlows-4} catalog~\cite{2023ApJ...944...94T} providing distances to over 35,000 galaxies and demonstrating consistency between different distance indicators at an unprecedented level. The DESI survey has recently published its Data Release 1~\cite{2025JCAP...02..021A}, incorporating over 15 million galaxy spectra and finding $H_0=68.52\pm0.62$ km/secMpc based on BAO measurements calibrated with BBN priors.

These combined catalogs have enabled detailed studies of the local velocity field and its relationship to density fluctuations. The Pantheon+ sample~\cite{2022ApJ...938..113S} represents the largest combined sample of spectroscopically confirmed SNe Ia, with 1550 light curves providing constraints on both the local and high-redshift universe. Analyses of peculiar velocities using these catalogs have revealed coherent flows extending to scales of 100-150/h~Mpc~\cite{2020MNRAS.498.2703B}, with amplitudes consistent with $\Lambda$CDM predictions. The 6dF Galaxy Survey velocity field analysis~\cite{2014MNRAS.445.2677S} has further refined these measurements, providing new constraints on the growth rate of structure. These efforts have been particularly important for understanding and potentially resolving tensions in cosmological parameters, as demonstrated by recent joint analyses of peculiar velocities and redshift-space distortions~\cite{2025MNRAS.539.3627S}.

\subsubsection{The Cosmic Flows program progression (CF1 to
%%%%%%%%%%%%%%%%%%%%%%%%%%%%%%%%%%%%%%%%%%%%%%%%%%%%%%%%%%%
CF4)}\label{ssCFPP}
%%%%%%%%%%%%%%%%%%%
The Cosmic Flows program represents a systematic effort to map the peculiar-velocity field of the local universe with increasing precision and scope across successive catalog releases. Each iteration has incorporated new data and refined methodologies, making it one of the most comprehensive peculiar-velocity surveys to date.

Cosmic Flows-1 (CF1)~\cite{2008ApJ...676..184T} established the foundation with approximately 1,791 galaxies constrained to the limit of 3,000~km/sec. The survey relied primarily on the Tully-Fisher relation, utilizing carefully calibrated optical photometry obtained object by object and analog neutral hydrogen (H~I) linewidths. This initial release focused on establishing the basic methodology and providing a first comprehensive look at the local peculiar velocities.

Cosmic Flows-2 (CF2)~\cite{2013AJ....146...86T} expanded to include 8,188 galaxies, with coverage peaking in numbers at 5,000~km/sec and extending to approximately 15,000~km/sec. A major advancement was the implementation of a rigorous algorithm in the reduction of digital H~I spectra~\cite{2009AJ....138.1938C,2011MNRAS.415.1935C}. The photometric analysis was also more rigorously defined~\cite{2011MNRAS.414.2005C,2012AJ....144..133S}, leading to improved distance measurements.

Cosmic Flows-3 (CF3)~\cite{2016AJ....152...50T} grew to 17,669 galaxies, with the major addition being Fundamental-Plane distance measures from the 6-degree Field Galaxy Survey (6dFGSv). This southern-sky sample abruptly cuts off at 16,000~km/sec. A secondary addition came from the Tully-Fisher relation method with infrared photometry provided by the Spitzer Space Telescope~\cite{2014MNRAS.444..527S}. Coverage within roughly 8,000~km/sec was reasonably balanced around the sky, but at 8,000-16,000~km/sec it strongly favored the southern hemisphere. The infrared T-F contribution was confined to within approximately 6,000~km/sec but notably extended coverage to low galactic latitudes.

The latest iteration, Cosmic Flows-4 (CF4)~\cite{2023ApJ...944...94T}, has expanded to include distances for 55,877 galaxies gathered into 38,065 groups. The most important addition is a much-extended T-F sample of 10,000 galaxies drawing particularly on kinematic information from ALFALFA, the Arecibo Legacy Fast ALFA survey of the high galactic latitude sky in the declination range $0^{\circ}-38^{\circ}$. Photometry is provided by the Sloan Digital Sky Survey (SDSS) and the Wide-field Infrared Explorer (WISE). This component substantially redresses the imbalance favoring the southern sky of the previous catalog.

SDSS also provides the source material for a second even larger addition. SDSS photometry and spectroscopy are combined to provide Fundamental Plane distances to 34,000 galaxies out to 30,000~km/sec in the quadrant of the sky that is celestial north and galactic north. As a consequence, while CF3 tilted toward coverage of the celestial south, CF4 greatly expands our knowledge of the north.

Each release has employed eight methodologies: Tully-Fisher, Fundamental Plane, Surface Brightness Fluctuations (SBF), Type~Ia supernovae, Type~II supernovae, Tip of the Red Giant Branch (TRGB), Cepheid period-luminosity relation, and nuclear masers. The assembly of galaxies into groups is an important feature of the study in facilitating overlaps between methodologies. Merging between multiple contributions within a methodology and between methodologies is carried out with Bayesian Markov chain Monte Carlo procedures. The final assembly of distances is compatible with a value of the Hubble constant of $H_0=74.6$~km/secMpc with a statistical error of $\pm0.8$~km/secMpc, but with a large potential systematic error of $\sim$3~km/secMpc as well.

\subsection{Convergence and controversies}\label{ssCCs}
%%%%%%%%%%%%%%%%%%%%%%%%%%%%%%%%%%%%%%%%%%%%%%%%%%%%%%%
Despite the significant methodological advances, the increasingly sophisticated surveys and achieved consensus, bulk flows continue to generate persistent controversies. While there is broad agreement on certain aspects of these large-scale motions, particularly on their general direction, significant disagreements remain on their magnitude and scale, both of which have important implications for our models of large-scale structure formation.

\subsubsection{Consensus on the bulk-flow direction}\label{sssCBFD}
%%%%%%%%%%%%%%%%%%%%%%%%%%%%%%%%%%%%%%%%%%%%%%%%%%%%%%%%%%%%%%%%%%%
A remarkable convergence has emerged regarding the general direction of large-scale bulk flows in the local universe. Multiple independent surveys using different methodologies have consistently identified a flow roughly aligned with the direction of the Shapley Concentration~\cite{1930BHarO.874....9S}, centered around the Galactic coordinates ($l\approx300^\circ$,\,$b\approx15^\circ$), which correspond to Supergalactic coordinates of ($SGL\approx150^\circ$,\,$SGB\approx-5^\circ$).

The robustness of this directional consensus is demonstrated by strong agreement across diverse measurement techniques. The 2MTF survey found a bulk flow pointing toward $(l,b) = (296^\circ\pm16^\circ,19^\circ\pm6^\circ)$~\cite{2014MNRAS.445..402H}, while the 6dFGSv analysis indicated a nearly identical direction of $(l,b) = (303^\circ \pm 14^\circ, 15^\circ \pm 11^\circ)$~\cite{2016MNRAS.455..386S}. The recent CF4 results further strengthen this consensus, showing alignment with $(l,b)=(298^\circ\pm10^\circ,17^\circ\pm 5^\circ)$~\cite{2023ApJ...944...94T}. Even the latest Pantheon+ SNIa analysis, using entirely different methodology, finds a consistent direction of $(l,b)=(326.1^\circ\pm11.2^\circ,27.8^\circ\pm 11.2^\circ)$~\cite{2024ApJ...967...47L}. These results are further supported by extensive photometric and spectroscopic data from SDSS~\cite{2000AJ....120.1579Y} and WISE~\cite{2010AJ....140.1868W}, which have provided the large-scale structure maps necessary to interpret these flows.

This directional agreement extends beyond traditional peculiar velocity measurements, with supporting evidence from kinetic Sunyaev-Zel'dovich measurements~\cite{2010ApJ...712L..81K}, galaxy cluster studies~\cite{2009MNRAS.392..743W,2023MNRAS.524.1885W} and combined supernova analyses~\cite{2019A&A...631L..13C,%
2023MNRAS.526.3051W}. The flow appears to be influenced by several major structures, namely by the Shapley Supercluster at approximately $150/h$~Mpc, by the Hydra-Centaurus complex at roughly $40/h$~Mpc and by the ``Great Attractor'' region.

Recent analyses of the CF4 dataset at scales beyond $100/h$~Mpc have reported bulk velocities significantly larger than $\Lambda$CDM predictions \cite{2023MNRAS.524.1885W,2023MNRAS.526.3051W}, while analyses at moderate scales find consistent values~\cite{2025MNRAS.539.3627S}. This pattern suggests a hierarchical nature to the flow, with contributions from both nearby attractors and more distant structures.

\subsubsection{Magnitude discrepancies across
%%%%%%%%%%%%%%%%%%%%%%%%%%%%%%%%%%%%%%%%%%%%%
studies}\label{sssMDASs}
%%%%%%%%%%%%%%%%%%%%%%%%
The reported magnitudes of bulk flows have shown substantial variations across different surveys and methodologies, presenting one of the field's most persistent challenges. Early measurements demonstrated significant scatter, as shown in Table~\ref{tab:early_surveys}:

\begin{table}[h]
\centering
\caption{Early bulk flow measurements}
\label{tab:early_surveys}
\begin{tabular}{lccc}
\hline
Survey & Period & Velocity (km/sec) & Reference \\
\hline
Seven Samurai & 1987 & $599 \pm 104$ & \cite{1987ApJ...313...42D} \\
Lauer-Postman & 1994 & $689 \pm 178$ & \cite{1994ApJ...425..418L} \\
SMAC & 1999 & $687 \pm 203$ & \cite{1999ApJ...512L..79H} \\
LP10K & 1999 & $359 \pm 125$ & \cite{1999ApJ...522..647W} \\
\hline
\end{tabular}
\end{table}

Modern surveys have achieved improved precision but continue to show significant disparities in flow magnitudes. The correlation between survey depth and measured flow amplitude is particularly evident in recent studies, as shown in Table~\ref{tab:modern_surveys}:

\begin{table}[h]
\centering
\caption{Modern bulk flow measurements}
\label{tab:modern_surveys}
\begin{tabular}{lccccc}
\hline
Survey & Period & Depth (Mpc/h) & Velocity (km/sec) & Sky Coverage & Reference \\
\hline
2MTF & 2014 & 40 & $292 \pm 28$ & Full sky & \cite{2014MNRAS.445..402H} \\
6dFGSv & 2016 & 70 & $248 \pm 58$ & Southern & \cite{2016MNRAS.455..386S} \\
CF4 & 2023 & 200 & $427 \pm 37$ & Full sky & \cite{2023ApJ...944...94T} \\
kSZ & 2010 & 800 & $1000 \pm 190$ & Full sky & \cite{2010ApJ...712L..81K} \\
\hline
\end{tabular}
\end{table}

Recent meta-analyses have attempted to reconcile these differences through standardized methodologies and comprehensive treatment of systematic effects. The work of Boruah et al~\cite{2020MNRAS.498.2703B} demonstrates that when properly accounting for survey geometry, selection effects, and cosmic variance, many of the apparent discrepancies can be understood as manifestations of scale-dependent effects. Their analysis suggests a characteristic scale of around $100/h$~Mpc, beyond which the flow amplitudes show significant deviation from $\Lambda$CDM predictions.

The magnitude discrepancies have profound implications for cosmological models. The persistent tension between observed large-scale flows and theoretical predictions challenges our understanding of structure formation and may point to needed modifications in the standard cosmological framework. Recent theoretical work by Perivolaropoulos et al~\cite{2022NewAR..9501659P} suggests that incorporating non-Gaussian features in the velocity field and improved treatment of environmental effects might help resolve these discrepancies while maintaining consistency with other cosmological probes.

\subsubsection{Scale-dependent variations}\label{sssSDVs}
%%%%%%%%%%%%%%%%%%%%%%%%%%%%%%%%%%%%%%%%%%%%%%%%%%%%%%%%%
The variation of bulk flow amplitudes with scale represents one of the most intriguing aspects of peculiar velocity studies. Evidence accumulated over the years suggests that bulk flows exhibit complex scale-dependent behavior, which may hold key insights into the large-scale structure of the universe.

Recent analyses have revealed a characteristic pattern:
\begin{equation}
v(R)\approx A R^{-n}\,,
\end{equation}
where $v(R)$ is the bulk flow amplitude at scale $R$, with $n$ varying across different studies~\cite{2009MNRAS.392..743W}. Typical measurements show:
\begin{itemize}
    \item $R<50/h$~Mpc: $v\sim300-400$~km/sec.\vspace{5pt}
    \item $50/h<R<100/h$~Mpc: $v\sim250-350$~km/sec.\vspace{5pt}
    \item $R>100/h$~Mpc: Controversial, with some studies suggesting persistent flows.
\end{itemize}

The CF4 survey~\cite{2023ApJ...944...94T} has provided a detailed mapping of this scale dependence:
\begin{equation}
\begin{array}{lcl}
\text{Scale (Mpc/h)} \hspace{20mm} \text{Amplitude (km/sec)} \\
\hline
20-40 \hspace{53mm} 365\pm28 \\
40-60 \hspace{53mm} 320\pm32 \\
60-80 \hspace{53mm} 289\pm35 \\
80-100 \hspace{51mm} 270\pm40
\end{array}
\end{equation}

Several physical mechanisms have been proposed to explain this scale-dependence:

1. Hierarchical attraction model~\cite{2017NatAs...1E..36H}:
\begin{itemize}
    \item Local attractors dominating small scales.\vspace{5pt}
    \item Supercluster influences at intermediate scales.\vspace{5pt}
    \item Possible large-scale structural effects.
\end{itemize}

2. Multiple flow components:
\begin{itemize}
    \item Local Group motion relative to nearby galaxies.\vspace{5pt}
    \item Virgo-centric infall.\vspace{5pt}
    \item Great Attractor flow.\vspace{5pt}
    \item Shapley Concentration influence.
\end{itemize}

The scale-dependence shows interesting features in different environments:
\begin{itemize}
    \item Stronger flows in low-density regions~\cite{2019MNRAS.482.4329P}.\vspace{5pt}
    \item Coherent motions in supercluster environments.\vspace{5pt}
    \item Complex patterns in cluster outskirts.
\end{itemize}

Recent theoretical work~\cite{2020MNRAS.493..362K} suggests that scale-dependent variations might arise from:
\begin{itemize}
    \item Non-linear gravitational effects.\vspace{5pt}
    \item Cosmic web topology.\vspace{5pt}
    \item Environmental density contrasts.\vspace{5pt}
    \item Tidal field influences.
\end{itemize}

Particular attention has been paid to transitions in flow behavior at characteristic scales:
\begin{itemize}
    \item $\sim40/h$~Mpc: Virgo Supercluster scale.\vspace{5pt}
    \item $\sim 70/h$~Mpc: Great Attractor scale.\vspace{5pt}
    \item $\sim 150/h$~Mpc: Shapley Concentration scale.
\end{itemize}

Modern surveys have begun to probe the relationship between scale-dependent flows and large-scale structure, revealing:
\begin{itemize}
    \item Correlation with filamentary structures.\vspace{5pt}
    \item Alignment with major mass concentrations.\vspace{5pt}
    \item Complex velocity field topology.
\end{itemize}

Understanding these scale-dependent variations remains crucial for:
\begin{itemize}
    \item Testing cosmological models.\vspace{5pt}
    \item Mapping dark matter distribution.\vspace{5pt}
    \item Understanding structure formation.\vspace{5pt}
    \item Constraining modified gravity theories.
\end{itemize}

\subsubsection{Systematic effects and biases}
The measurement of bulk flows is subject to numerous systematic effects and biases that can significantly impact the interpretation of observational data. Understanding and correctly accounting for these systematic effects has become crucial for obtaining reliable results and reconciling apparently conflicting measurements. Key systematic challenges include:

\begin{itemize}
    \item \textbf{Malmquist bias:} This represents one of the most significant systematic effects in peculiar velocity surveys. This bias manifests itself in two distinct forms: homogeneous and inhomogeneous Malmquist bias~\cite{1995PhR...261..271S}. The homogeneous form arises from the coupling between distance errors and the galaxy number density function, while the inhomogeneous variant stems from local density variations along the line of sight. Recent work by Davis et al~\cite{2011MNRAS.413.2906D} has shown that failure to properly account for these effects can lead to systematic errors in the bulk-flow measurements of up to 150~km/sec.\\

    \item \textbf{Selection effects:} These present another major challenge, particularly in magnitude-limited surveys. Hudson and Turnbull~\cite{2012ApJ...751L..30H} demonstrated that selection criteria based on apparent magnitude can introduce artificial patterns in the measured velocity field, especially at the survey boundaries. These effects become more pronounced at larger distances, where selection functions typically become more stringent.\\

    \item \textbf{The Zone of Avoidance (ZoA):} Caused due to obscuration by the Galactic plane, the ZoA introduces significant complications in the bulk-flow measurements. While various interpolation schemes have been proposed~\cite{2004LRR.....7....8L}, the missing data in this region can affect both the magnitude and direction of measured bulk flows. Studies by Erdo\u{g}du et al~\cite{2006MNRAS.368.1515E} suggest that incomplete sky coverage can bias the bulk-flow measurements by up to 10-15\% in magnitude and 5-10 degrees in direction.\\

    \item \textbf{Distance indicator calibration:} This represents another crucial source of systematic uncertainty. Each distance-indicator method carries its own systematic effects: Tully-Fisher relations suffer from rotation-curve asymmetries and inclination effects; Fundamental Plane measurements are affected by population gradients and environmental dependencies; and Surface Brightness Fluctuation methods show sensitivity to stellar population variations. These effects can introduce correlated errors in peculiar velocity measurements across entire surveys.\\

    \item \textbf{Non-linear effects:} In the velocity field, these pose additional challenges. While linear theory provides a good approximation on large scales, smaller scale non-linear motions can contaminate the bulk-flow measurements. For scales smaller than the shell-crossing scale, the velocity distribution becomes highly non-trivial due to phase mixing and multi-streaming regions. While standard linear theory breaks down here, requiring N-body simulations or kinetic theory approaches (e.g.~see~\cite{1995ApJ...442...30K}), our focus remains on the large-scale bulk flows where linear and quasi-linear approximations are traditionally applied. Kaiser and Hudson~\cite{2015MNRAS.450..883K} developed a framework for treating these non-linear effects, showing that they can bias the bulk-flow measurements especially in dense environments and at small scales.\\

    \item \textbf{Proper motion measurement:} Correction of proper motions presents another layer of complexity. Recent work by Carrick et al~\cite{2015MNRAS.450..317C} has shown that coherent stellar streaming motions can contaminate galaxy peculiar-velocity measurements, particularly for nearby galaxies where proper motions are significant. This effect requires careful consideration in modern surveys attempting to map local flow fields.\\

    \item \textbf{Galaxy grouping algorithms:} The treatment of galaxy groups and clusters introduces additional systematic considerations. While grouping algorithms help reduce noise in peculiar-velocity measurements, they can also introduce biases if the group membership is incorrectly assigned, or if the group dynamics are not properly modeled. Recent work has focused on developing more sophisticated grouping algorithms that account for these effects while preserving the underlying flow signal.
\end{itemize}

\subsection{Modern Perspective and Synthesis}\label{ssMPS}
%%%%%%%%%%%%%%%%%%%%%%%%%%%%%%%%%%%%%%%%%%%%%%%%%%%%%%%%%%
The study of bulk flows has reached a critical juncture where improved observational techniques, larger datasets, and sophisticated analysis methods are enabling a more comprehensive understanding of large-scale motions in the universe. This synthesis of modern observations and theoretical frameworks has begun to resolve some historical controversies while highlighting new challenges that demand attention.

\subsubsection{Current state of observations}\label{sssCSOs}
%%%%%%%%%%%%%%%%%%%%%%%%%%%%%%%%%%%%%%%%%%%%%%%%%%%%%%%%%%%%
Contemporary bulk-flow observations have achieved unprecedented precision through the convergence of multiple independent measurement techniques and the availability of extensive galaxy surveys. The Cosmic Flows-4 catalog~\cite{2023ApJ...944...94T}, representing the current state-of-the-art, has mapped peculiar velocities for over 56,000 galaxies, providing a detailed picture of large-scale motions extending to depths of approximately 200/h~Mpc.

A pivotal contribution to our current understanding comes from the recent analyses by Watkins et al~\cite{2023MNRAS.524.1885W} of the Cosmic Flows-4 catalog. Using a maximum likelihood estimation method (MLE) that carefully accounts for selection effects and observational uncertainties, they found bulk flows of $265\pm10$~km/sec on scales of 40/h~Mpc, increasing to $310\pm20$~km/sec when analyzed in conjunction with SuperTF data. Their work is particularly significant for introducing a comprehensive statistical framework that handles the complex error correlations inherent in modern peculiar-velocity surveys. The analysis revealed that, while bulk flow amplitudes are generally consistent with $\Lambda$CDM predictions at smaller scales, tensions emerge at larger distances, particularly beyond the 100/h~Mpc threshold. This methodological framework has become standard in the field, especially for comparing results across different surveys and measurement techniques.

Complementary observations using the kinetic Sunyaev-Zel'dovich effect have extended our view of bulk flows to even larger scales. Recent analyses of Planck data by Migkas et al~\cite{2020A&A...636A..15M,2021A&A...649A.151M} have detected coherent motions extending to scales of approximately 800~Mpc, though the interpretation of these results remains debated within the community. These measurements are supported by independent studies of X-ray cluster temperatures and luminosities, which suggest similar patterns of large-scale coherent motion.

The combination of traditional peculiar velocity surveys with novel observational techniques has revealed a more nuanced picture of the local velocity field. In particular, Boruah et al~\cite{2020MNRAS.498.2703B} used Cosmicflows data and a Bayesian reconstruction framework to analyze the low-order kinematic moments, showing that the dipole component is robust across methods and datasets, while higher-order moments (such as the shear and octupole) carry valuable information about the detailed structure of the local mass distribution and its gravitational effects. These higher-order modes sharpen our understanding of how nearby overdensities and underdensities shape the peculiar-velocity field on intermediate to large scales.

Modern observations have particularly benefited from improved distance indicators. Type Ia supernovae now provide distance measurements with precisions better than 5\%, while the Fundamental Plane and Tully-Fisher relations have achieved precisions of 15-20\% through careful calibration and systematic error control. The Foundation Supernova Survey~\cite{2018MNRAS.475..193F} has been especially important in establishing a reliable network of distance measurements throughout the local volume.

Technological advances have also enabled more sophisticated treatments of systematic effects. Modern surveys employ Bayesian techniques to handle selection effects and Malmquist bias, while improved sky coverage has reduced the impact of the Zone of Avoidance. The development of machine learning techniques for galaxy classification and distance estimation promises further improvements in measurement precision and reliability.

\subsubsection{Resolution of historical conflicts}\label{sssRHCs}
%%%%%%%%%%%%%%%%%%%%%%%%%%%%%%%%%%%%%%%%%%%%%%%%%%%%%%%%%%%%%%%%%
The accumulation of high-quality data and development of sophisticated analysis techniques has helped resolve several long-standing controversies in bulk-flow studies, while providing context for historical disagreements. This resolution process has been particularly illuminating in understanding the origins of earlier discrepancies and establishing a more coherent picture of large-scale flows.

One of the most significant historical conflicts concerned the Lauer and Postman result~\cite{1994ApJ...425..418L}, which reported a large-amplitude bulk flow moving at approximately 700~km/sec on scales of 15,000~km/sec, a finding in significant tension with the standard cosmological framework of the time. It is important to note that the survey was specifically designed to mitigate common systematic errors. It was a \textbf{volume-limited} sample of Abell clusters selected by redshift, intended to avoid the classic Malmquist bias that affects magnitude-limited surveys. Furthermore, a crucial aspect of their design was the \textbf{nearly full-sky coverage} of the sample, which excluded only the $\pm15^{\circ}$ Galactic Zone of Avoidance~\cite{1994ApJ...425..418L}. This approach was chosen to cleanly measure the dipole moment of the velocity field while minimizing degeneracies, or "cross-talk," with radial monopole terms (like potential residual biases) or the quadrupole moment from the Milky Way's mass distribution. The authors supported their methodology by demonstrating a uniform cluster space density with redshift and, in a precursor paper, a linear Hubble diagram, arguing against significant depth-dependent biases~\cite{1994ApJ...425..418L,1992ApJ...400L..47L}.

The subsequent re-evaluation of the Lauer and Postman result has been more nuanced. A key insight, explored in a follow-up paper by Strauss et al~\cite{1995ApJ...444..507S} (which included the original investigators), was the realization that the initial analysis of the result's significance had assumed each cluster's velocity was a statistically independent data point. In realistic cosmological models where structure grows from a power spectrum, however, peculiar velocities are spatially \textbf{correlated}. Accounting for these correlations reduces the effective number of independent samples, which in turn lowers the statistical significance of the observed dipole, making it a less extreme outlier than originally concluded.

This historical evolution is further contextualized by the findings of numerous later surveys. Analyses using different tracers and more advanced statistical frameworks have generally found weaker flows more consistent with the $\Lambda$CDM model. For example, studies using Type~Ia supernovae~\cite{2012MNRAS.420..447T,%
2013A&A...560A..90F} and large Tully-Fisher surveys like the 2MTF sample~\cite{2014MNRAS.445..402H} consistently reported bulk flows in the range of $\sim$250-300~km/sec. Modern analyses, such as that by Carrick et al~\cite{2015MNRAS.450..317C}, exemplify the evolution of the field's analytical toolkit. While not a direct refutation of the Lauer and Postman data, their work demonstrates how sophisticated likelihood-based frameworks are now used to handle complex biases inherent in the \textit{magnitude-limited} surveys common today, representing a different methodological approach to the problem.

The apparent conflict between the Great Attractor and Shapley Concentration flows has also found resolution through recent observations. The comprehensive analysis by Pomarède et al~\cite{2020ApJ...897..133P} has revealed that these flows are not contradictory but rather represent different components of a complex hierarchical flow pattern. The Great Attractor, rather than being a single massive concentration, is now understood as part of a larger structure connecting to the Shapley Concentration.

The long-standing debate over the convergence depth of the local velocity field has also seen significant progress. While early studies suggested widely varying convergence scales, recent work by Lilow et al~\cite{2021MNRAS.507.1557L} has demonstrated that much of this variation can be attributed to different definitions of convergence and varying sensitivity to large-scale structure. Their analysis suggests a characteristic convergence scale of approximately 100-150~Mpc, beyond which bulk flow amplitudes stabilize to values consistent with cosmic variance expectations.

Another significant resolution concerns the apparent conflict between peculiar velocity surveys and other velocity-sensitive probes. An analysis by Boruah et al~\cite{2021MNRAS.507.2697B} showed that, once sparse sampling, survey selection, and cosmic variance are properly accounted for within consistent modeling frameworks, independent probes of large-scale motions yield mutually consistent constraints on the velocity field. Furthermore, it has been argued (see~\cite{2020MNRAS.498.2703B}) that properly accounting for the survey window function and sparse sampling can reconcile the apparent conflict between bulk flow measurements and $\Lambda$CDM predictions, attributing many discrepancies to the specific realization of structure within our local volume.

\subsubsection{Outstanding discrepancies with
%%%%%%%%%%%%%%%%%%%%%%%%%%%%%%%%%%%%%%%%%%%%%
$\Lambda$CDM}\label{sssODLCDM}
%%%%%%%%%%%%%%%%%%%%%%%%%%%%%%
While many historical conflicts in bulk flow measurements have been resolved, several important discrepancies between observations and the predictions of the standard $\Lambda$CDM cosmological model persist. These tensions have become more significant as observational precision has improved, potentially pointing to gaps in our understanding of large-scale structure formation or possibly suggesting the need for modifications to the standard cosmological framework.

One of the most prominent discrepancies concerns the amplitude of bulk flows on large scales. Recent analyses of the CF4 catalog by Watkins et al~\cite{2023MNRAS.524.1885W} have found bulk flows that exceed $\Lambda$CDM predictions by approximately 2$\sigma$ at scales beyond 100/h~Mpc. This finding is particularly significant because it appears robust across different analysis methods and is supported by independent measurements using various distance indicators. The tension becomes more pronounced when considering the results of Migkas et al~\cite{2020A&A...636A..15M,2021A&A...649A.151M}, who found evidence for coherent flows extending to even larger scales using X-ray cluster observations.

The persistence of bulk flows at very large scales presents another challenge to $\Lambda$CDM. According to standard theory, flow amplitudes should decrease approximately as $1/R$ with increasing scale $R$, yet observations consistently find more gradual decline rates. Recent analyses based on Cosmicflows-4~\cite{2023MNRAS.524.1885W} have shown that the measured scale-dependence of bulk flows remains difficult to reconcile with standard $\Lambda$CDM expectations, even after accounting for realistic survey windows, cosmic variance, and selection effects. These results suggest that either subtle systematics remain to be fully understood, or that revisions to our understanding of structure formation on very large scales may be required.

The angular distribution of peculiar velocities has also shown unexpected features. Recent analyses of the low-redshift expansion field have identified statistically significant dipolar anisotropies aligned with the CMB dipole direction~\cite{2019A&A...631L..13C,2022PhRvD.105f3514K}. These results indicate departures from perfect statistical isotropy at the largest observable scales, and they suggest possible connections to other cosmological tensions, including the Hubble tension and reported dipole discordances across different tracers.

The correlation between bulk flows and large-scale structure poses additional challenges. Reconstructions and measurements have shown that the directions of large-scale flows tend to align with prominent nearby mass concentrations, including the Great Attractor and the Shapley Concentration. In particular, density–velocity reconstructions from 2M++~\cite{2015MNRAS.450..317C} and recent Cosmicflows-4 bulk-flow analyses~\cite{2023MNRAS.524.1885W} indicate a stronger-than-expected alignment between observed flow directions and the distribution of superclusters compared with simple $\Lambda$CDM expectations. This enhanced correlation, especially in connection with the Shapley region, suggests that our understanding of how large-scale structure sources peculiar velocities may be incomplete.

Another significant tension emerges from measurements of higher-order moments in the velocity field. The analysis of velocity shear and octupole moments by Boruah et al~\cite{2021MNRAS.507.2697B} reveals patterns that may deviate from $\Lambda$CDM expectations, particularly in their scale dependence and correlation structure. These findings suggest that the standard model might not fully capture the complexity of large-scale gravitational dynamics.

\subsubsection{Concluding remarks}\label{sssCRs}
%%%%%%%%%%%%%%%%%%%%%%%%%%%%%%%%%%%%%%%%%%%%%%%%
The study of bulk flows has evolved dramatically since its inception in the 1970s, progressing from contentious early measurements to the current era of precision cosmology. Modern surveys, particularly the Cosmic Flows program, have provided unprecedented detail in mapping the local velocity field, while improved analysis techniques have helped resolve many historical conflicts. However, the persistence of certain discrepancies with $\Lambda$CDM predictions, particularly regarding flow amplitudes at large scales and unexpected anisotropies in the velocity field, suggests that bulk flows may continue to play a crucial role in testing and potentially refining our cosmological models. As we enter an era of even more precise measurements and larger surveys, these peculiar velocity studies promise to remain a vital probe of large-scale structure and cosmological physics.

\section{Methodologies in measuring peculiar
%%%%%%%%%%%%%%%%%%%%%%%%%%%%%%%%%%%%%%%%%%%%
velocities}\label{sec:pec_vel_method}
%%%%%%%%%%%%%%%%%%%%%%%%%%%%%%%%%%%%%
The measurement of peculiar velocities--deviations from the uniform Hubble expansion--constitutes a fundamental probe of large-scale structure in the universe. At its core, this measurement requires disentangling the Hubble flow from the peculiar motion component, employing both distance indicators independent of redshift and precise redshift measurements~\cite{1994ARA&A..32..371D,%
1995PhR...261..271S}.

\subsection{Fundamental Principles of Peculiar Velocity
%%%%%%%%%%%%%%%%%%%%%%%%%%%%%%%%%%%%%%%%%%%%%%%%%%%%%%%
Measurement}\label{sub:pv methods}
%%%%%%%%%%%%%%%%%%%%%%%%%%%%%%%%%%
For a galaxy with redshift $z$ and independently measured distance ($d$), the radial peculiar velocity ($v_p$) in the simplest approximation is given by:
\begin{equation}
v_p= cz- H_0d\,,  \label{eq:pec_vel_basic}
\end{equation}
with $c$ being the speed of light and $H_0$ the Hubble constant~\cite{1988MNRAS.231..149K}. This approximation is valid only in the nearby universe where $z \ll 1$. For intermediate redshifts, the relationship between observed redshift, cosmological redshift, and peculiar velocity becomes more complex due to relativistic effects. The more accurate expression is:
\begin{equation}
v_p= \frac{c(z-z_H)}{1+z_H}\,,  \label{eq:pec_vel_relativistic}
\end{equation}
where $z_H$ is the redshift corresponding to the Hubble flow at the measured distance~\cite{2011MNRAS.413.2906D,1974ApJ...191L..51H}.

The fundamental limitation in peculiar velocity measurements stems from the propagation of distance errors. The uncertainty in peculiar velocity scales with distance is:
\begin{equation}
\sigma_{v_p}\approx H_0d\cdot\frac{\sigma_d}{d}\,,  \label{eq:pec_vel_error}
\end{equation}
This relationship reveals why peculiar velocity measurements become increasingly challenging at greater distances--even with a constant fractional distance error ($\sigma_d/d$), the absolute error in peculiar velocity grows linearly with distance. At 100 Mpc, a typical distance indicator with 10\% precision yields peculiar velocity uncertainties of approximately 700 km/sec, approaching the expected signal of large-scale flows~\cite{2013AJ....146...86T}.

Two primary methodological approaches have emerged for characterizing bulk flows. The first involves measuring individual galaxy peculiar velocities and then calculating their weighted average within specified volumes~\cite{2009MNRAS.392..743W,%
2010MNRAS.407.2328F}. The second employs likelihood methods to fit parametric models directly to the observed velocity field \cite{2011ApJ...736...93N,2013MNRAS.428.2017M}.

The maximum likelihood approach has become standard for estimating bulk flows, where the likelihood function is typically expressed as:
\begin{equation}
\mathcal{L}(\mathbf{B})\propto \exp\left(-\frac{1}{2}(\mathbf{v}_p- \mathbf{M}\mathbf{B})^T \mathbf{C}^{-1} (\mathbf{v}_p- \mathbf{M}\mathbf{B})\right)\,.
\label{eq:bulk_flow_likelihood}
\end{equation}
Here, $\mathbf{v}_p$ represents the vector of measured peculiar velocities, $\mathbf{B}$ is the bulk-flow vector to be estimated, $\mathbf{M}$ is a matrix encoding the spatial distribution of galaxies in the sample, and $\mathbf{C}$ is the covariance matrix incorporating measurement errors~\cite{2009MNRAS.392..743W,%
2010MNRAS.407.2328F}.

Advanced methodologies have been developed to address the challenges of sparse and nonuniform sampling. Kaiser \cite{1988MNRAS.231..149K} introduced optimal weighting schemes that minimize variance while maintaining sensitivity to large-scale flows. Building on this work, Watkins et al~\cite{2009MNRAS.392..743W} developed the Minimum Variance (MV) estimator, which constructs an optimal weighting scheme:
\begin{equation}
\mathbf{B_{MV}}= \mathbf{A}^{-1}\mathbf{g}\,,  \label{eq:MV_estimator}
\end{equation}
where $\mathbf{A}$ is a $3\times3$ matrix and $\mathbf{g}$ is a three-component vector, both derived from the data and window function properties.

More recent approaches employ Bayesian frameworks that can incorporate prior information about the velocity field. Carrick et al~\cite{2015MNRAS.450..317C} developed techniques to reconstruct the full 3-D velocity field from redshift surveys, providing a theoretical expectation to compare with direct peculiar velocity measurements:
\begin{equation}
\mathbf{v}(\mathbf{r})= \frac{H_0 f}{4\pi} \int_V \frac{(\mathbf{r}'-\mathbf{r})\delta(\mathbf{r}')} {|\mathbf{r}'-\mathbf{r}|^3} d^3\mathbf{r}'\,,
\label{eq:velocity_reconstruction}
\end{equation}
with $f \approx \Omega_m^{0.55}$ representing the growth rate of structure, and $\delta(\mathbf{r})$ the density-contrast field. This approach allows for direct comparison between predicted and observed peculiar velocities, providing constraints on cosmological parameters \cite{2014MNRAS.444.3926J,2017MNRAS.464.2517H}.

The increasing statistical power of modern surveys has necessitated more sophisticated approaches to error analysis. Correlation between measurements must be properly accounted for, particularly when combining different distance indicators or when galaxies are spatially clustered. Ma and Scott~\cite{2013MNRAS.428.2017M} demonstrated that neglecting these correlations can lead to significant underestimation of uncertainties in bulk flow measurements. Advances in computational methods, particularly Markov Chain Monte Carlo techniques, have enabled full exploration of the posterior probability distributions, providing robust uncertainty estimates~\cite{2012MNRAS.420..447T}.

As peculiar velocity surveys probe ever-larger volumes, the interpretation of results increasingly requires precise cosmological modeling. The expected amplitude of bulk flows on various scales is directly linked to the power spectrum of density fluctuations~\cite{2003ApJ...599..820F,2009MNRAS.392..743W}, making these measurements valuable probes of cosmological parameters, particularly when combined with other observations~\cite{2017MNRAS.464.2517H}.

\subsection{Distance indicator methods}\label{ssDIMs}
%%%%%%%%%%%%%%%%%%%%%%%%%%%%%%%%%%%%%%%%%%%%%%%%%%%%%
\subsubsection{Tully-Fisher elation}\label{sssTFR}
%%%%%%%%%%%%%%%%%%%%%%%%%%%%%%%%%%%%%%%%%%%%%%%%%%
The Tully-Fisher (TF) relation is an empirical scaling law for spiral galaxies that tightly correlates their luminosity ($L$) with rotational velocity ($v_{\mathrm{rot}}$). Typically expressed in logarithmic form, the relation is given as:
\begin{equation}
\log(L)= a\,\log(v_{\mathrm{rot}})+ b\,,  \label{eq:tully_fisher}
\end{equation}
where $a$ and $b$ are empirically determined calibration constants depending on the galaxy sample and the observational wavelength used~\cite{1977A&A....54..661T,2000ApJ...533..744T}.

Distances to galaxies can be derived using the T-F relation through the distance modulus, defined as:
\begin{equation}
\mu= m- M= m- [a\,\log(v_{\mathrm{rot}})+b]\,, \label{eq:tf_distance}
\end{equation}
with $m$ being the observed apparent magnitude and $M$ the absolute magnitude predicted by the T-F relation for a given rotational velocity~\cite{2000ApJ...533..744T,2007ApJS..172..599S}.

Current implementations of the TF relation primarily utilize infrared (IR) wavelengths, significantly improving precision over traditional optical bands. Near-infrared bands ($J$, $H$, $K$) offer reduced scatter due to decreased sensitivity to internal extinction, star formation regions, and stellar population variations~\cite{2008AJ....135.1738M,2014ApJ...792..129N}. Typical fractional uncertainties in distance measurements from modern IR TF studies range between 15-20\%~\cite{2013AJ....146...86T,2014ApJ...792..129N}.

Observational scatter in the T-F relation varies significantly with wavelength, with typical intrinsic scatter levels given by:
\begin{equation}
\sigma_{M} \approx
\begin{cases}
0.40-0.50~\mathrm{mag}, & \text{(optical B-band)} \\[4pt]
0.30-0.40~\mathrm{mag}, & \text{(optical I-band)} \\[4pt]
0.20-0.30~\mathrm{mag}, & \text{(near-infrared K-band)}
\end{cases}
\label{eq:tf_scatter}
\end{equation}
These values illustrate the advantage of infrared observations in reducing uncertainties and improving distance accuracy \cite{2008AJ....135.1738M,2012AJ....144..133S}.

Corrections for systematic effects such as galaxy inclination ($i$), internal extinction, and morphological peculiarities are essential. For instance, internal extinction corrections typically follow the empirical model:
\begin{equation}
    A_{\text{int}} = \gamma\,\log_{10}(\sec i),
    \label{eq:tf_extinction}
\end{equation}
where the coefficient $\gamma$ depends on wavelength and galaxy type, and $i$ is the inclination angle determined from galaxy axial ratios~\cite{2006ApJ...653..861M,2014ApJ...792..129N}.

Extensive calibration efforts, such as those from the Cosmicflows program and the 2MASS Tully-Fisher (2MTF) survey, have provided robust, self-consistent TF calibrations across multiple wavelengths, significantly enhancing the reliability and cosmological utility of TF-based peculiar velocity measurements~\cite{2013AJ....146...86T,2017MNRAS.471.3135H}.

Recent and upcoming surveys, notably employing radio H~I 21-cm measurements (e.g.~WALLABY, ASKAP, and SKA precursor surveys), promise further reductions in T-F scatter by obtaining precise galaxy rotation velocities unaffected by dust extinction, thereby refining peculiar velocity studies and bulk flow analyses~\cite{2020Ap&SS.365..118K,2023ApJ...944...94T}.

\subsubsection{Fundamental plane}\label{sssFP}
%%%%%%%%%%%%%%%%%%%%%%%%%%%%%%%%%%%%%%%%%%%%%%
The Fundamental Plane (FP) relation is an empirical scaling relation observed for elliptical and early-type galaxies, linking three observable parameters, namely the effective radius ($R_e$), the central stellar velocity dispersion ($\sigma_0$), and the mean effective surface brightness ($\langle I_e \rangle$) within $R_e$. Typically, the relation is expressed as:
\begin{equation}
\log R_ =\alpha\,\log\sigma_0+ \beta\,\log\langle I_e\rangle+ \gamma\,,  \label{eq:fundamental_plane}
\end{equation}
where $\alpha$, $\beta$, and $\gamma$ are empirically calibrated coefficients. The FP relation provides a robust distance indicator for galaxies out to intermediate redshifts~\cite{1987ApJ...313...59D,1987ApJ...313...42D,%
2003AJ....125.1817B}.

Under the assumption of virial equilibrium, theoretical expectations for the FP coefficients are $\alpha=2$ and $\beta=-1$. However, observational studies consistently find deviations from these predictions, typically measuring $\alpha\approx1.2$ and $\beta\approx-0.8$. These deviations indicate complexities beyond simple virial equilibrium, including varying stellar populations, non-homology in galaxy structures, and dark matter content variations among galaxies~\cite{2003AJ....125.1817B,2013MNRAS.432.1709C,%
2019A&A...622A..83S}.

Galaxy distances can be inferred from the FP relation by comparing the observed effective radius, $R_e$, with the radius predicted from velocity dispersion and surface brightness measurements, $R_{e,\text{FP}}$:
\begin{equation}
    \frac{d}{d_0} = \frac{R_e}{R_{e,\text{FP}}},
    \label{eq:fp_distance}
\end{equation}
where $d_0$ represents a reference or calibration distance established using galaxies with independently known distances~\cite{2003AJ....125.1817B,2012MNRAS.427..245M}.

The intrinsic scatter in the FP relation translates to typical fractional uncertainties in distance measurements of approximately:
\begin{equation}
\frac{\sigma_d}{d}\approx 18\%-22\%\,,
    \label{eq:fp_scatter}
\end{equation}
depending on observational methods, wavelengths, sample selection, and calibration strategies~\cite{2012MNRAS.427..245M,2020MNRAS.497.1275S}.

Environmental dependencies significantly impact FP precision. Studies have demonstrated that the FP coefficients exhibit measurable variations between galaxies residing in dense cluster environments and those in lower-density field regions. For instance, Bernardi et al~\cite{2003AJ....125.1817B} found systematic variations in the FP slopes of approximately $\Delta\alpha\sim0.05-0.1$ between low- and high-density environments. Such variations necessitate careful calibration of the FP relation within different cosmic environments, utilizing large homogeneous samples to minimize systematic biases~\cite{2003AJ....125.1817B,2015SSRv..193....1J,%
2019A&A...622A..83S}.

Recent large-scale galaxy surveys, such as the Sloan Digital Sky Survey (SDSS), the 6-degree Field Galaxy Survey (6dFGS), and more recent surveys like the Cosmicflows programs, have provided extensive FP datasets, enabling improved calibration, detailed studies of environmental dependencies, and more precise cosmological measurements~\cite{2012MNRAS.427..245M,2020MNRAS.497.1275S,%
2019A&A...622A..83S}. These surveys have significantly advanced the utility of the FP as a robust distance indicator for peculiar velocity analyses, bulk flow measurements, and cosmological parameter constraints.

\subsubsection{Surface brightness fluctuations}\label{sssSBFs}
%%%%%%%%%%%%%%%%%%%%%%%%%%%%%%%%%%%%%%%%%%%%%%%%%%%%%%%%%%%%%%
The Surface Brightness Fluctuations (SBF) method leverages statistical fluctuations in the integrated brightness of unresolved stellar populations within galaxies to measure precise distances~\cite{1988AJ.....96..807T,2021ApJ...911...65B}. These fluctuations arise because the number of stars per resolution element is finite and decreases with increasing distance, causing the observed variance in surface brightness to diminish predictably with galaxy distance.

The amplitude of these brightness fluctuations is quantified by the SBF magnitude, defined as:
\begin{equation}
\overline{M}= -2.5\,\log(\overline{f})+ \mathrm{ZP}\,,
    \label{eq:sbf_magnitude}
\end{equation}
with $\overline{f}$ being the luminosity-weighted variance of the stellar flux distribution, and $\mathrm{ZP}$ is the photometric zero-point determined by calibration observations~\cite{1988AJ.....96..807T,2009ApJ...694..556B}.

The SBF method provides exceptionally accurate distances for galaxies within the local universe, achieving fractional distance uncertainties typically on the order of:
\begin{equation}
    \frac{\sigma_d}{d} \approx
    \begin{cases}
        4\% - 5\%, & d < 20\,\text{Mpc}, \\[4pt]
        7\% - 8\%, & 20\,\text{Mpc} < d < 50\,\text{Mpc}, \\[4pt]
        9\% - 10\%, & 50\,\text{Mpc} < d < 100\,\text{Mpc}.
    \end{cases}
    \label{eq:sbf_precision}
\end{equation}
However, beyond approximately 100 Mpc, the fluctuation amplitude becomes comparable to observational noise, making distance measurements increasingly challenging and requiring deeper imaging data with higher spatial resolution \cite{2021ApJ...911...65B,2018ApJ...854L..31C}.

Crucially, SBF magnitudes exhibit strong dependence on the galaxy's stellar population characteristics, especially the galaxy color, which serves as a proxy for stellar age and metallicity. Consequently, modern calibrations of the SBF magnitude typically utilize a color-dependent relation of the form:
\begin{equation}
    \overline{M}= \alpha+ \beta\,(g-i)\,,
    \label{eq:sbf_calibration}
\end{equation}
where $(g-i)$ denotes the galaxy's optical color, and $\alpha$ and $\beta$ are calibration constants empirically determined from observations of galaxies with independently measured distances. Recent calibration studies in the $I$-band report slopes of $\beta\approx1.5-2.0$~\cite{2009ApJ...694..556B,2021ApJ...911...65B,%
2021ApJS..255...21J}.

SBF distance measurements have been extensively calibrated using Hubble Space Telescope observations and other high-resolution facilities, significantly refining their precision and reliability. These distances have notably contributed to peculiar velocity analyses and provided valuable constraints on local cosmological parameters, such as the Hubble constant and local bulk flows \cite{2021ApJ...911...65B,2021ApJS..255...21J,2021A&A...647A..72K}. Upcoming facilities, including the James Webb Space Telescope (JWST) and the Vera C. Rubin Observatory, promise further substantial improvements in the reach, precision, and cosmological utility of the SBF method \cite{2021ApJ...911...65B}.

\subsubsection{Standard candles}\label{sssSCls}
%%%%%%%%%%%%%%%%%%%%%%%%%%%%%%%%%%%%%%%%%%%%%%%
Standard candles represent a class of astrophysical objects with well-calibrated intrinsic luminosities, enabling robust distance measurements. Among these, Type Ia supernovae (SNe Ia) have proven particularly valuable for probing galaxy peculiar velocities at intermediate to large cosmological distances. After correcting observed brightness for light-curve shape and color variations, SNe Ia exhibit remarkably uniform standardized peak luminosities, empirically parameterized as~\cite{2018ApJ...859..101S,2022ApJ...938..110B}:
\begin{equation}
M_B= M_B^0+ \alpha\,x_1- \beta\,c+ \Delta_M\,,
    \label{eq:sne_standardization}
\end{equation}
where $M_B^0$ is a fiducial absolute magnitude, $x_1$ quantifies the light-curve stretch (decline rate), $c$ describes intrinsic color variations, and $\Delta_M$ represents host-galaxy-dependent luminosity corrections. Recent analyses from large SN compilations, such as Pantheon and Pantheon+, have achieved typical distance precisions at the 5-7\% level per object, making SNe Ia exceptional cosmological probes~\cite{2018ApJ...859..101S,2022ApJ...938..110B}.

At nearer distances ($d\lesssim20$~Mpc), the Tip of the Red Giant Branch (TRGB) method provides another powerful standard candle. The TRGB corresponds to a sharply defined cutoff in the luminosity function of evolved red giant branch stars, offering a reliable luminosity marker nearly independent of galaxy type or environment. The absolute magnitude of the TRGB in the I-band ($M_I^{\mathrm{TRGB}}$) is calibrated as a function of stellar color $(V-I)$~\cite{2019ApJ...882...34F,2021ApJ...906..125J}:
\begin{equation}
    M_I^{\mathrm{TRGB}} \approx -4.05 + 0.22\,[(V - I) - 1.6].
    \label{eq:trgb_calibration}
\end{equation}

Recent calibrations using the Hubble Space Telescope and the Gaia mission have confirmed that the TRGB method achieves exquisite accuracy of about 3--5\% in distance measurements for galaxies within approximately 10--20~Mpc, making it one of the most precise local distance indicators available~\cite{2019ApJ...882...34F,2022ApJ...932...15A}. The TRGB method's robustness against extinction and minimal dependence on stellar age and metallicity significantly reduces systematic uncertainties compared to other local distance indicators such as Cepheid variables~\cite{2021ApJ...906..125J,2021ApJ...919...16F}.

Both SNe Ia and TRGB methods are central to modern peculiar velocity surveys, providing complementary distance constraints across a broad cosmological scale. Future observational facilities and surveys, including the Vera C. Rubin Observatory, JWST, and dedicated SN surveys (e.g.~LSST), promise substantial improvements in the sample size, precision, and cosmological reach of distance measurements obtained from standard candles~\cite{2022ApJ...938..110B,%
2021ApJ...919...16F}.

\subsection{Advanced observation techniques}\label{ssAOTs}
%%%%%%%%%%%%%%%%%%%%%%%%%%%%%%%%%%%%%%%%%%%%%%%%%%%%%%%%%%
\subsubsection{Kinematic Sunyaev-Zel'dovich effect}\label{sec:kinematicSN}
%%%%%%%%%%%%%%%%%%%%%%%%%%%%%%%%%%%%%%%%%%%%%%%%%%%%%%%%%%%%%%%%%%%%%%%%%
The kinematic Sunyaev-Zel'dovich (kSZ) effect provides a powerful and unique method to probe peculiar velocities at cosmological distances, independent of traditional distance indicators~\cite{1980MNRAS.190..413S,1999PhR...310...97B}. This effect arises from the Doppler shift induced on cosmic microwave background (CMB) photons when scattering off free electrons moving with a peculiar velocity $v_r$ along the line-of-sight direction. The observed fractional temperature fluctuation is
\begin{equation}
\frac{\Delta T}{T_{\text{CMB}}}= -\tau_e \frac{v_r}{c}\,,  \label{eq:ksz_effect}
\end{equation}
where $\tau_e=\sigma_T\int n_e dl$ is the Thomson optical depth, typically $\tau_e\sim10^{-2}$-$10^{-3}$ for massive galaxy clusters, and $c$ is the speed of light~\cite{2002ARA&A..40..643C,2019SSRv..215...17M}.

The kSZ signal is directly proportional to the cluster's line-of-sight velocity and optical depth, independent of redshift, making it particularly attractive for probing velocity fields at large cosmological distances. Unlike thermal SZ (tSZ), which has a characteristic spectral dependence, the kSZ effect preserves the blackbody spectrum of the CMB, appearing only as a temperature shift. Thus, multi-frequency observations are crucial for reliable separation from the much stronger tSZ signal~\cite{1999PhR...310...97B,2019SSRv..215...17M}.

The amplitude of kSZ temperature fluctuations is typically very small, on the order of $\Delta T \sim 1$--$10\,\mu$K for clusters moving with velocities $v_r \sim 300$--$1000\,\text{km\,s}^{-1}$, significantly weaker than typical tSZ amplitudes ($\sim 100$--$1000\,\mu$K)~\cite{2019SSRv..215...17M}. The signal-to-noise ratio (SNR) for detecting the kSZ effect in an individual galaxy cluster can be approximated as:
\begin{equation}
\text{SNR} \approx \frac{\tau_e (v_r/c)}{\sigma_{\text{noise}}/\sqrt{N_{\text{pix}}}},
\label{eq:ksz_snr}
\end{equation}
where $\sigma_{\text{noise}}$ is the per-pixel instrumental noise level, and $N_{\text{pix}}$ is the number of pixels subtending the cluster~\cite{2016ApJ...823...98F}.

Given the small amplitude of the kSZ signal, several statistical methods have been developed to measure it robustly:

\begin{enumerate}
    \item \textbf{Pairwise kSZ estimator}: Leveraging pairs of galaxy clusters or groups, this method measures the mean pairwise velocity as a function of separation, exploiting opposite Doppler shifts from gravitational infall. The pairwise momentum estimator is typically defined as \cite{2012PhRvL.109d1101H,2016MNRAS.461.3172S}:
    \begin{equation}
    \hat{p}(r) = - \frac{\sum_{i<j}(T_i - T_j)c_{ij}}{\sum_{i<j} c_{ij}^2},
    \label{eq:ksz_pairwise}
    \end{equation}
    where $T_i$ is the measured CMB temperature fluctuation at object $i$, and $c_{ij}$ encodes geometry and projection effects.\\

    \item \textbf{Velocity reconstruction methods}: This approach reconstructs the velocity field from galaxy surveys and cross-correlates it with CMB maps, yielding constraints on optical depth and bulk flows \cite{2016A&A...594A..13P,2016PhRvD..93h2002S,%
        2020JCAP...12..011N}.\\

    \item \textbf{kSZ tomography}: This method utilizes large-scale structure tracers to reconstruct the electron density field, enabling direct cross-correlation with CMB temperature maps. Recent studies have demonstrated significant detections using this approach~\cite{2018arXiv181013423S,2021PhRvD.104d3502C}.
\end{enumerate}

These statistical techniques have yielded robust detections: for example, the Atacama Cosmology Telescope (ACT) team reported a $4.1\sigma$ detection using pairwise statistics~\cite{2012PhRvL.109d1101H}, while recent analyses using ACT and BOSS luminous red galaxies reached $6.5\sigma$ significance~\cite{2021PhRvD.104d3502C}. Individual cluster measurements remain challenging but have been demonstrated for massive merging clusters, such as MACS~J0717.5+3745 \cite{2013ApJ...778...52S,2017A&A...598A.115A}.

The kSZ effect offers distinct advantages for cosmic velocity studies:

\begin{itemize}
    \item \textbf{Redshift independence}: kSZ amplitude does not diminish at higher redshift, enabling measurements out to $z\sim 2$ or beyond~\cite{2016PhRvD..94l3526F}.\\

    \item \textbf{Baryon census}: Directly probes all free electrons, potentially resolving the ``missing baryon'' problem by tracing diffuse baryonic components in the intergalactic medium~\cite{2016PhRvL.117e1301H,2019A&A...624A..48D}.\\

    \item \textbf{Tests of gravity and cosmology}: The relationship between density and velocity fields measured via kSZ can test gravity theories beyond general relativity~\cite{2015ApJ...808...47M}.\\

    \item \textbf{Cluster astrophysics}: High-resolution kSZ mapping can reveal internal cluster motions, merger dynamics, and non-thermal pressure contributions~\cite{2019ApJ...880...45S}.
\end{itemize}

Future CMB experiments, such as the Simons Observatory \cite{2019JCAP...02..056A} and CMB-S4 \cite{2019arXiv190704473A}, combined with large-scale spectroscopic galaxy surveys (e.g., DESI, Euclid), are expected to improve the sensitivity of kSZ measurements dramatically. Forecasts suggest individual-cluster velocity measurements with uncertainties of $40$-$80$~km/sec for massive clusters ($M_{500}\sim10^{14}M_\odot$) will soon become feasible~\cite{2015ApJ...812..154B,2018JCAP...02..032M}, potentially revolutionizing our understanding of cosmological velocity fields and structure formation.

\subsubsection{Novel techniques and emerging
%%%%%%%%%%%%%%%%%%%%%%%%%%%%%%%%%%%%%%%%%%%%
methods}\label{sssNTEMs}
%%%%%%%%%%%%%%%%%%%%%%%%
Recent advances in machine learning (ML) techniques have opened new avenues for improving peculiar velocity estimations. Convolutional neural networks (CNNs), in particular, have demonstrated significant potential for accurately predicting galaxy distances and peculiar velocities even with limited or noisy observational data. For instance, CNN-based approaches have been shown to achieve comparable or superior precision compared to traditional distance indicators, significantly reducing required observational parameters and enabling efficient analyses of large galaxy surveys~\cite{2020MNRAS.498.2703B,2023EPJC...83..304G}. Such ML approaches are particularly promising for upcoming large-scale surveys, where traditional methods might become computationally prohibitive.

Another emerging and highly promising method for measuring peculiar velocities involves gravitational wave (GW) sources as ``standard sirens''. Binary neutron star mergers, accompanied by electromagnetic counterparts, provide precise luminosity distances ($d_L$) directly from GW waveform measurements. Combining these luminosity distances with spectroscopic redshift measurements ($z$) from electromagnetic follow-up observations allows for direct peculiar velocity measurements through the relationship~\cite{2017Natur.551...85A,2021PhRvD.103j3507P}:
\begin{equation}
    v_p = c \left[z - \frac{H_0 d_L}{c(1+z)}\right],
    \label{eq:gw_pec_vel}
\end{equation}
where $H_0$ is the Hubble constant, and $c$ is the speed of light. We note that this relationship is valid in the low-redshift limit (i.e.~for $z\ll1$).

The first multi-messenger GW detection (GW170817) demonstrated the feasibility of this method, yielding peculiar velocity constraints of approximately $v_p = 310 \pm 70\,\text{km\,s}^{-1}$ for the host galaxy NGC~4993~\cite{2017Natur.551...85A,2017ApJ...848L..31H}. Future GW observatories, including the Einstein Telescope, Cosmic Explorer, and LISA, combined with planned deep optical surveys (e.g., Vera C. Rubin Observatory and Roman Space Telescope), are expected to dramatically expand the number of GW standard sirens detected, enabling systematic peculiar velocity measurements with unprecedented precision and volume coverage~\cite{2021PhRvD.103j3507P,2024JCAP...05..031Y}. These standard sirens will provide highly accurate and independent cosmological distance measurements, complementing traditional distance indicators and significantly improving bulk flow determinations and cosmological parameter constraints.

\section{Theoretical framework}\label{sTF}
%%%%%%%%%%%%%%%%%%%%%%%%%%%%%%%%%%%%%%%%%%
This section provides the basic theoretical background of the 1+3 covariant approach to relativistic cosmology, which will be used in the subsequent sections to study the evolution of the observed large-scale peculiar motions. In the process, the relativistic results will also be directly compared to those of the available Newtonian and quasi-Newtonian studies.

\subsection{Covariant spacetime splitting}\label{ssCSS}
%%%%%%%%%%%%%%%%%%%%%%%%%%%%%%%%%%%%%%%%%%%%%%%%%%%%%%%
The covariant formalism dates back to the 1950s (e.g.~see~\cite{1955ZA.....38...95H,1957ZA.....43..161R}) and has since been used in numerous studies by many authors
(see~\cite{2008PhR...465...61T,2012reco.book.....E} for recent reviews with more details and references). The approach utilises a local 1+3 splitting of the spacetime into time and 3-dimensional space. This is achieved by introducing a group of observers, ``living'' along timelike worldlines with their associated 3-spaces orthogonal to them. Instead of using the metric tensor, the covariant formalism employs the Ricci and the Bianchi identities, together with the Einstein field equations~\cite{2008PhR...465...61T,2012reco.book.....E}.

Consider a group of observers with 4-velocity $u^a={\rm d}x^a/{\rm d}\tau$, where $\{x^a\}$ are their (generalised) coordinates and $\tau$ their proper time, so that $u_au^a=-1$. The $u_a$-field determines the observers' temporal direction and introduces a local 1+3 ``threading'' of the spacetime into time and 3-D space. Then, the tensor $h_{ab}=g_{ab}+u_au_b$ projects into the observers' instantaneous rest space and also acts as the 3-D metric. Note that the projector is symmetric, spacelike and also satisfies the constraints $h_a{}^a=3$ and $h_{ac}h^c{}_b=h_{ab}$.

Using $u_a$ and $h_{ab}$, the temporal and spatial covariant derivatives of a general tensor field $S_{ab\cdots}{}^{cd\cdots}$, relative to the aforementioned observers, are respectively given by
\begin{equation}
\dot{S}_{ab\cdots}{}^{cd\cdots}=
u^e\nabla_eS_{ab\cdots}{}^{cd\cdots} \hspace{5mm} {\rm and}
\hspace{5mm} {\rm D}_eS_{ab\cdots}{}^{cd\cdots}=
h_e{}^sh_a{}^fh_b{}^ph_q{}^ch_r{}^d\cdots
\nabla_sS_{fp\cdots}{}^{qr\cdots}\,.  \label{deriv}
\end{equation}

Note that we adopt spacetimes with metric ($g_{ab}$) of signature ($-,\,+,\,+,\,+$) and use geometrised units with $c=1=8\pi G$, unless stated otherwise. Also, Latin indices take values between $0$ and $3$, while their Greek counterparts run from $1$ to $3$.

\subsubsection{The gravitational field}\label{sssGrF}
%%%%%%%%%%%%%%%%%%%%%%%%%%%%%%%%%%%%%%%%%%%%%%%%%%%%%
In general relativity gravity is no longer a force but the manifestation of spacetime curvature. The presence of matter forces the spacetime to curve and the curvature dictates the motion of the matter. The whole interaction is monitored by Einstein's equations, namely by
\begin{equation}
G_{ab}\equiv R_{ab}- {1\over2}\,Rg_{ab}= T_{ab}- \Lambda g_{ab}\,.
\label{EFE1}
\end{equation}
Here, $G_{ab}$ is the Einstein tensor, $R_{ab}=R_{acb}{}^c$ is the Ricci tensor of the 4-D spacetime (with $R_{abcd}$ being the associated Riemann tensor), $R=R_a{}^a$ is the Ricci scalar, $T_{ab}$ is the stress-energy tensor of the matter and $\Lambda$ is the cosmological constant. Then, the (twice contracted) Bianchi identities ($\nabla^bG_{ab}=0$)
guarantee that $\nabla^bT_{ab}=0$ and energy-momentum
conservation.

The Ricci tensor and the Einstein equations describe the local gravitational field, whereas the action of gravity at a distance (i.e.~gravitational waves and tidal forces) is monitored by the Weyl tensor ($C_{abcd}$). These two tensors emerge after splitting the Riemann tensor, that is the total
gravitational field, as
\begin{equation}
R_{abcd}= C_{abcd}+ {1\over2}\left(g_{ac}R_{bd}+g_{bd}R_{ac}
-g_{bc}R_{ad}-g_{ad}R_{bc}\right)
-{1\over6}\,R\left(g_{ac}g_{bd}-g_{ad}g_{bc}\right)\,,
\label{Riemann}
\end{equation}
where $C_{abcd}$ has all the symmetries of $R_{abcd}$ and it is also trace-free. When timelike observers are introduced, the Weyl tensor decomposes into an electric and a magnetic part given by
(e.g.~see~\cite{2008PhR...465...61T,2012reco.book.....E})
\begin{equation}
E_{ab}= C_{acbd}u^cu^d \hspace{15mm} {\rm and} \hspace{15mm}
H_{ab}= {1\over2}\,\varepsilon_a{}^{cd}C_{cdbe}u^e\,.
\label{EabHab}
\end{equation}
respectively. In the above, $\varepsilon_{abc}=\eta_{abcd}u^d$ is the totally antisymmetric Levi-Civita tensor of the observer's 3-dimensional rest-space, with $\varepsilon_{abd}u^c=0$ by construction and with $\eta_{abcd}$ representing its 4-dimensional analogue. Also note that $E_{ab}$ has a Newtonian analogue but $H_{ab}$ has not (e.g.~see~\cite{1971glc..conf....104E}). For this reason the electric Weyl tensor is primarily associated with the tidal field and the magnetic with gravitational waves, although both Weyl tensors are necessary for the propagation of gravitational waves. In addition, both $E_{ab}$ and $H_{ab}$ are spacelike, symmetric and trace-free tensors.

\subsubsection{Matter fields}\label{sssMFs}
%%%%%%%%%%%%%%%%%%%%%%%%%%%%%%%%%%%%%%%%%%%
When decomposed relative to a family of timelike observers, the energy-momentum tensor of an imperfect medium splits as\footnote{In studies of multi-component media, or when allowing for peculiar velocities, it is necessary to introduce more than one 4-velocity fields in relative motion with each other (see~\S~\ref{sCPVT} below).}
\begin{equation}
T_{ab}= \rho u_au_b+ ph_{ab}+ 2q_{(a}u_{b)}+ \pi_{ab}\,.
\label{Tab1}
\end{equation}
where $\rho=T_{ab}u^au^b$ is the matter density, $p=T_{ab}h^{ab}/3$ is its isotropic pressure, $q_a=-h_a{}^bT_{bc}u^c$ is the energy-flux and $\pi_{ab}=h_{\langle a}{}^{c} h_{b\rangle}{}^dT_{cd}$ is the viscosity.\footnote{Hereafter, angled brackets will indicate the symmetric and traceless part of spatially projected second-rank tensors, as well as projected vectors. For example,
\begin{equation}
S_{\langle ab \rangle}= h_{\langle a}{}^{c}h_{b\rangle}{}^dS_{cd}=
h_{(a}{}^{c}h_{b)}{}^dS_{cd}- {1\over3}\,h^{cd}S_{cd}h_{ab} \hspace{10mm} {\rm
and} \hspace{10mm} {V}_{\langle a\rangle}= h_a{}^bV_b\,.
\label{angled}
\end{equation}} When dealing with a perfect fluid, $q_a$ and $\pi_{ab}$ vanish identically and (\ref{Tab1}) reduces to
\begin{equation}
T_{ab}= \rho u_au_b+ ph_{ab}\,.  \label{pfTab}
\end{equation}
Setting $p=0$ we have the simplest case of ``dust'', which refers to low-energy baryonic matter (after decoupling) and to Cold Dark Matter (CDM). Otherwise, we need an equation of state, which generally takes the form $p=p\,(\rho,s)$, with $s$ representing the specific entropy. Nevertheless, in most applications the cosmic medium is treated as barotropic with $p=p\,(\rho)$.

Starting from (\ref{Tab1}), while keeping in mind that $R=4\Lambda-T$ and $T=T_a{}^a$, the Einstein equations recast into
\begin{equation}
R_{ab}= T_{ab}- {1\over2}\,Tg_{ab}+ \Lambda g_{ab}\,.
\label{EFE2}
\end{equation}
Then, after successively contracting the above, we arrive at the algebraic relations
\begin{equation}
R_{ab}u^au^b= {1\over2}\,(\rho+3p)- \Lambda\,, \hspace{15mm} h_a{}^bR_{bc}u^c= -q_a  \label{EFE34}
\end{equation}
and
\begin{equation}
h_a{}^ch_b{}^dR_{cd}= {1\over2}\,(\rho-p)h_{ab}+ \Lambda h_{ab}+ \pi_{ab}\,,  \label{EFE5}
\end{equation}
between the Ricci field and the matter component (e.g.~see~\cite{2008PhR...465...61T,2012reco.book.....E}).

\subsection{Covariant relativistic cosmology}\label{ssCRC}
%%%%%%%%%%%%%%%%%%%%%%%%%%%%%%%%%%%%%%%%%%%%%%%%%%%%%%%%%%
Depending on the problem in hand, there are different choices for the timelike 4-velocity field ($u_a$). Typically, $u_a$ defines the frame where the CMB dipole vanishes, or occasionally the rest-frame of the matter. In most theoretical studies these two coordinate systems coincide, but this no longer holds when peculiar motions are involved (e.g.~see \S~\ref{sCPVT} below).

\subsubsection{Kinematics}\label{sssKs}
%%%%%%%%%%%%%%%%%%%%%%%%%%%%%%%%%%%%%%%
Covariantly, the motion of the observer is monitored by a set of four kinematic variables, which emerge after decomposing the 4-velocity gradient as follows (e.g.~see~\cite{2008PhR...465...61T,2012reco.book.....E})
\begin{equation}
\nabla_bu_a= {\rm D}_bu_a- A_au_b= {1\over3}\,\Theta h_{ab}+ \sigma_{ab}+ \omega_{ab}- A_au_b\,.  \label{Nbua}
\end{equation}
Here, $\Theta={\rm D}^au_a$ is the volume scalar, $\sigma_{ab}={\rm D}_{\langle b}u_{a\rangle}$ is the shear, $\omega_{ab}={\rm D}_{[b}u_{a]}$ is the vorticity and $A_a=\dot{u}_a=u^b\nabla_bu_a$ is the 4-acceleration. The last three are purely spacelike, since $\sigma_{ab}u^a=0= \omega_{ab}u^a=A_au^a$. The volume scalar describes expansion/contraction when positive/negative respectively. In the homogeneous and isotropic FLRW models, $\Theta$ is used to define the cosmological scale factor ($a$) by means of $\dot{a}/a=\Theta/3$. The same relation also defines the mean scale factor in the anisotropic Bianchi cosmologies. Given its trace-free nature, the shear monitors shape distortions under constant volume, while antisymmetric vorticity tensor describes rotation and also defines the rotational axis via the vorticity vector $\omega_a=\varepsilon_{abc}\omega^{bc}/2$. Finally, nonzero 4-acceleration indicates the presence of non-gravitational forces and ensures that the observers' worldlines are not geodesics.

The covariant kinematics are determined by three propagation and three constraint equations, obtained after applying the Ricci identities (e.g.~see~\cite{2008PhR...465...61T,2012reco.book.....E})
\begin{equation}
2\nabla_{[a}\nabla_{b]}u_c= R_{abcd}u^d\,,  \label{Ris}
\end{equation}
to the $u_a$-field. Substituting (\ref{Nbua}) into the right-hand side of the above, employing (\ref{Riemann}) and (\ref{EabHab}), as well as the auxiliary relations (\ref{EFE34}) and (\ref{EFE5}), the timelike part of the resulting expression leads to a set of three time-propagation formulae. These are the Raychaudhuri equation
\begin{equation}
\dot{\Theta}= -{1\over3}\,\Theta^2- {1\over2}\,(\rho+3p)-
2(\sigma^2-\omega^2)+ {\rm D}^aA_a+ A_aA^a+ \Lambda\,, \label{Ray}
\end{equation}
for the evolution of $\Theta$, the shear propagation formula
\begin{equation}
\dot{\sigma}_{\langle ab\rangle}= -{2\over3}\,\Theta\sigma_{ab}-
\sigma_{c\langle a}\sigma^c{}_{b\rangle}- \omega_{\langle
a}\omega_{b\rangle}+ {\rm D}_{\langle a}A_{b\rangle}+ A_{\langle
a}A_{b\rangle}- E_{ab}+ {1\over2}\,\pi_{ab}\,, \label{sigmadot}
\end{equation}
and the evolution equation of the vorticity
\begin{equation}
\dot{\omega}_{\langle a\rangle}= -{2\over3}\,\Theta\omega_a-
{1\over2}\,\curl A_a+ \sigma_{ab}\omega^b\,, \label{omegadot}
\end{equation}
where ${\rm curl}A_a= \varepsilon_{abc}{\rm D}^bA^c$, $\sigma^2=\sigma_{ab}\sigma^{ab}/2$ and
$\omega^2=\omega_{ab}\omega^{ab}/2=\omega_a\omega^a$ by definition.

Proceeding in an analogous way, the spacelike component of (\ref{Ris}) provides a set of three
constraints. These operate in the observer's 3-D rest-space and are given by
\begin{equation}
{\rm D}^b\sigma_{ab}= {2\over3}\,{\rm D}_a\Theta+ \curl \omega_a+
2\varepsilon_{abc}A^b\omega^c- q_a\,, \label{shearcon}
\end{equation}
\begin{equation}
{\rm D}^a\omega_a= A_a\omega^a\,,  \label{omegacon}
\end{equation}
and
\begin{equation}
H_{ab}= \curl \sigma_{ab}+ {\rm D}_{\langle a}\omega_{b\rangle}+
2A_{\langle a}\omega_{b\rangle}\,, \label{Hcon}
\end{equation}
with ${\rm curl}\sigma_{ab}=\varepsilon_{cd\langle a}{\rm D}^c\sigma^d{}_{b\rangle}$ by construction.

The Raychaudhuri equation (see~\cite{1957ZA.....43..161R} and
also~\cite{2005gr.qc....11123D,2006gr.qc....11123K}) is the key formula used to study the expansion/contraction of self-gravitating systems. Among others, expression (\ref{Ray}) determines the deceleration parameter ($q$) of the universe and it will play a central role in \S~\ref{sOSPMs} below. Until then, let us say that positive terms on the right-hand side of (\ref{Ray}) have the tendency to accelerate the expansion, whereas negative ones do the opposite.

\subsubsection{Conservation laws}\label{sssCLs}
%%%%%%%%%%%%%%%%%%%%%%%%%%%%%%%%%%%%%%%%%%%%%%%
Applying the twice contracted Bianchi identities to the Einstein field equations leads to $\nabla^bT_{ab}=0$, which in turn provides the conservation laws of the energy and the momentum densities. For an imperfect fluid, these read~\cite{2008PhR...465...61T,2012reco.book.....E}
\begin{equation}
\dot{\rho}= -\Theta(\rho+p)- {\rm D}^aq_a- 2A^aq_a-
\sigma^{ab}\pi_{ab}  \label{edcl}
\end{equation}
and
\begin{equation}
(\rho+p)A_a= -{\rm D}_ap- \dot{q}_{\langle a\rangle}-
{4\over3}\,\Theta q_a- (\sigma_{ab}+\omega_{ab})q^b- {\rm
D}^b\pi_{ab}- \pi_{ab}A^b\,,  \label{mdcl}
\end{equation}
respectively. In the case of a perfect medium (with $q_a=0=\pi_{ab}$) on the other hand, the above expressions simplify and reduce to
\begin{equation}
\dot{\rho}= -\Theta(\rho+p) \hspace{15mm} {\rm and} \hspace{15mm}
(\rho+p)A_a= -{\rm D}_ap\,,  \label{pfcls}
\end{equation}
respectively. Following (\ref{mdcl}), imperfect media have nonzero 4-acceleration even in the absence of pressure (isotropic or viscous). This fact will prove crucial in \S~\ref{ssRA} and later in \S~\ref{sOSPMs}.

\subsubsection{Spatial curvature}\label{sssSC}
%%%%%%%%%%%%%%%%%%%%%%%%%%%%%%%%%%%%%%%%%%%%%%
The geometry of the 3-space orthogonal to the observer's 4-velocity field is determined by the associated 3-dimensional Riemann tensor.\footnote{When $u_a$ is irrotational, the observers' rest-spaces form integrable spacelike surfaces, which define their hypersurfaces of simultaneity. In the presence of vorticity, however, Frobenius' theorem forbids the existence of such integrable 3-dimensional surfaces (e.g.~see~\cite{1984ucp..book.....W,2004rtmb.book.....P}).} The latter is given by
\begin{equation}
\clr_{abcd}= h_a{}^qh_b{}^sh_c{}^fh_d{}^pR_{qsfp}- v_{ac}v_{bd}+
v_{ad}v_{bc}\,, \label{3Riemann1}
\end{equation}
where $v_{ab}={\rm D}_bu_a$ (see Eq.~(\ref{Nbua}) in \S~\ref{sssKs}). Combining Eq.~(\ref{EFE1}) with decomposition (\ref{Riemann}) and definitions (\ref{EabHab}), one arrives at the expression
\begin{eqnarray}
\clr_{abcd}&=& -\varepsilon_{abq}\varepsilon_{cds}E^{qs}+
{1\over3}\left(\rho-{1\over3}\,\Theta^2+\Lambda\right)(h_{ac}h_{bd}
-h_{ad}h_{bc})\nonumber\\
&{}&+{1\over2}\left(h_{ac}\pi_{bd}+\pi_{ac}h_{bd}
-h_{ad}\pi_{bc}-\pi_{ad}h_{bc}\right)\nonumber\\
&{}&-{1\over3}\,\Theta\left[h_{ac}(\sigma_{bd}+\omega_{bd})
+(\sigma_{ac}+\omega_{ac})h_{bd}-h_{ad}(\sigma_{bc}+\omega_{bc})
-(\sigma_{ad}+\omega_{ad})h_{bc}\right] \nonumber\\
&{}&-(\sigma_{ac}+\omega_{ac})(\sigma_{bd}+\omega_{bd})+(\sigma_{ad}
+\omega_{ad})(\sigma_{bc}+\omega_{bc})\,,  \label{3Riemann2}
\end{eqnarray}
for the spatial Riemann tensor~\cite{2007PhR...449..131B}. Then, in direct analogy with 4-dimensions, the 3-Ricci tensor and 3-Ricci scalar are respectively given by
\begin{equation}
\clr_{ab}= h^{cd}\clr_{acbd}= \clr^c{}_{acb} \hspace{15mm} {\rm and}
\hspace{15mm} \clr= h^{ab}\clr_{ab}\,, \label{3Ricci}
\end{equation}
with
\begin{eqnarray}
\clr_{ab}&=& E_{ab}+ {2\over3}\left(\rho-{1\over3}\Theta^2
+\sigma^2-\omega^2+\Lambda\right)h_{ab}+ {1\over2}\,\pi_{ab}-
{1\over3}\,\Theta(\sigma_{ab}+\omega_{ab})+ \sigma_{c\langle
a}\sigma^c{}_{b\rangle} \nonumber\\
&{}&-\omega_{c\langle a}\omega^c{}_{b\rangle}+
2\sigma_{c[a}\omega^c{}_{b]}  \label{GC}
\end{eqnarray}
and
\begin{equation}
\clr= h^{ab}\clr_{ab}= 2\left(\rho-{1\over3}\,\Theta^2+\sigma^2
-\omega^2+\Lambda\right)\,.  \label{3R}
\end{equation}

It follows that, in the absence of rotation, $\mathcal{R}_{abcd}$ and $\mathcal{R}_{ab}$ have all the symmetries of their 4-dimensional counterparts. When $\omega_a\neq0$, on the other hand, $\clr_{abcd}\neq\clr_{cdab}$ and $R_{[ab]}\neq0$.

Finally, the commutation of the spatial covariant derivatives is governed by the 3-Ricci identities, which read (e.g.~see~\cite{2008PhR...465...61T,2012reco.book.....E})
\begin{equation}
{\rm D}_{[a}{\rm D}_{b]}f= -\dot{f}\omega_{ab}\,, \hspace{15mm} {\rm D}_{[a}{\rm D}_{b]}v_c= -\omega_{ab}\dot{v}_{\langle c\rangle}+ {1\over2}\,\mathcal{R}_{dcba}v^d  \label{3Ricci1}
\end{equation}
and
\begin{equation}
{\rm D}_{[a}{\rm D}_{b]}S_{cd}= -\omega_{ab}h_c{}^eh_d{}^fS_{ef}+ {1\over2}\,(\mathcal{R}_{ecba}S^e{}_d +\mathcal{R}_{edba}S_c{}^e)\,,  \label{3Ricci2}
\end{equation}
when applied to scalars, vectors and second-rank tensors respectively. Thus, in contrast to their 4-D counterparts, the spatial covariant derivatives of scalars do not generally commute.

\subsection{The Friedmann universes}\label{ssFU}
%%%%%%%%%%%%%%%%%%%%%%%%%%%%%%%%%%%%%%%%%%%%%%%
Our discussion so far refers to cosmological models that are inhomogeneous and anisotropic. Nevertheless, observations and our theoretical prejudice (the CMB and the Copernican principle) support the homogeneous and isotropic Friedmann-Robertson-Walker (FLRW) models.

\subsubsection{The FLRW metric}\label{sssFLRWM}
%%%%%%%%%%%%%%%%%%%%%%%%%%%%%%%%%%%%%%%%%%%%%
The simplest cosmological solution of the
Einstein equations is described by the Robertson-Walker line element. Written in spherical coordinates $(r,\theta,\phi)$, the latter takes the form
\begin{equation}
{\rm d}s^2=-{\rm d}t^2+ a^2(t)\left[{\rm d}r^2+f_K^2(r)({\rm
d}\theta^2+\sin^2\theta{\rm d}\phi^2)\right]\,,  \label{FLRWm1}
\end{equation}
where $a=a(t)$ is the cosmological scale factor and $f_K(r)$ is a function that depends on the geometry of the 3-dimensional spaces. More specifically,
\begin{equation}
f_K(r)=\left\{\begin{array}{l} \sin r \hspace{10mm} {\rm
when}~K=+1\,,\\ r \hspace{16mm} {\rm when}~K=0\,,\\ \sinh r
\hspace{8mm} {\rm when}~K=-1\,.\\ \end{array}\right.  \label{FLRWm2}
\end{equation}
The 3-curvature index ($K=0,\pm1$) is related to the 3-Ricci scalar by means of $\mathcal{R}=6K/a^2$. Then, the 3-space is closed with spherical geometry for $K=+1$, flat with Euclidean 3-D hypersurfaces when $K=0$ and open with hyperbolic geometry for $K=-1$.

\subsubsection{FLRW cosmologies}\label{sssFLRWCs}
%%%%%%%%%%%%%%%%%%%%%%%%%%%%%%%%%%%%%%%%%%%%%%%
The high symmetry of the Friedmann models ensures that time-dependent scalars are the only surviving variables. Hence, in covariant terms, the FLRW cosmologies have $\Theta=3H(t)\neq0$, where $H=\dot{a}/a$ is the familiar Hubble parameter,
$\sigma_{ab}=0=\omega_a=A_a$ and $E_{ab}=0=H_{ab}$. In addition, the isotropy of the Friedmann
universes ensures that they can only accommodate perfect fluids, with $\rho=\rho(t)$ and $p=p(t)$). Finally, due to spatial homogeneity, all the 3-gradients (e.g.~${\rm D}_a\rho$, ${\rm D}_ap$, etc) vanish by default. All this means that the only nontrivial equations are the FLRW versions of Raychaudhuri's formula, together with the continuity and the Friedmann equations. These follow from (\ref{Ray}), (\ref{pfcls}a) and (\ref{3R}) and are given by
\begin{equation}
\dot{H}= -H^2- {1\over6}\,(\rho+3p)+ {1\over3}\,\Lambda\,,
\hspace{20mm} \dot{\rho}= -3H(\rho+p)  \label{FLRWeqs1}
\end{equation}
and
\begin{equation}
H^2= {1\over3}\,\rho- {K\over a^2}+ {1\over3}\,\Lambda\,,
\label{FLRWeqs2}
\end{equation}
respectively. Also note that due to the high symmetry of the Friedmann models, Eq.~(\ref{3Riemann2}) reduces to $\clr_{abcd}= (K/a^2)(h_{ac}h_{bd}-h_{ad}h_{bc})$.

Introducing the $\Omega$-parameter $\Omega_{\rho}=\rho/3H^2$ for the matter, $\Omega_{\Lambda}=\Lambda/3H^2$ for the cosmological constant and $\Omega_K=-K/(aH)^2$ for the 3-curvature, the Friedmann equation (\ref{FLRWeqs2}) reads
\begin{equation}
1=\Omega_{\rho}+ \Omega_K+ \Omega_{\Lambda}\,.  \label{FLRWeqs3}
\end{equation}
Moreover, after employing some straightforward algebra, Raychaudhuri's formula (\ref{FLRWeqs1}a) assumes the alternative form
\begin{equation}
qH^2= {1\over6}\,(\rho+3p)- {1\over3}\,\Lambda\,, \label{FLRWeqs5}
\end{equation}
where $q=-\ddot{a}a/\dot{a}^2=-[1+(\dot{H}/H^2)]$ is the (scalar) deceleration parameter. By construction, negative values of $q$ indicate an accelerating universe, while positive ones mean deceleration. Thus, in exact FLRW cosmologies with vanishing $\Lambda$, one needs to violate the strong energy condition (i.e.~set $\rho+3p<0$) to achieve universal acceleration.

The expansion rate of the universe also defines a representative length scale, commonly referred to as the Hubble radius ($\lambda_H$), with
\begin{equation}
\lambda_H=H^{-1}\,.  \label{Hr}
\end{equation}
In FLRW universes with conventional matter, $\lambda_H$ effectively coincides with the particle horizon (i.e.~$\lambda_H\propto t$), in which case the Hubble radius determines the regions of causal contact as well.

The scale factor of a Friedmann cosmology also defines the model's curvature scale ($\lambda_K=a/|K|$), where any departures from Euclidean flatness start becoming important (e.g.~see~\cite{1995PhRvD..52.3338L}). Following (\ref{FLRWeqs2}) and in the absence of a
cosmological constant, the curvature scale and the Hubble radius are related by
\begin{equation}
\left({\lambda_K\over\lambda_H}\right)^2=
-{K\over1-\Omega_{\rho}}\,, \label{K/H}
\end{equation}
with $0<\Omega_{\rho}<1$ in the open model and $\Omega_{\rho}>1$ in the closed. Consequently, $\lambda_K>\lambda_H$ always for $K=-1$, with $\lambda_K\rightarrow\infty$ as $\Omega_{\rho}\rightarrow1$ and $\lambda_K\rightarrow\lambda_H$ when $\Omega_{\rho}\rightarrow0$. On the other hand, $\lambda_K>\lambda_H$ for $1<\Omega_{\rho}<2$ and $\lambda_K\leq\lambda_H$ when $\Omega_{\rho}\geq2$ in closed ($K=+1$) models.

\subsubsection{Luminosity distance}\label{sssLD}
%%%%%%%%%%%%%%%%%%%%%%%%%%%%%%%%%%%%%%%%%%%%%%%%
The luminosity distance of a source at redshift $z$ is $d_L=a_0(1+z)r_0$, with $a_0$ and $r_0$ being the present values of the scale factor and of the source's radial distance (e.g.~see~\cite{1983QB981.N3.......}). The latter follows by integrating the line element ${\rm d}t= (a/\sqrt{1-Kr^2}){\rm d}r$ of a null geodesic. In a spatially flat FLRW model, the integration gives
\begin{equation}
r_0= a_0^{-1}\int_0^zH^{-1}{\rm d}x\,.  \label{rd}
\end{equation}
With the supernovae observations in mind, it helps to express the above in terms of the deceleration parameter. Following~\S~\ref{sssFLRWCs} and recalling that ${\rm d}z=-(1+z)H{\rm d}t$, we find~\cite{2002ApJ...569...18T}
\begin{equation}
\int_H^{H_0}H^{-1}{\rm d}H= \ln\left({H_0\over H}\right)= -\int_0^z(1+q){\rm d}[\ln(1+x)]\,,  \label{int1}
\end{equation}
which substituted back into the right-hand side of (\ref{rd}) leads to
\begin{equation}
a_0r_0= H_0^{-1}\int_0^z{\rm e}^{-\int_0^x(1+q){\rm d} [\ln(1+y)]}{\rm d}x\,. \label{int2}
\end{equation}
So, finally, expressed in terms of the kinematical parameters ($H$ and $q$) of an exact Friedmann model with flat spatial sections, the luminosity distance of a source at redshift $z$ is given by
\begin{equation}
d_L= (1+z)H_0^{-1}\int_0^z{\rm e}^{-\int_0^x(1+q){\rm d}[\ln(1+y)]}{\rm d}x\,.  \label{DL2}
\end{equation}
Confronting the above with the measured luminosity distance of remote type Ia supernovae indicated that our universe had recently entered a phase of accelerating expansion~\cite{2002ApJ...569...18T,2004ApJ...607..665R}.

It should be noted that the above analysis applies to exact FLRW models free of peculiar-velocity perturbations, or of any other type of distortions for that matter. Alternative expressions of the luminosity distance, as well as generalisations of $d_L$ that incorporate the (linear) effects of peculiar motions, can be found in~\S~\ref{sssRLDMs}.

\subsubsection{Scale-factor evolution in FLRW
%%%%%%%%%%%%%%%%%%%%%%%%%%%%%%%%%%%%%%%%%%%%
cosmologies}\label{sssSfEFLRWCs}
%%%%%%%%%%%%%%%%%%%%%%%%%%%%%%%
In the following sections, the host universe will typically be a spatially flat FLRW model with $K=0$. Nevertheless, there will also be cases where the background spacetime will be allowed to have nonzero spatial curvature (see \S~\ref{sssPVFLRWUOm2} and \S~\ref{sssGFLRWC} below). It will therefore help to briefly go through the basic evolution laws of Friedmann universes with $K=\pm1$ as well.

\underline{\textit{The $K=0$ case:}} Let us consider perfect fluids with constant barotropic index $w=p/\rho$. Then, the continuity equation (\ref{FLRWeqs1}b) gives $\rho\propto a^{-3(1+w)}$. Substituting this result into the Friedmann equation and setting the cosmological constant to zero, we obtain
\begin{equation}
a= a_0\left({t\over t_0}\right)^{2/3(1+w)}\,,  \label{sfFLRWa}
\end{equation}
provided that $w\neq-1$. Therefore, for pressure-free ``dust'' (baryonic or not) with $w=0$, we arrive at the Einstein-de Sitter solution $a\propto t^{2/3}$. Alternatively, the above leads to $a\propto t^{1/2}$ for radiation and to $a\propto t^{1/3}$ for stiff matter (with $w=1/3$ and $w=1$ respectively).

Since $w\neq-1$, solution (\ref{sfFLRWa}) does not extend to an epoch of de Sitter type (i.e.~exponential) inflation. In that case, Eq.~(\ref{FLRWeqs1}) gives $\rho=\rho_0=$~constant, which combines with (\ref{FLRWeqs2}) to give $H=H_0=$~constant and eventually exponential inflation with $a\propto{\rm e}^{H_0(t-t_0)}$. In an additional special case, where the matter has $w=-1/3$ and therefore zero (effective) gravitational mass, we obtain ``coasting'' expansion with $a\propto t$.

\underline{\textit{The $K=+1$ case:}} In FLRW spacetimes with curved spatial sections it helps to parametrise the solutions in terms of the conformal time ($\eta$ -- defined so that $\dot{\eta}=1/a$). Then, for $K=+1$, $\Lambda=0$ and $w\neq-1/3$ relations (\ref{FLRWeqs1}) and (\ref{FLRWeqs2}) combine to the parametric solution
\begin{equation}
a= a_0\left\{{\sin[(1+3w)\eta/2]\over
\sin[(1+3w)\eta_0/2]}\right\}^{2/(1+3w)}\,,  \label{scFLRWa}
\end{equation}
with $(1+3w)\eta/2\in(0,\pi)$. Then, maximum expansion corresponds to $\eta=\pi/(1+3w)$, with
$a=a_{max}=a_0\{\sin[(1+3w)\eta_0/2]\}^{-2/(1+3w)}$. Also, the above solution reduces to $a\propto\sin^2(\eta/2)$ for dust and to $a\propto\sin\eta$ for radiation~\cite{1983QB981.N3.......}. Although we can use solution (\ref{scFLRWa}) when $w=-1/3$, Eq.~(\ref{FLRWeqs1}a) leads immediately to $a\propto t$ and the familiar coasting-expansion phase.

A special closed Friedmann cosmology is the Einstein static universe, with positive cosmological constant. There, there is no time-evolution, which means that expressions (\ref{FLRWeqs1}a) and (\ref{FLRWeqs2}) are no longer propagation equations, but reduce to the constraints
\begin{equation}
\rho_0+3p_0= 2\Lambda \hspace{15mm} {\rm and} \hspace{15mm} {1\over a_0^2}= {1\over3}\,(\rho_0+\Lambda)\,,  \label{Eeqs1}
\end{equation}
respectively.

\underline{\textit{The $K=-1$ case:}} In Friedmann universes with hyperbolic 3-geometry, zero cosmological constant and $w\neq-1/3$, the scale factor evolves as
\begin{equation}
a= a_0\left\{{\sinh[(1+3w)\eta/2]\over
\sinh[(1+3w)\eta_0/2]}\right\}^{2/(1+3w)}\,,  \label{soFLRWa}
\end{equation}
where now $(1+3w)\eta/2>0$. Assuming pressureless dust, we find $a\propto\sinh^2(\eta/2)$ and $a\propto\sinh\eta$ when the universe is dominated by relativistic species~\cite{1983QB981.N3.......}.

The simplest open FLRW cosmology is the vacuum Milne universe~\cite{1935MNRAS..95..663M}, which is often treated as the late-time limit of the open Friedmann models. There the Friedmann equation -- see expression (\ref{FLRWeqs2}) in \S~\ref{sssFLRWCs} -- guarantees coasting expansion with $a=t$.

\section{Cosmologies with a peculiar-velocity ``tilt''}\label{sCPVT}
%%%%%%%%%%%%%%%%%%%%%%%%%%%%%%%%%%%%%%%%%%%%%%%%%%%%%%%%%%%%%%%%%%%%
By default, the study of peculiar motions requires tilted cosmologies, equipped with two (at least) families of observers/frames moving relative to each other. Then, by construction, at every spacetime event there are two coordinate systems. This in turn guarantees that boosting from one frame to the other does not violate causality.

\subsection{Relatively moving observers}\label{ssRMOs}
%%%%%%%%%%%%%%%%%%%%%%%%%%%%%%%%%%%%%%%%%%%%%%%%%%%%%%
Peculiar velocities are typically defined and measured with respect to the reference frame of the universe, which has been largely identified with that of the cosmic microwave photons. By definition, the latter is the only coordinate system where the CMB dipole vanishes. While only idealised observers follow the cosmic microwave frame, all those living in typical galaxies (like our Milky Way) are expected to have some finite peculiar velocity with respect to the CMB. Typically, these realistic observers measure their own peculiar velocity by the magnitude and the direction of the observed CMB dipole. This way, for example, the peculiar velocity of our Local Group of galaxies has been estimated at approximately 600~km/sec (e.g.~see~\cite{1993ApJ...419....1K,1996ApJ...470...38L}). The reader is also referred to~\cite{1998ASSL..231..185H} for further discussion and references.

\begin{figure*}
\begin{center}
\includegraphics[height=3in,width=5in,angle=0]{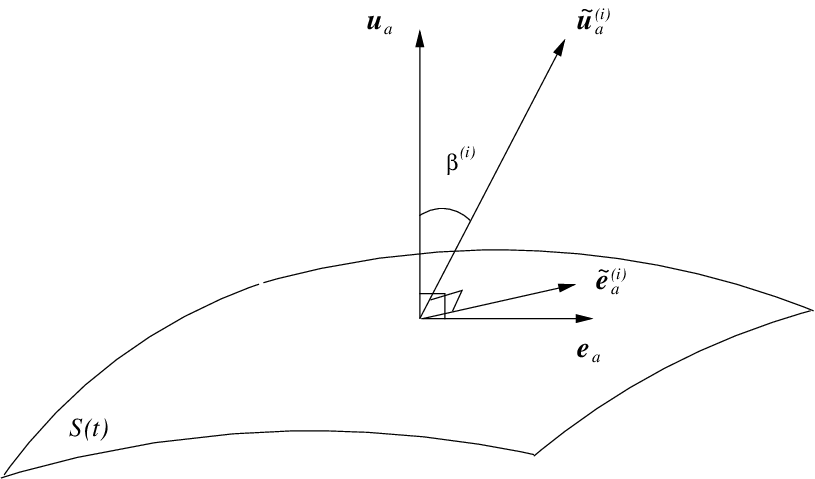}\quad
\end{center}
\caption{In a multi-component system, the 4-velocity $\tilde{u}_a^{(i)}$ of the $i$-th fluid makes a hyperbolic angle $\beta^{(i)}$ with the reference 4-velocity field $u_a$, normal to the hypersurfaces of homogeneity $S(t)$. The unit vectors $e_a$ and $\tilde{e}_a^{(i)}$ are orthogonal to $u_a$ and $\tilde{u}_a^{(i)}$ respectively. Following definition (\ref{Lboost}), the peculiar velocity of the $i$-th species is $\tilde{v}_a^{(i)}= \tilde{v}^{(i)}e_a$, with $\tilde{v}_{(i)}^2= \tilde{v}_a^{(i)}\tilde{v}_{(i)}^a$ being its squared magnitude~\cite{1973CMaPh..31..209K}.}  \label{mfluid}
\end{figure*}

Let us consider a tilted spacetime filled with a multi-component fluid and allow for several 4-velocity fields in relative motion at every spacetime event. Suppose also that $u_a$ is the reference 4-velocity of the CMB frame and $\tilde{u}_a^{(i)}$ that of the $i$-th fluid component (see Fg.~\ref{mfluid}), with $u_au^a=-1=\tilde{u}_a^{(i)}\tilde{u}_{(i)}^a$. Then, the symmetric 3-tensors
\begin{equation}
h_{ab}=g_{ab}+ u_au_b \hspace{15mm} \mathrm{and} \hspace{15mm}
\tilde{h}_{ab}^{(i)}=g_{ab}+ \tilde{u}_a^{(i)}\tilde{u}_b^{(i)}\,,  \label{h-hi}
\end{equation}
project orthogonal to $u_a$ and $\tilde{u}_a^{(i)}$ respectively. By construction, the above projectors satisfy the constraints $h_{ab}u^b=0= \tilde{h}^{(i)}_{ab}\tilde{u}_{(i)}^a$, $h_a{}^a=3= \tilde{h}_a^{(i)a}$, $h_{ac}h^c{}_b=h_{ab}$ and $\tilde{h}^{(i)}_{ac}\tilde{h}^{(i)c}{}_b=\tilde{h}^{(i)}_{ab}$.

The presence of more than one 4-velocity fields implies the existence of an equal number of temporal directions, together with their corresponding 3-dimensional spatial hypersurfaces. We therefore need to define the associated (covariant) derivative operators. These are
\begin{equation}
{}^{\cdot}= u^a\nabla_a \hspace{15mm} {\rm and} \hspace{15mm} {}_{(i)}^{\prime}= \tilde{u}_{(i)}^a\nabla_a\,,  \label{tder}
\end{equation}
denoting time-differentiation along the $u_a$ and the $\tilde{u}^{(i)}_a$ fields respectively. These differential operators are supplemented by their spatial counterparts
\begin{equation}
{\rm D}_a= h_a{}^b\nabla_b \hspace{15mm} {\rm and} \hspace{15mm} \tilde{\rm D}_a^{(i)}= \tilde{h}_a^{(i)b}\nabla_b\,,  \label{sder}
\end{equation}
employed in the 3-spaces orthogonal to $u_a$ and $\tilde{u}^{(i)}_a$ respectively.

The relation between the 4-velocities of the matter fields  ($\tilde{u}_a^{(i)}$) and that of the CMB frame ($u_a$) is given by the familiar Lorentz boost
\begin{equation}
\tilde{u}_a^{(i)}=\gamma^{(i)}\left(u_a+\tilde{v}_a^{(i)}\right)\,, \label{Lboost}
\end{equation}
where $u_a\tilde{v}^a_{(i)}=0$ and $\tilde{v}_a^{(i)}$ is the peculiar velocity of the $i$-th component relative to $u_a$ (see Fig.~\ref{mfluid}). Also, $\gamma_{(i)}= (1-\tilde{v}_{(i)}^2)^{-1/2}$ is the associated Lorentz-boost factor. Note that, for non-relativistic peculiar motions $\tilde{v}_{(i)}^2\ll1$ and therefore $\gamma_{(i)}\simeq1$.

One can recast the velocity boost (\ref{Lboost}) in terms of the hyperbolic ``tilt'' angle ($\beta_{(i)}$) between the two 4-velocity vectors (see Fig.~\ref{mfluid}). Indeed, given that $\cosh\beta^{(i)}=-\tilde{u}_a^{(i)}u^a=\gamma^{(i)}>1$ and that
$\sinh\beta^{(i)}=\gamma^{(i)}v^{(i)}$ (see Fig.~\ref{mfluid} above and also~\cite{1973CMaPh..31..209K}), we have
\begin{equation}
\tilde{u}_a^{(i)}= \cosh\beta^{(i)}u_a+ \sinh\beta^{(i)}e_a\,,  \label{ui}
\end{equation}
with $\tilde{v}_{(i)}=\tanh\beta_{(i)}$. All this means that, when the tilt angle is small (i.e.~for $\beta_{(i)}\ll1$), we have
$\tilde{v}_{(i)}\simeq\beta_{(i)}$ and non-relativistic peculiar velocities.

\subsection{Relations between the frames}\label{ssRBFs}
%%%%%%%%%%%%%%%%%%%%%%%%%%%%%%%%%%%%%%%%%%%%%%%%%%%%%%%
In a realistic universe, observers living in typical galaxies like our Milky Way move with 4-velocity $\tilde{u}_a$ relative to the $u_a$-field, which defines the cosmic rest-frame of the CMB photons. Then, following (\ref{Lboost}), we have $\tilde{u}_a=\gamma(u_a+\tilde{v}_a)$, with $u_a\tilde{v}^a=0$ and $\gamma=(1-\tilde{v}^2)^{-1/2}$. Also, the projection tensors and the spatial Levi-Civita tensors associated with the two coordinate systems are related by
\begin{equation}
\tilde{h}_{ab}= h_{ab}+ \gamma^2\left(v^2u_au_b
+2u_{(a}\tilde{v}_{b)}+\tilde{v}_a\tilde{v}_b\right)  \label{rels1a}
\end{equation}
and
\begin{equation}
\tilde{\varepsilon}_{abc}= \gamma\varepsilon_{abc}+
\gamma\left(2u_{[a}\varepsilon_{b]cd}
+u_c\varepsilon_{abd}\right)\tilde{v}^d\,,  \label{rels1b}
\end{equation}
respectively.

The kinematic variables, as measured in the reference ($u_a$) frame, are given by the irreducible decomposition of the gradient of the associated 4-velocity field  (see Eq.~(\ref{Nbua}) in \S~\ref{sssKs}). Written in the tilted coordinate system ($\tilde{u}_a$), the aforementioned decomposition reads
\begin{equation}
\nabla_b\tilde{u}_a=
{1\over3}\,\tilde{\Theta}\tilde{h}_{ab}+\tilde{\sigma}_{ab}
+\tilde{\varepsilon}_{abc}\tilde{\omega}^c-\tilde{A}_a\tilde{u}_b\,.
\label{tNbua}
\end{equation}
Combining (\ref{Nbua}) with the above and using the auxiliary relation $\nabla_a\gamma=\gamma^3v^b\nabla_av_b$, we arrive at the following (nonlinear) relations between the kinematic variables measured in the two frames. More specifically, we have~\cite{1998PhRvD..58l4006M}
\begin{equation}
\tilde{\Theta}= \gamma\Theta+ \gamma\left(\D_a\tilde{v}^a +A^a\tilde{v}_a\right)+
\gamma^3W\,,  \label{ab3}
\end{equation}
\begin{eqnarray}
\tilde{A}_a&=& \gamma^2A_a+
\gamma^2\left[\dot{\tilde{v}}_{\langle a\rangle} +{1\over3}\,\Theta\tilde{v}_a
+\sigma_{ab}\tilde{v}^b-\varepsilon_{abc}\omega^b\tilde{v}^c+
\left({1\over3}\,\Theta\tilde{v}^2+A^b\tilde{v}_b
+\sigma_{bc}\tilde{v}^b\tilde{v}^c\right)u_a\right. \nonumber\\ &&\left.
+{1\over3}\,(\D_b\tilde{v}^b)\tilde{v}_a +{1\over2}\,\varepsilon_{abc}\tilde{v}^b\curl\tilde{v}^c
+\tilde{v}^b\D_{\langle b}\tilde{v}_{a\rangle}\right] +\gamma^4W(u_a+\tilde{v}_a)\,,
\label{ab4}
\end{eqnarray}
\begin{eqnarray}
\tilde{\omega}_a&=& \gamma^2\left[\left(1-
{1\over2}\,\tilde{v}^2\right)\omega_a-{1\over2}\,\curl\tilde{v}_a
+{1\over2}\,\tilde{v}_b\left(2\omega^b-\curl\tilde{v}^b \right)u_a+
{1\over2}\,\tilde{v}_b\omega^b\tilde{v}_a \right. \nonumber\\ &&\left.
+{1\over2}\,\varepsilon_{abc}A^b\tilde{v}^c
+{1\over2}\,\varepsilon_{abc}\dot{\tilde{v}}^b\tilde{v}^c
+{1\over2}\, \varepsilon_{abc}\sigma^b{}_d\tilde{v}^c\tilde{v}^d\right]\,,
\label{ab5}
\end{eqnarray}
and
\begin{eqnarray}
\tilde{\sigma}_{ab}&=& \gamma\sigma_{ab}+
\gamma\left(1+\gamma^2\right)u_{(a}\sigma_{b)c}\tilde{v}^c+
\gamma^2A_{(a}\left[\tilde{v}_{b)}+\tilde{v}^2u_{b)}\right] \nonumber\\
&&+\gamma\D_{\langle a}\tilde{v}_{b\rangle}- {1\over3}\,h_{ab}
\left[A_c\tilde{v}^c +\gamma^2\left(W-\dot{\tilde{v}}_c\tilde{v}^c\right)\right] \nonumber\\
&&+\gamma^3u_au_b\left[\sigma_{cd}\tilde{v}^c\tilde{v}^d +{2\over3}\,\tilde{v}^2A_c\tilde{v}^c
-\tilde{v}^c\tilde{v}^d\D_{\langle c}\tilde{v}_{d\rangle}
+\left(\gamma^4-{1\over3}\,\tilde{v}^2\gamma^2-1\right)W\right] \nonumber\\
&&+\gamma^3u_{(a}\tilde{v}_{b)}\left[A_c\tilde{v}^c +\sigma_{cd}\tilde{v}^c\tilde{v}^d-\dot{\tilde{v}}_c\tilde{v}^c+
2\gamma^2\left(\gamma^2-{1\over3}\,\right)W\right] \nonumber\\
&&+{1\over3}\,\gamma^3\tilde{v}_a\tilde{v}_b\left[\D_c\tilde{v}^c -A_c\tilde{v}^c+\gamma^2\left(3\gamma^2-1\right)W\right] +\gamma^3\tilde{v}_{\langle a}\dot{\tilde{v}}_{b\rangle } +\tilde{v}^2\gamma^3u_{(a}\dot{\tilde{v}}_{\langle b\rangle)} \nonumber\\
&&+\gamma^3\tilde{v}_{(a}\sigma_{b)c}\tilde{v}^c -\gamma^3\omega^b\tilde{v}^c\varepsilon_{bc(a}
\left(\tilde{v}_{b)}+\tilde{v}^2u_{b)}\right)+ 2\gamma^3\tilde{v}^c\D_{\langle c}\tilde{v}_{(a\rangle} \left(\tilde{v}_{b)}+u_{b)}\right)\,,  \label{ab6}
\end{eqnarray}
where
\begin{equation}
W\equiv \dot{\tilde{v}}_c\tilde{v}^c+ {1\over3}\,\tilde{v}^2\D_c\tilde{v}^c+ \tilde{v}^c\tilde{v}^d\D_{\langle c}\tilde{v}_{d\rangle}\,. \label{W}
\end{equation}

Similarly, one can recast the decomposition of the energy-momentum tensor (see expression (\ref{Tab1}) in \S~\ref{sssMFs}) with respect to the $\tilde{u}_a$-frame. The result reads
\begin{equation}
T_{ab}=\tilde{\rho}\tilde{u}_a\tilde{u}_b+
\tilde{p}\,\tilde{h}_{ab}+ 2\tilde{q}_{(a}\tilde{u}_{b)}+
\tilde{\pi}_{ab}\,,  \label{tilTab}
\end{equation}
with $\tilde{\rho}$, $\tilde{p}$, $\tilde{q}_a$ and $\tilde{\pi}_{ab}$ being the density, the pressure, the energy flux and the viscosity of the matter in the tilted coordinate system. The two sets of variables are related by~\cite{1998PhRvD..58l4006M}
\begin{eqnarray}
\tilde{\rho}= \rho+ \gamma^2\left[(\rho+p)\tilde{v}^2 -2q_a\tilde{v}^a +\pi_{ab}\tilde{v}^a\tilde{v}^b\right]\,, \label{ab7}
\end{eqnarray}
\begin{eqnarray}
 \tilde{p}= p+
{1\over3}\,\gamma^2\left[(\rho+p)\tilde{v}^2- 2q_a\tilde{v}^a
+\pi_{ab}\tilde{v}^a\tilde{v}^b\right]\,, \label{ab8}
\end{eqnarray}
\begin{eqnarray}
\tilde{q}_a&=& \gamma q_a-
\gamma\pi_{ab}\tilde{v}^b- \gamma^3\left[(\rho+p) -2q_b\tilde{v}^b
+\pi_{bc}\tilde{v}^b\tilde{v}^c\right]\tilde{v}_a \nonumber\\ &&-\gamma^3\left[(\rho+p)\tilde{v}^2-
(1+\tilde{v}^2)q_b\tilde{v}^b +\pi_{bc}\tilde{v}^b\tilde{v}^c\right]u_a\,, \label{ab9}
\end{eqnarray}
and
\begin{eqnarray}
\tilde{\pi}_{ab}&=& \pi_{ab}+ 2\gamma^2\tilde{v}^c\pi_{c(a}\left[u_{b)}
+\tilde{v}_{b)}\right] -2\tilde{v}^2\gamma^2q_{(a}u_{b)} -2\gamma^2q_{\langle a}\tilde{v}_{b\rangle} \nonumber\\ &&-{1\over3}\,\gamma^2\left[(\rho+p)\tilde{v}^2
+\pi_{cd}\tilde{v}^c\tilde{v}^d\right]h_{ab} \nonumber\\
&&+{1\over3}\,\gamma^4\left[2(\rho+p)\tilde{v}^4 -4\tilde{v}^2q_c\tilde{v}^c
+(3-\tilde{v}^2)\pi_{cd}\tilde{v}^c\tilde{v}^d\right]u_au_b \nonumber\\
&&+{2\over3}\,\gamma^4\left[2(\rho+p)\tilde{v}^2 -(1+3v^2)q_c\tilde{v}^c +2\pi_{cd}\tilde{v}^c\tilde{v}^d\right]u_{(a}\tilde{v}_{b)} \nonumber\\
&&+{1\over3}\,\gamma^4\left[(3-\tilde{v}^2)(\rho+p) -4q_c\tilde{v}^c +2\pi_{cd}\tilde{v}^c\tilde{v}^d\right]\tilde{v}_a\tilde{v}_b\,. \label{ab10}
\end{eqnarray}
The transformation sets (\ref{ab3})-(\ref{ab6}) and (\ref{ab7})-(\ref{ab10}) are fully nonlinear and therefore apply to any tilted spacetime equipped with two families of observers/frames in (relativistic) relative motion. Next we will linearise the above around the two most familiar and best physically motivated cosmological spacetimes, namely the Friedmann and the Bianchi models. In both cases, we will also assume that the peculiar velocities are non-relativistic.

\subsection{Linear relations between the frames}\label{ssLRBFs}
%%%%%%%%%%%%%%%%%%%%%%%%%%%%%%%%%%%%%%%%%%%%%%%%%%%%%%%%%%%%%%%
Consider an FLRW background universe, where the only non-vanishing variables are scalars that depend solely on time. These are the density ($\rho$), the isotropic pressure ($p$), the volume expansion ($\Theta$, with $\Theta/3=H$) and the spatial Ricci scalar ($\mathcal{R}$, with $\mathcal{R}=6K/a^2$ and $K=0,\pm1$ -- see~\S~\ref{sssFLRWCs} earlier). Then, when dealing with non-relativistic peculiar velocities (with $\tilde{v}^2\ll1$ and $\gamma\simeq1$), the nonlinear relations (\ref{ab3})-(\ref{ab6}) linearise to
\begin{equation}
\tilde{\Theta}= \Theta+ \tilde{\vartheta}\,, \hspace{25mm} \tilde{A}_a= A_a+ \dot{\tilde{v}}_a+ H\tilde{v}_a\,, \label{lab34}
\end{equation}
and
\begin{equation}
\tilde{\omega}_a= \omega_a+ \tilde{\varpi}_a\,, \hspace{25mm} \tilde{\sigma}_{ab}= \sigma_{ab}+ \tilde{\varsigma}_{ab}\,,
\label{lab56}
\end{equation}
respectively. Recall that $\tilde{v}_a$ is the peculiar velocity of the tilted frame relative to that of the CMB (see Fig.~\ref{mfluid} in \S~\ref{ssRMOs} before). Then, $\tilde{\vartheta}={\rm D}^a\tilde{v}_a$ is the local volume scalar, with $\tilde{\vartheta}\gtrless0$ since the peculiar flow can expand or contract locally, $\tilde{\varpi}_a= -\curl\tilde{v}_a/2$ is the local votricity vector and $\tilde{\varsigma}_{ab}={\rm D}_{\langle a}\tilde{v}_{b\rangle}$ is the local shear tensor of the peculiar motion. The above follow from the spatial gradient of the peculiar-velocity field, which decomposes as~\cite{2013PhRvD..88h3501T}
\begin{equation}
{\rm D}_b\tilde{v}_a= {1\over3}\,\tilde{\vartheta}h_{ab}+ \varepsilon_{abc}\tilde{\varpi}^c+ \tilde{\varsigma}_{ab}\,,  \label{Dbtva}
\end{equation}
with respect to the reference (CMB) frame. Proceeding as with the kinematic variables before, the nonlinear relations between the matter variables measured in the two frames (see expressions (\ref{ab7})-(\ref{ab10}) above) linearise to
\begin{equation}
\tilde{\rho}= \rho\,, \hspace{10mm} \tilde{p}= p\,, \hspace{10mm} \tilde{q}_a= q_a- (\rho+p)\tilde{v}_a \hspace{10mm} {\rm and} \hspace{10mm} \tilde{\pi}_{ab}= \pi_{ab}\,,  \label{lab710}
\end{equation}
respectively. Note that, in deriving the above, we have also used the linear relation $\tilde{h}_{ab}=h_{ab}+2u_{(a}\tilde{v}_{b)}$ between the two projectors.

According to (\ref{lab34}) and (\ref{lab56}), the kinematics of the relatively moving observers/frames differ due to peculiar-motion effects even at the linear level. Of the above relations, the most important for our purposes are the two seen in Eq.~(\ref{lab34}). The former ensures that the expansion rates measured in the two frames differ, depending on the sign and the magnitude of the local volume scalar ($\tilde{\vartheta}$). Recall that the latter can take both positive and negative values, when the associated peculiar flow is respectively expanding or contracting locally. Also, in line with  (\ref{lab34}b), there is always a nonzero 4-acceleration vector in the system when peculiar velocities are present. Indeed, setting $A_a=0$ for example, implies that $\tilde{A}_a=\dot{\tilde{v}}_a+ H\tilde{v}_a\neq0$ in the tilted frame. Similarly, $\tilde{A}_a=0$ leads to $A_a=-\dot{\tilde{v}}_a-H\tilde{v}_a\neq0$ in the reference frame. In fact, expression (\ref{lab34}b) ensures that the 4-acceleration, namely non-gravitational forces, are the sources and the drivers of peculiar-velocity perturbations. Note that setting both 4-acceleration vectors to zero in (\ref{lab34}b) leads to a peculiar-velocity field that decays with the expansion (i.e.~$\tilde{v}_a\propto a^{-1}$). Such a result, is at direct odds with the plethora of observations confirming the ubiquitous presence of peculiar motions in the universe.

Following (\ref{lab710}), to first approximation, the density, the isotropic pressure and the viscosity of the matter remain unchanged. There is a difference, however, in the energy flux measured in the two frames due to relative-motion effects alone (see relation (\ref{lab710}c)). The latter implies that, in the presence of peculiar velocities, the cosmic medium can no longer be treated as a perfect fluid, with the imperfection taking the form of an effective energy-flux vector solely triggered by the peculiar flow (e.g.~see (\ref{lab710}c) here, as well as \S~5.2.1 in~\cite{2012reco.book.....E}). Indeed, even when one of the two flux vectors is zero, the other will take nonzero values just because of the observers' relative motion~\cite{2020EPJC...80..757T,2021Ap&SS.366....4F,%
2022PhRvD.106h3505M,2024PhRvD.110f3540M}. Note that we cannot set both fluxes to zero, because then we are left with no peculiar-velocity field to study. The only exception is when the background universe has a (de Sitter) inflationary equation of state with $p=-\rho$ (see~\cite{2022PhRvD.106h3505M} and~\S~\ref{ssPVI} here for further discussion on this special case).

In general relativity energy-fluxes ``gravitate'', since they also contribute to the stress-energy tensor of the matter. As we shall see in \S~\ref{sLPVs} next, it is the contribution of the \textit{peculiar flux} to the relativistic gravitational field which makes the difference between the Newtonian/quasi-Newtonian and the relativistic treatments of peculiar motions.

Assuming a Bianchi-type background universe (see \S~\ref{sssBC} below), namely maintaining the homogeneity of the 3-space but allowing for spatial  anisotropy, has no effect on the linear relations (\ref{lab710}) between the matter variables. There are changes, however, in two of the four kinematic relations. In particular expressions (\ref{lab34}b) and (\ref{lab56}b) respectively recast as
\begin{equation}
\tilde{A}_a= A_a+ \dot{\tilde{v}}_a+ H\tilde{v}_a+ \sigma_{ab}\tilde{v}^b \hspace{10mm} {\rm and} \hspace{10mm} \tilde{\sigma}_{ab}= \sigma_{ab}+ \tilde{\varsigma}_{ab}+ 2u_{(a}\sigma_{b)c}\tilde{v}^c\,.  \label{Blab1}
\end{equation}
Therefore, at the linear level, the background anisotropy has added two shear-related terms on the right-hand sides of the above. Nevertheless, following (\ref{Blab1}a), the 4-acceleration remains the only source of linear peculiar-velocity perturbations.

\section{Linear peculiar velocities}\label{sLPVs}
%%%%%%%%%%%%%%%%%%%%%%%%%%%%%%%%%%%%%%%%%%%%%%%%%
Bulk peculiar flows are commonplace in the universe, confirmed by several surveys over many years (e.g.~see~\cite{2014A&A...571A..27P} for the \textit{Planck} 2013/2015 constraints). Such large-scale motions are believed to have started as weak peculiar-velocity perturbations around the time of recombination, which then grew to the observed sizes and speeds driven by structure formation. There are open issues, however. The reported velocities generally lie within $\Lambda$CDM expectations on relatively small scales, roughly $\lesssim100/h$~Mpc~\cite{2011ApJ...736...93N,2011JCAP...04..015D,%
2012MNRAS.420..447T,2012MNRAS.424..472B,2013MNRAS.428.2017M,%
2013A&A...560A..90F,2014MNRAS.437.1996M,2014MNRAS.445..402H,%
2015MNRAS.450..317C,2016ApJ...827...60M,2016MNRAS.455..386S,%
2016MNRAS.456.1886S,2018MNRAS.477.5150Q,2019MNRAS.482.1920Q,%
2020MNRAS.498.2703B,2021MNRAS.505.2349S,2021ApJ...922...59Q}, but appear in excess of the typical $\Lambda$CDM predictions on larger scales~\cite{2004MNRAS.352...61H,2008MNRAS.387..825F,%
2009MNRAS.392..743W,2010MNRAS.407.2328F,2010ApJ...709..483L,%
2011MNRAS.414..264C,2015MNRAS.447..132W,2018MNRAS.481.1368P,%
2021MNRAS.504.1304S,2023MNRAS.524.1885W,%
2023MNRAS.526.3051W}.\footnote{The Planck constraints disfavour the so called ``dark flows''. The latter are extreme bulk flows, which sizes and speeds well in excess of the current $\Lambda$CDM limits (see \S~\ref{ssDFQ} below).} In this section we will review the theoretical studies of peculiar velocities, referring the reader to \S~\ref{sPMLCDM} for the observational surveys and reports.

\subsection{Newtonian analysis}\label{ssNA}
%%%%%%%%%%%%%%%%%%%%%%%%%%%%%%%%%%%%%%%%%%%
We will begin with an outline of the Newtonian covariant treatment of peculiar motions. This brief introduction will help us present the results of the available Newtonian and quasi-Newtonian studies, as well as compare them with those of the relativistic analysis.

\subsubsection{Newtonian peculiar kinematics}\label{ssNPKs}
%%%%%%%%%%%%%%%%%%%%%%%%%%%%%%%%%%%%%%%%%%%%%%%%%%%%%%%%%%%
In what follows, we will look at the Newtonian covariant approach to the linear evolution of peculiar velocities in cosmology during the Einstein-de Sitter epoch of the universe. This will also facilitate the direct comparison of the purely Newtonian results with those of the subsequent quasi-Newtonian and relativistic studies (see \S~\ref{ssQ-NA} and \S~\ref{ssRA} respectively). The linear nature of all these treatments confines them to sufficiently large scales, which typically lie close and beyond the 100~Mpc threshold. For a discussion on the available nonlinear studies, the reader is referred to \S~\ref{ssNRZA} - \S~\ref{ssRPSD}.

Consider a pair of relatively moving observers with velocities $u_{\alpha}$ and $\tilde{u}_{\alpha}$. In Newtonian theory, these two velocity fields are related by the Galilean transformation
\begin{equation}
\tilde{u}_{\alpha}= u_{\alpha}+ \tilde{v}_{\alpha}\,,  \label{Galilean}
\end{equation}
where $\tilde{v}_{\alpha}$ is the peculiar velocity of the $\tilde{u}_{\alpha}$-field relative to the (reference) $u_{\alpha}$-frame.\footnote{Typical Newtonian studies of peculiar motions use physical ($r^{\alpha}$) and comoving ($x^{\alpha}$) coordinates, with $r^{\alpha}=ax^{\alpha}$. The time derivative of the latter leads to $v=v_H+v_p$, where $v=\dot{r}^{\alpha}$, $v_H=Hr^{\alpha}$ and $v_p=a\dot{x}^{\alpha}$ are the total, the Hubble and the peculiar velocities respectively. On an FLRW background, the above velocity relation is equivalent to the one seen in Eq.~(\ref{Galilean}).} The irreducible kinematics of the $u_{\alpha}$-field are given by the decomposition
\begin{equation}
\partial_{\beta}u_{\alpha}= {1\over3}\,\Theta h_{\alpha\beta}+ \sigma_{\alpha\beta}+ \omega_{\alpha\beta}\,,  \label{Npbua}
\end{equation}
with an exactly analogous split for its tilded counterpart $\tilde{u}_{\alpha}$ (i.e.~$\partial_{\beta}\tilde{u}_{\alpha}= (\tilde{\Theta}/3)h_{\alpha\beta}+\tilde{\sigma}_{\alpha\beta}+ \tilde{\omega}_{\alpha\beta}$). Similarly, the spatial gradient of the peculiar-velocity field decomposes as
\begin{equation}
\partial_{\beta}\tilde{v}_{\alpha}= {1\over3}\,\tilde{\vartheta}h_{\alpha\beta}+ \tilde{\varsigma}_{\alpha\beta}+ \tilde{\varpi}_{\alpha\beta}\,,  \label{pbtva}
\end{equation}
with $\tilde{\vartheta}=\partial^{\alpha}\tilde{v}_{\alpha}$, $\tilde{\varsigma}_{\alpha\beta}= \partial_{\langle\beta}\tilde{v}_{\alpha\rangle}$ and $\tilde{\varpi}_{\alpha\beta}= \partial_{[\beta}\tilde{v}_{\alpha]}$ respectively representing the volume scalar, the shear tensor and the vorticity of the peculiar flow.\footnote{Comparing (\ref{Npbua}) to its relativistic counterpart (see decomposition (\ref{Nbua}) in~\S~\ref{sssKs}), one notices the absence of an acceleration term from the right-hand sides of the Newtonian expression. This reflects the fact that, in Newtonian physics, space and time are separate and independent entities.} As before (see Eq.~(\ref{Dbtva}) in \S~\ref{ssLRBFs}), positive/negative values for $\tilde{\vartheta}$ indicate (locally) expanding/contracting peculiar flows. Also, $\tilde{\varsigma}_{\alpha\beta}$ and $\tilde{\varpi}_{\alpha\beta}$ are respectively the shear and vorticity of the peculiar-velocity field. Starting from (\ref{Galilean}), one arrives at the following (nonlinear) relations
\begin{equation}
\tilde{\Theta}= \Theta+ \tilde{\vartheta}\,, \hspace{10mm} \tilde{\sigma}_{\alpha\beta}= \sigma_{\alpha\beta}+\tilde{\varsigma}_{\alpha\beta} \hspace{10mm} {\rm and} \hspace{10mm} \tilde{\omega}_{\alpha\beta}= \omega_{\alpha\beta}+\tilde{\varpi}_{\alpha\beta}\,,  \label{Nrels}
\end{equation}
between the three kinematic sets. In addition, taking the convective derivative of (\ref{Galilean}), with respect to the tilted frame, employing Eq.~(\ref{Galilean}) again and then assuming that our unperturbed background cosmology is described by the Newtonian version of the spatially flat Friedmann universe, we arrive at the linear expression
\begin{equation}
\tilde{u}_{\alpha}^{\prime}= \dot{u}_{\alpha}+ \tilde{v}_{\alpha}^{\prime}+ H\tilde{v}_{\alpha}= \dot{u}_{\alpha}+ \dot{\tilde{v}}_{\alpha}+ H\tilde{v}_{\alpha}\,,  \label{tEuler1}
\end{equation}
relating the acceleration vectors in the two coordinate systems. Note that in deriving the above we have used the fact that $\tilde{v}_{\alpha}^{\prime}= \dot{\tilde{v}}_{\alpha}$ to first order and the zero-order relation $\Theta=3H$. Also note that primes denote convective derivatives in the tilted frame (i.e.~$\tilde{u}_{\alpha}^{\prime}= \partial_t\tilde{u}_{\alpha}+ \tilde{u}^{\beta}\partial_{\beta}\tilde{u}_{\alpha}$), while overdots indicate convective differentiation in the reference $u_{\alpha}$-frame (i.e.~$\dot{u}_{\alpha}=\partial_tu_{\alpha}+ u^{\beta}\partial_{\beta}u_{\alpha}$). Following (\ref{tEuler1}), the presence of relative motion means that (generally) we cannot set $\dot{u}_{\alpha}$ and $\tilde{u}^{\prime}_{\alpha}$ to zero simultaneously. The same is also true for the rest of the kinematic variables (see Eqs.~(\ref{Nrels}a)-(\ref{Nrels}c) above). Finally, we should point out that, in contrast to the relativistic treatment (see relations (\ref{ab7})-(\ref{ab10}) in~\S~\ref{ssRBFs} earlier), the matter variables remain unchanged when transforming from one coordinate system to the other.

\subsubsection{Linear sources of peculiar %%%%%%%%%%%%%%%%%%%%%%%%%%%%%%%%%%%%%%%%%
velocities}\label{sssLSPVs}
%%%%%%%%%%%%%%%%%%%%%%%%%%%
In a perturbed almost-FLRW Newtonian universe (filled with a pressureless perfect fluid -- baryonic or/and CDM -- with $p=0= \pi_{\alpha\beta}$), the peculiar-velocity field is a linear perturbation. Then, by setting $\tilde{u}_{\alpha}^{\prime}=0$ in the matter frame, expression (\ref{tEuler1}) recasts to
\begin{equation}
\dot{\tilde{v}}_{\alpha}+ H\tilde{v}_{\alpha}= -\dot{u}_{\alpha}\,.  \label{ltEuler1}
\end{equation}
Hence, the driving force of linear peculiar velocities is the acceleration, which for zero pressure is given by $\dot{u}_{\alpha}= \partial_{\alpha}\Phi$ (with being the gravitational potential). As a result, Eq.~(\ref{ltEuler1}) reads
\begin{equation}
\dot{\tilde{v}}_{\alpha}+ H\tilde{v}_{\alpha}= -\partial_{\alpha}\Phi\,.  \label{ltv'}
\end{equation}
Thus, in Newtonian theory, linear peculiar velocities are generated and driven by gradients in the gravitational potential. Note that relation (\ref{ltv'}) is identical to the one obtained in typical (non-covariant) Newtonian studies, provided the equations of the latter are written in physical (rather than comoving) coordinates -e.g.~see~\cite{1976ApJ...205..318P,1980lssu.book.....P,%
1991ApJ...379....6N}).

\subsubsection{Linear evolution of peculiar
%%%%%%%%%%%%%%%%%%%%%%%%%%%%%%%%%%%%%%%%%%%
velocities}\label{sssLEPVs}
%%%%%%%%%%%%%%%%%%%%%%%%%%%
Taking the convective derivative of (\ref{ltv'}), employing the background relations $\dot{H}=-3H^2/2$, $\partial_{\beta}u_{\alpha}= Hh_{\alpha\beta}$ and keeping up to linear-order terms, leads to~\cite{2021Ap&SS.366....4F}
\begin{equation}
\ddot{\tilde{v}}_{\alpha}+ 2H\dot{\tilde{v}}_{\alpha}- {1\over2}\,H^2\tilde{v}_{\alpha}= -\partial_{\alpha}\dot{\Phi}\,,  \label{ltv''1}
\end{equation}
since $(\partial_{\alpha}\Phi)^{\cdot}=\partial_{\alpha}\dot{\Phi}- H\partial_{\alpha}\Phi$ to first approximation. Ignoring the term on the right-hand side of the above, namely assuming that $\dot{\Phi}$ has an nearly homogeneous spatial distribution, or/and that $\Phi$ varies slowly in time, the left-hand side of (\ref{ltv''1}) accepts the power-law solution
\begin{equation}
\tilde{v}= \mathcal{C}_1t^{1/3}+ \mathcal{C}_2t^{-2/3}= \mathcal{C}_3a^{1/2}+ \mathcal{C}_4a^{-1}\,,  \label{ltv1}
\end{equation}
since $H=2/3t$ in the background~\cite{2021Ap&SS.366....4F}. The above reproduces the familiar Newtonian growing mode ($\tilde{v}\propto t^{1/3}$), but not the decaying mode. One can recover the full Newtonian solution after taking the right-hand side of Eq.~(\ref{ltv''1}) into account as well. This is achieved by involving Poisson's formula, namely $\partial^2\Phi=\rho\delta/2$ (where $\delta$ is the familiar density contrast), the background relation $\rho=3H^2$ and the linear Newtonian result $\dot{\delta}=-\tilde{\vartheta}$. Then, a straightforward calculation leads to
\begin{equation}
\partial_{\alpha}\dot{\Phi}= {1\over2}\,H\partial_{\alpha}\Phi+ {3\over2}\,H\dot{\tilde{v}}_{\alpha}\,,  \label{graddotPhi}
\end{equation}
which substituting into the right-hand side of (\ref{ltv''1}) yields
\begin{equation}
\ddot{\tilde{v}}_{\alpha}+ 3H\dot{\tilde{v}}_{\alpha}- H^2\tilde{v}_{\alpha}= 0\,.  \label{ltddotv2}
\end{equation}
The latter solves immediately to give
\begin{equation}
\tilde{v}= \mathcal{C}_1t^{1/3}+ \mathcal{C}_2t^{-4/3}= \mathcal{C}_3a^{1/2}+ \mathcal{C}_4a^{-2}\,,  \label{ltv2}
\end{equation}
in complete agreement with~\cite{1976ApJ...205..318P,1980lssu.book.....P}. Therefore, in Newtonian cosmology, linear peculiar velocities grow as $\tilde{v}\propto t^{1/3}$ as long as pressure-free matter dominates the density of the universe. However, this rate is too slow to explain the fast bulk flows reported in~\cite{2004MNRAS.352...61H,2008MNRAS.387..825F,2009MNRAS.392..743W,%
2010MNRAS.407.2328F,2010ApJ...709..483L,2011MNRAS.414..264C,%
2015MNRAS.447..132W,2018MNRAS.481.1368P,2021MNRAS.504.1304S,%
2023MNRAS.524.1885W,2023MNRAS.526.3051W}.

\subsection{Quasi-Newtonian analysis}\label{ssQ-NA}
%%%%%%%%%%%%%%%%%%%%%%%%%%%%%%%%%%%%%%%%%%%%%%%%%%%
In the literature, there are also few quasi-Newtonian treatments that recover the Newtonian growth-rate~\cite{1998PhRvD..58l4006M,2001CQGra..18.5115E}. Nevertheless, despite their relativistic initial appearance, these studies reduce to Newtonian in practice. The reasons are multiple and they all stem from the severe mathematical restrictions imposed upon the perturbed spacetime, which inevitably compromise its relativistic nature (see below and also related comments in~\S~6.8.2 of~\cite{2012reco.book.....E}).

\subsubsection{The quasi-Newtonian setup}\label{sssQ-NS}
%%%%%%%%%%%%%%%%%%%%%%%%%%%%%%%%%%%%%%%%%%%%%%%%%%%%%%%%
The quasi-Newtonian approach is used to obtain Newtonian-like equations and solutions, by adopting a reference frame with zero linear vorticity and shear (i.e.~$\omega_{ab}=0=\sigma_{ab}$ at the linear level by default). This, however, imposes strict restrictions on the spacetime and compromises its relativistic nature. For instance, in the quasi-Newtonian models, the magnetic component of the Weyl field also vanishes, which in turn means that gravitational waves are switched off as well. Perturbed relativistic spacetimes without these features are rather unnatural and the problematic nature of the quasi-Newtonian approach has been noted in~\cite{2012reco.book.....E} (see \S~6.8.2 there for related discussion and ``warning'' comments). One should therefore be wary before adopting the quasi-Newtonian approximation in relativistic cosmological studies.

When it comes to the study of linear peculiar velocities, the downside of using the quasi-Newtonian approach is not the absence of gravitational waves, but the fact that the (purely relativistic) gravitational input of the \textit{peculiar flux} (see \S~\ref{ssLRBFs} earlier) is not accounted for. This is an additional side-effect of the zero vorticiy and shear constraints. The direct consequence is that the driving force of the linear peculiar-velocity field, namely the 4-acceleration, is given by the gradient of an ad hoc (Newtonian-like) scalar potential. Moreover, the potential has no expression to describe it and its evolution follows from an ansatz that is not uniquely determined. All this may simplify the calculations, but at the same time it seriously compromises the relativistic nature of the treatment and blurs the physics. As a result, the quasi-Newtonian analysis (inadvertently) reduces to Newtonian, which means that it cannot explain the fast and deep bulk peculiar flows reported in recent surveys, like those of~\cite{2023MNRAS.524.1885W,2023MNRAS.526.3051W} for example. In contrast, as will show in \S~\ref{ssRA} below, the relativistic analysis provides theoretical support to the aforementioned bulk flows, without imposing any constraints upon the perturbed spacetime. Al this makes the related quasi-Newtonian studies an interesting mathematical exercise but without any physical input.

\subsubsection{Linear quasi-Newtonian approach}\label{sssLQ-NA}
\begin{figure*}
\begin{center}
\includegraphics[height=2in,width=4in,angle=0]{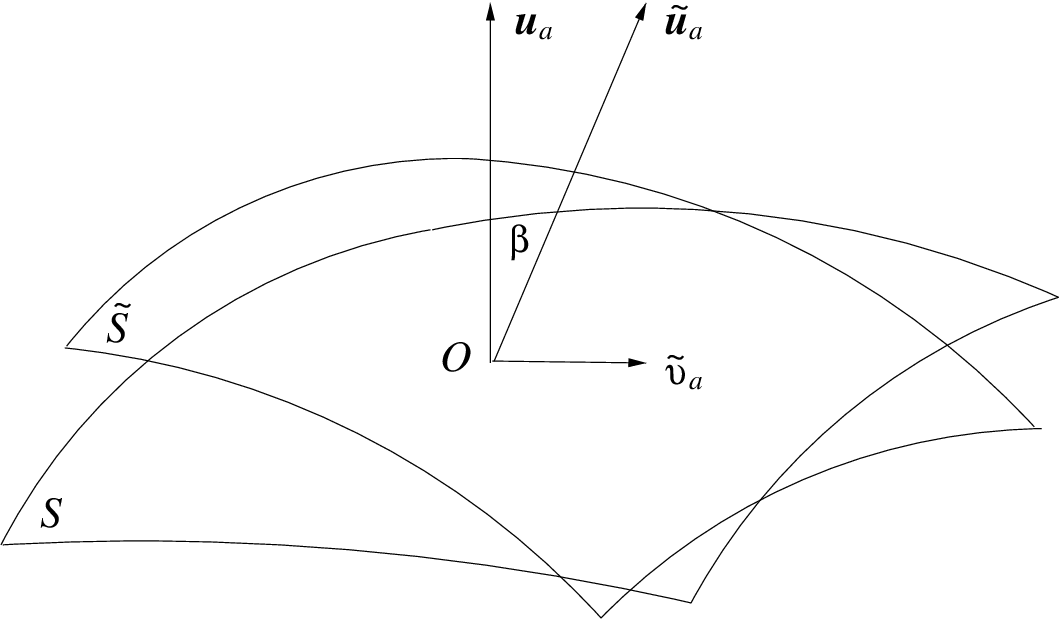}\quad
\end{center}
\caption{Typical tilted cosmologies equipped with two families of observers in relative motion, with peculiar velocity $\tilde{v}_a$. The associated 4-velocities ($u_a$ and $\tilde{u}_a$), which form a hyperbolic tilt angle ($\beta$) between them, are normal to their corresponding spatial hypersurfaces ($S$ and $\tilde{S}$).}  \label{fig:tilted}
\end{figure*}

%%%%%%%%%%%%%%%%%%%%%%%%%%%%%%%%%%%%%%%%%%%%%%%%%%%%%%%%%%%%%%%
The starting point is a tilted, perturbed Einstein-de Sitter background universe, with two families of observers in relative motion. These are the idealised (reference) CMB observers with 4-velocity $u_a$ and their realistic (tilted) counterparts with 4-velocity $\tilde{u}_a$. The latter are assumed to move with (non-relativistic) peculiar velocity $\tilde{v}_a$ with respect to the former, so that $\tilde{u}_a=u_a+\tilde{v}_a$ (see Fig.~\ref{fig:tilted}).\footnote{In the background spacetime, the peculiar-velocity field vanishes by default and the two 4-velocities coincide. This makes $\tilde{v}_a$ a gauge-invariant linear perturbation~\cite{1974RSPSA.341...49S}.} Assuming that the pressureless matter follows timelike geodesics, one may set $\tilde{A}_a=0=\tilde{q}_a$ in the tilted frame. Then, in line with (\ref{lab34}b) and (\ref{lab710}c), we have
\begin{equation}
\dot{\tilde{v}}_a+ H\tilde{v}_a= -A_a \hspace{15mm} {\rm and} \hspace{15mm} q_a= \rho\tilde{v}_a\,,  \label{lqNrels}
\end{equation}
in the coordinate system of the CMB~\cite{1998PhRvD..58l4006M,%
2001CQGra..18.5115E}.

When studying peculiar motions, the apparent advantage of the quasi-Newtonian approach is that, without vorticity, one can appeal to a scalar potential ($\varphi$) and express the 4-acceleration as the spatial gradient of the latter, namely write~\cite{1998PhRvD..58l4006M,2001CQGra..18.5115E}
\begin{equation}
A_a= {\rm D}_a\varphi\,.  \label{qNlA}
\end{equation}
There is a serious downside, however, because $\varphi$ is an ad hoc potential with no expression for it. In addition, the time evolution of the quasi-Newtonian potential is not uniquely determined and follows from the ansatz $\dot{\varphi}=-\Theta/3$, which requires setting the shear to zero as well~\cite{1998PhRvD..58l4006M}. Moreover, when it comes to the study of peculiar velocities, the key disadvantage is that the 4-acceleration adopted in Eq.~(\ref{qNlA}) does not account for the gravitational input of the peculiar flux and this severely compromises the relativistic nature of the analysis. After all this, the linear evolution of the quasi-Newtonian 4-acceleration is given by~\cite{1998PhRvD..58l4006M,%
2001CQGra..18.5115E}
\begin{equation}
\dot{A}_a= -2HA_a- {3\over2}\,H^2\tilde{v}_a= 2H\dot{\tilde{v}}_a+{1\over2}\,H^2\tilde{v}_a\,.  \label{lqNdA}
\end{equation}
Note that in deriving the above, one also needs to use the linear shear constraint $3{\rm D}^b\sigma_{ab}=2{\rm D}_a\Theta+3{\rm curl}\omega_a-3q_a$ (see Eq.~\ref{shearcon} in \S~\ref{sssKs} for the nonlinear expression). In the absence of shear and rotation, the above reduces to ${\rm D}_a\Theta=3q_a/2$ and subsequently leads to ${\rm D}_a\Theta=9H^2\tilde{v}_a/2$, given that $q_a=\rho\tilde{v}_a$ to linear order and $\rho=3H^2$ in the Einstein-de Sitter background.

Finally, the time derivative of (\ref{lqNrels}a) combines with (\ref{lqNdA}) to provide the quasi-Newtonian  differential equation governing the linear evolution of peculiar velocities~\cite{1998PhRvD..58l4006M,%
2001CQGra..18.5115E}
\begin{equation}
\ddot{\tilde{v}}_a= -3H\dot{\tilde{v}}_a+ H^2\tilde{v}_a= -{2\over t}\,\dot{\tilde{v}}_a+ {4\over9t^2}\,\tilde{v}_a\,,  \label{lq-Nddotv}
\end{equation}
given that $H=2/3t$ after equipartition. Since the above differential formula is identical to its purely Newtonian counterpart (compare to Eq.~(\ref{ltddotv2}) in \S~\ref{sssLEPVs} previously), it comes to no surprise that its solution simply reproduces the Newtonian result (see~(\ref{ltv2}) in \S~\ref{sssLEPVs})
\begin{equation}
\tilde{v}= \mathcal{C}_1t^{1/3}+ \mathcal{C}_2t^{-4/3}\,.  \label{lqNv}
\end{equation}
Before proceeding to compare with the proper relativistic analysis, it is important to note that the zero shear and zero vorticity constraints of the quasi-Newtonian treatment, as well as the ``damaging'' subsequent introduction of the scalar potential ($\varphi$) in Eq.~(\ref{qNlA}), are not necessary, at least when studying the linear evolution of peculiar velocities (see below).

As we shall show in \S~\ref{ssRA} next, the key behind the considerable difference between the Newtonian/quasi-Newtonian results and the fully relativistic one stems from the different forms of the acceleration/4-acceleration. The latter is critical because it is the only driving force of the linear peculiar-velocity field in all these studies. In Newtonian physics and in the absence of matter pressure, the acceleration is the gradient of the gravitational potential and leads to a growth-rate of $v\propto t^{1/3}$ between recombination and the onset of the accelerated expansion (e.g.~see~\cite{1980lssu.book.....P} and \S~\ref{ssNA} here). This growth is too weak to explain fast bulk flows, like those reported in~\cite{2023MNRAS.524.1885W,2023MNRAS.526.3051W} for example, without introducing new parameters. In the quasi-Newtonian studies, the 4-acceleration is also given by the gradient of a Newtonian-like scalar potential. It therefore comes to no surprise that the result is the same fairly mediocre ($v\propto t^{1/3}$) growth-rate for the peculiar-velocity field~\cite{1998PhRvD..58l4006M,2001CQGra..18.5115E}.

To the unsuspecting reader, the agreement between the Newtonian and the quasi-Newtonian studies of linear peculiar velocities can be misleading, because it simply reflects the fact that the latter analysis is also Newtonian (for all practical purposes). This happens because both approaches bypass (albeit for different reasons) the gravitational input of the peculiar flux (see \S~\ref{ssLRBFs} previously). In Newtonian physics this is unavoidable, since only the density of the matter gravitates. In the quasi-Newtonian studies, however, the effect of the peculiar flux is not properly accounted for because of the entirely unnecessary zero vorticity and shear constraints and the subsequent introduction of an (also unnecessary) scalar potential for the 4-acceleration. All this blurs the physics and diverts the attention from the key role of the peculiar flux, so that its contribution to the relativistic gravitational field and subsequently to the linear evolution of peculiar velocities are inadvertently bypassed. As a result, the driving force of the peculiar velocity field is practically identical to its purely Newtonian counterpart, which explains why the two studies arrive at the same result.

Before closing this section, we should point out that the general problems of the quasi-Newtonian studies, namely of adopting a reference frame with zero linear shear and vorticity (and the rest of the constraints that follow), have been known and they are clearly stated in \S~6.8.2 of~\cite{2012reco.book.....E}. However, the extent and the depth of the problem was not realised because there was no direct comparison with a proper relativistic treatment that did not employ the quasi-Newtonian frame. Next, we will provide such a comparison by avoiding the quasi-Newtonian zero vorticity and shear restrictions, when studying linear peculiar velocities. In the process, we will also identify the pivotal role of the peculiar flux and show how deeply damaging it can be to bypass it. Without accounting for the gravitational input of the flux, peculiar-motion studies that may appear relativistic reduce to Newtonian.\footnote{In the literature there additional examples of studies involving peculiar motions, which start relativistically but reduce to Newtonian when the flux-input of the moving matter to the gravitational field is bypassed for one reason or another (see \S~\ref{ssMSE} and \S~\ref{ssNRZA} below).} Therefore, incorporating the peculiar-flux input to gravity is what distinguishes the relativistic treatments from the rest and in so doing it can also provide a solution to the bulk-flow puzzle.

\subsection{Relativistic analysis}\label{ssRA}
%%%%%%%%%%%%%%%%%%%%%%%%%%%%%%%%%%%%%%%%%%%%%%
When coming to the study of cosmological peculiar motions, the purely Newtonian and the quasi-Newtonian treatments are essentially identical and very different from the relativistic analysis. This is a direct consequence of the fact that the Newtonian/quasi-Newtonian studies do not account for the gravitational input of the peculiar flux. In addition, the relativistic studies allow one to investigate the implications of spatial curvature for the evolution of the peculiar-velocity field, as well as to probe cosmological epochs prior to recombination.

\subsubsection{Two alternative perspectives}\label{sssTAPs}
\begin{figure}[!tbp]
  \begin{subfigure}[b]{0.475\textwidth}
    \includegraphics[width=\textwidth]{pmotion1.eps}
    \caption{CMB-frame perspective.}
    \label{fig:f1}
  \end{subfigure}
  \hfill
  \begin{subfigure}[b]{0.475\textwidth}
    \includegraphics[width=\textwidth]{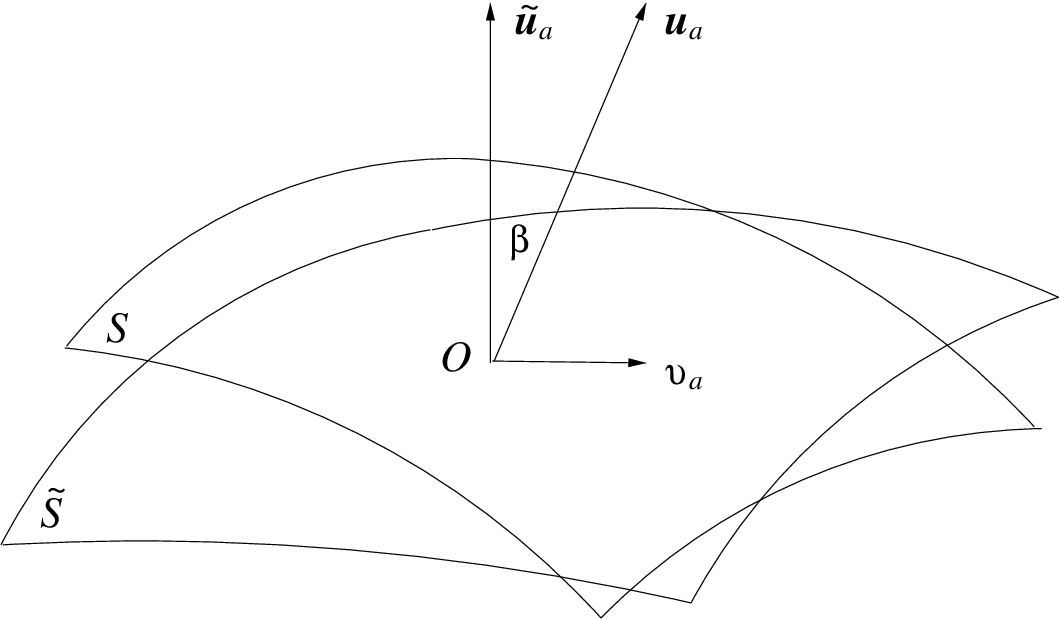}
    \caption{Tilted-frame perspective.}
    \label{fig:f2}
  \end{subfigure}
  \caption{Tilted spacetimes allow for two families of relatively moving observers, with 4-velocities $u_a$ and $\tilde{u}_a$, at every event ($O$). Assuming that the $u_a$-field defines the reference frame of the universe, $\tilde{u}_a$ is the 4-velocity of the tilted (the realistic) observers, ``drifting'' with peculiar velocity $\tilde{v}_a$ relative to the CMB (see Eq.~(\ref{4vels4vels}a) and Fig.~\ref{fig:f1} above). Alternatively, one may turn to Fig.~\ref{fig:f2} and adopt the $\tilde{u}_a$-field as their frame of reference. In that case, the idealised CMB observers are assumed to ``move'' relative to the tilted frame of a typical galaxy with (effective) ``peculiar'' velocity $v_a$. It goes without saying that both approaches are physically equivalent and both lead to the same results. Also note that in either case $\beta$ (with $\cosh\beta=-u_a\tilde{u}^a$) is the hyperbolic (tilt) angle between $u_a$ and $\tilde{u}_a$, while $S$ and $\tilde{S}$ are the 3-D rest-spaces of the aforementioned two groups of observers.}  \label{fig:pmotion}
\end{figure}

%%%%%%%%%%%%%%%%%%%%%%%%%%%%%%%%%%%%%%%%%%%%%%%%%%%%%%%%%%%
Let us consider a family of realistic observers moving relative to the reference (CMB) frame during the Enstein-de Sitter epoch of the universe. Assuming that $\tilde{u}_a$ and $u_a$ are the 4-velocities associated with the former and the latter group of observers respectively, we have
\begin{equation}
\tilde{u}_a= u_a+ \tilde{v}_a \hspace{15mm} {\rm and} \hspace{15mm} u_a= \tilde{u}_a+ v_a\,.  \label{4vels4vels}
\end{equation}
In the above, $\tilde{v}_a$ is the peculiar velocity of the tilted (the realistic) observers with respect to the CMB frame, whereas $v_a$ is the velocity of the latter relative to the former (with $u_a\tilde{v}^a=0=\tilde{u}_av^a$ and $\tilde{v}^2=v^2$ by construction -- see Fig.~\ref{fig:pmotion}). We remind the reader that both the CMB and the ``tilted'' frames are located at the same spacetime event by construction (e.g.~see~\cite{1973CMaPh..31..209K} and \S~\ref{ssRMOs} here). This guarantees that a Lorentz boosting between these two coordinate systems is local and does not violate causality. Instead, the transformation relates the quantities measured by the realistic (tilted) observers to those measured by their idealised (CMB) parters. This in turn allows the realistic observers to quantify the kinematic ``contamination'' induced by their own peculiar motion. Also, when the peculiar motion is not relativistic (as it happens in our case), we have $\tilde{v}^2,v^2\ll1$ and $\tilde{v}_a=-v_a$. When studying the effects of relative motion, one is free to choose any of the above approaches. Here, we will adopt the perspective of (\ref{4vels4vels}a), but it goes without saying that nothing would have changed if we were to adopt that of (\ref{4vels4vels}b).

We also remind the reader that all the variables describing perturbations, including the peculiar-velocity field ($\tilde{v}_a$), vanish identically in the FLRW background. This satisfies the Stewart \& Walker criterion for gauge invariance~\cite{1974RSPSA.341...49S} and guarantees that our analysis is independent of the gauge-choice ``connecting'' the background and the perturbed spacetime.

\subsubsection{The key role of the peculiar flux}\label{sssKRPF}
%%%%%%%%%%%%%%%%%%%%%%%%%%%%%%%%%%%%%%%%%%%%%%%%%%%%%%%%%%%%%%%%
We will start by drawing the reader's attention to the fundamental differences between the Newtonian/quasi-Newtonian treatments of peculiar velocities and the relativistic studies. These stem from the fundamentally different way the two theories treat the gravitational field, as well as its sources. The reader is referred to~\cite{2020EPJC...80..757T,2021Ap&SS.366....4F,2022PhRvD.106h3505M,%
2024PhRvD.110f3540M} for the original fully relativistic studies of peculiar-velocity fields, which employ both the tilted-frame and the CMB-frame approaches depicted in Fig.~\ref{fig:pmotion} here. A streamlined mathematical treatment and a physical summary of the subject was recently given in~\cite{2026ApJ...997...25T}.

Peculiar motions are nothing else but matter in motion. In relativity, as opposed to Newtonian gravity, matter fluxes ``gravitate'' since they also contribute to the energy-momentum tensor. The gravitational input of this \textit{peculiar flux} feeds into the field equations, then into the relativistic conservation laws and eventually emerges into the equations governing the peculiar-velocity field and its implications. Therefore, in a sense, one could say that bulk peculiar flows ``gravitate''~\cite{2020EPJC...80..757T}. Accounting for this flux contribution to the gravitational field is what distinguishes the relativistic treatments of peculiar motions from the rest (see~\cite{2020EPJC...80..757T,2021Ap&SS.366....4F,%
2022PhRvD.106h3505M,2024PhRvD.110f3540M} for the original studies and~\cite{2026ApJ...997...25T} for an updated summary presentation).

In what follows, we will demonstrate the implications of the above arguments in perturbed FLRW universes, starting with the Einstein-de Sitter background and then generalising to Friedmann models with nonzero spatial curvature. The latter studies will also allow us to probe the linear evolution of peculiar velocities in locally underdense and/or overdense regions of a nearly flat FLRW universe.

Adopting the CMB-frame perspective (see Eq.~(\ref{4vels4vels}a) and Fig.~\ref{fig:f1}), we will set the flux and the 4-acceleration to zero in the frame of the matter, so that $\tilde{q}_a=0$ and $\tilde{A}_a=0$. Put another way, we will assume that there is no peculiar flux in the coordinate system of the pressureless matter, which moves along timelike geodesics. Then, to linear order in the reference $u_a$-frame, we have (see Eqs.~(\ref{lab710}c) and (\ref{lab34}b) in \S~\ref{ssLRBFs} earlier)
\begin{equation}
q_a= \rho\tilde{v}_a \hspace{15mm} {\rm and} \hspace{15mm} \dot{\tilde{v}}_a+ H\tilde{v}_a= -A_a\,,  \label{lrels1}
\end{equation}
with zero linear pressure in both coordinate systems (i.e.~$\tilde{p}=p=0$ -- see (\ref{lab710}b) in \S~\ref{ssLRBFs}). It is important to note that, according to Eq.~(\ref{lrels1}a), the cosmic medium cannot be treated as perfect when peculiar motions are present. The ``imperfection'' appears as a nonzero \textit{peculiar flux} vector ($q_a$), solely triggered by the moving matter (e.g.~see \S~5.2.1 in~\cite{2012reco.book.....E} for related comments and also for the generalised linear version of (\ref{lrels1}a)). Overall, while the pressureless dust follows a geodesic frame, where the 4-acceleration ($\tilde{A}_a$) vanishes, there is nonzero 4-acceleration ($A_a$) in the CMB frame, which acts as the driving force of the linear peculiar-velocity field (see Eq.~(\ref{lrels1}b) above).\vspace{5pt}

At this point, before proceeding any further, it is important to note that the linear expression (\ref{lrels1}b) is also the staring point of the quasi-Newtonian analysis (see Eq.~(\ref{lqNrels}) in \S~\ref{sssLQ-NA} previously). However, the agreement between the quasi-Newtonian and the relativistic approaches to the question of the peculiar-velocity evolution stops here.

In a proper relativistic study no zero-vorticity and zero-shear constraints are imposed upon the perturbed spacetime. Also, there is no need to express the 4-acceleration as the gradient of an ad hoc scalar potential ($\varphi$ - see expression (\ref{qNlA}) in \S~\ref{sssLQ-NA}) and no need to use the ansatz $\dot{\varphi}=-\Theta/3$ to determine its temporal evolution. Most importantly, the gravitational input of the peculiar flux is fully accounted for throughout the analysis.

Via Einstein's equations, the flux contribution to relativistic gravitational field feeds into the conservation laws and eventually emerges in a host of relativistic relations (see~\cite{2026ApJ...997...25T} for a recent streamlined presentation). Among them are the formulae governing the linear evolution of inhomogeneities in the density and in the expansion. These are respectively monitored by the spatial gradients $\Delta_a=(a/\rho){\rm D}_a\rho$ and $\mathcal{Z}_a=a{\rm D}_a\Theta$, which in the presence of peculiar motions evolve as
\begin{equation}
\dot{\Delta}_a= -\mathcal{Z}_a+ 3aH\dot{\tilde{v}}_a+ 3aH^2\tilde{v}_a- a{\rm D}_a\tilde{\vartheta}  \label{ldotDel1}
\end{equation}
and
\begin{eqnarray}
\dot{\mathcal{Z}}_a&=& -2H\mathcal{Z}_a- {3\over2}\,H^2\Omega\Delta_a+ 3aH^2\left(1+{1\over2}\,\Omega\right)\dot{\tilde{v}}_a+ 3aH^3\left(1+{1\over2}\,\Omega\right)\tilde{v}_a- a{\rm D}_a\dot{\tilde{\vartheta}} \nonumber\\ &&-2aH{\rm D}_a\tilde{\vartheta}\,,  \label{ldotcZ1}
\end{eqnarray}
respectively~\cite{2026ApJ...997...25T}.\footnote{In~%
\cite{2026ApJ...997...25T} Eqs.~(\ref{ldotDel1}) and (\ref{ldotcZ1}) were obtained by taking the spatial gradients of the energy-density conservation law and of the Raychaudhuri equation (see (\ref{edcl}) and (\ref{Ray}) respectively), both linearised in the CMB frame. There are more  ways of obtaining (\ref{ldotDel1}) and (\ref{ldotcZ1}) however. The faster is by linearising the nonlinear formulae (2.3.1) and (2.3.2) of~\cite{2008PhR...465...61T}, or equivalently expressions (10.101) and (10.102) of~\cite{2012reco.book.....E}, while using the linear relations (\ref{lrels1}a) and (\ref{lrels1}b) as well.} Note that, without the peculiar-velocity terms seen on the right-hand side of the above differential equations, expressions (\ref{ldotDel1}) and (\ref{ldotcZ1}) coincide with those used in studies free of peculiar-motion effects~\cite{2008PhR...465...61T,2012reco.book.....E}. These velocity terms reflect the general-relativistic input of the peculiar flux to the gravitational field (see \S~\ref{sssKRPF} before).

Differentiating (\ref{ldotDel1}) in time, combining the resulting expression with (\ref{ldotcZ1}) and using the linear commutation law $({\rm D}_a\vartheta)^{\cdot}={\rm D}_a \dot{\vartheta}-H{\rm D}_a\vartheta$, leads to the (non-homogeneous) differential equation~\cite{2026ApJ...997...25T}
\begin{equation}
\ddot{\tilde{v}}_a+ 2\left(1-{1\over2}\,\Omega\right)H\dot{\tilde{v}}_a- {3\over2}\,\Omega H^2\tilde{v}_a= {1\over3aH}\left(\ddot{\Delta}_a+ 2H\dot{\Delta}_a- {3\over2}\,\Omega H^2\Delta_a\right)\,,  \label{lddotv1}
\end{equation}
since $\dot{H}=-H^2(1+\Omega/2)$ in the Friedmann background (with $\Omega=\kappa\rho/3H^2$ representing the density parameter). Before proceeding to the solution of (\ref{lddotv1}), we should clarify to the reader that the terms homogeneous/non-homogeneous refer to the nature of the differential equation and have noting to do with the homogeneity/inhomogeneity of the host spacetime. The latter remains both inhomogeneous and anisotropic at the linear perturbative level.

The differential equation (\ref{lddotv1}) was introduced in~\cite{2026ApJ...997...25T} and couples the linear evolution of peculiar-velocity perturbations to those in the density distribution of the matter. Also note that (\ref{lddotv1}) is not confined to the Einstein-de Sitter background, but includes Friedmann models with nonzero spatial curvature as well. The only proviso is that the matter component is pressure-free, which means that (\ref{lddotv1}) applies to baryonic ``dust'' after recombination and to low-energy CDM species after equipartition.

\subsubsection{The relativistic 4-acceleration}\label{sssR4A}
%%%%%%%%%%%%%%%%%%%%%%%%%%%%%%%%%%%%%%%%%%%%%%%%%%%%%%%%%%%%%
Before proceeding to the solutions of (\ref{lddotv1}), it is worth deriving the linear form of the relativistic 4-acceleration in the presence of peculiar motions and in the absence of matter pressure. In such an environment, the relativistic energy conservation law (see Eq.~(\ref{edcl}) in \S~\ref{sssCLs} previously for the nonlinear expression) linearises to
\begin{equation}
\dot{\rho}= -\Theta\rho- {\rm D}^aq_a\,.  \label{ledcl}
\end{equation}
Note the extra flux-related term on the right-hand side of the above, which results form the purely general relativistic contribution of the peculiar flux ($q_a=\rho\tilde{v}_a$) to the gravitational field (see \S~\ref{sssKRPF} before). As we will show next, accounting for the implications of the aforementioned flux-term is what separates the relativistic studies of peculiar motions from the rest.

Taking the spatial gradient of (\ref{ledcl}), employing the linear commutation law $({\rm D}_a\rho)^{\cdot}={\rm D}_a\dot{\rho}-H{\rm D}_a\rho+\dot{\rho}A_a$, recalling that $q_a=\rho\tilde{v}_a$ and keeping up to first-order terms, one arrives at the following linear expression for the 4-acceleration by~\cite{2024PhRvD.110f3540M}
\begin{equation}
A_a= -{1\over3H}\,{\rm D}_a\tilde{\vartheta}- {1\over3aH}\left(\dot{\Delta}_a+\mathcal{Z}_a\right)\,.  \label{rAa}
\end{equation}
The profound difference between the above and its quasi-Newtonian analogue (compare to Eq.~(\ref{qNlA}) in \S~\ref{sssLQ-NA} before), predisposes for an analogous difference between the results of the two approaches (see \S~\ref{sssPVFLRWUOm1} next).

We would also like to draw the reader's attention to the density and the expansion gradients seen on the right-hand side of (\ref{rAa}) and represented by $\Delta_a$ and $\mathcal{Z}_a$ respectively. The presence of these inhomogeneity gradients, in connection with Eq.~(\ref{lrels1}b), ensures that linear peculiar velocities are driven by structure formation. Finally, combining the above with Eq.~(\ref{lrels1}b) leads to expression (\ref{ldotDel1}).

\subsubsection{Peculiar velocities in FLRW universes with
%%%%%%%%%%%%%%%%%%%%%%%%%%%%%%%%%%%%%%%%%%%%%%%%%%%%%%%%
$\Omega=1$}\label{sssPVFLRWUOm1}
%%%%%%%%%%%%%%%%%%%%%%%%%%%
Since the observations support a universe with almost flat spatial sections, let us first confine to an Einstein de Sitter background. Then, after setting $\Omega=1$, Eq.~(\ref{lddotv1}) reduces to
\begin{equation}
\ddot{\tilde{v}}_a+ H\dot{\tilde{v}}_a- {3\over2}\,H^2\tilde{v}_a= {1\over3aH}\left(\ddot{\Delta}_a+ 2H\dot{\Delta}_a- {3\over2}\,H^2\Delta_a\right)\,.  \label{lfddotv1}
\end{equation}
The most straightforward way to show that the above leads to linear growth rates stronger than the Newtonian/quasi-Newtonian $\tilde{v}_a\propto t^{1/3}$-rate (see \S~\ref{ssNA} and \S~\ref{ssQ-NA} earlier), is by solving its homogeneous component. Since $a\propto t^{2/3}$ and $H=2/3t$ after equipartition, the latter reads
\begin{equation}
\ddot{\tilde{v}}_a+ {2\over3t}\,\dot{\tilde{v}}_a- {2\over3t^2}\,\tilde{v}_a= 0\,,  \label{lddotv2}
\end{equation}
and accepts the power-law solution
\begin{equation}
\tilde{v}= \mathcal{C}_1t+ \mathcal{C}_2t^{-2/3}= \mathcal{C}_3a^{3/2}+ \mathcal{C}_4a^{-1}\,,  \label{lv1}
\end{equation}
on all scales~\cite{2026ApJ...997...25T}. Therefore, the linear peculiar-velocity field grows as $\tilde{v}_a\propto t$, a rate considerably stronger than the $\tilde{v}\propto t^{1/3}$ growth of the Newtonian and the quasi-Newtonian treatments (see \S~\ref{ssNA} and \S~\ref{ssQ-NA} earlier). The reason is the aforementioned general relativistic input of the peculiar flux, which changes the local gravitational field and, by so doing, modifies the linear evolution of the peculiar-velocity field as well (compare the differential equation (\ref{lfddotv1}) to (\ref{ltddotv2}) and (\ref{lq-Nddotv}) in \S~\ref{ssNA} and \S~\ref{ssQ-NA}  respectively).

It is also quite important to realise that solution (\ref{lv1}) provides the minimum linear growth of peculiar velocities in tilted Einstein-de Sitter universes. Indeed, according to the mathematical theory of differential equations, the solution of the non-homogeneous formula (\ref{lfddotv1}) contains the general solution of its homogeneous part (in our case (\ref{lv1})) plus a partial solution of the full equation. Hence, solving (\ref{lfddotv1}) in full, will make physical difference only if the partial solution grows stronger than the strongest growing mode of its homogeneous counterpart. Put another way, $\tilde{v}\propto t$ is the minimum growth-rate of linear peculiar velocities. Indeed, by incorporating some of the density effects seen on the right-hand side of~(\ref{lfddotv1}) the relativistic analysis has led to the growth-rate of $\tilde{v}\propto t^{4/3}$ for the linear peculiar-velocity field~\cite{2020EPJC...80..757T,%
2021Ap&SS.366....4F,2022PhRvD.106h3505M,%
2024PhRvD.110f3540M}.\footnote{It is straightforward to show that the right-hand side of (\ref{lfddotv1}) also goes to zero when the standard evolution law $\Delta_a=\mathcal{C}_1t^{2/3}+ \mathcal{C}_2t^{-1}$ is adopted for the linear density contrast. However, the latter law holds in perturbed Einstein-de Sitter universes without peculiar velocities (e.g.~see~\cite{2008PhR...465...61T,2012reco.book.....E}). Therefore, even in the minimalist scenario where the linear density perturbations evolve unaffected by the existing peculiar-velocity field, the latter still grows at the $\tilde{v}\propto t$-rate.}

The above demonstrate how, as well as why, general relativity supports stronger growth rates for linear peculiar velocities than the Newtonian and quasi-Newtonian studies. This in turn supports faster and deeper peculiar flows, perhaps like those reported by several bulk-flow surveys~~\cite{2004MNRAS.352...61H,%
2008MNRAS.387..825F,2009MNRAS.392..743W,2010MNRAS.407.2328F,%
2010ApJ...709..483L,2011MNRAS.414..264C,2015MNRAS.447..132W,%
2018MNRAS.481.1368P,2021MNRAS.504.1304S}, including the very recent reports of~\cite{2023MNRAS.524.1885W,2023MNRAS.526.3051W} using the \textit{CosmicFlows-4} database. Overall, general relativity offers a natural explanation to the existence of high-amplitude bulk flows, which could bridge the gap between theory and observation without requiring any new physics beyond the standard model. Then, the tension between theory and observation is no longer a sign of a flawed model, but instead it reflects the limitations of the Newtonian approximations employed to predict the growth of peculiar velocities in cosmology. The debate thus shifts, from questioning the standard model itself, to scrutinizing the validity of the approximations used in its application to cosmic dynamics.

\subsubsection{The role of the peculiar
%%%%%%%%%%%%%%%%%%%%%%%%%%%%%%%%%%%%%%%
4-acceleration}\label{sssRP4A}
%%%%%%%%%%%%%%%%%%%%%%%%%%%%%%
A complementary, less technical but physically intuitive approach to the peculiar-velocity growth utilises the 4-acceleration, or (if the study is purely Newtonian) the acceleration. In either form the latter plays the key role, because it acts as the driving force of linear peculiar flows (see expressions (\ref{ltv'}), (\ref{lqNrels}a) and (\ref{lrels1}b) previously). In the Newtonian/quasi-Newtonian studies, where the gravitational contribution of the peculiar flux is not accounted for,  the linear peculiar velocity field grows at the mediocre rate of $\tilde{v}\propto t^{1/3}$. In both cases, the linear expressions (\ref{ltv'}) and (\ref{lqNrels}a) ensure that the corresponding driving force, namely the acceleration/4-acceleration, decays as $\partial_{\alpha}\Phi$, $A_a\propto t^{-2/3}$ -- see also solution (43) in~\cite{1998PhRvD..58l4006M}.

In the relativistic analysis of \S~\ref{sssPVFLRWUOm1} the peculiar-flux input to gravity increased the linear growth of the peculiar-velocity field from $\tilde{v}\propto t^{1/3}$ to the (minimum) rate of $\tilde{v}\propto t$. Following (\ref{lrels1}b), the aforementioned boost reflects the fact that the 4-acceleration driving the peculiar-velocity field no longer decays but remains constant in time. However, in \S~\ref{sssPVFLRWUOm1} we did not account for the additional effects of the density and the expansion gradients. The latter were (partially) accounted for in~\cite{2020EPJC...80..757T,2021Ap&SS.366....4F,2022PhRvD.106h3505M,%
2024PhRvD.110f3540M} and led to the faster growth-rate of $\tilde{v}\propto t^{4/3}$ for the linear peculiar-velocity field. According to (\ref{lrels1}b), this happens because the effective 4-acceleration was no longer constant but grew as $A_a\propto t^{1/3}$ (see also Eq.~(17) in~\cite{2020EPJC...80..757T}).

One can therefore explain the faster growth-rates of the relativistic treatments by looking at the driving forces of the linear peculiar-velocity field. In the Newtonian/quasi-Newtonian treatments, the latter is driven by a time-decaying acceleration/4-acceleration. This decay, which reflects the fact that the gravitational input of the peculiar flux has not been accounted for, is the reason for the mediocre $\tilde{v}\propto t^{1/3}$ Newtonian/quasi-Newtonian growth. The relativistic study incorporates the gravitational contribution of the peculiar flux and also brings the density gradients into play. Then, the 4-acceleration driving the linear peculiar velocities no longer decays, but remains constant (or even increases) in time. As a result, the velocity growth-rate goes up to $\tilde{v}\propto t$~\cite{2026ApJ...997...25T}, or even higher~\cite{2020EPJC...80..757T,2021Ap&SS.366....4F,%
2022PhRvD.106h3505M,2024PhRvD.110f3540M}.

In summary, according to the theoretical analyses of~\cite{2020EPJC...80..757T,2021Ap&SS.366....4F,2022PhRvD.%
106h3505M,2024PhRvD.110f3540M,2026ApJ...997...25T}, general relativity supports a considerably stronger linear growth for the peculiar-velocity fields, than the Newtonian theory. This in turn could relax the current $\Lambda$CDM limits, which are all based on Newtonian studies, to accommodate the faster and deeper bulk flows reported in~\cite{2004MNRAS.352...61H,2008MNRAS.387..825F,2009MNRAS.392..743W,%
2010MNRAS.407.2328F,2010ApJ...709..483L,2011MNRAS.414..264C,%
2015MNRAS.447..132W,2018MNRAS.481.1368P,2021MNRAS.504.1304S,%
2023MNRAS.524.1885W,2023MNRAS.526.3051W}. It is therefore conceivable that general relativity could resolve the persistent bulk-flow puzzle without the need for introducing any new free parameters, or any exotic new physics.

\subsubsection{Peculiar velocities in FLRW universes with
%%%%%%%%%%%%%%%%%%%%%%%%%%%%%%%%%%%%%%%%%%%%%%%%%%%%%%%%
$\Omega\neq1$}\label{sssPVFLRWUOm2}
%%%%%%%%%%%%%%%%%%%%%%%%%%%%%%%%%
Having tested the case of the Einstein-de Sitter background, let us move on to Friedmann universes with non-Euclidean spatial geometry. Allowing for nonzero spatial curvature in the background, makes it possible to study the evolution of peculiar velocities in the underdense and overdense environments of the open and the closed FLRW universes respectively. The downside is that the complexity of the analytic treatments increases considerably. To begin with, in Friedmann universes with nonzero spatial curvature, it is necessary to express the evolution equations in terms of the conformal time ($\eta$, with $\dot{\eta}=1/a$), instead of proper time (e.g.~see~\cite{1983QB981.N3.......} for further discussion). This is achieved by means of the transformation formulae
\begin{equation}
\dot{\tilde{v}}= {1\over a}{{\rm d}\tilde{v}\over{\rm d}\eta} \hspace{15mm} {\rm and} \hspace{15mm} \ddot{\tilde{v}}= {1\over a^2} \left({{\rm d}^2\tilde{v}\over{\rm d}\eta^2}- \mathcal{H}{{\rm d}\tilde{v}\over{\rm d}\eta}\right)\,,  \label{tls}
\end{equation}
where $\mathcal{H}=aH$ is the ``rescaled'' Hubble parameter. On using the above, the homogeneous component of (\ref{lddotv1}) takes the form
\begin{equation}
{{\rm d}^2\tilde{v}_a\over{\rm d}\eta^2}+ (1-\Omega)\mathcal{H} {{\rm d}\tilde{v}_a\over{\rm d}\eta}- {3\over2}\,\Omega\mathcal{H}^2\tilde{v}_a= 0\,.  \label{K1ddotv1}
\end{equation}
For negative spatial curvature and pressureless matter (baryonic or not), we have $a\propto\sinh^2(\eta/2)$ with $\eta>0$ (see solution (\ref{soFLRWa}) in \S~\ref{sssSfEFLRWCs}). This leads to $\mathcal{H}=\coth(\eta/2)$ and recasts Eq.~(\ref{K1ddotv1}) as
\begin{equation}
{{\rm d}^2\tilde{v}_a\over{\rm d}\eta^2}+ (1-\Omega)\coth
\left({\eta\over2}\right){{\rm d}\tilde{v}_a\over{\rm d}\eta}- {3\over2}\,\Omega\coth^2\left({\eta\over2}\right)\tilde{v}_a= 0\,,  \label{-1ddotv1}
\end{equation}
with $0<\Omega<1$ at all times. On the other hand, the scale factor of a closed FLRW universe with dust is given by $a\propto\sin^2(\eta/2)$ (see solution (\ref{scFLRWa}) in \S~\ref{sssSfEFLRWCs}) and the rescaled Hubble parameter is $\mathcal{H}=\cot(\eta/2)$, with $0\leq\eta\leq2\pi$. In this environment, Eq.~(\ref{K1ddotv1}) reads
\begin{equation}
{{\rm d}^2\tilde{v}_a\over{\rm d}\eta^2}+ (1-\Omega)\cot
\left({\eta\over2}\right){{\rm d}\tilde{v}_a\over{\rm d}\eta}- {3\over2}\,\Omega\cot^2\left({\eta\over2}\right)\tilde{v}_a= 0\,,  \label{+1ddotv1}
\end{equation}
where $\Omega>1$ always.

Before closing this section, it worth noticing that close to the Euclidean limit, we have $\Omega\rightarrow1^{\pm}$ and $\eta\rightarrow0^+$.\footnote{In the presence of nonzero spatial curvature, the Friedmann equation recasts as $|1-\Omega|=|K|/\mathcal{H}$. After equipartition, when $\mathcal{H}=\coth(\eta/2)$ in an open model and $\mathcal{H}=\cot(\eta/2)$ in a closed one, we have $|1-\Omega|=\tanh^2(\eta/2)=\tan(\eta/2)$. Therefore, $\Omega\rightarrow1^{\pm}$ leads to $\eta\rightarrow0^+$ and vice versa.} Then, a simple Maclaurin-series expansion gives $\coth(\eta/2)=2/\eta=\cot(\eta/2)$ to first approximation and reduces Eqs.~(\ref{-1ddotv1}) and (\ref{+1ddotv1}) to
\begin{equation}
{{\rm d}^2\tilde{v}\over{\rm d}\eta^2}- {6\over\eta^2}\,\tilde{v}= 0\,,  \label{mK1ddotv1}
\end{equation}
which accepts the power-law solution
\begin{equation}
\tilde{v}= \mathcal{C}_1\eta^3+ \mathcal{C}_2\eta^{-2}\,.  \label{mKv1}
\end{equation}
Since $t\propto\eta-\sinh(\eta)$ in the open Friedmann models and $t\propto\eta-\sin(\eta)$ in their closed counterparts (e.g.~see~\cite{1983QB981.N3.......}), we have $t\propto\eta^3$ at the $\eta\rightarrow0^+$ limit in both cosmologies. Then, solution (\ref{mKv1}) transforms into
\begin{equation}
\tilde{v}= \mathcal{C}_3t+ \mathcal{C}_4t^{-2/3}= \mathcal{C}_5a^{3/2}+ \mathcal{C}_6a^{-1}\,,  \label{mKv2}
\end{equation}
in agreement with (\ref{lv1}) here and with~\cite{2026ApJ...997...25T} as well. Note that $a\propto1-\cosh(\eta)$ in open Friedmann universes and $a\propto1-\cos(\eta)$ in the closed models (e.g.~see~\cite{1983QB981.N3.......}), both of which lead to $a\propto\eta^2\propto t^{2/3}$ when $\eta\rightarrow0^+$ (as expected).

\underline{\it The case of an open FLRW background:} The hyperbolic geometry of the open FLRW background makes it essentially impossible to obtain analytic linear solutions for the whole evolution of these models. The existing analytic solutions usually cover the late-time evolution, when $\Omega\ll1$, $\eta\gg1$ and $\coth(\eta/2)\rightarrow1$. At this limit, Eq.~(\ref{-1ddotv1}) simplifies to
\begin{equation}
{{\rm d}^2\tilde{v}_a\over{\rm d}\eta^2}+ (1-\Omega){{\rm d}\tilde{v}_a\over{\rm d}\eta}- {3\over2}\,\Omega\tilde{v}_a= 0\,.  \label{-1ltlddottv2}
\end{equation}
Given that $\Omega$ is (nearly) constant during the late-time period under consideration, the above accepts the scale-independent power-law solution of the form
\begin{equation}
\tilde{v}= \mathcal{C}_1{\rm e}^{\alpha_{_1}\eta}+  \mathcal{C}_2{\rm e}^{\alpha_{_2}\eta}\,,  \label{-1ltlv1}
\end{equation}
where
\begin{equation}
\alpha_{1,2}= {1\over2}\left(\Omega-1\pm\sqrt{\Omega^2+4\Omega+1}\right)  \label{alphas1}
\end{equation}
and $0<\Omega\ll1$. Consequently, solution (\ref{-1ltlv1}) contains one growing and one decaying mode (with $\alpha_1>0$ and $\alpha_2<0$ respectively). In addition, since $t\propto\sinh(\eta)-\eta$ in open Friedmann universes and $\sinh(\eta)=({\rm e}^{\eta}-{\rm e}^{-\eta})/2$ always, we have $t\propto{\rm e}^{\eta}$ at late times (i.e.~when $\eta\gg1$). Then, in terms of proper time, solution (\ref{-1ltlv1}) reads
\begin{equation}
\tilde{v}= \mathcal{C}_3t^{\alpha_{_1}}+  \mathcal{C}_4t^{\alpha_{_2}}\,,  \label{-1ltltv2}
\end{equation}
with $\alpha_{1,2}$ still given by (\ref{alphas1}). The growing mode of the above corresponds to $\alpha_1= (\Omega-1+\sqrt{\Omega^2+4\Omega+1})/2$, which suggests that the lower the value of $\Omega$, the slower the linear growth of the peculiar-velocity field. It is also straightforward to show that $\alpha_1<1$ as long as $\Omega\lesssim4/5$, while keeping in mind that $\Omega\ll1$ at late times. Recall that the relativistic analysis has led to $\tilde{v}\propto t$ on spatially flat FLRW backgrounds (see~\S~\ref{sssPVFLRWUOm1} here).\footnote{At late times, the open FLRW model approaches the empty Milne universe. Then, taking the $\Omega\rightarrow0^+$ limit of (\ref{alphas1}), we obtain $\alpha_1\simeq3\Omega/2\simeq0$ and $\alpha_2\simeq -1$, leading to the solution $\tilde{v}=C_3+C_4t^{-1}$ for the linear evolution of the peculiar-velocity field. Therefore, near the Milne limit, linear peculiar-velocities remain essentially constant, which agrees with solution (\ref{-1ltltv2}) at the $\Omega\rightarrow0^+$ limit.}

Although the above solutions cover limiting periods in the evolution of Friedmann universes with hyperbolic spatial geometry, they do suggest that linear peculiar velocities grow slower on open FLRW backgrounds than on their spatially flat counterparts. In fact, the lower the background density parameter, the slower the resulting peculiar growth-rate.

\underline{\it The case of a closed FLRW background:} In contrast to their flat and open counterparts, Friedmann models with spherical spatial geometry are finite both in space and in time (conformal as well as proper). Another key difference is that the second term on the left-hand side of (\ref{+1ddotv1}) changes sign when the closed FLRW background enters its contracting phase. In particular, the coefficient of the aforementioned term is negative throughout the expanding phase (i.e.~when $0<\eta<\pi$), but turns positive once the universe has started to collapse (i.e.~for $\pi<\eta<2\pi$). We therefore expect the peculiar velocities to evolve differently during these two phases. Next, we will put this expectation to the test by looking for analytic solutions at characteristic moments in the lifetime of a spatially closed Friedmann universe.

Focusing on the expanding phase and setting $\eta=\pi/2$, gives $\cot(\eta/2)=1$ at the mid-expansion point. This in turn reduces Eq.~(\ref{+1ddotv1}) to
\begin{equation}
{{\rm d}^2\tilde{v}_a\over{\rm d}\eta^2}+ (1-\Omega){{\rm d}\tilde{v}_a\over{\rm d}\eta}- {3\over2}\,\Omega\tilde{v}_a= 0\,, \label{+1ddotv2}
\end{equation}
with $\Omega>1$. Since we are considering a brief period near the middle of the expansion phase, we may treat $\Omega$ as almost constant. Then, the above accepts the power-law solution
\begin{equation}
\tilde{v}= \mathcal{C}_1{\rm e}^{\alpha_{_1}\eta}+  \mathcal{C}_2{\rm e}^{\alpha_{_2}\eta}\,,  \label{+1ltv1}
\end{equation}
where
\begin{equation}
\alpha_{1,2}= {1\over2}\left(\Omega-1\pm\sqrt{\Omega^2+4\Omega+1}\right)\,,  \label{alphas2}
\end{equation}
guaranteeing one growing and one decaying mode (with $\alpha_1>0$ and $\alpha_2<0$ respectively). The growing mode ensures that the higher the value of the density parameter, the stronger the linear growth of the $\tilde{v}$-field. Although (\ref{+1ltv1}) and (\ref{+1ltv1}) are formally identical to the late-time solution obtained in open FLRW universes, there is a key difference between the two results, because $\Omega<1$ in the $K=-1$ case and $\Omega>1$ here. This allows for growth rates stronger than those obtained in flat and open FLRW models. In fact, when $\Omega\gg1$, we find that $\alpha_1\simeq(1+2\Omega)/2\gg1$. The latter argues for an arbitrarily strong linear growth of the peculiar velocity, if the background density parameter is allowed to increase arbitrarily as well.

Half the way down the contracting phase of a closed Friedmann universe, we have $\eta=3\pi/2$. There, $\cot(\eta/2)=-1$ and Eq.~(\ref{+1ddotv1}) reads
\begin{equation}
{{\rm d}^2\tilde{v}_a\over{\rm d}\eta^2}- (1-\Omega){{\rm d}\tilde{v}_a\over{\rm d}\eta}- {3\over2}\,\Omega\tilde{v}_a= 0\,, \label{+1tv''3}
\end{equation}
giving
\begin{equation}
\tilde{v}= \mathcal{C}_1{\rm e}^{\alpha_{_1}\eta}+  \mathcal{C}_2{\rm e}^{\alpha_{_2}\eta}\,,  \label{+1ltv2}
\end{equation}
where now
\begin{equation}
\alpha_{1,2}= {1\over2}\left(1-\Omega\pm\sqrt{\Omega^2+4\Omega+1}\right)\,.  \label{alphas3}
\end{equation}
As in the expanding phase before, $\alpha_1>0$ and $\alpha_2<0$ mark the growing and decaying modes respectively. Also, the larger/smaller the value of $\Omega$, the stronger/weaker the growth of the peculiar velocities. However, comparing the sets (\ref{+1ltv1}), (\ref{alphas2}) and (\ref{+1ltv2}), (\ref{alphas3}) closer, shows that peculiar-velocity perturbations grow faster during the expanding phase than when the universe is contracting. The reason is that the density parameter of the closed Friedmann models increases when they expand, but decreases when they contract (recall that $\dot{\Omega}=-H(1-\Omega)\Omega$ for dust). This means that, unlike solution (\ref{+1ltv1})-(\ref{alphas2}), the growing mode of  (\ref{+1ltv2})-(\ref{alphas3}) has a finite upper bound. Indeed, in a highly dense FLRW background with $\Omega\gg1$, we find $\alpha_1\leq3/2$  near the middle of its contraction.

At the moment of maximum expansion $\eta=\pi$, $\cot(\eta/2)=0$ and $\mathcal{H}=0$. Close to this ``turning'' point, Eq.~(\ref{+1ddotv1}) gives $\tilde{v}=\tilde{C}_1+\tilde{C}_2\eta$, suggesting linear growth (in terms of conformal time) for the $\tilde{v}$-field. This solution is independent of the value of $\Omega$ and also holds when the closed FLRW background is replaced by the static Einstein model.\footnote{On the Einstein-static background, the homogeneous part of Eq.~(\ref{lddotv1}) reduces to $\ddot{v}_a=0$, which leads to $v\propto t\propto\eta$ on all scales (recall that $\eta\propto t$ in the Einstein-static universe).} This confirms the instability of the Einstein universe against linear peculiar-velocity perturbations as well~\cite{2024PhRvD.110f3540M}.\footnote{The Einstein static universe has long been known to be unstable against linear homogeneous and isotropic density perturbations as well~\cite{1930MNRAS..90..668E}, though its instability is less straightforward in the case of conformal metric and inhomogeneous distortions~\cite{1967RvMP...39..862H,%
1987NuPhB.292..784G,2003CQGra..20L.155B}}

Finally, we should remind the reader that all the linear solutions in this section solve the homogeneous part of differential equation (\ref{lddotv1}). Therefore, for the reasons explained in \S~\ref{sssPVFLRWUOm1}, the growth rates of the peculiar-velocity field are the minimum possible~\cite{2024PhRvD.110f3540M}.

In summary, linear peculiar velocities can grow faster in closed FLRW universes than in their flat counterparts and even faster than in the spatially open ones~\cite{2024PhRvD.110f3540M}. This is still a tentative claim, however, since our analytic solutions do not cover the whole lifetime of Friedmann models with $\Omega\lessgtr1$. Having said that, there is a persistent pattern in all the solutions, which supports the aforementioned claim. Also note that, even if the universe is close to the Euclidean limit globally, our results suggest faster peculiar velocities in locally overdense regions (e.g.~in galaxy clusters) than in underdense domains (e.g.~in the so-called voids)~\cite{2024PhRvD.110f3540M}.

\subsubsection{Peculiar velocities before %%%%%%%%%%%%%%%%%%%%%%%%%%%%%%%%%%%%%%%%%
recombination}\label{sssPVBR}
%%%%%%%%%%%%%%%%%%%%%%%%%%%%%
Prior to recombination perturbations in the baryonic distribution cannot really grow. During the radiation epoch the Jeans length of the (tightly coupled) photon-baryon system essentially equals that of the horizon. Between equilibrium and decoupling, Silk damping essentially erases any variations in the baryon density that may happened to exist (see~\cite{1968ApJ...151..459S} as well as~\cite{1990eaun.book.....K,1995coec.book.....C}). Consequently, since peculiar velocities are sourced by inhomogeneities in the matter, one does not expect appreciable velocity perturbations in the baryonic component. On the other hand, typical CDM species interact neither with the photons nor with the baryons and become non-relativistic in the radiation ere. More specifically, the heavier the dark matter particle, the earlier it ceases being relativistic. In addition, the Landau damping (or free-streaming) scale of typical CDM particles is too small to be of cosmological relevance. Nevertheless, CDM perturbations are not expected grow during most of the radiation era because of the Mezsaros stagnation effect (see~\cite{1974A&A....37..225M}, as well as~\cite{2008PhR...465...61T,2012reco.book.....E} and \S~\ref{ssMSE} here). Only as the universe approaches the time of matter-radiation equality, linear inhomogeneities in the dark matter can start to grow and thus trigger peculiar-velocity perturbations in their sector. Having said that, we should remind the reader that all the aforementioned theoretical studies are essentially Newtonian and there is no fully relativistic analysis in the literature as yet.\footnote{A relativistic study of the peculiar velocities in the pressureless matter component (baryonic or/and CDM) prior to equipartition, which accounts for the gravitational input of the peculiar flux, has been given in~\cite{2022PhRvD.106h3505M}. According to the latter, the associated linear peculiar-velocity field can grow during the late radiation era as well, thought at a rate slower than that of the subsequent dust epoch. Nevertheless, a refined two-fluid approach is required to confirm this result.}

After recombination, any peculiar-velocity fields that might have already developed in the dark matter could enhance those in the baryonic component, in the same way that ``gravitational wells'' in the CDM amplify density perturbations in the baryons.\footnote{Studies involving peculiar velocities in the CDM component, though without focusing on their linear evolution and growth, have also appeared in the recent literature. For instance, drift motions in the CDM sector, within Szekeres-type regions matched on a $\Lambda$CDM background, have been investigated in an attempt to alleviate the Hubble-tension problem~\cite{2021EPJC...81..374N}. Also, studies of dark-matter decay in cosmic voids, involving tilted cosmological models, were recently pursued in~\cite{2021PhRvD.104l3540L}.}

\subsection{Peculiar velocities and inflation}\label{ssPVI}
%%%%%%%%%%%%%%%%%%%%%%%%%%%%%%%%%%%%%%%%%%%%%%%%%%%%%%%%%%%
Large-scale peculiar motions are believed to be a recent addition to the universal kinematics and a byproduct of structure formation. Having said that, it has been speculated that asymmetries, perhaps of dipolar nature, suggesting the existence of a preferred direction analogous to that induced by a large-scale velocity field, might instead have primordial origin~\cite{1991PhRvD..44.3737T,%
2007PhRvD..75h3502A,2008PhRvD..78l3520E,2011PhRvD..83j3002M}.

\subsubsection{The dynamics of de Sitter inflation}\label{sssDdSI}
%%%%%%%%%%%%%%%%%%%%%%%%%%%%%%%%%%%%%%%%%%%%%%%%%%%%%%%%%%%%%%%%%%
Inflationary scenarios are based on the assumption that the very early universe is dominated by scalar fields. The inflationary paradigm, which was introduced in the early 1980s and has been developing ever since, can provide answers to questions regarding the magnetic-monopole, the horizon, the flatness, as well as the initial-perturbations problems (see~\cite{1980ApJ...241L..59K,1980PhLB...91...99S,%
1981PhRvD..23..347G,1982PhLB..108..389L,1982PhRvL..48.1220A,%
1982JETP...56..258M} for an incomplete, though historically representative list). When the scalar fields are minimally coupled to gravity, they behave as effective perfect fluids with energy-momentum tensor~\cite{1988CQGra..5.627M}
\begin{equation}
T_{ab}= \bar{\rho}u_au_b+ \bar{p}h_{ab}\,,  \label{Tab}
\end{equation}
relative to the timelike 4-velocity $u_a$. Then, the effective energy and pressure of the inflaton field ($\phi$) are respectively $\rho= \dot{\phi}^2/2+V$ and $p=\dot{\phi}^2/2-V$, where $V=V(\phi)$ is the potential and the overdots denote derivatives along the $u_a$-field. A key relation is the Klein-Gordon equation
\begin{equation}
\ddot{\phi}+ 3H\dot{\phi}+ \hat{V}= 0\,,  \label{KG}
\end{equation}
which monitors the time evolution of $\phi$ (with $\hat{V}={\rm d}V/{\rm d}\phi$) and coincides with the continuity equation of a minimally coupled scalar field~\cite{2008PhR...465...61T,%
2012reco.book.....E}.

In order to achieve exponential (de Sitter-type) inflation, one needs to impose the slow-rolling conditions. This means demanding that $\dot{\phi}\simeq0$ and $3H\dot{\phi}+\hat{V}\simeq0$ simultaneously. The former makes sure that the inflaton rolls very slowly down its potential, which in turn guarantees that $\bar{p}\simeq-\bar{\rho}\simeq-V$ and consequently exponential expansion. Following (\ref{KG}), the latter condition ensures that $\ddot{\phi}\simeq0$ and that the slow-rolling phase lasts long enough to bring the primordial universe (arbitrarily) close to the homogeneous, isotropic and spatially flat Friedmann model.

\subsubsection{Tilted inflationary models}\label{sssTIMs}
%%%%%%%%%%%%%%%%%%%%%%%%%%%%%%%%%%%%%%%%%%%%%%%%%%%%%%%%%
Consider a perturbed inflationary model and allow for a pair of relatively moving frames as described in \S~\ref{ssRMOs}. When the velocity between the two coordinate systems is not relativistic, the linear relations (\ref{lrels1}) still apply. However, there is a key difference between tilted inflationary models and those filled with conventional matter, because $\bar{p}=-\bar{\rho}$ during the de Sitter regime. Applied to (\ref{lrels1}c), the latter ensures that $\tilde{q}_a=q_a$ throughout the exponential phase. Together with (\ref{lrels1}d), this result implies that, if the cosmic fluid appears perfect to one family of observers, it will look the same to any other relatively moving group of observers as well. Given that minimally coupled scalar fields correspond to perfect fluids, we deduce that in typical inflationary models the peculiar flux vanishes (at the linear level) by default.

The absence of a peculiar flux vector in all frames means that the energy-momentum tensor has no flux-related terms, despite the presence of nonzero peculiar velocities. Put another way, during de Sitter inflation, any peculiar-velocity field that may exist has zero contribution to the stress-energy tensor and therefore to Einstein's equations and to the local gravitational field~\cite{2022PhRvD.106h3505M}. This fact was also observed in the analysis of an unperturbed tilted Bianchi-type cosmology pursued in~\cite{2023EPJC...83..874K}. In a sense, the de Sitter environment has decoupled the peculiar-velocity field from the gravitational field. This has a profound effect on the evolution of peculiar velocities during de Sitter inflation, at least at the linear level (see~\cite{2022PhRvD.106h3505M} and also \S~\ref{sssPVDI} next).

In the presence of pressure, perturbed inflationary universes have nonzero 4-acceleration. More specifically, following~\cite{2008PhR...465...61T,2012reco.book.....E}, we have $A_a=-({\rm D}_a\dot{\phi)}/\dot{\phi}$ in the reference CMB frame and $\tilde{A}_a=-(\tilde{\rm D}_a\phi^{\prime})/ \phi^{\prime}$ in its tilted counterpart.\footnote{Neither $\phi$ nor $V=V(\phi)$ change after a Lorentz boost between frames, since they are both spacetime quantities. Put another way, $\tilde{\phi}=\phi$ and $\tilde{V}=V$ at all perturbative levels.} Then, the linear relation (\ref{lab34}b) guarantees that the two 4-acceleration vectors (and their associated non-gravitational forces) are the sources of peculiar-velocity perturbations during inflation. The latter can therefore survive a phase of exponential expansion as long as $\tilde{A}_a-A_a\neq0$. Let us take a closer look at this possibility.

\subsubsection{Peculiar velocities during inflation}\label{sssPVDI}
%%%%%%%%%%%%%%%%%%%%%%%%%%%%%%%%%%%%%%%%%%%%%%%%%%%%%%%%%%%%%%%%%%%
Throughout the de Sitter phase, linear peculiar-velocity perturbations (if any) evolve according to (\ref{lab34}b). In our inflationary environment, the latter acquires the form
\begin{equation}
\dot{\tilde{v}}_a+ H\tilde{v}_a= \tilde{A}_a- A_a= {1\over\phi^{\prime}}\,\tilde{\rm D}_a\phi^{\prime}- {1\over\dot{\phi}}\,{\rm D}_a\dot{\phi}\,.  \label{dStv'1}
\end{equation}
with $\phi$ and $V=V(\phi)$ being invariant (see footnote~18) under frame changes. However, this is not necessarily the case for their time derivatives and their spatial gradients, since $\phi^{\prime}\neq\dot{\phi}$ and $\tilde{\rm D}_a\phi^{\prime}\neq{\rm D}_a\dot{\phi}$ in general. Having said that, at the linear level, we have
\begin{equation}
\phi^{\prime}= \tilde{u}^a\nabla_a\phi= \dot{\phi}+ \tilde{v}^a{\rm D}_a\phi= \dot{\phi}\,,  \label{phis1}
\end{equation}
given that $\tilde{v}^a{\rm D}_a\phi$ is a second-order perturbation. Using this result and keeping in mind that $\tilde{h}_{ab}=h_{ab}+2u_{(a}\tilde{v}_{b)}$ to first approximation (see \S~\ref{ssLRBFs} earlier) leads to
\begin{equation}
\tilde{\rm D}_a\phi^{\prime}= {\rm D}_a\dot{\phi}+ \ddot{\phi}\tilde{v}_a= {\rm D}_a\dot{\phi}\,,  \label{phis2}
\end{equation}
since $\ddot{\phi}=0$ during the slow-rolling phase (see \S~\ref{sssTIMs} and discussion therein).\footnote{Strictly speaking, the slow-roll conditions are $\dot{\phi}\simeq0$ and $\ddot{\phi}\simeq0$. These ensure that $p\simeq\rho$ throughout the de Sitter phase and that the terms $(p-\rho)\tilde{v}_a$ and $\ddot{\phi}\tilde{v}_a$ seen in Eqs.~(\ref{lrels1}c) and (\ref{phis2}) are second-order perturbations. All this guarantees the consistency and the generality of our linear analysis.}

Combining relations (\ref{dStv'1})-(\ref{phis2}), the former of the three reduces to the source-free linear propagation formula~\cite{2022PhRvD.106h3505M}
\begin{equation}
\tilde{v}_a^{\prime}+ H\tilde{v}_a= 0\,,  \label{dStv'2}
\end{equation}
which solves immediately to give
\begin{equation}
\tilde{v}_a\propto a^{-1}\propto {\rm e}^{-H_0t}\,.  \label{dStv}
\end{equation}
Note that $H=H_0=$~constant all along de Sitter inflation and that the zero suffix marks the start of the exponential expansion. Following (\ref{dStv}), during de Sitter inflation linear peculiar velocities decay exponentially, so that at the end of that era
\begin{equation}
\tilde{v}_f= \tilde{v}_0{\rm e}^{-\mathcal{N}}\,,  \label{dSvf}
\end{equation}
where $\mathcal{N}$ is the number of e-folds~\cite{2022PhRvD.106h3505M}. Assuming the minimum required period of inflation, namely setting $\mathcal{N}\simeq60$, the peculiar-velocity field drops by about 26 orders of magnitude. This goes up to roughly 43 orders of magnitude for a 100 e-fold inflation, and so on.

In summary, there are no linear sources of peculiar velocity perturbations during de Sitter inflation. Moreover, any 4-velocity ``tilt'' that might had been present at the onset of inflation, will quickly decay away by the end of the exponential expansion era, leaving the post-inflationary universe free of peculiar velocities. The reason is traced to the fact that $\tilde{q}_a=q_a$ throughout the de Sitter phase. This linear result, which is a direct consequence of the inflationary equation of state (i.e.~$p=-\rho$), ensures that the energy-flux vector vanishes in all relatively moving frames, if it happens to be zero in one of them (see relation (\ref{lrels1}c) in \S~\ref{ssLRBFs}). When $\tilde{q}_a=0=q_a$, the momentum conservation law (see Eq.~(\ref{mdcl}) in \S~\ref{sssCLs} and recall that $\tilde{p}=p$ to first approximation) guarantees that $\tilde{A}_a=A_a$. This ensures the source-free nature of (\ref{dStv'1}) and the exponential decay of linear peculiar-velocity perturbations during de Sitter inflation.

\subsection{Peculiar velocities and dark-energy}\label{ssPVDE}
%%%%%%%%%%%%%%%%%%%%%%%%%%%%%%%%%%%%%%%%%%%%%%%%%%%%%%%%%%%%%%
In accordance with the current cosmological model, namely with the $\Lambda$CDM paradigm, the Einstein-de Sitter period of the post-recombination universe is followed by a late phase of accelerated expansion. The latter is triggered by the dominance of an as yet unknown and elusive matter field, commonly referred to as dark energy. Intuitively speaking, once the $\Lambda$-dominated phase of the universe starts, one expects the accelerated expansion to suppress the growth of the peculiar-velocity perturbation, as it does with essentially all types of distortions.

A proper analysis of the peculiar-velocity evolution during the recent dark-energy epoch requires a two-fluid approach, analogous to that given in \S~\ref{ssLRT} below. More specifically, the material component of the late-time universe should be treated a mixture of the dominant dark-energy component and of the subdominant baryonic matter and/or CDM. Although, the detailed analysis is still missing, the numerical study of~\cite{2016MNRAS.463..512D} reached the conclusion that dark energy inhibits the growth of peculiar velocities, when compared to the preceding Einstein-de Sitter era. This result is also corroborated by the theoretical analysis of~\cite{2022PhRvD.106h3505M}, as outlined in \S~\ref{sssPVDI} here, which found exponential decay for the linear peculiar-velocity field during a period of de Sitter inflation. On these grounds, one would expect to see the linear peculiar velocities to grow at a slower pace relative to the Einstein-de Sitter epoch, if not to decay.

When combining this expectation with the faster growing rates for the linear peculiar-velocity field of the relativistic analysis (see \S~\ref{ssRA} before), one expects to measure bulk-flow velocities faster than anticipated, but also to see them drop at relatively low redshifts, as the accelerated expansion starts to dominate the kinematics of our cosmos~\cite{2026ApJ...997...25T}. Interestingly, bulk-flow profiles with a late-time velocity drop have been reported in~\cite{2011MNRAS.414..264C} and more recently in~\cite{2023MNRAS.524.1885W} (see Fig.~1 in~\cite{2026ApJ...997...25T}, as well as Fig.~\ref{fig:Colin} and Fig.~\ref{fig:Watkins} in \S~\ref{ssBFELCDMLs} below). Although more work is required, it is conceivable that the bulk-velocity decay detected in~\cite{2011MNRAS.414..264C,2023MNRAS.524.1885W} could reflect/mark the onset of the recent dark-energy era. This possibility may be also supported by the fact that, so far at least, the surveys that report slower bulk velocities seem confined to relatively low redshifts (corresponding to scales smaller than $\sim100$~Mpc), whereas those reporting faster bulk flows appear to extend beyond the aforementioned threshold (see related discussion in \S~\ref{sPMLCDM} below).

We will close this section with a brief reference to the numerical treatments that aim to simulate the evolution structure formation. In the recent literature one can find a number of such works that use relativistic techniques in their study. Nevertheless, to the best of our knowledge, the available simulations do not focus on the evolution and the implications of the peculiar velocities reviewed here. Perhaps the only exception are a couple of works that simulated the rotational component of the peculiar-velocity field (e.g.~see~\cite{2018JCAP...09..006J,2021MNRAS.501.5697B}). However, the peculiar vorticity, as well as the peculiar shear, grow slower that the peculiar velocity itself~\cite{2020EPJC...80..757T}, so their effects are considerably weaker. Moreover, almost all of the simulations operate during the recent $\Lambda$-dominated phase of universe, where the accelerated expansion inhibits the growth of the peculiar-velocity field~\cite{2018JCAP...09..006J,%
2021MNRAS.501.5697B,2019PhRvD.100b1301A,2019PhRvD..99f3522M,%
2020JCAP...01..007B,2025LRR....28....5A}. Having said that, the advent of the relativistic numerical simulations is a very recent one and the codes are still under development. So, there is considerable room for improvement. In this respect, the exact analytical results reviewed in this section can provide a valuable testing ground for the existing and the upcoming simulations. The fact that the aforementioned analytical studies are also gauge invariant is an additional advantage, since their solutions are free of spurious gauge-depended modes. Recall that the numerical studies rely on particular gauge-choices to address the problems in hand.

\subsection{Cosmologies with a generic ``tilt''}\label{ssCGT}
%%%%%%%%%%%%%%%%%%%%%%%%%%%%%%%%%%%%%%%%%%%%%%%%%%%%%%%%%%%%%
So far our discussion has been about cosmologies where the velocity tilt is a relatively recent addition to the universal kinematics, triggered by the ongoing process of structure formation and reflecting the resulting peculiar-velocity field. In the literature there are also generically tilted cosmological models, equipped with two non-comoving 4-velocity fields ab initio. These cosmologies, which typically form a special subgroup of the spatially homogeneous Bianchi spacetimes, allow for two 4-velocity fields with a hyperbolic angle between them (e.g.~see~\cite{1979PhR....56...65C,1997dsc..book.....W}). One of the 4-velocities is normal to the hypersurfaces of homogeneity (i.e.~of constant time), while the other is not.

In the related studies, the aforementioned tilt is typically treated as a generic feature of the host spacetime and does not necessarily represent a physical peculiar velocity. Having said that, one can always associate to the tilt angle ($\beta$) an effective peculiar-velocity field ($\tilde{v}_a$, or $v_a$) by means of the Lorentz transformations (\ref{Lboost}) seen in \S~\ref{ssRMOs}. Then, the ``size'' of the tilt is monitored by the Lorentz-boost factor between the 4-velocities involved. More specifically, in a effective two-fluid system we have $\cosh\beta=\gamma\geq1$, which immediately implies that
\begin{equation}
\beta= \cosh^{-1}(\gamma)= \ln\left(\gamma+\sqrt{\gamma^2-1}\right)\geq 0\,.  \label{beta1}
\end{equation}
Recalling that $\gamma^2=1/(1-v^2)$ and therefore that $\gamma^2-1=\gamma^2v^2$, with $v^2=\tilde{v}^2$, one may successively recast the above as
\begin{equation}
\beta= \ln\left[\gamma(1+v)\right]= \ln\gamma+ \ln(1+v)= {1\over2}\left[\ln(1+v)-\ln(1-v)\right]\,.  \label{beta2}
\end{equation}
Then, a simple Taylor-expansion of the logarithmic functions leads to
\begin{equation}
\beta= v+ {1\over3}\,v^3+ {1\over5}\,v^5+ \cdots\,,  \label{beta3}
\end{equation}
which diverges at the relativistic limit (i.e.~for $v=1$). When $v\ll1$, on the other hand, we have $\beta\simeq v$ and the tilt angle grows in tune with the associated (effective) linear peculiar-velocity field (see \S~\ref{ssRA} and \S~\ref{ssPVI} earlier). To be precise, the time derivative of (\ref{beta3}), gives
\begin{equation}
\dot{\beta}= \left(1+v^2+v^5+\cdots\right)\dot{v}\,.  \label{dotbeta}
\end{equation}
Consequently, as expected, the tilt angle grows in time as long as the corresponding (effective or physical) peculiar velocity does the same. Note that the above relations are fully nonlinear and hold in a general spacetime equipped with an arbitrary 4-velocity tilt. In other words, the associated effective peculiar-velocity field does not need to be weak. Clearly, when $v\ll1$, expression (\ref{dotbeta}) reduces to $\dot{\beta}\simeq\dot{v}$ at the linear level.

Tilted cosmological models are naturally associated with the presence of dipolar anisotropies, which may reflect the generic anisotropy of the host spacetine, or they can be mere artifacts of the observers' peculiar motion. In the former case, one can use the tilted models to investigate (among others) potential violations of the cosmological principle. This possibility has received considerable interest from recent observations reporting dipole-like anisotropies in the number counts of distant astrophysical sources (e.g.~see~\cite{2021ApJ...908L..51S,2022MNRAS.511.1819S,%
2022ApJ...937L..31S,2025RvMP...97d1001S} and references therein). Theoretical studies of this nature were recently performed in the context of the \textit{dipole-cosmology} scenario (e.g.~see~\cite{2023JCAP...07..020K,2023EPJC...83..874K,%
2024CQGra..41n5007E}). The latter generalises the spatially homogeneous and isotropic Friedmann models, by allowing for a preferred spatial direction. In practice, the dipole-cosmology scenarios employ spatially homogeneous but anisotropic spacetimes of the Bianchi type, equipped with up to three scale factors and a 4-velocity tilt ($\beta$) like the one defined above. It has been claimed that the hyperbolic tilt-angle ($\beta$), as well as the dipolar asymmetries related to it, can grow even when the model's global anisotropy fades away~\cite{2023JCAP...07..020K}.

Assuming, on the other hand, that the dipolar anisotropies are deceptive side-effects of the observers' peculiar flow relative to the cosmic rest-frame, one can analyse their implications within the realm of the \textit{tilted universe} scenario (see \S~\ref{sOSPMs} next).

\section{Observational signatures of peculiar motions}\label{sOSPMs}
%%%%%%%%%%%%%%%%%%%%%%%%%%%%%%%%%%%%%%%%%%%%%%%%%%%%%%%%%%%%%%%%%%%%
Relative motions have long been known to ``contaminate'' the observations and ``interfere'' with their interpretation. When looking back into the history of Astronomy, we can find a number of incidents where relative-motion effects have led us to gross misinterpretations of reality, some of which lasted for centuries, if not millennia (see \S~\ref{ssHMRMEs} below). Here, we will consider the possibility that analogous misinterpretations can occur in nowadays cosmology.

\subsection{Relative-motion effects on the deceleration %%%%%%%%%%%%%%%%%%%%%%%%%%%%%%%%%%%%%%%%%%%%%%%%%%%%%%%
parameter}\label{ssRMEDP}
\begin{figure*}
\begin{center}
\includegraphics[height=2in,width=4in,angle=0]{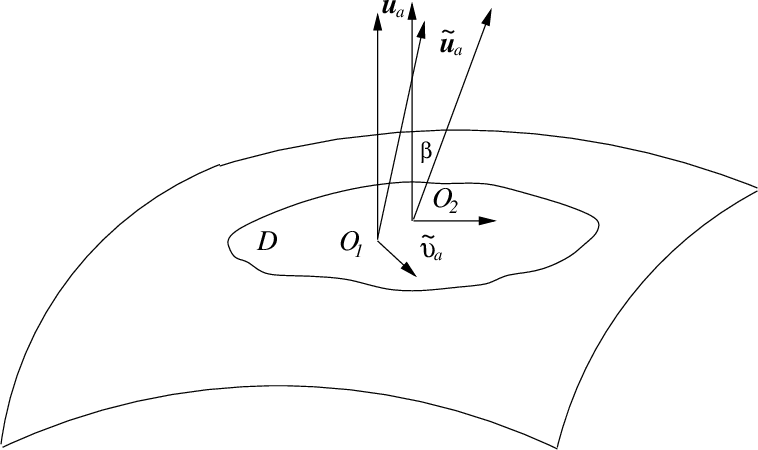}\quad
\end{center}
\caption{Realistic observers ($O_1$, $O_2$) inside a bulk-flow domain ($D$), moving with peculiar velocity $\tilde{v}_a$ relative to their idealised CMB counterparts.  The 4-velocities $u_a$ and $\tilde{u}_a$, with a hyperbolic ``tilt'' angle ($\beta$) between them, define the reference (CMB) frame of the universe and that of the bulk peculiar motion respectively (see Eqs.~(\ref{4vels4vels})).}  \label{fig:bflow}
\end{figure*}

%%%%%%%%%%%%%%%%%%%%%%%%%
We are moving observers in the universe. Our Milky Way and the local group of galaxies move at approximately 600~km/sec relative to the CMB frame. The latter is believed to define the cosmic rest-frame, namely the only coordinate system where the CMB dipole vanishes. Moreover, bulk peculiar flows, with sizes varying between few hundred and several hundred Mpc and speeds ranging from few hundred to several hundred km/sec, are commonplace in the universe. It is therefore fair to say that no realistic observer follows the idealised CMB frame, but we all have some finite peculiar velocity relative to it (see Fig.~\ref{fig:bflow}).

Following the discussion given in \S~\ref{ssLRBFs}, observers in relative motion generally disagree on their measurements, even at the linear level. In particular, the expansion and the deceleration/acceleration rates of the universe, as measured by the idealised (CMB) observers and by their real counterparts living in a typical galaxy and following the tilted frame of the matter are related by (see Eq.~(\ref{lab34}a) in \S~\ref{ssLRBFs})
\begin{equation}
\tilde{\Theta}= \Theta+ \tilde{\vartheta} \hspace{15mm} {\rm and} \hspace{15mm} \tilde{\Theta}^{\prime}= \dot{\Theta}+ \dot{\tilde{\vartheta}}\,,  \label{lab11}
\end{equation}
respectively. Recall that $\tilde{\Theta}=3\tilde{H}$ and $\Theta=3H$, where $\tilde{H}$ and $H$ are the values of the Hubble parameter measured in the tilted and the CMB frame respectively. Also, $\tilde{\vartheta}={\rm D}^a\tilde{v}_a$ is the local volume scalar of the peculiar flow, with $|\tilde{\vartheta}|\ll\Theta$ throughout the linear regime. Note that $\tilde{\vartheta}$ can take positive (as well as negative) values, depending on whether the flow is locally expanding (or contracting). Finally, we remind the reader that overdots indicate time derivatives in the coordinate system of the CMB and primes in that of the tilted observers.

Relations (\ref{lab11}) show that the kinematic differences between the two frames are entirely due to their relative motion. The same expressions also guarantee that the deceleration parameters measured in the CMB and the tilted frame will differ as well. Indeed, recalling that
\begin{equation}
\tilde{q}= -\left(1+{3\tilde{\Theta}^{\prime}\over\tilde{\Theta}^2}\right) \hspace{15mm} {\rm and} \hspace{15mm} q= -\left(1+{3\dot{\Theta}\over\Theta^2}\right)\,,  \label{lqsdef}
\end{equation}
by definition and using the linear relations (\ref{lab11}), we arrive at~\cite{2010MNRAS.405..503T,2011PhRvD..84f3503T}
\begin{equation}
1+\tilde{q}= \left(1+q-{\dot{\tilde{\vartheta}}\over3H^2}\right) \left(1+{\tilde{\vartheta}\over3H}\right)^{-2}\,,  \label{lqs0}
\end{equation}
which ensures that $\tilde{q}\neq q$ because of relative-motion effects alone. Given that $|\tilde{\vartheta}/H|\ll1$ at the linear perturbative level, the above reduces to~\cite{2015PhRvD..92d3515T,2021EPJC...81..753T}
\begin{equation}
\tilde{q}= q- {\dot{\tilde{\vartheta}}\over3H^2}\,. \label{lqs1}
\end{equation}
Therefore, the difference between the deceleration parameter ($\tilde{q}$) measured locally by the bulk-flow observer and the one ($q$) measured in the CMB frame, namely the deceleration parameter of the universe as a whole, is determined by the time derivative of the local volume scalar ($\tilde{\vartheta}$). The latter is given by the ``peculiar'' Raychaudhuri equation, which in a tilted almost-FLRW universe takes the form~\cite{2002PhRvD..66l4015E,%
2013PhRvD..88h3501T}
\begin{equation}
\dot{\tilde{\vartheta}}= -H\tilde{\vartheta}+ {\rm D}^a\dot{\tilde{v}}_a\,,  \label{lpRay1}
\end{equation}
to leading order. On using the linear relation (\ref{lab34}b), while keeping in mind that in our approach we have set $\tilde{q}_a=0=\tilde{A}_a$, the above recasts as
\begin{equation}
\dot{\tilde{\vartheta}}= -2H\tilde{\vartheta}+ {\rm D}^aA_a\,,  \label{lpRay2}
\end{equation}
which substituted back into the right-hand side of Eq.~(\ref{lqs1}) gives
\begin{equation}
\tilde{q}= q+ {1\over3H^2}\,{\rm D}^aA_a\,,  \label{lqs2}
\end{equation}
since $|\tilde{\vartheta}|/H\ll1$ throughout the linear phase. Consequently, the ``correction term'' separating two two deceleration parameters depends entirely on the peculiar 4-acceleration, which is nonzero even in the absence of pressure (see expression (\ref{lcls}b) in \S~\ref{sssKRPF}). As we shall see next, the general form of the 4-Acceleration, and therefore that of the correction term on the right-hand side of the above, depend on the specifics of the background cosmology. Nevertheless, as long as one confines to subhorizon scales, the Einstein-de Sitter picture prevails.

\subsubsection{The key role of the peculiar
%%%%%%%%%%%%%%%%%%%%%%%%%%%%%%%%%%%%%%%%%%%
4-acceleration}\label{sssKRP4A}
%%%%%%%%%%%%%%%%%%%%%%%%%%%%%%%
As stated in \S~\ref{sssKRPF} earlier, in the presence of peculiar motions the cosmic medium can no longer be treated as perfect. The imperfection, which appears in the form of the peculiar flux ($q_a=\rho\tilde{v}_a$ -- see (\ref{lrels1}a)), also guarantees a nonzero 4-acceleration even in the absence of pressure. This \textit{peculiar 4-acceleration} follows naturally from the energy and the momentum conservation laws (see Eqs.~(\ref{edcl}) and (\ref{mdcl}) in~\S~\ref{sssCLs} here). On our Einstein-de Sitter background these formulae linearise to
\begin{equation}
\dot{\rho}= -\Theta\rho- {\rm D}^aq_a \hspace{15mm} {\rm and} \hspace{15mm} \rho A_a= -\dot{q}_a- 4Hq_a\,,  \label{lcls}
\end{equation}
respectively.\footnote{Recalling that $q_a=\rho\tilde{v}_a$ (see expression (\ref{lrels1}a) above), it is straightforward to demonstrate the consistency between Eqs.~(\ref{lrels1}b) and (\ref{lcls}b).} The flux terms on the right-hand sides of the above reflect the purely general-relativistic contribution of the peculiar flux to the (perturbed) energy-momentum tensor (noted in \S~\ref{ssLRBFs} and then discussed in \S~\ref{sssKRPF} earlier). The latter feeds into the Einstein field equations and leads to the flux-related terms in the linear conservation laws seen above.

The momentum conservation law (\ref{lcls}b) ensures a nonzero peculiar 4-acceleration and the conservation law of the energy provides its form. Indeed, taking the gradient of (\ref{lcls}a) and using the linear commutation law ${\rm D}_a\dot{\rho}=({\rm D}_a\rho)^{\cdot}- \dot{\rho}A_a+H{\rm D}_a\rho$, together with the background relation $\dot{\rho}= -3H\rho$, leads to the linear evolution formula
\begin{equation}
\dot{\Delta}_a= -\mathcal{Z}_a- 3aHA_a- a{\rm D}_a\tilde{\vartheta}\,,  \label{ldDel}
\end{equation}
of the density inhomogeneities.\footnote{Alternatively, one can obtain Eq.~(\ref{ldDel}) by linearising around the Einstein-de Sitter background the nonlinear expression (2.3.1) of~\cite{2008PhR...465...61T}, or Eq.~(10.101) of~\cite{2012reco.book.....E}, while keeping in mind that the 4-acceleration is now given by (\ref{lcls}b) above.} The latter are described by the dimensionless spatial gradient $\Delta_a=(a/\rho){\rm D}_a\rho$, while the gradient $Z_a=a{\rm D}_a\Theta$ monitors inhomogeneities in the universal expansion. Solving the above for the 4-acceleration, immediately gives
\begin{equation}
A_a= -{1\over3H}\,{\rm D}_a\tilde{\vartheta}- {1\over3aH}\left(\dot{\Delta}_a +\mathcal{Z}_a\right)\,.  \label{lA}
\end{equation}
Therefore, the momentum conservation law (\ref{lcls}b) guarantees the presence of a nonzero peculiar 4-acceleration and the energy conservation law (\ref{lcls}a) provides the relativistic expression (\ref{lA}) for it. Crucially, the latter holds even in the absence of pressure.

\subsubsection{The Einstein-de Sitter case}\label{sssE-dSC}
%%%%%%%%%%%%%%%%%%%%%%%%%%%%%%%%%%%%%%%%%%%%%%%%%%%%%%%%%%%
After equipartition the universe is believed to be close to the Einstein-de Sitter model, which is a flat FLRW cosmology with zero pressure. Perturbing the above spacetime and allowing for non-relativistic peculiar velocities, the associated 4-acceleration is given by the linear expression (\ref{lA}) above. Substituting the latter into the right-hand side of (\ref{lqs2}) gives
\begin{equation}
\tilde{q}= q- {1\over9H^3}\,{\rm D}^2\tilde{\vartheta}- {1\over9a^2H^2}\left({\dot{\Delta}\over H}+{\mathcal{Z}\over H}\right)\,,  \label{lqs3}
\end{equation}
with ${\rm D}^2={\rm D}^a{\rm D}_a$ being the 3-D covariant Laplacian operator. Also, $\Delta=a{\rm D}^a\Delta_a$ monitors scalar perturbations in the matter distribution and $\mathcal{Z}=a{\rm D}^a\mathcal{Z}_a$ describes scalar perturbations in the universal expansion (see~\cite{2008PhR...465...61T,%
2012reco.book.....E} for details). The Laplacian on the right-hand side of (\ref{lqs3}) implies scale dependence. Indeed, after a simple harmonic splitting the above reads
\begin{equation}
\tilde{q}= q+ {1\over9}\left({\lambda_H\over\lambda}\right)^2 {\tilde{\vartheta}\over H}- {1\over9}\left({\lambda_H\over\lambda_K}\right)^2 \left({\dot{\Delta}\over H}+{\mathcal{Z}\over H}\right)\,,  \label{lqs4}
\end{equation}
where $\lambda_H=1/H$ is the Hubble radius, $\lambda$ is the bulk-flow scale and $\lambda_H=a/|K|$ is the curvature scale of the universe. According to current observations $\lambda_H/\lambda_K\ll1$, which makes the last term on the right-hand side of the above redundant. Then, (\ref{lqs4}) reduces to~\cite{2015PhRvD..92d3515T,%
2021EPJC...81..753T}
\begin{equation}
\tilde{q}= q+ {1\over9}\left({\lambda_H\over\lambda}\right)^2 {\tilde{\vartheta}\over H}\,.  \label{lqs5}
\end{equation}
This is the linear relation between the local deceleration parameter ($\tilde{q}$) measured by the tilted (the bulk-flow) observers and the deceleration parameter ($q$) of the universe itself. The correction term is entirely due to the relative motion of the two frames, since $\tilde{\vartheta}=0$ when there are no peculiar velocities. Note that the overall effect of the correction term is more sensitive to the scale ratio ($\lambda_H/\lambda$), rather that to the velocity ratio ($\tilde{\vartheta}/H$). This ensures that even bulk flows with $\tilde{\vartheta}/H\ll1$ can have a strong effect on the locally measured deceleration parameter ($\tilde{q}$) on subhorizon scales with $\lambda_H/\lambda\gg1$. By looking at the form of (\ref{lqs5}), one can also notice the analogy between the linear effect of peculiar-velocity perturbations on the deceleration parameter and that of pressure-gradient perturbations on linear density inhomogeneities, which leads to the familiar Jeans length (see~\cite{2021EPJC...81..753T} for further discussion).

The scale dependence of the correction term on the right-hand side of Eq.~(\ref{lqs5}) also shows that the relative-motion effect on $\tilde{q}$ decreases as one moves out to progressively larger lengths. This is to be expected, since peculiar velocities are expected to fade away on increasing length scales. An additional, potentially very important, conclusion following from (\ref{lqs5}) is that the overall effect on $\tilde{q}$ also depends on the sign of the local volume scalar ($\tilde{\vartheta}$). The latter takes positive values in locally expanding bulk flows, but turns negative when the latter is contracting. This means that $\tilde{q}<q$ inside contracting bulk peculiar flows. Therefore, according to (\ref{lqs5}), observers in (slightly) contracting bulk flows with sizes much smaller than the horizon, namely with $\lambda_H/\lambda\gg1$, could assign negative values to their deceleration parameter, while the host universe is globally decelerating. Such unsuspecting observers are likely to erroneously believe that their universe has recently entered a phase of accelerated expansion, by misinterpreting a local effect as a recent global event~\cite{2015PhRvD..92d3515T,%
2021EPJC...81..753T}.\footnote{The term ``unsuspecting'' refers to observers who are unaware of their peculiar motion, or do not account for it. By default, these observers are oblivious to the fact that, among others, the deceleration parameter ($\tilde{q}$) measured inside their bulk flow has been ``contaminated'' by relative-motion effects. At the opposite end of the spectrum are the ``informed'' observers, who are aware of their peculiar motion and also account for it. The former are prone to misinterpret the local contraction of their bulk flow as acceleration of the surrounding universe. The latter are likely to realise that their unsuspecting partners were deceived by the own relative motion. In an everyday analogy, the unsuspecting observers are like passengers on a train, or in car, who confuse the deceleration of their own vehicle as acceleration of those next to them. For a discussion on a range of kinematic deceptions triggered by peculiar motions on cosmological scales, the interested reader is referred to~\cite{2025arXiv250104680T}.}

\subsubsection{The transition length}\label{sssTL}
\begin{figure*}
\begin{center}
\includegraphics[height=2in,width=4in,angle=0]{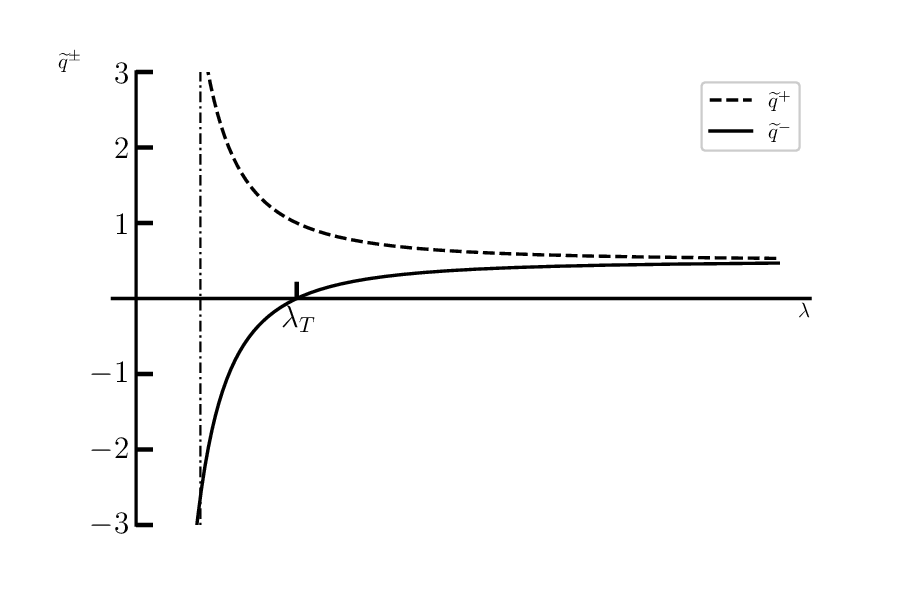}\quad
\end{center}
\caption{The transition length $\lambda_T$ on an Einstein-de Sitter background with $q=1/2$. On scales much larger than $\lambda_T$ the relative-motion effects are negligible and $\tilde{q}^{\,\pm}\rightarrow1/2$. Scales close and inside the transition scale, on the other hand, are heavily contaminated by the observers peculiar motion. There, $\tilde{q}^{\,+}$ becomes increasingly more positive (dashed curve), while $\tilde{q}^{\,-}$ turns negative at $\lambda_T$ and keeps decreasing on progressively smaller wavelengths (solid curve). The vertical line marks the nonlinear cutoff, where the linear approximation is expected to break down~\cite{2021EPJC...81..753T}. In both cases the (unsuspecting) observer is located at the $\lambda=0$ point.}  \label{fig:tq+-}
\end{figure*}

%%%%%%%%%%%%%%%%%%%%%%%%%%%%%%%%%%%%%%%%%%%%%%%%%%
The correction term dominates the right-hand side of (\ref{lqs5}) when it equals (in absolute value) the deceleration parameter ($q$) of the universe. This happens at the critical length~\cite{2021EPJC...81..753T}
\begin{equation}
\lambda_T= {1\over3}\sqrt{{|\tilde{\vartheta}|\over qH}} \lambda_H\,,  \label{lambdaT1}
\end{equation}
keeping in mind that $\tilde{\vartheta}\gtrless0$ in general and that $q>0$ in Friedmann universes with conventional matter and no cosmological constant. On using the above, the linear relation (\ref{lqs5}) reads
\begin{equation}
\tilde{q}= q\left[1\pm\left({\lambda_T\over\lambda}\right)^2\right]\,.  \label{eq:lqs6}
\end{equation}
with the $+/-$ sign corresponding to locally expanding/contracting bulk flows. Accordingly, $\tilde{q}>q$ when $\tilde{\vartheta}>0$ and $\tilde{q}<q$ when $\tilde{\vartheta}<0$. More specifically, following (\ref{eq:lqs6}), we may distinguish between three very characteristic cases, namely
\begin{equation}
\tilde{q}^{\pm}\rightarrow q\,, \hspace{5mm} {\rm when} \hspace{5mm} \lambda\gg\lambda_T\,,  \label{lqEdS}
\end{equation}
\begin{equation}
\tilde{q}^+> 2q\,, \hspace{5mm} {\rm when} \hspace{5mm} \lambda<\lambda_T \hspace{5mm} {\rm and} \hspace{5mm} \tilde{\vartheta}> 0  \label{lq+}
\end{equation}
and
\begin{equation}
\tilde{q}^-< 0\,, \hspace{5mm} {\rm when} \hspace{5mm} \lambda<\lambda_T \hspace{5mm} {\rm and} \hspace{5mm} \tilde{\vartheta}< 0\,.  \label{lq-}
\end{equation}
where $\tilde{q}^+$ is associated with a locally expanding bulk flow and $\tilde{q}^-$ with a contracting one. Note that, in the third case, the critical length ($\lambda_T$) also marks the \textit{transition length}, that is the scale where the deceleration parameter measured by the unsuspecting bulk-flow observers appears to turn negative (e.g.~see Fig.~\ref{fig:tq+-} here and also~\cite{2021EPJC...81..753T}). Clearly, this is the most intriguing of the three cases, since it opens the possibility the recent accelerated expansion of the universe to be a mere illusion triggered by our galaxy's peculiar motion.

\subsubsection{Estimating $\tilde{q}$ and
%%%%%%%%%%%%%%%%%%%%%%%%%%%%%%%%%%%%%%%%%
$\lambda_T$}\label{sssEqlT}
%%%%%%%%%%%%%%%%%%%%%%%%%%%
The available surveys provide the size and the mean velocity of the bulk flow. Currently, there seems to be a general agreement on the direction of the reported flows, but not on their sizes and speeds. In what follows, we will use data from a representative sample of these surveys to estimate the values of $\tilde{q}$ and $\lambda_T$. According to (\ref{lqs5}) and (\ref{lambdaT1}), the latter are determined by the local volume scalar, which provides the spatial divergence of the bulk flow velocity (recall that $\tilde{\vartheta}={\rm D}^a\tilde{v}_a$). Here, we will use standard dimensional analysis arguments to approximate $\tilde{\vartheta}$ with the ratio $\pm\sqrt{3}\langle\tilde{v}\rangle/\lambda$, where $\langle\tilde{v}\rangle$ and $\lambda$ are respectively the reported mean velocity and scale~\cite{2015PhRvD..92d3515T,2021EPJC...81..753T}. Then, allowing for both locally expanding and contracting bulk flows, expressions (\ref{lqs5}) and (\ref{lambdaT1}) recast as
\begin{equation}
\tilde{q}^{\pm}\simeq q\pm {\sqrt{3}\over9}\left({\lambda_H\over\lambda}\right)^2 {\langle\tilde{v}\rangle\over v_H} \hspace{15mm} {\rm and} \hspace{15mm} \lambda_T\simeq {1\over3}\sqrt{{\sqrt{3}\langle\tilde{v}\rangle\over qv_H}} \lambda_H\,,  \label{estqlT}
\end{equation}
respectively. Note that the $+/-$ signs in the former of the above indicate expanding/contracting bulk flows, the mean peculiar velocity ($\langle\tilde{v}\rangle$) takes always positive values and $v_H=\lambda H$ is the Hubble velocity on the scale ($\lambda$) of the reported bulk flow.

One can now use (\ref{estqlT}) to estimate the values of $\tilde{q}^{\pm}$ and $\lambda_T$ by turning to the observations (see Table~\ref{tab1}). These claim velocities between 250 and 430~km/sec over regions with radius ranging from 150 to 280~Mpc. Assuming slightly expanding bulk flows, we find that the deceleration parameters measured by observers located at the centre of these bulk flows lie in the range $+1.25\lesssim \tilde{q}^+\lesssim+5.90$ (3rd column in Table~\ref{tab1}). For slightly contracting peculiar motions, on the other hand, Eq.~(\ref{estqlT}a) assigns negative values to the local deceleration parameter, with $-4.90\lesssim \tilde{q}^-\lesssim-0.25$ (4th column in Table~\ref{tab1}). In the latter case, $\lambda_T$ also marks the transition length, where the deceleration parameter crosses the $\tilde{q}=0$ divide. These vary between 300~Mpc and 490~Mpc (5th column in Table~\ref{tab1}). Note that, in all cases, the universe is assumed to decelerate globally, with $q=+0.5$ in the (reference) CMB frame. Therefore, the over-deceleration, or the acceleration, seen in Table~\ref{tab1} are mere local artifacts of the observers' relative motion. Nevertheless, the affected scales are large enough ($\lambda_T$ is of the order of few to several hundred Mpc) to create the false impression of a recent global change in the expansion of the universe as a whole.

\begin{table}
\caption{Representative estimates of the deceleration parameter ($\tilde{q}^{\pm}$ -- see Eq.~(\ref{estqlT}a)) measured in the rest-frame of some of the reported bulk flows (see \S~\ref{sPMLCDM} below), with $\tilde{q}^+$ corresponding to slightly expanding and $\tilde{q}^-$ to slightly contracting bulk motions. In the latter case, $\lambda_T$ (see last column) marks the transition length where the sign of $\tilde{q}^-$ turns negative (see Eqs.~(\ref{eq:lqs6}), (\ref{lq-}) and also Fig.~\ref{fig:tq+-}). Note that in all cases the background universe is assumed to decelerate with $q=+0.5$. Finally, we have set $H\simeq70$~km/sec\,Mpc and $\lambda_H=1/H\simeq4\times10^3$~Mpc today.}\vspace{5pt}
\begin{center}\begin{tabular}{cccccccc}
\hline \hline & \hspace{-20pt} Survey & $\lambda$ & $\langle\tilde{v}\rangle$ & $\tilde{q}^+$ & $\tilde{q}^-$ & $\lambda_T$ &\\ \hline \hline & $\begin{array}{c} \hspace{-20pt} {\rm Nusser\,\&\,Davis}~\cite{2011ApJ...736...93N} \\ \hspace{-20pt} {\rm Feldman,\,et\,al}~\cite{2010MNRAS.407.2328F} \\ \hspace{-20pt} {\rm Colin,\,et\,al}~\cite{2011MNRAS.414..264C} \\ \hspace{-20pt} {\rm Watkins,\,et\,al}~\cite{2023MNRAS.524.1885W} \\ \hspace{-20pt} {\rm Watkins,\,et\,al}~\cite{2023MNRAS.524.1885W} \\ \hspace{-20pt} {\rm Whitford,\,et\,al}~\cite{2023MNRAS.526.3051W} \end{array}$ & $\begin{array}{c} 150~{\rm Mpc} \\ 150~{\rm Mpc} \\  250~{\rm Mpc} \\ 210~{\rm Mpc} \\ 280~{\rm Mpc} \\ 250~{\rm Mpc} \end{array}$ & $\begin{array}{c} 250~{\rm km/sec} \\ 410~{\rm km/sec} \\ 260~{\rm km/sec} \\ 395~{\rm km/sec} \\ 425~{\rm km/sec} \\ 430~{\rm km/sec} \end{array}$ & $\begin{array}{c} +3.20 \\ +5.90 \\ +1.25 \\ +2.70 \\ +1.30 \\ +1.50 \end{array}$ & $\begin{array}{c} -2.20 \\ -4.90 \\ -0.25 \\ -1.70 \\ -0.30 \\ -0.50 \end{array}$ & $\begin{array}{c} 380~{\rm Mpc} \\ 490~{\rm Mpc} \\ 300~{\rm Mpc} \\ 415~{\rm Mpc} \\ 370~{\rm Mpc} \\ 390~{\rm Mpc}\end{array}$\\ [2.5truemm] \hline \hline
\end{tabular}\end{center}  \label{tab1}\vspace{10pt}
\end{table}

\subsubsection{Refining $\tilde{q}^{-}$}\label{sssRtq-}
%%%%%%%%%%%%%%%%%%%%%%%%%%%%%%%%%%%%%%%%%%%%%%%%%%%%%%%
Expressions (\ref{estqlT}) highlight the key role played by the local volume scalar ($\tilde{\vartheta}$) when it comes to determine the values of $\tilde{q}$ and $\lambda_T$. To this point, we have adopted the simplifying assumption that $\tilde{\vartheta}=\langle\tilde{\vartheta}\rangle$ and assigned an average value to the local volume element, which remains constant inside the whole of the bulk-flow domain. In reality, however, the local volume scalar is expected to
vary with scale. One therefore needs to consider physically motivated profiles for $\tilde{\vartheta}$, which can then be tested against the observations.

Since peculiar motions fade away with increasing scale, $\tilde{\vartheta}$ is expected to decrease as one moves out to progressively larger wavelengths, that is $\tilde{\vartheta}\rightarrow0$ as $\lambda\rightarrow\infty$. In addition, contracting bulk flows are expected to contract faster near the outskirts of their domain and slower near its centre, where the stabilising effects of the pressure gradients are expected to take over. Such a qualitative profile for $\tilde{\vartheta}$ is well parametrized by the scale-dependent functional form~\cite{2022MNRAS.513.2394A}
\begin{equation}
\tilde{\vartheta}= \tilde{\vartheta}(\lambda)= \frac{m\lambda^2}{\alpha+\beta\lambda^3}\,,  \label{eq:vthetalambda}
\end{equation}
where $m$, $\alpha$, and $\beta$ are free parameters to be constrained by observational data. The next step involves deriving the deceleration parameter in the tilted frame, using Eq.~(\ref{eq:vthetalambda}) as input. For scales of several hundred Mpc - well within the cosmic horizon - it is reasonable to treat the Hubble parameter as approximately constant ($H\simeq H_0$) over the relevant timescales.

To compare with observational data, particularly those of the Type~Ia supernovae, the deceleration parameter must be expressed as a function of redshift. In an Einstein–de Sitter background, the physical scale ($\lambda$) of the bulk flow is identified with the line-of-sight comoving distance ($\chi$). Substituting Eq.~(\ref{eq:vthetalambda}) into expression (\ref{lqs5}) and appropriately redefining the parameters $m$, $\alpha$, and $\beta$, leads to the redshift-dependent form
\begin{equation}
\tilde{q}(z)=\frac{1}{2} \left[1-\frac{1}{a + b\,d_r^{3}} \right]\,,  \label{eq:qzfull}
\end{equation}
with $a$ and $b$ being dimensionless parameters fitted to the \textit{Pantheon} supernova dataset. Also, $d_r$ is connected to the comoving distance through the relation $d_r=d_r(z)\equiv H_0\chi/c$. Further details regarding the derivation and parametrization can be found in~\cite{2022MNRAS.513.2394A}.

\begin{figure*}
\begin{center}
\includegraphics[height=2in,width=4in,angle=0]{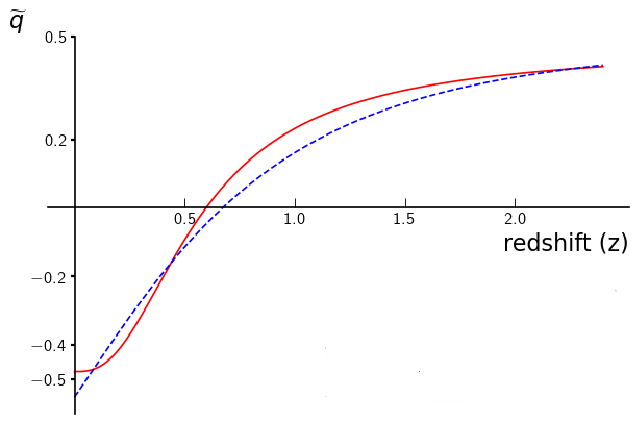}
\end{center}
\caption{The redshift profile of the deceleration parameter $\tilde{q}$ (red solid line) in accord with the Pantheon dataset (see also Fig.~3 in~\cite{2022MNRAS.513.2394A}), which refines that of the black solid line in Fig.~\ref{fig:tq+-} earlier. Here, the red solid line has been drawn after setting $q=1/2$ in the CMB frame and by using the best-fit parameters for the tilted cosmological model with an Einstein-de Sitter comoving distance. The dashed blue line compares to the best-fit $\Lambda$CDM universe. In both models the deceleration parameter starts positive at high redshifts, turns negative at $z\simeq0.5$ and ends up close to $-0.5$ today. The above figure demonstrates how local bulk motions can closely mimic the phenomenology of the $\Lambda$CDM paradigm and, in so doing, explain the recent universal acceleration without the need of dark energy, of a cosmological constant, or of any new physics, but as an artifact of our galaxy's peculiar motion.}
\label{fig:qteds}
\end{figure*}

It is important to note that the negative signin Eq.~\eqref{eq:qzfull} applies to contracted bulk flows. Also, the prefactor $1/2$ indicates that, in the absence of peculiar motions, the global universe is decelerating at the Einstein-de Sitter pace. However, below the critical transition scale ($\lambda_T$), the locally inferred deceleration parameter $\tilde{q}$ becomes negative, even though the global deceleration parameter remains positive (with $q=1/2$). The scale/redshift dependence seen in (\ref{eq:qzfull}) highlights the significance of bulk flows in interpreting low-redshift cosmological observables. In contrast, on large enough scales (i.e.~at sufficiently high redshifts), $\tilde{q}\rightarrow q=1/2$ and the deceleration parameter measured in the tilted frame approaches its CMB counterpart. This behavior is consistent with the general expectation that the impact of the peculiar velocities should fade away on progressively larges scales. The redshift evolution of $\tilde{q}$ in the tilted Einstein-de Sitter model, together with a comparison to the corresponding $\Lambda$CDM profile (see Fig.~\ref{fig:qteds} here), was given in~\cite{2022MNRAS.513.2394A}. There, the Pantheon data were used to provide the best-fit values of the parameters. Here, we illustrate the results of that work in Fig.~\ref{fig:qteds}, referring the reader to~\cite{2022MNRAS.513.2394A} for the details.

Overall, the statistical analysis of~\cite{2022MNRAS.513.2394A} indicates that the tilted Einstein-de Sitter model performs comparably to the $\Lambda$CDM scenario, when evaluated using model selection criteria such as the Akaike and Bayesian Information Criteria. Put another way, within the ``tilted universe'' paradigm, the recent accelerated expansion can be explained as a linear effect of our galaxy's peculiar flow relative to the CMB frame. If so, the perceived universal acceleration can be an artifact of our local motion rather than a recent global event

\subsubsection{The case for a contracting bulk flow}\label{sssCCBF}
%%%%%%%%%%%%%%%%%%%%%%%%%%%%%%%%%%%%%%%%%%%%%%%%%%%%%%%%%%%%%%%%%%%
Let us now consider the likelihood that we may actually live within a locally contracting bulk flow. So far, the related surveys provide the mean bulk velocity on a given scale, but not its divergence. Assuming that there is no natural bias in favour of contracting, or expanding, bulk flows on cosmologically relevant scales (i.e.~typically greater than 100~Mpc), the chances of an observer to reside within either of them should be around 50\%.

\begin{table*}[ht]
\centering
%\begin{ruledtabular}
\begin{tabular}{@{}lccc@{}}
$\hspace{20pt}\lambda$ & $\hspace{-25pt}\tilde{\vartheta}$ & $\hspace{5pt}\tilde{q}$ \\
\hline \hline
%\midrule
\textbf{Finite Difference}\\
\hline
70~{\rm Mpc/h}  & \hspace{-30pt} $-3.36\substack{+0.16\\-0.16}$ & $-6.36\substack{+0.33\\-0.33}$ \\
100~{\rm Mpc/h}  & \hspace{-30pt} $-2.77\substack{+0.13\\-0.13}$ & $-2.27\substack{+0.13\\-0.13}$ \\
125~{\rm Mpc/h}  & \hspace{-30pt} $-1.48\substack{+0.07\\-0.07}$ & $-0.45 \substack{+0.05\\-0.05}$ \\
150~{\rm Mpc/h}  & \hspace{-30pt} $-0.65\substack{+0.03\\-0.03}$ & $+0.21\substack{+0.01\\-0.01}$ \\
200~{\rm Mpc/h} & \hspace{-30pt} $-0.24\substack{+0.01\\-0.01}$ & $+0.44\substack{+0.005\\-0.005}$ \\
\hline
\textbf{Integral Approximation}\\
\hline
70~{\rm Mpc/h}  & \hspace{-30pt} $-2.45\substack{+0.12\\-0.12}$ & $-4.50 \substack{+0.24\\-0.24}$ \\
100~{\rm Mpc/h}  & \hspace{-30pt} $-1.99\substack{+0.10\\-0.10}$ & $-1.49 \substack{+0.10\\-0.10}$ \\
125~{\rm Mpc/h}  & \hspace{-30pt} $-1.13\substack{+0.06\\-0.06}$ & $-0.22\substack{+0.04\\-0.04}$ \\
150~{\rm Mpc/h}  & \hspace{-30pt} $-0.47\substack{+0.02\\-0.02}$ & $+0.29\substack{+0.01\\-0.01}$ \\
200~{\rm Mpc/h}  & \hspace{-30pt} $-0.21 \substack{+0.01\\-0.01}$ & $+0.45\substack{+0.005\\-0.005}$ \\
\hline
\textbf{Discrete Density Integration}\\
\hline
70~{\rm Mpc/h}  & \hspace{-30pt} $-3.94\substack{+0.19\\-0.19}$ & $-7.54 \substack{+0.39\\-0.39}$ \\
100~{\rm Mpc/h}  & \hspace{-30pt} $-3.17\substack{+0.15\\-0.15}$ & $-2.67\substack{+0.15\\-0.15}$ \\
125~{\rm Mpc/h}  & \hspace{-30pt} $-1.66\substack{+0.08\\-0.08}$ & $-0.56\substack{+0.05\\-0.05}$ \\
150~{\rm Mpc/h}  & \hspace{-30pt} $-0.76\substack{+0.04\\-0.04}$ & $+0.16\substack{+0.02\\-0.02}$ \\
200~{\rm Mpc/h}  & \hspace{-30pt} $-0.29 \substack{+0.01\\-0.01}$ &
$+0.43\substack{+0.005\\-0.005}$ \\
\end{tabular}
\caption{The divergence ($\tilde{\vartheta}$) of the reconstructed peculiar-velocity field, estimated by means of three different methods, takes consistently negative values across a range of scales. This indicates that we happen to live inside a locally contracting bulk peculiar flow. As result of this, the locally measured deceleration parameter ($\tilde{q}$) appears to take negative values, despite the fact that the universe is globally decelerating (with $q=0.5$ and $h\simeq0.7$). Note that the effect of the observer’s peculiar motion becomes stronger as we go down to smaller scales and fades away on progressively longer lengths (extract from~\cite{2024PDU....4301385P} - see also Table~1 there for more details and further discussion).}
\label{tabla1}
%\end{ruledtabular}
\end{table*}

To the best of our knowledge, the first systematic attempt to reconstruct the local peculiar-velocity field and, in so doing, extract its divergence was made by Pasten et al~\cite{2024PDU....4301385P}, using data from the \textit{2M++} galaxy survey~\cite{2011MNRAS.416.2840L}. The authors' employed three different methods, namely Finite Difference, Integral Approximation and Discrete Density Integration, to estimate the divergence of the reconstructed local velocity field on different scales. In all cases, the divergence was found to take negative values, thus indicating a locally contracting velocity field (see Table~\ref{tabla1} here, or Table~1 in~\cite{2024PDU....4301385P}). Pasten et al then estimated the values of the locally measured deceleration, as predicted by the ``tilted universe'' scenario (see Eq.~(\ref{lqs5}) in  \S~\ref{sssE-dSC} previously). The numerical results assigned negative average values to the local deceleration parameter on a scales reaching out to approximately $150/h$~Mpc, in support of the tilted scenario as a natural alternative to dark energy~\cite{2024PDU....4301385P}. Note that the aforementioned work of Pasten et al, followed that by Pasten and Cardenas~\cite{2023PDU....4001224P}, which identified tensions at the $2\sigma-3\sigma$ level in the deceleration parameter across various redshift bins.

Before closing this section, we should mention that the results of~\cite{2024PDU....4301385P}, claiming a contracting local velocity field, are corroborated by the recent report of~\cite{2024JCAP...12..003S}. The latter argues for an infalling local velocity field, which extends out to $z\simeq0.4$ and corresponds to a radius of approximately $120/h$~Mpc.

\subsubsection{The Newtonian case}\label{sssNtC}
%%%%%%%%%%%%%%%%%%%%%%%%%%%%%%%%%%%%%%%%%%%%%%%%
In Newtonian theory the peculiar flux of the matter, although nonzero, has no gravitational contribution and this has a drastic effect on the results. More specifically, whereas the linear relation between the two deceleration parameters,
\begin{equation}
\tilde{q}= q- {\dot{\tilde{\vartheta}}\over3H^2}\,,  \label{lNqs1}
\end{equation}
is formally identical to its relativistic counterpart, the evolution of the local volume scalar is given by the divergence of Eq.~(\ref{ltv'}), namely by
\begin{equation}
\dot{\tilde{\vartheta}}= -2H\tilde{\vartheta}- \partial^2\Phi\,,  \label{lNvthetadot}
\end{equation}
instead of (\ref{lpRay2}). Substituting the above into the right-hand side of Eq.~(\ref{lNqs1}), while keeping in mind that $\partial^2\Phi=\rho\delta/2$ (where $\delta=\delta\rho/\rho$ is the familiar density contrast) and that $\vartheta/H\ll1$ throughout the linear regime, we find that~\cite{2021Ap&SS.366...90T}
\begin{equation}
\tilde{q}= q+ {1\over2}\,\delta\simeq q\,,  \label{lNqs2}
\end{equation}
to first order (where $\delta\ll1$). Therefore, within the limits of Newtonian gravity, the linear relative-motion effects leave the deceleration parameter essentially unaffected on all scales. Put another way, the deceleration parameter ($\tilde{q}$) measured locally in the tilted frame coincides (for all practical purposes) with the deceleration parameter ($q$) of the whole universe. The reason for the considerable difference between the results of the Newtonian and the relativistic studies is traced to the fundamentally different role played by the peculiar flux in these two  theories. This in turn reflects the fundamentally different way Newton's and Einstein's theories treat both the gravitational field and its sources (see~\cite{2021Ap&SS.366...90T} for further discussion).

\subsubsection{The generalised FLRW case}\label{sssGFLRWC}
%%%%%%%%%%%%%%%%%%%%%%%%%%%%%%%%%%%%%%%%%%%%%%%%%%%%%%%%
Allowing for nonzero pressure and spatial curvature changes the linear relation between the deceleration parameters measured in the tilted frame of the matter and in that of the CMB. Instead of (\ref{lqs1}), the two parameters are now related by~\cite{2022EPJC...82..521T}
\begin{equation}
\tilde{q}= q+ {\dot{\tilde{\vartheta}}\over6\dot{H}} \left[2+(1+3w)\Omega\right]\,,  \label{FLRWlqs1}
\end{equation}
where $w=p/\rho$ is the barotropic index of the matter (with $w\neq-1$) and $\Omega=\rho/3H^2$ is the density parameter of the background universe. Note that generalising the Friedmann host does not change the linear time evolution of the local volume scalar ($\tilde{\vartheta}$), which is still monitored by Eq.~(\ref{lpRay2}). However, in the presence of pressure, the linearised energy conservation law is no longer given by Eq.~(\ref{lcls}a) of \S~\ref{sssKRP4A}, but by
\begin{equation}
\dot{\rho}= -\Theta(\rho+p)- {\rm D}^aq_a\,.  \label{FLRWecl}
\end{equation}
Taking the spatial gradient of the above and employing the linear commutation law ${\rm D}_a\dot{\rho}=({\rm D}_a\rho)^{\cdot}- \dot{\rho}A_a+H{\rm D}_a\rho$, together with the background relation $\dot{\rho}=-3H(\rho+p)$, we arrive at the following expression for the peculiar 4-acceleration
\begin{equation}
A_a=-{1\over3H}\,{\rm D}_a\tilde{\vartheta}- {1\over3aH}\left[\dot{\Delta}_a-3wH\Delta_a +(1+w)\mathcal{Z}_a\right]\,.  \label{FLRWlA}
\end{equation}
instead of (\ref{lA}). Combining all of the above and setting $c_s^2={\rm d}p/{\rm d}\rho$ as the square of the sound speed, the linear relation between the two deceleration parameters reads~\cite{2022EPJC...82..521T}
\begin{equation}
\tilde{q}= q+ {2\over3}\left[1-{3\over2}\,c_s^2 +{1\over6}\left({\lambda_H\over\lambda}\right)^2\right] {\tilde{\vartheta}\over H}- {|1-\Omega|\over9(1+w)} \left[{\dot{\tilde{\Delta}}\over H} -3w\tilde{\Delta}+(1+w){\tilde{\mathcal{Z}}\over H}\right]\,,  \label{FLRWlqs2}
\end{equation}
since $(\lambda_H/\lambda_K)^2=|1-\Omega|$ to zero order. The above holds on all FLRW backgrounds, irrespective of the their spatial curvature and equation of state (with the exception of the $w=-1$ models). Clearly, on subhorizon scales, where $\lambda_H/\lambda\gg1$, the dominant (correction) term on the right-hand side of (\ref{FLRWlqs2}) is the one containing the aforementioned ratio. The rest of the terms have coefficients of order unity or less, given that $w\neq-1$ and provided the density parameter ($\Omega$) does not take unrealistically large values.\footnote{Given that the perturbations $\tilde{\vartheta}/H$, $\Delta$ and $\tilde{\mathcal{Z}}/H$ are all very small during the linear regime, Eq.~(\ref{FLRWlqs2}) suggests that $\tilde{q}\rightarrow q$ on scales close and beyond the Hubble horizon.} On these grounds, well inside the Hubble radius, the two deceleration parameters are still related by (\ref{lqs5}) and the transition length is still given by expression (\ref{lambdaT1}) of the Einstein-de Sitter background (see \S~\ref{sssE-dSC} previously). Therefore, on subhorizon scales, the linear relative-motion effects on the deceleration parameter are the same on all physically realistic FLRW backgrounds~\cite{2022EPJC...82..521T}. In other words, observers living inside (slightly) contracting bulk peculiar flows in any observationally realistic Friedmann universe are likely to misinterpret their slower local expansion rate as recent acceleration of the surrounding universe.

\subsubsection{The Bianchi case}\label{sssBC}
%%%%%%%%%%%%%%%%%%%%%%%%%%%%%%%%%%%%%%%%%%%%%
Adopting a spatially homogeneous but anisotropic background, of the Bianchi type, allows one to investigate the changes the anisotropy introduces to the peculiar-motion effects discussed in the previous sections.

Typically, the Bianchi spacetimes are classified into the non-tilted and the tilted models (e.g.~see~\cite{1997dsc..book.....W} for the details). The former class has the worldlines of the matter orthogonal to the hypersurfaces of homogeneity, while in the latter this is not the case. Here, we will perturb a non-tilted Banchi background and introduce the tilt as a linear velocity perturbation due to the peculiar motion of the real observers with respect to the CMB frame. These non-tilted Bianchi universes are a group of ten cosmological models, five of which contain the Friedmann universes (with all three types of spatial curvature) as special cases.

The anisotropy of the Bianchi spacetimes means that they can naturally support kinematic, dynamic and geometric anisotropies. These respectively include, the shear, the viscosity and the Weyl field, with the latter monitoring tidal forces and gravitational waves. Here, we will confine to ideal pressure-free media and set both the pressure and the viscosity to zero.

In a perturbed Bianchi universe with a peculiar-velocity tilt the linear relations (\ref{lab34}a), (\ref{lab56}a) and (\ref{lab710}) still hold. On the other hand, the relations between the 4-acceleration vectors and between the shear tensors change. More specifically, even when the cosmic medium is a pressureless perfect fluid, the linear expressions (\ref{lab34}b) and (\ref{lab56}b) change to
\begin{equation}
\tilde{A}_a= A_a+ \dot{\tilde{v}}_a+ H\tilde{v}_a+ \sigma_{ab}\tilde{v}^b \hspace{10mm} {\rm and} \hspace{10mm} \tilde{\sigma}_{ab}= \sigma_{ab}+ 2u_{(a}\sigma_{b)c}\tilde{v}^c+ \tilde{\varsigma}_{ab}\,,  \label{Blrels}
\end{equation}
respectively~\cite{2024EPJC...84.1061T}. As before, we have switched off the flux and the 4-acceleration in the tilted frame of the matter, by setting $\tilde{q}_a=0=\tilde{A}_a$. Then, written in the coordinate system of the CMB, the momentum conservation law linearises to
\begin{equation}
\rho A_a= -\dot{q}_a- 4Hq_a- \sigma_{ab}q^b\,,  \label{Bmcl}
\end{equation}
which reproduces (\ref{Blrels}a) once the expression $q_a=\rho\tilde{v}_a$ is used for the peculiar flux. To linear order, the Bianchi background does not change the energy conservation law, which in the absence of pressure retains its almost-FLRW form (see Eq.~(\ref{lcls}a) in \S~\ref{sssKRP4A} previously). Taking the spatial gradient of (\ref{lcls}a), using the linear commutation law ${\rm D}_a\dot{\rho}=({\rm D}_a\rho)^{\cdot}+H{\rm D}_a\rho+3H\rho A_a+\sigma_{ab}{\rm D}^b\rho$ and solving the resulting expression for the 4-acceleration yields~\cite{2024EPJC...84.1061T}
\begin{equation}
A_a= -{1\over3H}\,{\rm D}_a\vartheta- {1\over3aH}\left(\dot{\Delta}_a+\mathcal{Z}_a+ \sigma_{ab}\Delta^b\right)\,.  \label{BlA}
\end{equation}
This is the relativistic linear peculiar 4-acceleration in a tilted, almost-Bianchi universe filled with pressureless matter in the presence of peculiar velocities. Comparing the above to its almost-FLRW counterpart (see Eq.~(\ref{lA}) in \S~\ref{sssKRP4A} previously), one notices that the Bianchi background has added a shear-induced term to the right-hand side.

The assumption of an anisotropic Bianchi cosmology also changes the linear relation between the deceleration parameters measured in the tilted and the CMB frames. In particular, expression (\ref{lqs2}) recasts into
\begin{equation}
\tilde{q}= q+ {2\over3H^2}\,\sigma_{ab}\tilde{\varsigma}^{ab}+ {1\over3H^2}\,{\rm D}^aA_a\,,  \label{Blqs1}
\end{equation}
where $\tilde{\varsigma}_{ab}={\rm D}_{\langle b} \tilde{v}_{a\rangle}$ represents the peculiar shear~\cite{2024EPJC...84.1061T}. Substituting the spatial divergence of (\ref{BlA}) into the right-hand side of the above, the linear relation between the two deceleration parameters acquires the form
\begin{equation}
\tilde{q}= q+ {2\over3H^2}\,\sigma_{ab}\tilde{\varsigma}^{ab}- {1\over9H^3}\,{\rm D}^2\tilde{\vartheta}- {1\over9a^2H^2} \left({\dot{\Delta}\over H} +{\mathcal{Z}\over H} +{2\over H}\, \sigma_{ab}\Sigma^{ab}\right)\,,  \label{Blqs2}
\end{equation}
with two extra shear-related terms due to the background anisotropy (compare to Eq.~(\ref{lqs3}) in \S~\ref{sssE-dSC}). Note that the symmetric and trace-free tensor $\Sigma_{ab}=a{\rm D}_{\langle b}\Delta_{a\rangle}$ monitors linear anisotropies (shape distortions) in the density distribution of the matter~\cite{2008PhR...465...61T,%
2012reco.book.....E}.

Without compromising the physics, we may treat the peculiar velocity as a shear eigenvector, in which case $\sigma_{ab}\tilde{v}^b=\zeta\tilde{v}_a$, with $\zeta$ being the associated eigenvalue. Then, $\sigma_{ab}\tilde{\varsigma}^{ab}= {\rm D}^a(\sigma_{ab}\tilde{v}^b)=\zeta\tilde{\vartheta}$ to linear order. Also, since $1/a^2H^2=(\lambda_H/\lambda_K)^2\ll1$ according to the observations, we may ignore the last term on the right-hand side of (\ref{Blqs2}). In that case, the latter reduces to~\cite{2024EPJC...84.1061T}
\begin{equation}
\tilde{q}= q+ {2\over3}\left[{\zeta\over H} +{1\over6}\left(\lambda_H\over\lambda\right)^2\right] {\tilde{\vartheta}\over H}\,,  \label{Blqs3}
\end{equation}
after a simple harmonic splitting. The impact of the Bianchi background is carried by the $\zeta/H$ ratio, which also provides a measure of the model's anisotropy. Assuming realistically anisotropic backgrounds we may set $|\zeta|/H\ll1$. Then, on subhorizon scales where $\lambda_H/\lambda\gg1$, the two deceleration parameters are still related by Eq.~(\ref{lqs5}) and the transition length is still given by expression (\ref{lambdaT1}) of the Einstein-de Sitter case (see \S~\ref{sssE-dSC} previously).\footnote{The Bianchi transition length, marking the $\tilde{q}=0$ threshold when $\tilde{\vartheta}<0$, is generally given by~\cite{2024EPJC...84.1061T}
\begin{equation}
\lambda_T= {1\over3}\sqrt{{|\tilde{\vartheta}|\over[q +(2\zeta\vartheta/3H^2)]H}}\,\lambda_H  \label{Btl}
\end{equation}
and reduces to its Einstein de Sitter counterpart (\ref{lambdaT1}) when $\zeta=0$ (as expected). There, as well as here, the transition length is a fraction of the Hubble scale and closely analogous to the Jeans length~\cite{2021EPJC...81..753T}. Here, the background anisotropy increases $\lambda_T$ when $\zeta>0$ and decreases it if $\zeta<0$ (recall that $\tilde{\vartheta}<0$). In either case, the effect is weak in realistically anisotropic models (with $|\zeta|/H\ll1$).} Hence, as in the generalised FLRW case (see \S~\ref{sssGFLRWC} before), on subhorizon scales the linear relative-motion effects on the deceleration parameter remain unaffected, unless the Bianchi background has unacceptably high anisotropy and/or unacceptably strong spatial curvature.\footnote{In the observationally unrealistic case of high background anisotropy with $|\zeta|/H\gg1$, the relative-motion effects on $\tilde{q}$ are dominated by the aforementioned ratio. For example, confining to super-Hubble lengths, where $\lambda_H/\lambda\ll1$, and allowing for high anisotropy, Eq.~(\ref{Blqs3}) reduces to
\begin{equation}
\tilde{q}= q+ {2\zeta\tilde{\vartheta}\over3H^2}\,.  \label{ltq5}
\end{equation}
The effect is now solely determined by the sign of $\zeta\vartheta$ and by the magnitude of the correction term on the right-hand side of the above, which combines the kinematic anisotropy of the Bianchi background with that of the bulk peculiar flow. Then, when  $\zeta\tilde{\vartheta}<0$, the local value of the deceleration parameter decreases (i.e.~$\tilde{q}<q$) and it can even drop below zero (see~\cite{2024EPJC...84.1061T} for further discussion).} Therefore, observers living inside (slightly) contracting bulk peculiar flows in observationally realistic Bianchi universes, are still likely to misinterpret their slower local expansion rate as recent acceleration of the surrounding universe.

In summary, qualitatively speaking, the physical mechanism responsible for the relative-motion effects on the locally measured deceleration parameter seem generic, as it appears independent of the background model. In fact, the close analogies between the transition length ($\lambda_T$) and the Jeans length ($\lambda_J$ - see~\cite{2022EPJC...82..521T} for further discussion) suggest that the former is as generic to peculiar-velocity perturbations as the letter is to density perturbations. Also note that the scale dependence of the linear relative-motion effects on $\tilde{q}$, raises the intriguing possibility that they may even dominate the nonlinear regime of structure formation. After all, on the typical nonlinear scales, with $\lambda\ll100$~Mpc, the ratio $({\lambda_H/\lambda})^2$ takes very large values.

\subsubsection{The null case}\label{sssNC}
\begin{figure*}
\begin{center}
\includegraphics[height=2in,width=4in,angle=0]{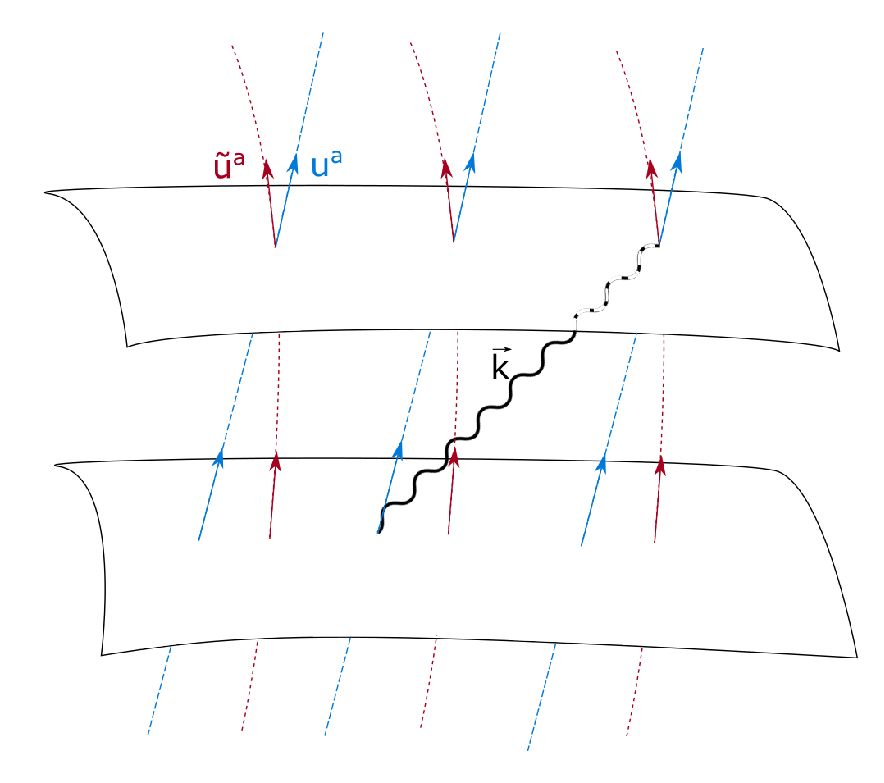}
\end{center}
\caption{Radiation signal emitted and received between two different spacetime events. At reception the signal in measured simultaneously by the idealised (CMB) observers and by their realistic (tilted) partners, associated with the blue and the red worldlines respectively. Note that the sources are assumed to reside at sufficiently high redshits to be treated as comoving with the CMB frame.}
\label{fig:null}
\end{figure*}

%%%%%%%%%%%%%%%%%%%%%%%%%%%%%%%%%%%%%%%%%%
Assume a family of observers with 4-velocity $u_{a}$ and introduce the unitary basis $\{{e}_1, {e}_2, {e}_3\}$ spanning the 3-D space orthogonal to the $u_a$-field, so that $u_ae^a=0$, $h_a{}^be_b=e_a$ and $e_ae^a=1$. Suppose also that $e_a$ is the spatial direction the observers collect their incoming data. Then, the photon 4-momentum ($k_{a}$) decomposes as
\begin{equation}
k_a = E(u_a -e_a)\,,  \label{k}
\end{equation}
where $E=-k_au^a$ is the energy of the photons. The latter travel along null geodesics tangent to the $k_a$-vector (with $k_ak^a=0$) and they are received along the spacelike $e_a$-direction by observers following timelike worldlines tangent to the $u_a$-field (e.g.~see Fig.~\ref{fig:null}).

In what follows, we will assume that the radiation source is sufficiently distant, namely it has sufficiently high redshift, to be treated as comoving with the universal expansion. We may then ignore the peculiar velocity of the source at the point of emission and align its 4-velocity there with that of the CMB frame ($u_a$). The signal is received by the realistic (tilted) observer (red worldline) at the same future event, where it is causally boosted to the CMB frame (blue worldline). Then, any difference in the measurements (e.g.~like that seen in Eq.~(\ref{lQs1}) below) will be attributed to the peculiar motion of the realistic observer (see also Fig.~\ref{fig:null}).

The luminosity distance of a radiation source is typically Taylor-expanded in terms of redshift. When dealing with timelike worldlines, the Taylor-series is expressed in terms of the familiar Hubble ($H$) and deceleration ($q$) parameters, as well as in terms of higher-order derivatives of the scale factor (e.g.~see~\cite{2004CQGra..21.2603V}). When the Taylor expansion takes place along null rays, the standard cosmological parameters are replaced by their effective ``null'' counterparts~\cite{2021JCAP...05..008H}. For example, the familiar Hubble and deceleration parameters ($H$ and $q$) are replaced by their effective null analogues ($\mathfrak{H}$ and $\mathfrak{Q}$). These are respectively given by
\begin{equation}
	\label{fancyh}
	\mathfrak{H}\equiv -\frac{1}{E^2}\frac{{\rm d}E}{{\rm d}s} \hspace{15mm} {\rm and} \hspace{15mm}
	\mathfrak{Q}\equiv -1 - \frac{1}{E\mathfrak{H}^2}\frac{{\rm d}\mathfrak{H}}{{\rm d}s}\,,
\end{equation}
where $s$ is the null affine parameter and ${\rm d}E/{\rm d}s<0$ in an expanding universe. Comparing $\mathfrak{H}$ and $\mathfrak{Q}$ to their timelike counterparts $H$ and $q$, shows that $E$ and $s$ act as effective cosmological scale factor ($a$) and conformal time ($\eta$) respecyively. Recall that $H=({\rm d}a/{\rm d}\eta)/a^2$ and $q=-1-({\rm d}H/{\rm d}\eta)/aH^2$, with $\dot{a}/a=H$ and $\dot{\eta}=1/a$.

In an almost-FLRW universe with peculiar-velocity perturbations, the null and the timelike deceleration parameters, as measured in the CMB and the tilted frames simultaneously at the reception event (see Fig.~\ref{fig:null}), are related by
\begin{equation}
\mathfrak{Q}= q- {1\over3H^2}\,{\rm D}^aA_a \hspace{15mm} {\rm and} \hspace{15mm} \tilde{\mathfrak{Q}}= \tilde{q}- {1\over3H^2}\,\tilde{{\rm D}}^a\tilde{A}_a\,,  \label{lQs1}
\end{equation}
to first approximation~\cite{2022arXiv220301126S}. Therefore, the null deceleration parameters measured by the real (the tilted) and by the idealised (the CMB) observers generally differ (i.e.~$\tilde{\mathfrak{Q}}\neq\mathfrak{Q}$), since
\begin{equation}
\tilde{\mathfrak{Q}}- \mathfrak{Q}= \tilde{q}- q- {1\over3H^2}\left(\tilde{\rm D}^a\tilde{A}_a-{\rm D}^aA_a\right)\,.  \label{Q-tQ1}
\end{equation}
To proceed one linearises the spatial divergence of (\ref{lab34}b) and then combines it with (\ref{lqs1}) and the linear commutation law ${\rm D}^a\dot{\tilde{v}}_a=({\rm D}^a\tilde{v}_a)^{\cdot}+H{\rm D}^a\tilde{v}_a$. The result reads
\begin{equation}
\tilde{q}- q= -{1\over3H^2}\left(\tilde{\rm D}^a\tilde{A}_a-{\rm D}^aA_a\right)  \label{q-tq2}
\end{equation}
and recasts expression (\ref{Q-tQ1}) into
\begin{equation}
\tilde{\mathfrak{Q}}- \mathfrak{Q}= 2(\tilde{q}-q)\,.  \label{Q-tQ2}
\end{equation}
Accordingly, the linear relative-motion effect on the null deceleration parameter is twice stronger than on its timelike counterpart. Indeed, on using (\ref{lqs5}), we arrive at~\cite{2022arXiv220301126S}
\begin{equation}
\tilde{\mathfrak{Q}}= \mathfrak{Q}+ {2\over9}\left({\lambda_H\over\lambda}\right)^2 {\tilde{\vartheta}\over H}\,, \label{dtQ1}
\end{equation}
with the relative-motion effect being twice stronger than in Eq.~(\ref{lqs5}). A direct consequence of this result is that the null transition length ($\lambda_\mathfrak{T}$) associated with $\tilde{\mathfrak{Q}}$, is larger than its timelike counterpart ($\lambda_T$). More specifically, the null transition length, marking the scale where $\tilde{\mathfrak{Q}}$ appears to turn negative is no longer given by (\ref{lambdaT1}), but by the linear expression~\cite{2022arXiv220301126S}
\begin{equation}
\lambda_\mathfrak{T}= \sqrt{{2\over9q}\,{|\tilde{\vartheta}|\over H}}\,\lambda_H= \sqrt{2}\,\lambda_T\,,  \label{tQlambdaT}
\end{equation}
since $\tilde{\mathfrak{Q}}=q$ in our FLRW background. This means that the transition length  for $\mathfrak{Q}$ is approximately $1.4$ times larger than the one associated with its timelike counterpart ($q$)~\cite{2022arXiv220301126S}.

\subsection{Doppler-like dipoles in the universal
%%%%%%%%%%%%%%%%%%%%%%%%%%%%%%%%%%%%%%%%%%%%%%%%%
 expansion}\label{ssD-LDUE}
%%%%%%%%%%%%%%%%%%%%%%%%%%%
Relative motions have a trademark signature, which is nothing else but an apparent (Doppler-like) dipolar anisotropy, like that seen in the CMB spectrum. Hence, if the recent universal acceleration is an artefact of our galaxy's peculiar flow, we should ``see'' a dipole is the sky distribution of the deceleration parameter. The universe should appear to accelerate faster along one direction on the celestial sphere and equally slower along the antipodal~\cite{2010MNRAS.405..503T,%
2011PhRvD..84f3503T}.

\subsubsection{The deceleration tensor}\label{sssDT}
%%%%%%%%%%%%%%%%%%%%%%%%%%%%%%%%%%%%%%%%%%%%%%%%%%%%
The deceleration parameter ($q$) is a scalar monitoring the isotropic deceleration/acceleration of the universal expansion. The latter is not fully isotropic, however, mainly due to structure-formation effects. There are more than one ways of describing anisotropies in the deceleration/acceleration of the universe. Here, we will do so by means of the deceleration tensor
\begin{equation}
Q_{ab}= -\left(h_{ab}+{9\over\Theta^2}\, h_a{}^ch_b{}^d\dot{\Theta}_{cd}\right)\,,  \label{Qab}
\end{equation}
first introduced in~\cite{2025glc..conf....104T} and subsequently used in~\cite{2025EPJC...85..596S}. In the above,
\begin{equation}
\Theta_{ab}= {1\over3}\,\Theta h_{ab}+ \sigma_{ab}\,,  \label{Thetaab}
\end{equation}
is the more familiar expansion tensor~\cite{2008PhR...465...61T,%
2012reco.book.....E}. By construction, both $\Theta_{ab}$ and $Q_{ab}$ are symmetric and spacelike tensors, with traces given by
\begin{equation}
Q= 3q \hspace{15mm} {\rm and} \hspace{15mm} \Theta=3H\,, \label{Q}
\end{equation}
where $q=-[1+(3\dot{\Theta}/\Theta^2)]$ and $H=\Theta/3$ are the  deceleration and Hubble parameters respectively. Therefore, the expansion and the deceleration tensors are 3$\times$3 matrices, the non-diagonal components of which describe anisotropies in the universal expansion.

Assuming that $n_a$, with $n_an^a=1$ and $u_an^a=0$, is the unit vector along a spatial direction, the deceleration/acceleration of the expansion along $n_a$ is given by the scalar $Q_{ab}n^an^b$. On using definition (\ref{Qab}), the latter reads~\cite{2025glc..conf....104T}
\begin{equation}
Q_{ab}n^an^b= q- {9\over\Theta^2}\,\dot{\sigma}_{ab}n^an^b\,,  \label{Qn}
\end{equation}
given that $\dot{h}_{ab}n^an^b= (A_au_b+u_aA_b)n^an^b=0$. Hence, the deceleration parameter along a selected spatial direction consists of its mean value plus/minus shear-like corrections. As expected, in the spatially isotropic FLRW universes, definitions (\ref{Qab}) and (\ref{Qn}) reduce to
\begin{equation}
Q_{ab}= qh_{ab} \hspace{15mm} {\rm and} \hspace{15mm} Q_{ab}n^an^b= q\,.  \label{FLRWQ}
\end{equation}

\subsubsection{The deceleration tensor in tilted
%%%%%%%%%%%%%%%%%%%%%%%%%%%%%%%%%%%%%%%%%%%%%%%%
universes}\label{sssDTTUs}
%%%%%%%%%%%%%%%%%%%%%%%%%%
Consider a group of observers living in typical galaxies like our Milky Way and moving relative to the CMB frame with peculiar velocity $\tilde{v}_a$ (see Fig.~\ref{fig:bflow} in \S~\ref{ssRMEDP} earlier). In the tilted frame of the matter, definition (\ref{Qab}) reads
\begin{equation}
\tilde{Q}_{ab}= -\left(\tilde{h}_{ab}+{9\over\tilde{\Theta}^2}\, \tilde{h}_a{}^c\tilde{h}_b{}^d\tilde{\Theta}^{\prime}_{cd}\right)\,,  \label{tQab}
\end{equation}
with $\tilde{\Theta}_{ab}=(\tilde{\Theta}/3)\tilde{h}_{ab}+ \tilde{\sigma}_{ab}$ and the primes indicating time derivatives along the $\tilde{u}_a$-field~\cite{2025glc..conf....104T}.

Substituting the linear relations (\ref{lab34}a) and (\ref{lab56}b) into the right-hand side of (\ref{tQab}), recalling that $\tilde{h}_{ab}=h_{ab}+2u_{(a}\tilde{v}_{b)}$ to first approximation (see (\ref{rels1a}) in \S~\ref{ssRBFs}) and employing definitions (\ref{Qab}) and (\ref{Thetaab}), we obtain
\begin{equation}
\tilde{Q}_{ab}- Q_{ab}= 2qu_{(a}\tilde{v}_{b)}- {1\over H^2} \left({1\over3}\,\dot{\tilde{\vartheta}}h_{ab} +\dot{\tilde{\varsigma}}_{ab}\right)\,,  \label{Qs}
\end{equation}
at the linear level. The above relates the deceleration tensor ($\tilde{Q}_{ab}$) measured in the tilted frame of the real observers to the one measured in the CMB frame of their idealised counterparts ($Q_{ab}$). Not surprisingly, the two tensors coincide when there are no peculiar motions.\footnote{As expected, the trace of expression (\ref{Qs}) leads immediately to the linear relation (\ref{lqs1}) between the (scalar) deceleration parameters ($\tilde{q}$ and $q$) measured in the two frames.}

\subsubsection{Doppler-like dipole in the deceleration
%%%%%%%%%%%%%%%%%%%%%%%%%%%%%%%%%%%%%%%%%%%%%%%%%%%%%%
parameter}\label{sssDDDP}
%%%%%%%%%%%%%%%%%%%%%%%%%
Projecting (\ref{Qs}) twice along $n_a$, using definitions (\ref{Qab}) and (\ref{tQab}), while keeping in mind that $(\tilde{\vartheta}/3)\tilde{h}_{ab}+\tilde{\varsigma}_{ab}= {\rm D}_{(b}\tilde{v}_{a)}$, yields~\cite{2025glc..conf....104T}
\begin{equation}
\tilde{Q}_{ab}n^an^b- Q_{ab}n^an^b= -{1\over H^2}\left(\tilde{\rm D}_b\tilde{v}_a\right)^{\cdot}n^an^b\,,  \label{tQn1}
\end{equation}
to first order. Then employing the linear commutation law $\left({\rm D}_b\tilde{v}_a\right)^{\cdot}={\rm D}_b \dot{\tilde{v}}_a-H{\rm D}_b\tilde{v}_a$, we obtain
\begin{eqnarray}
\tilde{Q}_{ab}n^an^b- Q_{ab}n^an^b&=& {1\over H}\,n^a{\rm D}_a \left(\tilde{v}_bn^b\right)- {1\over H^2}\,n^a{\rm D}_a \left(\dot{\tilde{v}}_bn^b\right) \nonumber\\ &=&{1\over H}\, n^a{\rm D}_a\left(\tilde{v}\cos\phi\right)- {1\over H^2}\,n^a{\rm D}_a \left(\dot{\tilde{v}}\cos\psi\right)\,,  \label{tQn2}
\end{eqnarray}
where $\tilde{v}$ and $\dot{\tilde{v}}$ are the magnitudes of the two vectors, while $\phi$ and $\psi$ are the respective trigonometric angles between $\tilde{v}_a$, $\dot{\tilde{v}}_a$ and $n_a$ (with $0\leq\phi,\,\psi\leq\pi$). The right-hand side of the above ensures that $\tilde{Q}_{ab}n^an^b\neq Q_{ab}n^an^b$ when peculiar motions are present. Put another way, the measurements of the deceleration parameter made by the realistic and the idealised observers along the direction of $n_a$ differ entirely due to the peculiar flow of the former observer.

\begin{figure*}
\begin{center}
\includegraphics[height=2.25in,width=2.25in,angle=0]{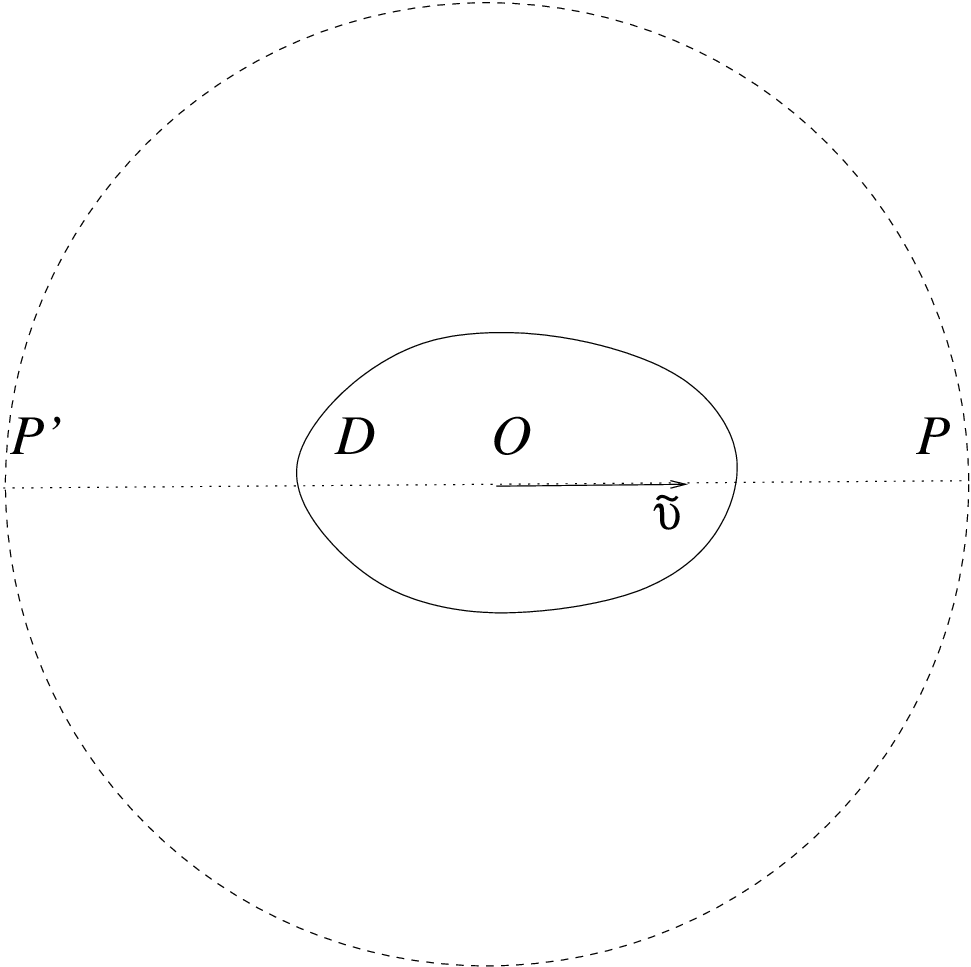}\quad
\end{center}
\caption{Consider an isotropic distribution of identical distant (comoving) sources, say of supernovae \textit{I\,{\rm a}} (circular dashed line). Consider also an observer ($O$) in a typical galaxy inside a bulk flow ($D$), moving with peculiar velocity $\tilde{v}$ relative to the sources. Following (\ref{DtQ1}), observers ``approaching'' point $P$ on the celestial sphere, and therefore simultaneously ``moving away'' from the antipodal point $P^{\prime}$, they will ``see'' an apparent (Doppler-like) dipole in the sky-distribution of the deceleration parameter forming along the direction of their peculiar motion (dotted line)~\cite{2025glc..conf....104T}.}  \label{fig:Doppler}
\end{figure*}

Unless the two terms on the right-hand side of (\ref{tQn2}) ``conspire'' to cancel each other out, they induce an apparent (Doppler-like) dipolar anisotropy in the sky distribution of the deceleration parameter. To demonstrate this, we take the standard approach that the universe appears isotropic in the idealised CMB frame and set $Q_{ab}n^an^b=q$ in Eq.~(\ref{tQn2}). Suppose also (for simplicity) that the first term dominates the right-hand side of (\ref{tQn2}). Then, the latter reads
\begin{equation}
\tilde{Q}_{ab}n^an^b= q+ {1\over H}\,n^a{\rm D}_a \left(\tilde{v}\cos\phi\right)\,.  \label{tQn3}
\end{equation}
Finally, keeping the direction vector ($n_a$) fixed, we arrive at~\cite{2025glc..conf....104T}
\begin{equation}
\tilde{Q}_{ab}n^an^b= q+ {1\over H}\,n^a{\rm D}_a\tilde{v}\,, \hspace{10mm} {\rm and} \hspace{10mm} \tilde{Q}_{ab}n^an^b= q- {1\over H}\,n^a{\rm D}_a\tilde{v}  \label{DtQ1}
\end{equation}
when $\tilde{v}_a\uparrow\uparrow n^a$ and $\tilde{v}_a\uparrow \downarrow n^a$ respectively (or equivalently for $\phi=0$ and $\phi=\pi$). According to the above, observers moving towards a certain point on the sky will assign an increased/decreased value to the deceleration parameter in that direction, compared to the value measured in the CMB frame. In contrast, observers moving away from the aforementioned point will assign an equally decreased/increased value to $\tilde{Q}_{ab}n^an^b$. It goes without saying that any observer moving towards a certain point in the sky also moves away from the antipodal. All this means that the observer in question will measure an increased/decreased deceleration parameter in the direction of their motion and an equally decreased/increased one in the direction they are moving away from. In other words, the bulk-flow observers should ``experience'' faster acceleration along a certain celestial direction and equally slower in the opposite. As a result, the sky distribution of the deceleration parameter should exhibit an apparent (Doppler-like) dipole solely triggered by their peculiar motion (e.g.~see Fig.~\ref{fig:Doppler}).

Following (\ref{DtQ1}), the magnitude of the induced dipole depends on that of the projected gradient $n^a\tilde{\rm D}_a\tilde{v}$. The latter ensures that the dipolar anisotropy should appear stronger on smaller scales and weaker on progressively larger lengths (i.e.~closer and away from the observer respectively). This is also corroborated by the fact that the peculiar velocities themselves are also expected to fade away with scale/redshift~\cite{2025glc..conf....104T}.

According to (\ref{tQn2}), the induced dipolar anisotropy generally depends on the angle ($\psi$) between $\dot{\tilde{v}}_a$ and $n_a$ as well. In that case, expression (\ref{tQn3}) generalises to
\begin{equation}
\tilde{Q}_{ab}n^an^b= q+ {1\over H}\,n^a{\rm D}_a \left(\tilde{v}\cos\phi\right)- {1\over H^2}\,n^a{\rm D}_a \left(\dot{\tilde{v}}\cos\psi\right)\,.  \label{tQn4}
\end{equation}
The last term on the right-hand side of the above also leads to an Doppler-like dipole, though this time the axis is not necessarily collinear with the peculiar velocity vector ($\tilde{v}_a$). Clearly, the overall $q$-dipole is the linear combination of the aforementioned two. This means that the dipole axis in the $q$-distribution should not necessarily coincide with that of the CMB dipole, assuming that the latter is purely kinematical and therefore collinear to $\tilde{v}_a$~\cite{2025glc..conf....104T}.\footnote{In all of the above, the observer has been assumed to reside close to the centre of a nearly spherical bulk flow, which is moving coherently relative to a largely isotropic distribution of distant sources. Violating any of these assumptions should inevitably lead to additional types of anisotropy on top of the dipolar. For example, observers residing well away from the centre of the dark flow should also see a hemisphere anisotropy in the sky-distribution of their deceleration parameter. Anisotropies in the bulk-flow kinematics are likely to cause quadrupolar distortions in the $q$-distribution. Accounting for the individual peculiar motions of the sources could further complicate the overall picture, and so on. Nevertheless, of all the different types of anisotropy, the dipolar one is the ``trademark'' signature of relative motion and the reason we have focused on it.}

Dipole-like anisotropies in the sky distribution of the deceleration have been reported in a number of surveys, with~\cite{2010MNRAS.401.1409C} being the first one (to the best of our knowledge). Nevertheless, it was not until the analysis of the JLA catalogue by~\cite{2019A&A...631L..13C} and of the $\Omega_{\Lambda}$ distribution by~\cite{2024ApJ...971...19C} that the measured $q$-dipole was attributed to our peculiar motion relative to the CMB frame. Subsequent analysis of the recent Pantheon+ dataset has confirmed the initial results and also verified that the magnitude of the aforementioned dipole decays with increasing scale/redshift~\cite{2025EPJC...85..596S}, as predicted by the ``tilted universe'' scenario. The reader is referred to~\cite{2019A&A...631L..13C,2024ApJ...971...19C,2025EPJC...85..596S} for the details,  as well as to \S~\ref{sssq-D} here for a discussion of the highlights.

\subsection{Doppler-like dipoles in the Hubble
%%%%%%%%%%%%%%%%%%%%%%%%%%%%%%%%%%%%%%%%%%%%%%
parameter}\label{ssDDHP}
%%%%%%%%%%%%%%%%%%%%%%%%
The close relation between the Hubble and the deceleration parameters - the latter is essentially the time-derivative of the former - suggests that peculiar motions should also trigger an analogous Doppler-like dipole in the sky-distribution of the former. Anisotropies in the cosmic expansion are typically monitored by the expansion tensor, defined in Eq.~(\ref{Thetaab}). Employing the latter, setting $\Theta_{ab}n^an^b=H$ in the CMB frame and proceeding as in \S~\ref{sssDDDP}, gives~\cite{2025glc..conf....104T}
\begin{equation}
\tilde{\Theta}_{ab}n^an^b= H+ n^b{\rm D}_b \left(\tilde{v}_an^a\right)=  H+ n^a{\rm D}_a \left(\tilde{v}\cos\phi\right)\,,  \label{tHn}
\end{equation}
with $0\leq\phi\leq\pi$. The above ensures that the observer's peculiar motion leads to a Doppler-like dipolar anisotropy in the Hubble parameter, exactly analogous to the one induced in the deceleration parameter. Indeed, keeping the direction of $n_a$ fixed, we obtain
\begin{equation}
\tilde{\Theta}_{ab}n^an^b= H+ n^a\tilde{\rm D}_a\tilde{v} \hspace{10mm} {\rm and} \hspace{10mm} \tilde{\Theta}_{ab}n^an^b= H- n^a\tilde{\rm D}_a\tilde{v}  \label{DtH1}
\end{equation}
when $\tilde{v}_a\uparrow\uparrow n^a$ and $\tilde{v}_a\uparrow\downarrow n^a$ respectively (alternatively for $\phi=0$ and for $\phi=\pi$)~\cite{2025glc..conf....104T}.

The presence of the spatial gradient terms on the right-hand side of both (\ref{DtH1}a) and (\ref{DtH1}b) ensures that the magnitude of the induced dipole should decrease on progressively higher redshifts, where both the peculiar-velocity field and its gradient are also expected to drop. As shown in the previous section, the same also applies to the magnitude of the $\tilde{q}$-dipole. The latter, however, carries additional effects coming from the gradient of the ``peculiar acceleration''  ($\dot{\tilde{v}}$ -- see Eq.~(\ref{tQn4}) in \S~\ref{sssDDDP}). This is to be expected, since the deceleration parameter is essentially the time derivative of the Hubble parameter. This also means that, to a larger or lesser degree, the two dipoles should differ both in magnitude and in direction~\cite{2025glc..conf....104T}.\footnote{It is worth pointing out that, although peculiar motions can have a strong effect on the local value of the deceleration parameter (even at the linear level -- see \S~\ref{ssRMEDP} earlier), their impact on the Hubble parameter is minimal. Indeed, in line with the linear expression (\ref{lab34}a), the expansion rates measured in the tilted and the CMB frames are related by $\tilde{\Theta}=\Theta+\tilde{\vartheta}$. Therefore, the ``correction'' to the local Hubble value due to the tilted observer's peculiar flow is determined by the ratio $|\tilde{\vartheta}|/\Theta=\tilde{\vartheta}/3H$, which is much smaller than unity on scales of cosmological interest (i.e.~of the order of 100~Mpc and beyond). Indeed, based on the current observations, the difference the above correction term will make to the local value ($\tilde{H}$) of the Hubble parameter will appear in the second decimal point at best.}

A dipolar asymmetry in the observed distribution of the Hubble parameter was (to the best of our knowledge) first reported in~\cite{2020A&A...636A..15M,2021A&A...649A.151M}. There, the authors suggested that a bulk flow with size close to 500~Mpc and speed around 800~km/sec could explain the measured dipole. A subsequent  independent analysis of the Pantheon+ data also reported an $H$-dipole, this time with a redshift-decaying magnitude~\cite{2025EPJC...85..596S}, as predicted by the ``tilted universe'' paradigm. The reader is also referred to \S~\ref{sssH-D} here for an outline of the studies and their results.

\subsection{Doppler-like dipoles in the number
%%%%%%%%%%%%%%%%%%%%%%%%%%%%%%%%%%%%%%%%%%%%%%
counts}\label{ssD-LDNCs}
%%%%%%%%%%%%%%%%%%%%%%%%
Observationally speaking, the high isotropy of the CMB seems to provide irrefutable evidence that our universe is highly uniform on large enough scales. Following the theoretical arguments of~\cite{1984MNRAS.206..377E}, however, this belief has been recently put under question. The reason can be traced to number of reports claimed dipolar anisotropies in the number counts of distant sources that appeared in disagreement with their CMB counterpart both in magnitude and in direction. It was the differences in the dipole amplitudes, in particular, that raised the major questions, which went as far as to doubt the validity of the Cosmological Principle itself. Such scenarios have also been proposed in the past, though primarily as theoretical alternatives. The reader is referred to \S~\ref{sssNCDs} here for a brief discussion of the observations and to the recent review of~\cite{2025RvMP...97d1001S} for details and more references.

The question of the Cosmological Principle lies outside the scope of this review. We will therefore outline here the results of the simple, though informative, study of~\cite{1992ApJ...395...34B} on the peculiar-velocity effects on the number counts of distant astrophysical sources, referring the reader to studies, such as those of~\cite{2011PhRvD..84d3516C,2022JCAP...01..061C} for example, for more extensive treatments.

Depending on their aims, typical cosmological studies employ two alternative reference frames. These are the energy (or Landau-Lifshitz) frame and the particle (or Eckart) frame, with associated 4-velocity vectors $u_a^{(E)}$ and $u_a^{(N)}$ respectively. In the former, the energy-flux vector is zero by default, while in the latter it is the particle-flux vector that vanishes (see decompositions (\ref{NatNa}) next). In a general spacetime the aforementioned coordinate systems are distinct, but in our homogeneous and isotropic FLRW background, where all physical phenomena obey the (strict) cosmological principle, $u_a^{(E)}$ are $u_a^{(N)}$ are assumed to coincide.

Let us consider a distribution of astrophysical sources (e.g.~supernovae, radio galaxies, etc) with 4-current density $\mathcal{N}_a$. As in \S~\ref{sssNC} before, the sources are at sufficiently high redshifts to ignore their individual peculiar velocities and treat them as comoving. Also, the number counts are measured locally (i.e.~at the same spacetime event) by the tilted and the CMB observers. Expressed relative to the CMB and the tilted frames, the 4-current density decomposes as
\begin{equation}
\mathcal{N}_a= nu_a+ \mathcal{J}_a \hspace{15mm} {\rm and} \hspace{15mm} \mathcal{N}_a= \tilde{n}\tilde{u}_a+ \tilde{\mathcal{J}}_a\,,  \label{NatNa}
\end{equation}
respectively~\cite{2008PhR...465...61T,2012reco.book.....E}. In the above $n=-\mathcal{N}_au^a$ and $\tilde{n}= -\mathcal{N}_a\tilde{u}^a$ are the number densities of the sources, while $\mathcal{J}_a=h_a{}^b\mathcal{N}_b$ and $\tilde{\mathcal{J}}_a=\tilde{h}_a{}^b\mathcal{N}_b$ are their source/particle flux vectors, as measured by the associated observers (with $\mathcal{J}_au^a=0= \tilde{\mathcal{J}}_a\tilde{u}^a$).

Applying Eqs.~(\ref{NatNa}) to a tilted almost-FLRW universe and linearising, the two number densities and the corresponding flux vectors are related by
\begin{equation}
\tilde{n}= n \hspace{15mm} {\rm and} \hspace{15mm} \tilde{\mathcal{J}}_a= \mathcal{J}_a- n\tilde{v}_a\,,  \label{nlrels1}
\end{equation}
to first approximation~\cite{1992ApJ...395...34B}. Therefore, although the number densities measured in the two frames are the same (at the linear level), the fluxes of the sources differ due to relative-motion effects alone. Indeed, when $\tilde{v}_a=0$, expression (\ref{nlrels1}b) ensures that $\tilde{\mathcal{J}}_a= \mathcal{J}_a$. Since the universe appears isotropic in the idealised frame of the CMB, we may set $\mathcal{J}_a=0$ there. Then, relation (\ref{nlrels1}b) reduces to
\begin{equation}
\tilde{\mathcal{J}}_a= -n\tilde{v}_a\,,  \label{nlrels2}
\end{equation}
ensuring that the tilted (i.e.~the realistic) observer sees a nonzero (effective) flux of sources entirely due to their peculiar motion.

According to Eq.~(\ref{nlrels1}a), for a given type of sources (e.g.~radio galaxies), all typical observers in the universe should measure the same number density at redshifts large enough for the linear approximation to hold. Moreover, the aforementioned number density should agree with that measured by their idealised counterparts in the CMB frame. Following (\ref{nlrels2}), however, all real observers ``experience'' a nonzero source-drift, when their CMB associates ``see'' none. What is most important, is that the (apparent) flux measured in the tilted frame depends on the number density of the corresponding sources. More specifically, the magnitude of the flux vector associated with a certain type of astrophysical sources is proportional to their number density. Indeed, expression (\ref{nlrels2}) immediately leads to $\tilde{\mathcal{J}}= n\tilde{v}$, with $\tilde{\mathcal{J}}= \sqrt{\tilde{\mathcal{J}}_a\tilde{\mathcal{J}}^a}$ and $\tilde{v}= \sqrt{\tilde{v}_a\tilde{v}^a}$ by definition.

The presence of an effective nonzero source-flux vector in the tilted frame of the real observers, implies a dipolar anisotropy in their number counts. This dipole is clearly apparent, since it is triggered solely by relative-motion effects. Indeed, even when the sources are isotropically distributed in space and simply comoving with the expansion, observers moving with peculiar velocity $\tilde{v}_a$ ``see'' an effective source-flux $\tilde{\mathcal{J}}_a^+=-n\tilde{v}_a$ coming towards them and a source-flux $\tilde{\mathcal{J}}_a^-=-\tilde{\mathcal{J}}_a^+= n\tilde{v}_a$ moving away from them along the opposite direction. As a result, the tilted observers will measure more sources in the direction of their peculiar flow and equally fewer in the antipodal. The amplitudes of such apparent dipoles should differ from source to source, since they depend on their number densities. For instance, assuming that the number density of the supernovae differs from that of, say, the radio galaxies, the apparent dipoles seen in the two distributions should have different amplitudes as well.

Provided that the reported number-count dipoles and their CMB counterpart are all kinematically induced, one expects their axes to essentially coincide. This expectation is supported by Eq.~(\ref{nlrels2}), according to which the spacelike vectors $\tilde{\mathcal{J}}_a$ and $\tilde{v}_a$ are (anti) parallel. Nevertheless, expression (\ref{nlrels2}) alone does not provide the full picture, since the aforementioned 3-vectors evolve differently in time. Indeed, taking the temporal derivative of (\ref{nlrels2}) and then using the background time-evolution law $\dot{n}=-3Hn$ for the number density, we arrive at the linear expression
\begin{equation}
\dot{\tilde{\mathcal{J}}}_a= -3H\tilde{\mathcal{J}}_a- n\dot{\tilde{v}}_a\,,  \label{ltcJ'1}
\end{equation}
ensuring that $\dot{\tilde{\mathcal{J}}}_a\neq\dot{\tilde{v}}_a$. Therefore, even if $\tilde{\mathcal{J}}_a$ and $\tilde{v}_a$ were along the same direction initially, their alignment should change as time goes by.

The fact that large-scale relative motions can trigger (apparent) dipolar anisotropies in the number counts of distant astrophysical sources has been largely known (e.g.~see~\cite{1992ApJ...395...34B}). What is not so well known perhaps, is that the amplitudes of the aforementioned dipoles vary because they are proportional to the number densities of the associated sources. Such differences may come to lie within our observational capabilities soon, if not already. In addition, the directions of the induced dipolar axes differ from source to source, as well as from that of their CMB counterpart, due to evolution effects. All this should be taken into account when considering the inconsistencies reported in the number counts of distant astrophysical sources and their potential implications for the Cosmological Principle.

\section{Peculiar velocities and structure formation}\label{sPVSF}
%%%%%%%%%%%%%%%%%%%%%%%%%%%%%%%%%%%%%%%%%%%%%%%%%%%%%%%%%%%%%%%%%%
Peculiar velocities can in principle affect the process of structure formation during the post-recombination era. The peculiar-velocity field can play a role during both the linear and the nonlinear regimes of structure formation, but the strength and the outcome of its effect seems to depend on the underlying theory of gravity (i.e.~Newtonian or relativistic).

\subsection{Linear Newtonian and quasi-Newtonian
%%%%%%%%%%%%%%%%%%%%%%%%%%%%%%%%%%%%%%%%%%%%%%%%
treatments}\label{ssLNQ-NTs}
%%%%%%%%%%%%%%%%%%%%%%%%%%%%
At the nonlinear perturbative level, the Newtonian evolution of density inhomogeneities are monitored by the system (e.g.~see~\cite{1990MNRAS.243..509E})
\begin{equation}
\dot{\Delta}_{\alpha}= -\mathcal{Z}_{\alpha}- (\sigma_{\beta\alpha}+\omega_{\beta\alpha})\Delta^{\beta}  \label{NnlDeldot}
\end{equation}
and
\begin{equation}
\dot{\mathcal{Z}}_{\alpha}= -{2\over3}\,\Theta\mathcal{Z}_{\alpha}- {1\over2}\,\rho\Delta_{\alpha}+ a\partial_{\alpha}\mathcal{A}- 2a\partial_{\alpha} \left(\sigma^2-\omega^2\right)- (\sigma_{\beta\alpha}+\omega_{\beta\alpha})\mathcal{Z}^{\beta}\,,  \label{NnlcZdot}
\end{equation}
with the gradients
\begin{equation}
\Delta_{\alpha}= {a\over\rho}\,\partial_{\alpha}\rho \hspace{15mm} {\rm and} \hspace{15mm} \mathcal{Z}_{\alpha}=a\partial_{\alpha}\Theta\,,  \label{NDeltacZ}
\end{equation}
describing inhomogeneities in the matter density and in the universal expansion respectively (see \S~\ref{sssKRPF} for their relativistic counterparts). Also note that $\mathcal{A}=\partial^{\alpha}A_{\alpha}$ by construction, where the vector $A_{\alpha}=\dot{u}_{\alpha}+\partial_{\alpha}\Phi$ carries the combined effect of the kinematic and the gravitational acceleration and provides the Newtonian analogue of the relativistic 4-acceleration. In what follows, we will linearise the above set around the Newtonian version of the Einstein-de Sitter universe, while allowing for peculiar-velocity perturbations at the same time.

In the absence of matter pressure, Euler's equation ensures that $\dot{u}_{\alpha}=-\partial_{\alpha}\Phi$ always. This means that $A_{\alpha}=0$ and consequently that $\mathcal{A}=0$ irrespective of the presence of peculiar motions. On these grounds, the linear evolution of density inhomogeneities on our (Newtonian) Einstein-de Sitter background, is monitored by the system
\begin{equation}
\dot{\Delta}_{\alpha}= -\mathcal{Z}_{\alpha} \hspace{15mm} {\rm and} \hspace{15mm} \dot{\mathcal{Z}}_{\alpha}= -2H\mathcal{Z}_{\alpha}- {1\over2}\rho\Delta_{\alpha}\,,  \label{lNDeltacZdot}
\end{equation}
familiar from Newtonian studies of linear density perturbations in cosmologies free of peculiar motions. Combined with the background relation $\rho=3H^2$, the system of (\ref{lNDeltacZdot}) leads to
\begin{equation}
\ddot{\Delta}_{\alpha}+ 2H\dot{\Delta}_{\alpha}- {3\over2}\,H^2\Delta_{\alpha}= 0\,,  \label{lNddotDelta}
\end{equation}
The above differential equation, which is identical to the one governing the evolution of linear density inhomogeneities in the absence of peculiar motions (e.g.~see~\cite{1990MNRAS.243..509E}), accepts the familiar power-law solution
\begin{equation}
\Delta_{\alpha}= \mathcal{C}_1t^{2/3}+ \mathcal{C}_2t^{-1}= \mathcal{C}_3a+ \mathcal{C}_4a^{-3/2}\,,  \label{lNDelta}
\end{equation}
on all scales (recall that $a\propto t^{2/3}$ in the Einstein-de Sitter universe). Accordingly, in Newtonian theory, peculiar velocities leave the linear growth-rate of density inhomogeneities unaffected.\footnote{In typical Newtonian studies, like that pursued in~\cite{1991ApJ...379....6N} for example, the peculiar-velocity field is treated as a perturbation of the Hubble velocity and it is given by $v_{\alpha}=\dot{x}_{\alpha}$ in comoving coordinates ($\left\{x^{\alpha}\right\}$), or by $v_{\alpha}=a\dot{x}_{\alpha}$ in physical coordinates ($\left\{r^{\alpha}\right\}$, with $r^{\alpha}=ax^{\alpha}$). At the same time, density perturbations are described by the density contrast $\delta=\delta\rho/\rho$. The linear evolution of the latter is monitored by the homogeneous differential equation $\ddot{\delta}+2H\dot{\delta}-(3H^2/2)\delta=0$ (e.g.~see~\cite{1991ApJ...379....6N}).}

Looking back to the discussion given in \S~\ref{ssQ-NA}, it should come to no surprise that the quasi-Newtonian treatment also finds that peculiar motions have no real effect on the linear evolution of density perturbations (e.g.~see~\cite{1998PhRvD..58l4006M}). In order to demonstrate this, let us set $\tilde{q}_a=0=\tilde{A}_a$ in the coordinate system of the matter and thus switch the peculiar-velocity effects off in the tilted frame. This means that, to linear order,
\begin{equation}
\tilde{\Delta}_a= \mathcal{C}_1t^{2/3}+ \mathcal{C}_2t^{-1}\,,  \label{tDelta}
\end{equation}
where the spatial gradient
$\tilde{\Delta}_a=(a/\rho)\tilde{\rm D}_a\rho$
describes density inhomogeneities in the matter frame. The situation changes in the coordinate system of the CMB, where $q_a=\rho\tilde{v}_a$ and $A_a=-\dot{\tilde{v}}_a-H\tilde{v}_a$ (see expressions (\ref{lqNrels}a) and (\ref{lqNrels}b) in \S~\ref{sssLQ-NA} earlier). There, the effects of the peculiar motion on the evolution of density perturbations are still active. Indeed, starting from the definition of $\tilde{\Delta}_a$ given above, employing the background continuity equation and keeping up to linear-order terms, we obtain~\cite{1998PhRvD..58l4006M}
\begin{equation}
\tilde{\Delta}_a= {a\over\rho}\,\tilde{h}_a{}^b\nabla_b\rho= \Delta_a- 3aH\tilde{v}_a\,,  \label{tDelta2}
\end{equation}
since $\tilde{h}_{ab}=h_{ab}+2u_{(a}\tilde{v}_{b)}$ to first approximation. Also, the gradient $\Delta_a=(a/\rho){\rm D}_a\rho$ monitors density perturbations in the CMB frame by definition. Recalling that $a\propto t^{2/3}$, $H\propto t^{-1}$ and using solution (\ref{tDelta}), the above recasts into
\begin{equation}
\Delta_a= \mathcal{C}_1t^{2/3}+ \mathcal{C}_2t^{-1}+ 3a_0H_0\left({t_0\over t}\right)^{1/3}\tilde{v}_a\,,  \label{Delta1}
\end{equation}
Therefore, the linear growth-rate of the density contrast, depends on that of the peculiar velocity. According to the quasi-Newtonian analysis~\cite{1998PhRvD..58l4006M,2001CQGra..18.5115E}, we have $\tilde{v}_a=C_1t^{1/3}+ C_2t^{-4/3}$, the growing mode of which leads to~\cite{1998PhRvD..58l4006M}
\begin{equation}
\Delta_a= \mathcal{C}_1t^{2/3}+ \mathcal{C}_2t^{-1}+ \mathcal{C}_3\,,  \label{Delta2}
\end{equation}
when substituted back into solution (\ref{Delta1}). Clearly, the growth-rate of the density contrast remains unaffected by the peculiar-velocity field, in full agreement with the Newtonian study. In retrospect, this null result should not come to our surprise, since the Newtonian and quasi-Newtonian treatments arrive at exactly the same evolution law for linear peculiar velocities (see~\cite{1998PhRvD..58l4006M,2001CQGra..18.5115E}, or compare solutions (\ref{ltv2}) and (\ref{lqNv}) in \S~\ref{sssLEPVs} and \S~\ref{sssLQ-NA} respectively).

Solution (\ref{Delta2}) and the conclusions derived from it, confirm that the quasi-Newtonian analysis of peculiar velocities reduces to purely Newtonian, despite its initially relativistic setup. The reason is the strict constraints that the quasi-Newtonian approximation imposes upon the perturbed spacetime, which compromise its relativistic nature and thus lead to Newtonian-like equations and results (see discussion in \S~\ref{ssQ-NA} earlier and \S~6.8.2 in~\cite{2012reco.book.....E} for warning comments).

\subsection{Linear relativistic treatment}\label{ssLRT}
%%%%%%%%%%%%%%%%%%%%%%%%%%%%%%%%%%%%%%%%%%%%%%%%%%%%%%%
During the evolution of the universe, there are periods when the cosmic medium corresponds to a mixture of several fluids, for example to a composition of baryons, dark matter, radiation and neutrinos, rather than to a single component. The relativistic treatment of such multi-fluid systems requires the use of multiple frames in relative motion with each other.

\subsubsection{Multiple perfect fluids}\label{sssMPFs}
%%%%%%%%%%%%%%%%%%%%%%%%%%%%%%%%%%%%%%%%%%%%%%%%%%%%%%
The aforementioned species may interact, or not, with each other and they generally have their individual 4-velocities, instead of sharing the same one. In the latter case, assuming that $u_a$ defines the reference (typically the CMB) frame and that the $i$-th species have non-relativistic peculiar velocities ($\tilde{v}_a^{(i)}$) with respect to $u_a$, their 4-velocity ($\tilde{u}_a^{(i)}$) is given by
\begin{equation}
\tilde{u}_a^{(i)}= u_a+ \tilde{v}_a^{(i)}\,,  \label{lui}
\end{equation}
where $u^a\tilde{v}_a^{(i)}=0$, $h_{ab}^{(i)}=g_{ab}+u_a^{(i)}u_b^{(i)}$ by construction and $\tilde{v}_{(i)}^2\ll1$ at the non-relativistic limit (e.g.~see~\cite{2008PhR...465...61T,2012reco.book.....E} and also \S~\ref{ssRMOs} here).

Suppose also that the individual matter components are perfect fluids with energy density $\rho^{(i)}$ and (isotropic) pressure $p^{(i)}$. Then, the associated stress-energy tensor, written in the coordinate system of the species in question, reads
\begin{equation}
T_{ab}^{(i)}= \rho^{(i)}u_a^{(i)}u_b^{(i)}+ p^{(i)}h_{ab}^{(i)}\,.  \label{Ti1}
\end{equation}
Alternatively, one could express the energy-momentum tensor of the $i$-th component relative to the reference $u_a$-frame. Keeping in mind that the latter coordinate system has ``peculiar'' velocity $v_a^{(i)}=-\tilde{v}_a^{(i)}$ with respect to each one of the species, the above recasts as
\begin{equation}
T_{ab}^{(i)}= \hat{\rho}u_au_b+ \hat{p}h_{ab}+ 2u_{(a}\hat{q}_{b)}^{(i)}+ \hat{\pi}_{ab}^{(i)}\,, \label{Ti2}
\end{equation}
with
\begin{equation}
\hat{\rho}^{(i)}= \rho^{(i)}\,, \hspace{10mm} \hat{p}^{(i)}= p^{(i)}\,, \hspace{10mm} \hat{q}_a^{(i)}= \left(\rho^{(i)}+p^{(i)}\right)\tilde{v}_a^{(i)} \hspace{7.5mm} {\rm and} \hspace{7.5mm} \hat{\pi}_{ab}^{(i)}= 0\,,  \label{lhatvars}
\end{equation}
to first approximation (e.g.~see~\cite{2008PhR...465...61T,2012reco.book.....E} and also \S~\ref{ssRBFs} here). Following Eqs.~(\ref{Ti2}) and (\ref{lhatvars}), despite the fact that the individual matter components are perfect fluids in their individual rest frames, relative to the reference $u_a$-field there is a nonzero linear effective energy-flux vector ($\hat{q}_a^{(i)}$) solely triggered  by the relative motion of the species.

Setting $\rho=\Sigma_i\hat{\rho}^{(i)}$, $p=\Sigma_i\hat{p}^{(i)}$ and $q_a=\Sigma_i\hat{q}_a{(i)}$ as the respective sums of the energy densities, the isotropic pressures and the energy fluxes of the matter components, the stress-energy tensor of the total fluid, written in the reference coordinate system, acquires the form
\begin{equation}
T_{ab}= \rho u_au_b+ p h_{ab}+ 2u_{(a}q_{b)}  \label{Tt}
\end{equation}
and satisfies the familiar conservation law $\nabla^bT_{ab}=0$. The above, together with the linear relations (\ref{Ti2}) and (\ref{lhatvars}), confirms that in the presence of peculiar velocities, the cosmic medium can no longer be treated as perfect. The ``imperfection'' appears as an effective energy flux, which is induced by relative-motion effects alone (see also discussion in \S~\ref{ssLRBFs} and \S~\ref{sssKRPF}, with related references therein).

Let us proceed by assuming that the total fluid has an effective equation of state of the barotropic form, namely set $p=w\rho$, where $w$ is the associated (effective) barotropic index. Then, the conservation laws for the energy and the momentum densities (see expressions (\ref{edcl}) and (\ref{mdcl}) in \S~\ref{sssCLs}) linearise to
\begin{equation}
\dot{\rho}= -\Theta(1+w)\rho- {\rm D}^aq_a  \label{ltecl}
\end{equation}
and
\begin{equation}
a(1+w)\rho A_a= -c_s^2\rho\Delta_a- \dot{q}_a- 4Hq_a\,,  \label{ltmcl}
\end{equation}
respectively. Note that the spatial gradients $\Delta_a=(a/\rho){\rm D}_a\rho$ and $\mathcal{Z}_a=a{\rm D}_a\Theta$ carry the effects of the total fluid. Also, in deriving the latter of the above, we have used the barotropic relation $a{\rm D}_ap=c_s^2\rho\Delta_a$, with $c_s^2={\rm d}p/{\rm d}\rho$ representing the square of the sound speed.

\subsubsection{Multiple non-interacting perfect
%%%%%%%%%%%%%%%%%%%%%%%%%%%%%%%%%%%%%%%%%%%%%%%
fluids}\label{sssMN-IPFs}
%%%%%%%%%%%%%%%%%%%%%%%%%
Provided the individual matter components do not interact with each other, their stress energy tensors are separately conserved, that is $\nabla^bT_{ab}^{(i)}=0$. Then, on using (\ref{Ti2}) and the linear relations (\ref{lhatvars}), we arrive at the linear conservation laws
\begin{equation}
\dot{\rho}^{(i)}= -\Theta\left(1+w^{(i)}\right)\rho^{(i)}- {\rm D}^aq_a^{(i)}  \label{iledcl}
\end{equation}
and
\begin{equation}
a\left(1+w^{(i)}\right)\rho^{(i)}A_a= -c^{2(i)}_s\rho^{(i)}\Delta_a^{(i)}- a\left(\dot{q}_a^{(i)}+4Hq^{(i)}\right)\,,  \label{ilmcl}
\end{equation}
for the energy and the momentum of the $i$-th species respectively. Here, $w^{(i)}=p^{(i)}/\rho^{(i)}$, $c_s^{2(i)}={\rm d}p^{(i)}/{\rm d}\rho^{(i)}$ and $\Delta_a^{(i)}=(a/\rho^{(i)}){\rm D}_a\rho^{(i)}$ by definition. The latter variable describes inhomogeneities in the density distribution of the $i$-th matter component and obeys the linear evolution law
\begin{equation}
\dot{\Delta}_a^{(i)}= 3Hw^{(i)}\Delta_a^{(i)}- \left(1+w^{(i)}\right)\mathcal{Z}_a+ {3aH\over\rho^{(i)}}\left(\dot{q}_a^{(i)}+4Hq_a^{(i)}\right)- {a\over\rho^{(i)}}\,{\rm D}_a{\rm D}^bq_b^{(i)}\,,  \label{liDeldot1}
\end{equation}
On using (\ref{ilmcl}), one can replace the third  term on the right-hand side of the above differential equation, which then recasts into
\begin{equation}
\dot{\Delta}_a^{(i)}= 3H\left(w^{(i)}-c_s^{2(i)}\right)\Delta_a^{(i)}- \left(1+w^{(i)}\right)\mathcal{Z}_a- 3aH\left(1+w^{(i)}\right)A_a- {a\over\rho^{(i)}}\,{\rm D}_a{\rm D}^bq_b^{(i)}\,.  \label{liDeldot2}
\end{equation}
Finally, employing the momentum conservation law of the total fluid (see Eq.~(\ref{ltmcl}) above), one arrives at
\begin{eqnarray}
\dot{\Delta}_a^{(i)}&=& 3H\left(w^{(i)}-c_s^{2(i)}\right)\Delta_a^{(i)}- \left(1+w^{(i)}\right)\mathcal{Z}_a+ {3Hc_s^2\left(1+w^{(i)}\right)\over1+w}\,\Delta_a \nonumber\\ &&+{3aH\left(1+w^{(i)}\right)\over\rho(1+w)} \left(\dot{q}_a+4Hq_a\right)- {a\over\rho^{(i)}}\,{\rm D}_a{\rm D}^bq_b^{(i)}\,.  \label{liDeldot3}
\end{eqnarray}

It should be noted that the second-last term on the right-hand side of (\ref{liDeldot3}) involves the total flux of the multi-fluid system, whereas the last term contains the peculiar flux of the $i$-th species only. Also note that the gradient $\mathcal{Z}_a=a{\rm D}_a\Theta$ carries the contribution of the total fluid to the universal expansion and satisfies the linear evolution law
\begin{equation}
\dot{\mathcal{Z}}_a= -2H\mathcal{Z}_a- {1\over2}\,\left(1+3c_s^2\right)\rho\Delta_a- {3\over2}\,\rho(1+w)aA_a+ a{\rm D}_a{\rm D}^bA_b\,.  \label{lcZdot1}
\end{equation}
On using the momentum conservation law of the total fluid (see Eq.~(\ref{ltmcl}) previously), employing the 3-Ricci identities (see the commutation law (\ref{3Ricci1}b) in \S~\ref{sssSC} earlier), while keeping in mind the spatial flatness of our Friedmann background, the above recasts into
\begin{eqnarray}
\dot{\mathcal{Z}}_a&=& -2H\mathcal{Z}_a- {1\over2}\,\rho\Delta_a- {c_s^2\over1+w}\,{\rm D}^2\Delta_a+ {3\over2}\,a(\dot{q}_a+4Hq_a)\nonumber\\ &&-{a\over\rho(1+w)}\,{\rm D}_a{\rm D}^b(\dot{q}_b+4Hq_b)\,.  \label{lcZdot2}
\end{eqnarray}

The differential formulae (\ref{liDeldot3}) and (\ref{lcZdot2}) monitor the linear evolution of a mixture of non-comoving, non-interacting barotropic perfect fluids on an spatially flat FLRW background.\footnote{For the system governing interacting, non-barotropic ideal media on a general Friedmann background, as well as for further discussion and additional references, the interested reader is referred to \S~3.3.1, \S~3.3.2 and \S~3.3.3 of~\cite{2008PhR...465...61T}, or/and to \S~10.3.2 of~\cite{2012reco.book.....E}.} In order to close the system, one needs to specify the form of the peculiar fluxes ($q_a^{(i)}$ and $q_a$). Next, we will do this within a perturbed spatially flat FLRW background cosmology that contains a composition of radiation and pressure-free dust.

\subsection{The Meszaros ``stagnation'' effect}\label{ssMSE}
%%%%%%%%%%%%%%%%%%%%%%%%%%%%%%%%%%%%%%%%%%%%%%%%%%%%%%%%%%%%
Consider a mixture of radiation and non-relativistic species in a perturbed Friedmann universe with flat spatial sections. This model could correspond to the later radiation era, where the relativistic species are photons and their non-relativistic counterparts are primarily cold dark matter and secondarily baryons. In such a case, the total energy density and pressure are $\rho=\rho^{(r)}+\rho^{(d)}$ and $p=p^{(r)}=\rho^{(r)}/3$ respectively. Then, the effective barotropic index of the radiation-dust mixture is $w=\rho^{(r)}/3(\rho^{(r)}+\rho^{(d)})$, while the effective sound speed is given by $c_s^2=4\rho^{(r)}/[3(4\rho^{(r)}+3\rho^{(d)})]$ (e.g.~see~\cite{2008PhR...465...61T,2012reco.book.....E} and references therein).

During the radiation era, when photons dominate the energy density of the universe, we may assume that $\rho^{(d)}/\rho^{(r)}\ll1$. At the same time the dark component is believed to dominate over its baryonic counterpart. Then, since dark matter does not interact with neither photons not baryons and neglecting any photon-baryon interactions, we may use the system (\ref{liDeldot3}) and (\ref{lcZdot2}) to monitor linear perturbations in the density of the pressureless matter. Finally, let us also assume that the radiation field is homogeneously distributed (i.e.~set $\Delta_a^{(r)}\simeq0$) on the scales of interest. Then, applied to linear inhomogeneities in the density of the dust, the system (\ref{liDeldot3}) and (\ref{lcZdot2}) reduces to\footnote{Prior to equipartition, radiation dominates the energy density of the universe, which means that $\rho^{(d)}/\rho^{(r)}\ll1$. On these grounds, we may write $w\simeq1/3\simeq c_s^2$ for the total effective barotropic index and sound speed. Moreover, by definition we have $\Delta_a\simeq\Delta_a^{(r)}+ (\rho^{(d)}/\rho^{(r)})\Delta_a^{(d)}\simeq0$ to linear order, since $\Delta_a^{(r)}\simeq0$ and $\rho^{(d)}/\rho^{(r)}\ll1$. Also $\rho\Delta_a=\rho^{(r)}\Delta_a^{(r)}+\rho^{(d)}\Delta_a^{(d)}\simeq \rho^{(d)}\Delta_a^{(d)}$, given that $\Delta_a^{(r)}\simeq0$. For a more rigorous derivation of the above linear results, which also involves effective entropy perturbations, the reader is referred to~\cite{1992ApJ...395...54D} and/or to \S~3.3.3 and \S~3.3.4 of~\cite{2008PhR...465...61T}).}
\begin{equation}
\dot{\Delta}_a^{(d)}= -\mathcal{Z}_a- a{\rm D}_a{\rm D}^b\tilde{v}_b^{(d)} \hspace{15mm} {\rm and} \hspace{15mm} \dot{\mathcal{Z}}_a= -2H\mathcal{Z}_a- {1\over2}\,\rho^{(d)}\Delta_a^{(d)}\,,  \label{lrd}
\end{equation}
where the spatial gradient $\Delta_a^{(d)}=(a/\rho^{(d)}{\rm D}_a\rho^{(d)}$ describes inhomogeneities in the density distribution of the pressureless component and $\tilde{v}_a^{(d)}$ is the latter's peculiar velocity relative to the reference (CMB) frame (e.g.~see~~\cite{2008PhR...465...61T,2012reco.book.....E}).

It is important to note that the above system has been written in the so-called Landau-Lifshitz (or energy) frame, where  the total flux to zero (i.e.~$q_a=\Sigma_iq_a^{(i)}=0$) by default~\cite{2008PhR...465...61T,2012reco.book.....E}.\footnote{The energy, or Landau-Lefshitz, frame is sometimes also called the ``center of mass'' frame~\cite{2019JCAP...06..041C}.} The peculiar fluxes of the individual species, on the other hand, can take nonzero values. This means that the relativistic contribution of the (total) peculiar flux to the gravitational field (see related discussion in \S~\ref{sssKRPF} earlier) is unaccounted for. The implications of bypassing this purely general relativistic effect will become clear as we proceed. To begin with, the momentum conservation of the total fluid (see Eq.~(\ref{ltmcl}) in \S~\ref{sssMPFs}) is now given by the expression $A_a=-\Delta_a/4a$, which implies that the 4-acceleration is zero to linear order (since $\Delta_a=0$ at the same perturbative level - see footnote~31) and explains its absence from the right-hand side of Eq.~(\ref{lrd}b).

The next step is to close the system of (\ref{lrd}a) and (\ref{lrd}b). In order to do so, we need the linear propagation formula of the peculiar velocity field. Applying the momentum-conservation of the $i$-th species (see Eq.~(\ref{ilmcl} in \S~\ref{sssMN-IPFs} before) to pressureless matter and keeping in mind that $A_a=0$, we arrive at
\begin{equation}
\dot{q}_a^{(d)}+ 4q_a^{(d)}= \rho^{(d)}\dot{\tilde{v}}_a^{(d)}+ H\rho^{(d)}\tilde{v}_a^{(d)}= 0\,, \label{intmed1}
\end{equation}
given that $q_a^{(d)}=\rho^{(d)}\tilde{v}_a^{(d)}$ to linear order and $\dot{\rho}^{(d)}=-3H\rho^{(d)}$ in the FLRW background. The above immediately leads to
\begin{equation}
\dot{\tilde{v}}_a^{(d)}= -H\tilde{v}_a^{(d)}\,,  \label{lrdtvdot}
\end{equation}
which describes linear peculiar velocities that decay with the expansion as $\tilde{v}^{(d)}\propto1/a$. A decaying $\tilde{v}$-field is clearly cosmologically unacceptable, however, since it cannot explain the numerous reports of bulk peculiar flows, several of which are in excess of those predicted by the $\Lambda$CDM cosmological model. Indeed, given that peculiar velocities start very weak around recombination, there should be no bulk flows to observe today, if the velocity fields were allowed to decay with time. Moreover, we remind the reader that, instead of decaying, linear peculiar-velocity perturbations grow with the expansion even in the Newtonian limit (as $\tilde{v}\propto t^{1/3}$ - see~\cite{1976ApJ...205..318P,1980lssu.book.....P} and also \S~\ref{ssNA} here). As mentioned above, this inconsistency stems from the choice of the Landau-Lifshitz frame where the total peculiar flux is zero.

Let us leave the drawback of selecting the Landau-Lifshitz frame aside for the moment and concentrate on the system of (\ref{lrd}a), (\ref{lrd}b) and (\ref{lrdtvdot}). Taking the time derivative of (\ref{lrd}a), substituting (\ref{lrd}b), (\ref{lrdtvdot}) and using the linear commutation laws $({\rm D}_af)^{\cdot}={\rm D}_a\dot{f}-H{\rm D}_af$ and $({\rm D}^av_a)^{\cdot}={\rm D}^a\dot{v}_a-H{\rm D}^av_a$, we arrive at (e.g.~see~\cite{2008PhR...465...61T,2012reco.book.....E})
\begin{equation}
\ddot{\Delta}_a^{(d)}+ 2H\dot{\Delta}_a^{(d)}- {1\over2}\,\rho^{(d)}\Delta_a^{(d)}= 0\,.  \label{lrdDelddot}
\end{equation}
This homogeneous differential equation is formally identical to the one governing the linear evolution of inhomogeneities in the dust on an Einstein-de Sitter background. Here, Eq.~(\ref{lrdDelddot}) holds on a spatially flat FLRW background filled with a non-interacting radiation-dust mixture and monitors the linear evolution of inhomogeneities in the dust component, assuming that the radiation field is homogeneously distributed in space. We may therefore use (\ref{lrdDelddot}) to infer the linear evolution of $\Delta_a^{(d)}$ both after and prior to equipartition.

Before proceeding to the solution of (\ref{lrdDelddot}), we should remind the readers that, although Eq.~(\ref{lrdDelddot}) may have been derived by means of an otherwise relativistic approach, it holds in the Landau-Lifshitz frame. There the energy flux vanishes by default, which means that the purely general-relativistic contribution of the (total) peculiar flux to the gravitational field has been bypassed (see \S~\ref{sssKRPF} earlier). As we will show next, switching the flux input off renders the study Newtonian for all practical purposes.

Indeed, recasting (\ref{lrdDelddot}) in terms of the scale factor, leads to~\cite{1974A&A....37..225M}
\begin{equation}
\Delta_a^{(d)\prime\prime}+ {2+3a\over2a(1+a)}\,\Delta_a^{(d)\prime}- {3\over2a(1+a)}\,\Delta_a^{(d)}= 0\,,  \label{lrdDel''}
\end{equation}
with the primes indicating differentiation with respect to the scale factor and with the scale factor normalised so that $a=1$ corresponds to the moment of matter-radiation equality~\cite{1992ApJ...395...54D}. The above differential equation accepts the solution\footnote{The first mode of solution (\ref{lrdDeld}) was originally obtained in~\cite{1974A&A....37..225M}, with the second mode been added subsequently in~\cite{1975A&A....41..143G}. Both solutions are Newtonian in their nature.}
\begin{equation}
\Delta^{(d)}= \mathcal{C}_1\left(1+{3\over2}\,a\right)- \mathcal{C}_2\left[\left(1+{3\over2}\,a\right) \ln\left({\sqrt{1+a}+1\over\sqrt{1+a}-1}\right)-3\sqrt{1+a}\,\right]\,.  \label{lrdDeld}
\end{equation}
Accordingly, we find that $\Delta^{(d)}\propto a$ at late times (i.e.~deep into the dust epoch when $a\gg1$). This agrees with the linear evolution of the density gradients in an Einstein-de Sitter universe free of peculiar motions. Earlier in the radiation era, on the other hand, we have $a\ll1$ and $\Delta^{(d)}$ remains essentially constant (it grows only logarithmically). This ``stagnation'' of the matter perturbations prior to equipartition is also referred to as the ``Meszaros effect''~\cite{1974A&A....37..225M}.

Both (\ref{lrdDel''}) and its associated solution (\ref{lrdDeld}) are identical to their Newtonian counterparts (e.g.~see~\cite{1974A&A....37..225M,1975A&A....41..143G} for the respective expressions). However, this agreement could misleading, since the relativistic analysis presented here degenerated (for all practical purposes) to Newtonian once the Landau-Lifshitz frame was adopted and the gravitational input of the peculiar flux was bypassed.\footnote{In cosmological perturbation theory two are the typical reference coordinate systems: the Landau-Lifshitz frame and the Eckart frame~\cite{2008PhR...465...61T,2012reco.book.....E}. The former sets the energy flux to zero, while the latter switches the particle flux off. This means that adopting the Landau-Lifshitz frame when studying a perturbed multi-component medium may set the energy flux to zero, but triggers a nonzero effective particle flux. Using the Eckart frame, on the other hand does the exact opposite. In either case, a system of several non-comoving perfect fluids cannot be treated as perfect.} In view of this, it would be interesting to see whether the Newtonian picture reproduced in Eqs.~(\ref{lrdDel''}) and (\ref{lrdDeld}) changes when the aforementioned flux-input is properly included in the relativistic studies of the peculiar-velocity effects on the evolution of density perturbations.

Although such a relativistic analysis is still missing, a recent essay has raised the possibility that structure formation could proceed more efficiently when the effects of peculiar motions are accounted for~\cite{2025IJMPD..3444010T}. This could in turn lead to fully formed galaxies at redshifts higher than it is generally anticipated. Perhaps like those reported on the basis of the James Webb Space Telescope (JWST) data~\cite{2022ApJ...938L..15C,2023ApJ...942L..27S,2023ApJ...946L..13F,2023ApJS..265....5H,2025NatAs...9..729H}, which seems at odds with the $\Lambda$CDM model~\cite{2022ApJ...939L..31H,2023NatAs...7..731B,2023MNRAS.518.2511L}.

\subsection{The nonlinear regime and the Zel'dovich
%%%%%%%%%%%%%%%%%%%%%%%%%%%%%%%%%%%%%%%%%%%%%%%%%%%
approximation}\label{ssNRZA}
%%%%%%%%%%%%%%%%%%%%%%%%%%%%
Most of the analytic studies looking into the evolution of density inhomogeneities in the universe are linear. Put another way, the common underlying assumption is that the inhomogeneities are small perturbations on a homogeneous expanding background. Nonlinear studies were possible in cases of high, more specifically spherical, symmetry. An important advance over these idealisations and approximations came through the seminal work of Zel'dovich~\cite{1970Afz.....6..319Z}, who addressed the problem of anisotropic gravitational collapse.

\subsubsection{The Zel'dovich ansatz}\label{sssZA}
%%%%%%%%%%%%%%%%%%%%%%%%%%%%%%%%%%%%%%%%%%%%%%%%%%
The Zel'dovich approximation is not restricted to spherical symmetry. It is a kinematical, Lagrangian approach that
addresses the issue of anisotropic collapse. What Zel'dovich showed was that, at least within the realm of Newtonian gravity, any generic (i.e.~nonspherical) overdensity will undergo a phase of anisotropic, effectively one-dimensional, collapse leading to the formation of two-dimensional flattened structures that are widely known as ‘‘pancakes’’~\cite{1970Afz.....6..319Z}. Over the years, the Zel'dovich approximation has provided a great deal of insight into the initial nonlinear evolution of density fluctuations (e.g.~see~\cite{1989RvMP...61..185S,1993CQGra..10...79B,%
1996ASPC...94...31B,2002PhRvD..66l4015E,2003MNRAS.338..785B,%
2013MNRAS.430L..54R} for an incomplete though representative list).

By construction, the approximation applies to scales well within the Hubble radius and addresses the mildly nonlinear regime of structure formation, when the protogalaxy has decoupled from the background expansion and ``turned around''. The Zel'dovich ansatz extends to this early nonlinear epoch a result of the linear phase, according to which the peculiar flow of the dust component is both irrotational and acceleration free~\cite{1970Afz.....6..319Z}. This result was originally obtained within the purely Newtonian framework, but it was later extended to the quasi-Newtonian studies. Given that the two approaches are identical for all practical purposes, they both arrive at the same conclusions. On these grounds, in what follows, we will provide a unified (Newtonian and quasi-Newtonian) treatment of the issue.

Let us start by recalling that the source of the linear peculiar-velocity field in the Newtonian studies is the gravitational potential, which is replaced by an effective gravitational potential in the quasi-Newtonian treatments (see expression (\ref{ltv'}) and Eqs.~(\ref{lqNrels}), (\ref{qNlA}) in \S~\ref{sssLSPVs} and \S~\ref{sssLQ-NA} respectively). As a result, both approaches arrive at the same growing mode $\tilde{v}\propto t^{1/3}\propto a^{1/2}$ for the linear peculiar-velocity perturbation (see solutions (\ref{ltv2}) and (\ref{lqNv}) in \S~\ref{sssLEPVs} and \S~\ref{sssLQ-NA} respectively).

With this result in hand, it helps to introduce the rescaled peculiar velocity $\tilde{V}_a=a^{1/2}\tilde{v}_a$. The latter has the advantage of being constant (i.e.~$\dot{\tilde{V}}=0$) within both the Newtonian and the quasi-Newtonian approximations, as one can easily verify (e.g.~see~\cite{1996ASPC...94...31B,2002PhRvD..66l4015E}). In addition, the rescaled peculiar velocity can be expressed as the spatial gradient of a (rescaled) gravitational potential, which makes the $\tilde{V}_a$-field irrotational as well~\cite{1996ASPC...94...31B,2002PhRvD..66l4015E}. The Zel'dovic ansatz extends these linear results into the mildly nonlinear regime of protogalactic collapse and, in so doing, forms the foundations of the Zel'dovich approximation.

\subsubsection{The rescaled second-order equations}\label{sssRS-OEs}
%%%%%%%%%%%%%%%%%%%%%%%%%%%%%%%%%%%%%%%%%%%%%%%%%%%%%%%%%%%%%%%%%%%%
For all practical purposes, the Zel'dovich approximation proceeds identically in both the purely Newtonian and the quasi-Newtonian approaches (see~\cite{1996ASPC...94...31B} and~\cite{2002PhRvD..66l4015E} for the respective covariant studies). We will therefore employ here a unified treatment for both, while pointing out their occasional (minor) differences.

In both types of study the background universe is of the Einstein-de Sitter type and the material component is irrotational pressure-free dust. Also, given that the protogalactic cloud has decoupled from the background expansion and started to collapse, we may assume that $\Theta\ll\tilde{\vartheta}$. Then, $\tilde{\Theta}\simeq\tilde{\vartheta}$ and the key variables are the local volume scalar ($\tilde{\vartheta}$) and the local shear tensor ($\tilde{\varsigma}_{ab}$) of the flow  (with $\tilde{\sigma}_{ab}=\tilde{\varsigma}_{ab}$). Accordingly, the mildly nonlinear evolution of the collapsing pressureless matter is monitored by the system~\cite{1996ASPC...94...31B,2002PhRvD..66l4015E}.
\begin{equation}
\dot{\tilde{\vartheta}}= -{1\over3}\,a^{1/2}\tilde{\vartheta}^2- 2a^{1/2}\tilde{\varsigma}^2+ {1\over2}\,a^{1/2}\rho \tilde{V}^2  \label{Zel1}
\end{equation}
and
\begin{equation}
\dot{\tilde{\varsigma}}_{ab}= -{2\over3}\,a^{1/2}\tilde{\vartheta}\tilde{\varsigma}_{ab}- a^{1/2}\tilde{\varsigma}_{c\langle a}\tilde{\varsigma}^c{}_{b\rangle}\,.  \label{Zel2}
\end{equation}
Note that the effect of the last term on the right-hand side of (\ref{Zel1}) decays away with time, since $\rho\propto a^{-3}$ in the background and $V=$~constant throughout the collapse.\footnote{Expression (\ref{Zel1}) is the Raychaudhuri equation monitoring the (peculiar) local volume scalar ($\tilde{\vartheta}$) of the collapsing protogalactic cloud. Intriguingly, the sign of the matter term at the end of the right-hand side is positive, in direct contrast to the standard version of Raychaudhuri's formula  (e.g.~compare to Eq.~(\ref{Ray}) in \S~\ref{sssKs}). This means that in (\ref{Zel1}) the role of the matter has been reversed, since it tends to decelerate the collapse and to accelerate the expansion. Nevertheless, this quite counterintuitive relative-motion effect is subdominant, unless the peculiar-velocity field is relativistic. For the fully nonlinear version of the (peculiar) Raychaudhuri equation and related comments, the interested reader is also referred to Eq.~(16) in~\cite{2002PhRvD..66l4015E} and to Eq.~(15) in~\cite{2013PhRvD..88h3501T}.} This means that the latter is increasingly dominated by the kinematics, as the role of gravity becomes progressively less important. With this in mind, from now on, we well remove the matter term from the right-hand side of Eq.~(\ref{Zel1}).

Let us now introduce the new ``time'' variable $\tau$, defined so that $\tau=-a^{1/2}\tilde{\vartheta}$, where the minus sign compensates for the fact that we are considering a collapsing protogalaxy (with $\tilde{\vartheta}<0$) and also ensures that $\dot{\tau}>0$ always. Then, the system of (\ref{Zel1}) and (\ref{Zel2}) recasts into
\begin{equation}
\tilde{\vartheta}^{\prime}= {1\over3}\,\tilde{\vartheta}+ 2\tilde{\vartheta}^{-1}\tilde{\varsigma}^2 \hspace{15mm} {\rm and} \hspace{15mm}
\tilde{\varsigma}^{\prime}_{ab}= {2\over3}\,\tilde{\varsigma}_{ab}+ \tilde{\vartheta}^{-1}\tilde{\varsigma}_{c\langle a}\tilde{\varsigma}^c{}_{b\rangle}\,,  \label{Zel34}
\end{equation}
where the primes denote differentiation with respect to $\tau$. Our next step is to introduce the shear eigenframe in which the associated tensor takes the diagonal form $\tilde{\varsigma}_{ab}={\rm diag}(\tilde{\varsigma}_1,\tilde{\varsigma}_2,\tilde{\varsigma}_3)$, with $\tilde{\varsigma}_1+\tilde{\sigma}_2+\tilde{\varsigma}_3=0$. In this coordinate system, the expressions (\ref{Zel34}) read as
\begin{equation}
\tilde{\vartheta}^{\prime}= {1\over3}\,\tilde{\vartheta}+ 2\tilde{\vartheta}^{-1}\left(\tilde{\varsigma}_1^2 +\tilde{\varsigma}_1^2 +2\tilde{\varsigma}_1\tilde{\varsigma}_2\right)\,,  \label{Zeleigen1}
\end{equation}
\begin{equation}
\tilde{\varsigma}_1^{\prime}= {2\over3}\,\tilde{\varsigma}_1+ {1\over3}\,\tilde{\vartheta}^{-1}\tilde{\varsigma}_1^2- {2\over3}\,\tilde{\vartheta}^{-1} (\tilde{\varsigma}_1+\tilde{\varsigma}_2)\tilde{\varsigma}_2  \label{Zeleigen2}
\end{equation}
and
\begin{equation}
\tilde{\varsigma}_2^{\prime}= {2\over3}\,\tilde{\varsigma}_2+ {1\over3}\,\tilde{\vartheta}^{-1}\tilde{\varsigma}_2^2- {2\over3}\,\tilde{\vartheta}^{-1} (\tilde{\varsigma}_1+\tilde{\varsigma}_2)\tilde{\varsigma}_1\,,  \label{Zeleigen3}
\end{equation}
since the evolution of $\tilde{\varsigma}_3$ is determined by that of $\tilde{\varsigma}_1$ and $\tilde{\varsigma}_2$. The last three nonlinear differential equations formulate the Zel'dovich approximation in covariant terms, within both the purely Newtonian and the quasi-Newtonian treatments~\cite{1996ASPC...94...31B,%
2002PhRvD..66l4015E}.

\subsubsection{The Zel'dovich ``pancakes''}\label{sssZPs}
\begin{figure*}
\begin{center}
\includegraphics[height=4in,width=4in,angle=0]{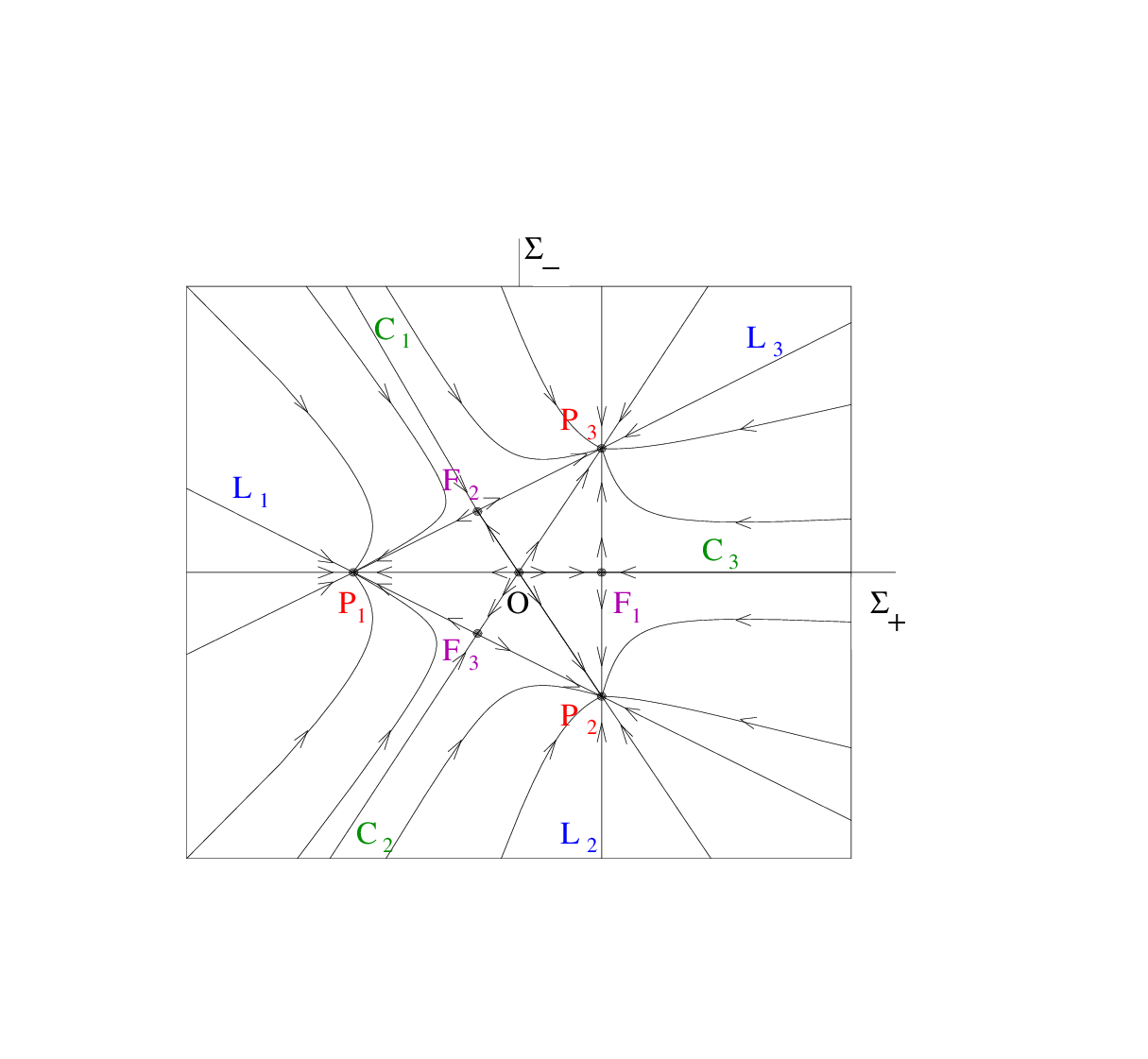}
\end{center}
\caption{Phase plane with $\Sigma_+=X$  and $\Sigma_-=Y$. The lines $L_1$, $L_2$ and $L_3$ forming the central triangle correspond to $\tilde{\varsigma}_{\alpha}=-\tilde{\vartheta}/3$ ($\alpha=1,2,3$), with the three pancakes located at $P_1$, $P_2$ and $P_3$ where these lines intersect. The points $F_1$, $F_2$, $F_3$ represent filamentary solutions and spindle-like singularities, while $O$ corresponds to spherically symmetric, isotropic collapse (see also~\cite{2002PhRvD..66l4015E} and~\cite{1996ASPC...94...31B}).}  \label{fig:Zeldovich}
\end{figure*}

%%%%%%%%%%%%%%%%%%%%%%%%%%%%%%%%%%%%%%%%%%%%%%%%%%%%%%%%%
The final step, in order to demonstrate that the Zel'dovich approximation leads to one-dimensional collapse and therefore to two-dimensional pancake-like structures, is to introduce the dimensionless variables~\cite{1996ASPC...94...31B,2002PhRvD..66l4015E}
\begin{equation}
\Sigma_+= {3\over2}\,\tilde{\vartheta}^{-1} (\tilde{\sigma}_1+\tilde{\varsigma}_2) \hspace{15mm} {\rm and} \hspace{15mm} \Sigma_-= {3\over2}\,\tilde{\vartheta}^{-1} (\tilde{\varsigma}_a-\tilde{\varsigma}_2)\,,  \label{Sigma+-}
\end{equation}
which provide a measure of the anisotropy of the collapse. On using the above, the system (\ref{Zeleigen1})-(\ref{Zeleigen3}) acquires the form
\begin{equation}
\tilde{\vartheta}^{\prime}= {1\over3}\,\tilde{\vartheta}+ 2\tilde{\vartheta}^{-1}\left(\Sigma_+^2+\Sigma_-^2\right)\,,  \label{Zeleigen4}
\end{equation}
\begin{equation}
\Sigma_+^{\prime}= {1\over3}\left[1-\Sigma_+-2\left(\Sigma_+^2 +\Sigma_-^2\right)\right]\Sigma_++ {1\over3}\,\Sigma_-^2  \label{Zeleigen5}
\end{equation}
and
\begin{equation}
\Sigma_-^{\prime}= {1\over3}\left[1+2\Sigma_+-2\left(\Sigma_+^2 +\Sigma_-^2\right)\right]\Sigma_-\,.  \label{Zeleigen6}
\end{equation}
Accordingly, the evolution of $\Sigma_+$ and $\Sigma_-$ has decoupled from that of $\tilde{\vartheta}$ and the final shape of the collapsing protogalactic cloud is now monitored by the subsystem (\ref{Zeleigen5}) and (\ref{Zeleigen6}). In technical terms, the problem has been reduced to the study of the planar dynamical system depicted in Fig.~\ref{fig:Zeldovich}. Physically speaking, the dimensional reduction means that the shape of the collapsing
dust cloud is independent of the collapse timescale.

Referring the reader to Fig.~\ref{fig:Zeldovich} for illustration purposes, the vertices $P_1$, $P_2$ and $P_3$ of the central triangle are stationary points of the system (\ref{Zeleigen5}) and (\ref{Zeleigen6}) and act as attractors. Each one of these vertices represents an one-dimensional pancake solution, which is stationary in two shear eigendirections and collapses along the third. Therefore, generic solutions are asymptotic to one of these pancakes (one along each eigendirection). The bisecting lines $C_1$, $C_2$ and $C_3$ intersect at the central stationary point $O$, corresponding to shear-free spherically symmetric collapse. Finally, the stationary points $F_1$, $F_2$ and $F_3$ represent filamentary solutions. The pancakes are stable nodes, the filaments are saddle points and the spherically symmetric collapse is an unstable node. All this means that, once the collapse sets in, the pancakes are the natural attractors for a generic overdensity, at least within the limits of the Newtonian and the quasi-Newtonian treatments.

\subsection{Beyond the Zel'dovich approximation: Radiation and the
%%%%%%%%%%%%%%%%%%%%%%%%%%%%%%%%%%%%%%%%%%%%%%%%%%%%%%%%%%%%%%%%%%
Sachs-Wolfe effect}\label{ssRESW}
%%%%%%%%%%%%%%%%%%%%%%%%%%%%%%%%%
While the Zel'dovich approximation provides a powerful framework for understanding structure formation in the matter-dominated era, a comprehensive treatment requires considering the radiation effects as well, particularly when connecting peculiar velocities to observable quantities like CMB temperature anisotropies. The connection between large-scale peculiar velocities and CMB anisotropies was established in the seminal work of Sachs and Wolfe~\cite{1967ApJ...147...73S}, who demonstrated how metric perturbations and velocity fields imprint on the observed radiation temperature.

The Sachs-Wolfe effect provides a natural extension to our discussion of peculiar velocities, as it explicitly demonstrates how these motions contribute to observable temperature fluctuations in the CMB. In the conformal Newtonian gauge, where the perturbed metric takes the form
\begin{equation}
ds^2 = a^2(\tau)[-(1+2\Phi)d\tau^2 + (1-2\Psi)dx^idx_i]\,,
\end{equation}
the total observed temperature fluctuation in direction $\hat{\mathbf{n}}$ can be decomposed as
\begin{equation}
    \frac{\Delta T}{T}(\hat{\mathbf{n}}) = \left[\Phi + \hat{\mathbf{n}}\cdot\mathbf{v}\right]_{\tau_{\rm ls}} + \int_{\tau_{\rm ls}}^{\tau_0} (\dot{\Phi} + \dot{\Psi})d\tau\,.
\end{equation}
Here, the first term represents the ordinary Sachs-Wolfe effect evaluated on the last scattering surface ($\tau_{\rm ls}$), which combines the gravitational potential ($\Phi$) and the Doppler contribution from peculiar velocities ($\hat{\mathbf{n}}\cdot\mathbf{v}$). The second term represents the integrated Sachs-Wolfe (ISW) effect, arising from the time evolution of potentials along the photon path.

The Doppler term $\hat{\mathbf{n}}\cdot\mathbf{v}$ directly connects the peculiar-velocity field discussed in previous sections to observable temperature anisotropies. In the linear regime, this term contributes primarily to the dipole and quadrupole moments of the CMB, while at smaller scales, the velocity field induces higher-order multipoles that carry information on the growth of structure.

This connection is particularly significant for understanding the evolution of peculiar velocities beyond the (mild) nonlinear regime addressed by the Zel'dovich approximation. While the Zel'dovich formalism describes the formation of pancake structures in collapsing regions, the Sachs-Wolfe effect allows us to connect these peculiar velocity fields to observable radiation patterns, providing a crucial observational window into the physics of structure formation.

The importance of this connection extends to modern precision cosmology, where accurate modeling of the peculiar velocity contributions to CMB anisotropies is essential for extracting cosmological parameters. For instance, the kinematic Sunyaev-Zel'dovich (kSZ) effect—a secondary anisotropy arising from the scattering of CMB photons off the free electrons moving with bulk peculiar velocities—provides a direct probe of the peculiar velocity field at various redshifts~\cite{1980MNRAS.190..413S}. This effect complements the pancake formation picture by revealing how the velocity field evolves across cosmic time and in different environments.

\subsection{Relativistic phase-space description}\label{ssRPSD}
%%%%%%%%%%%%%%%%%%%%%%%%%%%%%%%%%%%%%%%%%%%%%%%%%%%%%%%%%%%%%%%
The Zel'dovich approximation, while powerful, operates within the framework of Newtonian or quasi-Newtonian gravity. For a more comprehensive treatment that accounts for relativistic effects — particularly important on scales approaching the horizon, or in the early universe - one could turn to a fully relativistic phase-space description. Such an approach provides the natural framework for going beyond the limits of the Newtonian/quasi-Newtonian treatments.

Ma and Bertschinger~\cite{1995ApJ...455....7M} developed a comprehensive relativistic treatment using the distribution function formalism, which describes the phase space density of particles. The evolution of the distribution function $f(x^i, P^j, \tau)$ is governed by the relativistic Boltzmann equation:
\begin{equation}
    \frac{\partial f}{\partial \tau} + \frac{dx^i}{d\tau}\frac{\partial f}{\partial x^i} + \frac{dP^j}{d\tau}\frac{\partial f}{\partial P^j} = C[f]\,,
\end{equation}
where $P^j$ is the momentum and $C[f]$ represents collision terms. The above naturally incorporates both the metric perturbations and peculiar velocities in a unified relativistic framework.

The relativistic approach of~\cite{1995ApJ...455....7M} reveals several effects absent in the earlier Newtonian and quasi-Newtonian treatments of peculiar velocities:

\begin{itemize}
    \item \textbf{Frame-dragging effects:} Vector perturbations in the metric induce gravitomagnetic effects that can influence peculiar-velocity fields in ways not captured by scalar potentials.\\

    \item \textbf{Mode coupling:} Scalar, vector and tensor modes couple in the nonlinear regime, affecting the evolution of peculiar velocities in ways not captured by the Zel'dovich approximation.\\

    \item \textbf{Horizon-scale corrections:} On scales approaching the Hubble radius, where our earlier Newtonian/quasi-Newtonian treatments begins to break down, the relativistic framework remains valid and captures horizon-scale physics.
\end{itemize}

This relativistic phase space description provides a natural bridge between the Zel'dovich approximation discussed in \S~\ref{sssZPs} and the fully relativistic treatment of peculiar velocities presented in earlier sections of this review (particularly in \S~\ref{ssRA}). It demonstrates how the ``pancake'' structures predicted by Zel'dovich emerge naturally in the nonlinear regime of a more general relativistic framework when appropriate limits are taken.

The phase space approach also connects with our earlier discussion of the ``tilted universe'' (see \S~\ref{sCPVT} - \S~\ref{sPVSF}), as it provides a formal mathematical foundation for describing the relative motion between different components of the cosmic fluid. In particular, the moments of the distribution function relate directly to the peculiar-velocity field and the higher-order moments (like the anisotropic stress) encountered in our covariant 1+3 treatment.

Recent work has used the aforementioned phase-space approach in N-body simulations~\cite{2016NatPh..12..346A}. The study demonstrated that, while Newtonian calculations (including the Zel'dovich approximation) capture the dominant physics of structure formation on sub-horizon scales, relativistic corrections become important for precision cosmology, particularly when considering effects like the gravitational redshift and light propagation through nonlinear structures.

This completes our theoretical journey from linear peculiar velocities to nonlinear structure formation, providing a unified framework that encompasses both the Zel'dovich approximation's prediction of pancake structures and the relativistic effects that become important on the largest scales and earliest times in cosmic history.

\section{Peculiar motions and the $\Lambda$CDM}\label{sPMLCDM}
%%%%%%%%%%%%%%%%%%%%%%%%%%%%%%%%%%%%%%%%%%%%%%%%%%%%%%%%%%%%%%
The current cosmological model, the $\Lambda$CDM paradigm, was launched at the turn of the millennium after the SNI${\rm a}$ observations and the advent of the universal acceleration \cite{1998AJ....116.1009R,1999ApJ...517..565P}. Since then, there has been several observations in agreement with the model's predictions, thus strengthening its position as the concordance scenario. However, in recent years, an increasing number of independent observations seem at odds with the $\Lambda$CDM. Here, we will go through those observations that are, or could be, related to large-scale peculiar velocities.

\subsection{The bulk-flow question}\label{ssB-FQ}
%%%%%%%%%%%%%%%%%%%%%%%%%%%%%%%%%%%%%%%%%%%%%%%%%
The cornerstone of the concordance $\Lambda$CDM model is well known as the Cosmological Principle (CP), an idea that is simple yet powerful. It is based on the assertion that our universe is both spatially homogeneous and isotropic on cosmological scales (e.g.~see~\cite{1972gcpa.book.....W}), implying the absence of any preferred locations or directions. However, the real universe is neither perfectly homogeneous nor perfectly isotropic. At best, there is a length scale -- beyond the reach of the structures that we see locally -- where matter is distributed in a way that appears isotropic about every point. In other words, the universe is considered to be uniform only on sufficiently large scales.

Strong observational support for the CP comes from the near isotropy of the CMB. The latter exhibits temperature fluctuations of only one part in one hundred thousand, which is consistent with a highly uniform early universe. The most prominent deviation from this isotropy is the CMB dipole, which has been largely treated as purely kinematic and it is attributed to the peculiar motion of our Solar System relative to the rest-frame of the microwave photons. The amplitude of the dipole was measured with high accuracy by the \textit{Planck~2018} collaboration and corresponds to a velocity of $v=369\pm0.9$~km/sec towards the direction $(l,b)= (263.99^{\circ}\pm0.14^{\circ}, 48.26^{\circ}\pm0.03^{\circ}$) in Galactic coordinates~\cite{2020A&A...641A...6P}. Beyond the Solar System, the motion of the Local Group of galaxies relative to the CMB frame has also been established, with an estimated velocity of approximately 600~km/sec. This motion reflects the cumulative influence of large-scale inhomogeneities in the nearby universe.

In addition, over the last decades, there have been repeated reports of coherent large-scale peculiar motions - commonly referred to as \textit{bulk flows} - with sizes between few hundred and several hundred Mpc and speeds ranging from few hundred to several hundred km/sec. The amplitude and the scales of these bulk flows remain topics of active investigation, as they may provide critical insights into the underlying matter distribution and also prove potential challenges to the standard $\Lambda$CDM model. To these, one should probably add reports of extreme bulk flows with sizes and speeds close to thousand Mpc and thousand km/sec respectively~\cite{2008ApJ...686L..49K,2009ApJ...691.1479K,%
2010ApJ...712L..81K,2011ApJ...732....1K,2012MNRAS.419.3482A}. These are the controversial \textit{dark flows} (see \S~\ref{ssDFQ} for a brief presentation), which are currently at odds with the \textit{Planck~2013} results~\cite{2014A&A...561A..97A}, although the issue may not be over yet~\cite{2013A&A...557A.116A,2015ApJ...810..143A}.

According to the CP the peculiar velocities of the individual galaxies, as well as any bulk peculiar flows, should fade away and essentially vanish on sufficiently large scales (beyond those traced by the large-scale structure of the distribution of galaxies). In other words, on large enough lengths, the net peculiar motion of the galaxies should become statistically consistent with zero. Nevertheless, the precise depth at which this convergence occurs remains an open question, with no clear consensus as yet. As observational surveys grow in number and quality and probe progressively deeper parts of the universe, it is expected that this question will be addressed with greater accuracy in the near future.

To date, numerous studies have been performed to estimate the amplitude, the direction and the convergence depth of these bulk flows using various methodologies, as discussed in \S~\ref{sec:pec_vel_method} earlier. Observational constraints derived from different probes -- including galaxies, galaxy clusters, SNIa, quasars and gamma-ray bursts -- have yielded conflicting results. Some measurements generally agree with the predictions of the $\Lambda$CDM model, while others report significant departures from it. Intriguingly, the surveys that agree with the expectations of the concordance model report bulk flows on scales less than approximately $100/h$~Mpc (e.g.~see~\cite{2011ApJ...736...93N,2011JCAP...04..015D,%
2012MNRAS.420..447T,2012MNRAS.424..472B,2013MNRAS.428.2017M,%
2013A&A...560A..90F,2014MNRAS.437.1996M,2014MNRAS.445..402H,%
2015MNRAS.450..317C,2016ApJ...827...60M,2016MNRAS.455..386S,%
2016MNRAS.456.1886S,2018MNRAS.477.5150Q,2019MNRAS.482.1920Q,%
2020MNRAS.498.2703B,2021MNRAS.505.2349S,2021ApJ...922...59Q}), while those that disagree go beyond the aforementioned threshold (e.g.~see~\cite{2004MNRAS.352...61H,2008MNRAS.387..825F,%
2009MNRAS.392..743W,2010MNRAS.407.2328F,2010ApJ...709..483L,%
2011MNRAS.414..264C,2015MNRAS.447..132W,2018MNRAS.481.1368P,%
2021MNRAS.504.1304S,2023MNRAS.524.1885W,2023MNRAS.526.3051W}.). Furthermore, one should always keep in mind that the $\Lambda$CDM expectations are entirely based on Newtonian theoretical studies, which lead to peculiar-velocity magnitudes considerably smaller than the relativistic treatments (see~\cite{2020EPJC...80..757T,2021Ap&SS.366....4F,%
2022PhRvD.106h3505M,2024PhRvD.110f3540M,2026ApJ...997...25T} and also \S~\ref{ssRA} here). In the rest of this section, we will review a number of characteristic peculiar-velocity surveys and discuss their methodologies and results.

\subsection{Bulk flows within the $\Lambda$CDM limits}\label{ssBFWLCDMLs}
%%%%%%%%%%%%%%%%%%%%%%%%%%%%%%%%%%%%%%%%%%%%%%%%%%%%%%%%%%%%%%%%%%%%%%%%%
Over the years, many peculiar-velocity surveys have reported bulk-flow measurements that are consistent with the predictions of the concordance $\Lambda$CDM model (see \S~\ref{ssPCTBFLCDM} earlier). In this section, we will go through a representative (to the best of our knowledge) number of them.

The work of Nusser $\&$ Davis~\cite{2011ApJ...736...93N} provides a detailed investigation into the observed bulk flow in the local universe and into its implications for cosmological parameters within the framework of the $\Lambda$CDM paradigm. The authors assess the consistency of the bulk flow with the predictions of the standard cosmological model, by specifically focusing on how the flow constrains the amplitude of density fluctuations ($\sigma_8$) and the growth factor of structure ($f$).

The aforementioned work employs a range of techniques to estimate the bulk flow, including redshift surveys and peculiar-velocity data, in conjunction with statistical approaches such as the Minimum Likelihood Estimate (MLE) and the All Space Constrained Estimate (ASCE), as described in detail in~\ref{sub:pv methods} earlier. The analysis focuses on the determination of the growth rate of large-scale structures, using the measured bulk flow to place constraints on cosmological parameters like $\sigma_8$ and $f$. Then, by fitting these parameters to the observed velocity fields, one can in principle differentiate between the various cosmological models.

\begin{figure*}
\begin{center}
\includegraphics[height=2.75in,width=4.5in,angle=0]{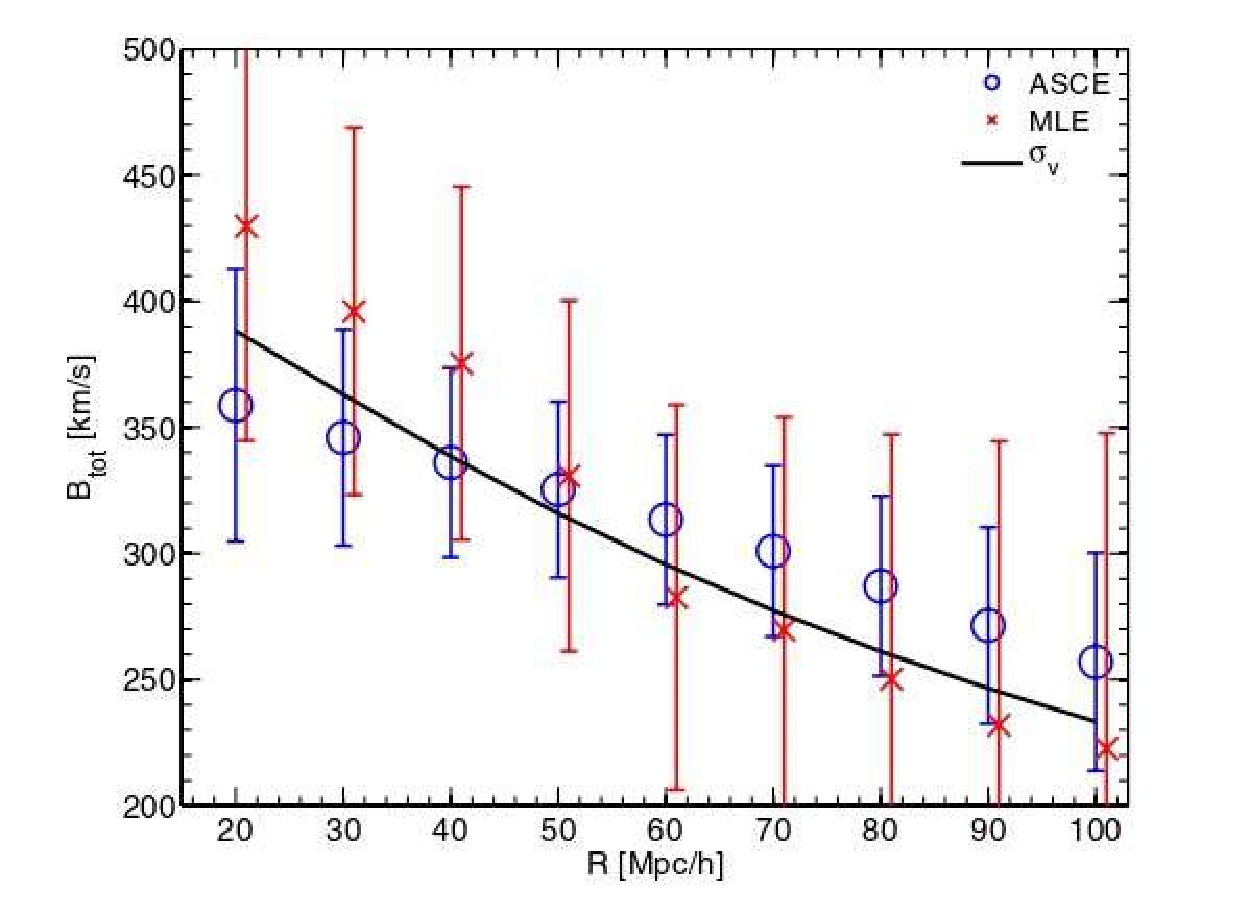}\quad
\end{center}
\caption{The amplitude of the bulk as a function of the bulk-flow radius for ASCE and MLE (in blue and red respectively). The solid curve shows the rms velocity of the bulk flow as expected in a flat $\Lambda$CDM model, with $\Omega_m=0.266$, $h=0.71$, and $\sigma_8=0.85$ (see also Fig.~6 in~\cite{2011ApJ...736...93N}).}  \label{fig:Nusser}
\end{figure*}

Among the key findings of~\cite{2011ApJ...736...93N} is a bulk flow centered to the Milky Way with radius $100/h$~Mpc and velocity around 250~km/sec, which increases on progressively smaller scales and aligns with the predictions of the $\Lambda$CDM scenario (see Fig.~\ref{fig:Nusser} and also Fig.~6 in~\cite{2011ApJ...736...93N}). In addition, the direction of the bulk motion, towards $(\ell,b)=(279^{\circ},10^{\circ}$),  agrees with the motion of our Local Group, after correcting for the Virgocentric infall, thus reinforcing the consistency of these measurements with the concordance model. However, the results of~\cite{2011ApJ...736...93N} are in contrast with those of Feldman~et~al~\cite{2009MNRAS.392..743W,2010MNRAS.407.2328F}, for example, who around the same time reported a significant large-scale bulk flow at distances beyond 100/h~Mpc by using galaxy surveys (see \S~\ref{ssBFELCDMLs} below for further discussion). This discrepancy highlights the ongoing debate over the origin and the evolution of the large-scale bulk peculiar motions.

Bulk flows on scales close to $50/h$~Mpc where investigated by Ma and Scott in~\cite{2013MNRAS.428.2017M}, utilizing a Bayesian framework and multishell likelihood analysis. The study uses various peculiar velocity surveys to reconstruct the bulk flow, trimming the data sets at $80/h$~Mpc in order to reconstruct the bulk-flow moments accurately on the $50/h$~Mpc scale. The analysis incorporates a Bayesian hyper-parameter method to assess the amplitude and direction of the bulk flow. This approach allows for more flexibility and efficiency in accounting for possible systematic errors and uncertainties in the data. The multishell likelihood analysis involves partitioning the universe into multiple concentric shells at varying distances, allowing one to study the bulk-flow effects across different redshift ranges and assess how the flow evolves with distance. This method provides a detailed view of the bulk flow across different scales, offering insights into its redshift-dependence.

Reconstructing the bulk-flow moments on $50/h$~Mpc scales by means of the Minimum Variance (MV) method showed consistency with the $\Lambda$CDM predictions. In particular, after correcting for the inhomogeneous Malmquist bias and properly selecting the samples,
all the used survey catalogs showed a coherent flow of about $310$~km/sec~\cite{2013MNRAS.428.2017M}. These findings support earlier studies, like that of~\cite{2011ApJ...736...93N} for example, which also suggested that the bulk flow observed on intermediate scales (around $50/h$~Mpc) is in agreement with the standard model.

As mentioned earlier in this section, other surveys, such as those by Watkins et~al~\cite{2009MNRAS.392..743W} and by Feldman et~al~\cite{2010MNRAS.407.2328F}, have reported a more significant bulk flow on larger scales (beyond the 100/h~Mpc threshold -- see also \S~\ref{ssBFELCDMLs} below). This discrepancy could be due to the distinct statistical methods employed across these studies, or it may reflect differences in the bulk-flow behavior between different cosmological scales. Recall that the studies of~\cite{2009MNRAS.392..743W,2010MNRAS.407.2328F} extent beyond $100/h$~Mpc, whereas that of~\cite{2013MNRAS.428.2017M} is centered on scales around $50/h$~Mpc.

Another notable example is the work of Turnbull~et~al~\cite{2012MNRAS.420..447T}, who employed SNIa to estimate the amplitude and direction of the bulk flow as a function of distance. Utilizing two complementary statistical techniques - the Minimum Variance (MV) weighting scheme (originally introduced in~\cite{2009MNRAS.392..743W}) and the Maximum Likelihood Estimate (MLE) method - they carefully accounted for the sparse spatial sampling of the SNIa dataset. To this end, the authors compiled the \textit{First Amendment} dataset, consisting of 245 SNIa with redshifts $z<0.06$. The analysis of~\cite{2012MNRAS.420..447T} utilized the MLE approach to yield a bulk-flow velocity of $330\pm120$~km/sec. However, their most robust results were obtained using the MV method, which is specifically designed to suppress contributions from small-scale velocity noise and survey geometry effects. This approach resulted in a bulk flow of $249\pm76$~km/sec, directed toward Galactic coordinates ($l=319^{\circ}\pm18^{\circ}$, $b=7^{\circ}\pm14^{\circ}$), within a Gaussian-weighted volume of radius $50/h$~Mpc. These measurements are also consistent with the statistical predictions of the $\Lambda$CDM model.

It is important to emphasize again that the work of~\cite{2012MNRAS.420..447T} is confined to the local Universe and relies exclusively on nearby SNIa with redshifts $z<0.06$. Consequently, while the results align with the $\Lambda$CDM expectations on small scales, they do not constrain bulk flows on larger, cosmological scales. Note that the above study also highlights the critical role of the measurement methodologies, particularly in the comparison of bulk-flow statistics with theoretical models. Proper accounting for sparse sampling, incomplete sky coverage, and cosmic variance is essential, as these technical factors are likely to account for much of the apparent discrepancies between studies that aim to measure the same physical quantity.

Building upon this foundation, subsequent work has extended the investigation of bulk flows to other tracers and larger datasets. For example, Hong~et~al~\cite{2014MNRAS.445..402H} focused on determining the bulk-flow amplitude and direction in the local universe using data from the \textit{2MASS Tully-Fisher} Survey (2MTF). The latter provides distances and peculiar velocities for over 2,000 nearby bright spiral galaxies, with a homogeneous sky distribution, thus making it a powerful resource for probing cosmic flows in the nearby universe. Peculiar velocities were estimated by comparing observed redshifts to distances derived from the Tully-Fisher relation, and a dipole model was fitted to infer the amplitude and the direction of the bulk-flow velocity. To further refine their constraints across a range of depths, the authors adopted the Gaussian density profile of Watkins~et~al~\cite{2009MNRAS.392..743W} and applied three Gaussian window functions centered at depths of $20/h$, $30/h$, and $40/h$~Mpc. The resulting analysis yielded bulk-flow velocities around $310$~km/sec, $280$~km/sec, and $290$~km/sec at the corresponding depths, all of which are consistent with the $\Lambda$CDM model. Furthermore, their findings are in good agreement with other independent studies utilizing SNIa and galaxy redshift surveys~\cite{2013MNRAS.428.2017M,2011ApJ...736...93N,%
2012MNRAS.420..447T}, reinforcing the emerging consensus on the amplitude and the scale of the local bulk flows.

In their seminal work, Branchini et al~\cite{2012MNRAS.424..472B} utilized the flux-limited \textit{2MASS Redshift Survey} (2MRS) to reconstruct the linear peculiar velocity field in the local universe and to place constraints on the amplitude and direction of the local bulk flow. The methodology hinges on the fact that peculiar velocities introduce systematic biases in the estimated absolute magnitudes of galaxies. More specifically, in a flux-limited sample an underlying bulk flow modifies the inferred luminosity distribution of galaxies, with those moving away from us appearing dimmer than expected and those moving toward us looking brighter. By exploiting this effect, the authors infer the amplitude and direction of the bulk flow directly from the statistical properties of the luminosity function.

To ensure robustness, the study placed significant emphasis on the accurate modeling of the galaxy luminosity and selection functions, both of which are crucial for mitigating systematic uncertainties. The authors found that the resulting bulk-flow estimate is consistent with previous independent measurements based on alternative tracers of the velocity field. In particular, the components (90,-230,50)~km/sec of the bulk velocity reported at $60/h$~Mpc are in excellent agreement with the results of Nusser $\&$ Davis~\cite{2011ApJ...736...93N} (see above). Additionally, the derived bulk flow is compatible with that obtained from SNIa-based peculiar velocity measurements by Turnbull et al~\cite{2012MNRAS.420..447T} (see above as well), thus providing further support to $\Lambda$CDM.

Additionally, Feindt~et~al~\cite{2013A&A...560A..90F} set out to measure the cosmic bulk flow using a sample of SNIa observed by the \textit{Nearby Supernova Factory}~\cite{2002SPIE.4836...61A}, which provides a collection of 117 spectro-photometrically calibrated supernovae in the redshift range $0.03< z<0.08$ (corresponding to distances of approximately 90-240/h~Mpc). To improve statistical power and sky coverage, the authors supplemented their sample with SNIa from other compilations, such as the \textit{Union2} dataset. Motivated by earlier claims of large-scale coherent motion by Colin et al~\cite{2011MNRAS.414..264C} and Kashlinsky et al~\cite{2011ApJ...732....1K}, the work of~\cite{2013A&A...560A..90F} was the first attempt to probe bulk flows beyond the Shapley Supercluster using conventional distance indicators derived from SNIa. The analysis involved dividing the SNIa dataset into distinct redshift shells and perform a dipole fit based on~\cite{2006PhRvL..96s1302B}, to determine the bulk-flow velocity in every redshift shell. As a cross-check, the authors reimplemented the smoothed residuals method, used in~\cite{2011MNRAS.414..264C} earlier, which allowed them to estimate the anisotropy in the local Hubble flow as a function of scale. The methodology offered a robust test of large-scale isotropy and indicated the consistency of the bulk-flow amplitudes with the $\Lambda$CDM predictions, especially at intermediate cosmological distances.

The findings of the study reveal that the observed bulk flow amplitude is in excellent agreement with the expectations of the $\Lambda$CDM model. Using the dipole fit method, the authors report a bulk-flow velocity of approximately 250~km/sec for the combined SNIa dataset. The direction of this flow is consistent with both the location of the Shapley Supercluster and the CMB dipole, and it aligns well with results from independent analyses employing different methodologies and datasets, such as the \textit{2MASS Redshift Survey} and other SNIa-based studies. A complementary smoothed residual analysis yields a bulk flow direction that corroborates the dipole-fit results, further supporting the reliability of the measurements. Finally, in the highest redshift shell ($0.06<z<0.1$), the analysis of~\cite{2013A&A...560A..90F} finds no significant evidence of a bulk flow, with the velocity consistent with zero and a statistical uncertainty of approximately 240~km/sec when limiting the direction to that of the CMB
dipole. This result strongly disfavors the detection of a ``dark flow'' as reported by Kashlinsky et al (e.g.~see~\cite{2008ApJ...686L..49K,2011ApJ...732....1K} and also \S~\ref{ssDFQ} below), ruling it out at a confidence level of around $4\sigma$. Overall, the study supports the view that the peculiar motion of the Local Group is primarily driven by gravitational interactions with nearby large-scale structures — most notably the Shapley Supercluster—within a cosmological context that remains fully consistent with the $\Lambda$CDM paradigm.

The above cited results are consistent with other bulk-flow surveys that align with the $\Lambda$CDM model, such as that of~\cite{2012MNRAS.420..447T}, who also used SNIa to estimate the peculiar velocity of the Local Group. Both studies support the view that the bulk flow observed in the local universe should be attributed to the gravitational pull of large-scale structures, rather than to a breakdown of standard cosmology. The work of~\cite{2013A&A...560A..90F}, alongside these other measurements, underscores the importance of using multiple datasets and methodologies to obtain consistent results.

\begin{figure*}
\begin{center}
\includegraphics[height=2.75in,width=3.5in,angle=0]{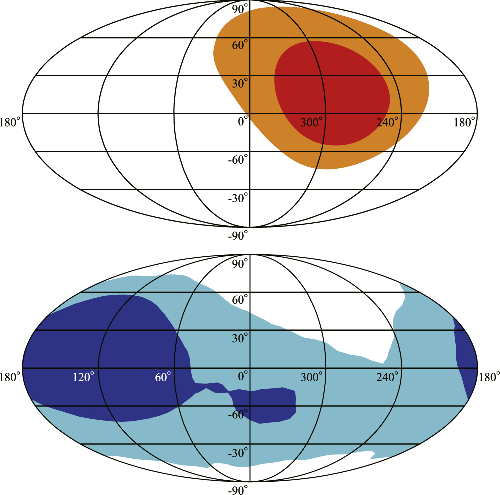}
\end{center}
\caption{Angular distribution of the bulk velocities derived from a Bayesian MCMC analysis of the SNIa \textit{Union2.1} compilation. The upper plot shows the inferred bulk-flow direction and magnitude based on low-redshift SNIa data ($z<0.05$) and the lower plot from high-redshift SNIa ($z>0.05$). The darker areas indicate the $1\sigma$ confidence region, while the lighter ones the $2\sigma$ region (see also Fig.~2 in~\cite{2016ApJ...827...60M}).}
\label{fig:mathews16}
\end{figure*}

The use of SNIa to probe the bulk flow out to larger distances was pursued in the study of Mathews et al~\cite{2016ApJ...827...60M}. That work analyzed observational evidence for the magnitude and direction of the bulk flow, using the SNIa redshift-distance relationship from the \textit{Union2.1} dataset, as well as from the \textit{SDSS-II Supernova Survey} catalog. The authors explored the potential detectability of a large-scale bulk flow using two different methodologies. First, they  implemented a Bayesian MCMC search of the parameter space of the three Cartesian velocity components, using $\Lambda$CDM parameters from the \textit{WMAP} survey. The analysis separates the data into low-redshift ($z<0.05$) and high-redshift ($z >0.05$) subsets, corresponding to distances smaller/greater than $145/h$~Mpc, to assess their scale-dependent detectability (see Fig.~\ref{fig:mathews16} here or Fig.~2 in~\cite{2016ApJ...827...60M}). The upper plot shows the $1\sigma$ and $2\sigma$ confidence intervals of the bulk-flow direction estimated from the low-redshift data. The analysis reveals a well defined bulk-flow velocity of approximately 270~km/sec towards galactic coordinates $(l, b)\simeq(295^{\circ},10^{\circ})$, consistent with the $\Lambda$CDM predictions. These results are in agreement with those of~\cite{2012MNRAS.420..447T}.

At higher redshifts, however, the analysis of~\cite{2016ApJ...827...60M} is rather inconclusive, suggesting a possible larger bulk flow, but with wide confidence intervals and low statistical significance due to increased measurement uncertainties and limited sky coverage. The authors conclude that current SNIa data are insufficient to confirm, or refute, a high-redshift bulk flow, but that future large-scale surveys (with improved precision and broader sky-coverage) may be able to look for departures from isotropy at cosmic scales with greater accuracy.

The study by Qin et al~\cite{2021ApJ...922...59Q} is a comprehensive analysis of cosmic flows using the CF4 catalog~\cite{2023ApJ...944...94T}. The authors measure the low-order kinematic moments of the cosmic-flow field, specifically the bulk flow and shear moments, to test the consistency of cosmological models with the observed nearby density field. To achieve accurate cosmological inferences, they develop a mock sampling algorithm that accurately replicates the survey geometry and luminosity selection function of the CF4 sample. These mocks are instrumental in exploring how systematics affect the measurements and can be further utilized to estimate the covariance matrix and errors of power spectrum and two-point correlation functions in future works.

The study reports a measured bulk flow in the local universe of $376\pm23$~km/sec at a depth of $35/h$ Mpc, directed towards Galactic coordinates $(l,b)=(298^{\circ}\pm3^{\circ},-6^{\circ}\pm3^{\circ})$. Both the measured bulk and shear moments are consistent with the predictions of the concordance cosmological model, indicating that the observed cosmic flows do not deviate significantly from the theoretical expectations.

\begin{figure*}
\begin{center}
\includegraphics[height=2.5in,width=4.5in,angle=0]{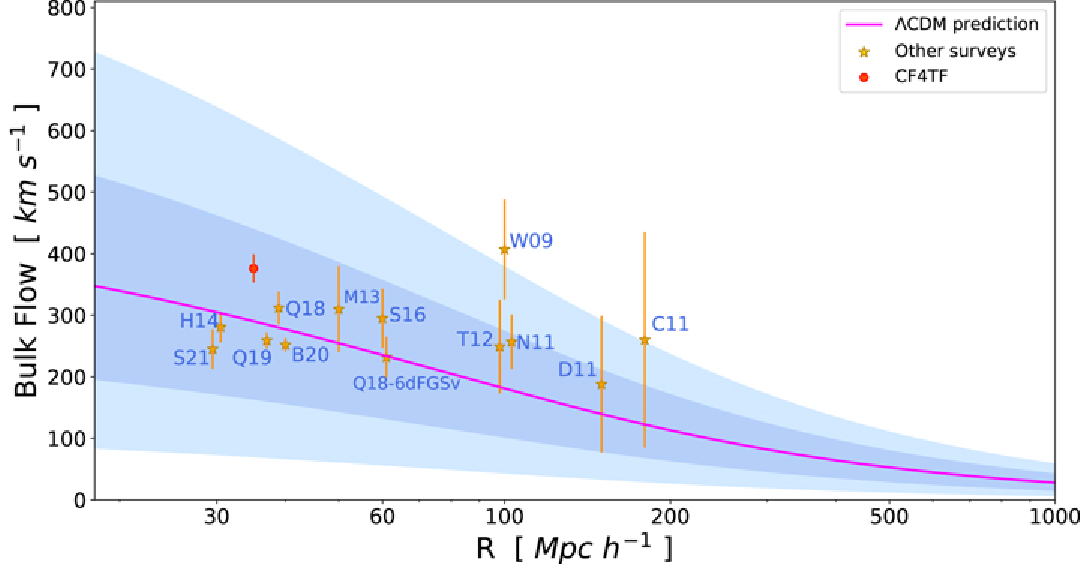}
\end{center}
\caption{Comparison of reconstructed bulk flows in our local neighborhood (see also Fig.~9 in~\cite{2021ApJ...922...59Q}). The pink curve is the $\Lambda$CDM prediction calculated from a spherical top-hat window function (rather than the gaussian window function). The shaded areas indicate the $1\sigma$ and $2\sigma$ cosmic variance. The yellow stars represent the bulk measurements from different surveys, namely W09:~\cite{2009MNRAS.392..743W}, C11:~\cite{2011MNRAS.414..264C}, D11:~\cite{2011JCAP...04..015D}, N11:~\cite{2011ApJ...736...93N}, T12:~\cite{2012MNRAS.420..447T}, M13:~\cite{2013MNRAS.428.2017M}, H14:~\cite{2014MNRAS.445..402H}, S16:~\cite{2016MNRAS.455..386S}, Q18:~\cite{2018MNRAS.477.5150Q}, Q18:-6dFGSv, Q19:~\cite{2019MNRAS.482.1920Q}, B20:~\cite{2020MNRAS.498.2703B} and S21:~\cite{2021MNRAS.505.2349S}. The red dot is the bulk flow measurement of~\cite{2021ApJ...922...59Q} using CF4 data.}
\label{fig:qin}
\end{figure*}

The numerical results of different bulk-flow studies are depicted in Fig.~\ref{fig:qin}, where the pink curve represents the $\Lambda$CDM prediction. Note that the $\Lambda$CDM expectation (pink curve) has been calculated from the spherical top-hat window function, which gives higher predictions than the gaussian window function (e.g.~see Fig.~1 in~\cite{2016IAUS..308..336M}).

We close this section with the work of~\cite{2020MNRAS.498.2703B} which reports a bulk flow with amplitude of $239\pm65$~km/sec within a sphere of radius $50/h$~Mpc, using the MLE statistical method and an amplitude of $253\pm44$~km/sec employing the MV technique. However, the bulk flow direction is not well aligned with the CMB dipole. The authors also investigated the scale dependence of the bulk flow by varying the depth of the sample. The results showed a bulk-flow amplitude that decreases with scale, as expected by the $\Lambda$CDM model. Nevertheless, the uncertainties are large, highlighting the need for obtaining larger and more accurate peculiar velocity samples to further test the standard model and constrain the cosmological parameters.

\subsection{Bulk flows in excess of the $\Lambda$CDM
%%%%%%%%%%%%%%%%%%%%%%%%%%%%%%%%%%%%%%%%%%%%%%%%%%%%
limits}\label{ssBFELCDMLs}
%%%%%%%%%%%%%%%%%%%%%%%%%%
Over the past two decades or so, a number of independent studies have reported bulk flows significantly faster and deeper than those predicted by the standard model. These results, obtained using improved distance indicators and sophisticated statistical techniques, have collectively deepened the tension between observations and theoretical expectations. An early notable analysis by Hudson et al~\cite{2004MNRAS.352...61H}, using 56 \textit{SMAC} clusters within $120/h$~Mpc, reported a bulk flow of $687\pm203$~km/sec directed toward $(l=260^{\circ}\pm13^{\circ}$,$b=0^{\circ}\pm11^{\circ}$). Notably, this flow did not diminish with increasing depth, suggesting it could not be solely attributed to the gravitational influence of the Great Attractor. Instead, they proposed that the Shapley Concentration, one of the most massive superclusters in the nearby universe, might be responsible, albeit with marginal statistical significance. Furthermore, their analysis indicated that convergence to the CMB rest frame does not occur within $60/h$~Mpc, and that at depths between $60/h$ and $120/h$~Mpc, the bulk flow amplitude is limited to approximately 600~km/sec.

Building on the work of~\cite{2004MNRAS.352...61H}, a novel method for optimally weighting peculiar-velocity measurements, known as the ``Minimum Variance'' (MV) weighting scheme was introduced by Watkins et al in~\cite{2009MNRAS.392..743W}. This technique is designed to minimize the variance of the estimated bulk flow, by calculating the weighted average of the peculiar velocities of all galaxies in the sample. Consequently, the MV method ensures that galaxies with more precise velocity measurements contribute more significantly to the bulk-flow estimate, thereby improving the precision of the measurements. The authors compiled data from all major peculiar-velocity surveys, including SMAC and other major galaxy surveys combining FP, TF, SBF and SNIa distance indicators. The findings of~\cite{2009MNRAS.392..743W} showed a consistent bulk flow of $407\pm81$~km/sec within an approximate radius of $100/h$~Mpc, pointing toward ($l=287^{\circ}\pm9^{\circ}$,$b=8^{\circ}\pm6^{\circ}$). Intriguingly, the results also suggested that nearly $50\%$ of the Local Group (LG) motion relative to the CMB frame is generated by structures beyond this depth.

Extending the analysis of~\cite{2009MNRAS.392..743W}, Feldman~et~al reported a similar bulk flow with $v=416\pm78$~km/sec directed toward ($l,b)=(282^{\circ}\pm11^{\circ},6^{\circ}\pm 6^{\circ}$)~\cite{2010MNRAS.407.2328F}. The latter study employed a novel method, known as the ``Standardized Minimum Variance'' (SMV) estimator, to analyze the bulk flow, extending their previous work \cite{2009MNRAS.392..743W} to include the next higher elements in the expansion, namely the shear and the octopole moments of the peculiar-velocity field. This method accounts for the sparse and non-uniform spatial distribution of galaxy samples that had plagued previous surveys. The technique effectively compares measurements across different studies by calculating the flow within an idealized survey geometry - essentially a Gaussian window function with a characteristic depth. Analyzing a total sample of more than 4000 data covering the whole sky outside the Galactic plane and using the method mentioned above, the authors confirmed that their findings substantially exceeded the predictions of the standard cosmological model, being approximately $2\sigma$-$3\sigma$ higher than expected. The same study also revealed that the sources responsible for the bulk flow were at an effective distance greater than $300/h$~Mpc and thus too far to identify in the existing all-sky redshift surveys. It should be noted that the above cited results agree with those from the studies of~\cite{2010ApJ...709..483L,2008MNRAS.387..825F}, which used different catalogs and methodologies.

A significant contribution to the study of large-scale peculiar flows is the work of Watkins and Feldman~\cite{2015MNRAS.447..132W}, which utilises the \textit{CF2} catalog. The latter contains most of the data used in~\cite{2010MNRAS.407.2328F} along with a considerable amount of more recent data, reaching a total size of more than 8000 objects with good coverage beyond the $100/h$~Mpc mark. The catalog combines multiple distance measurement techniques, including TF relations, FP measurements, SBF, SNIa, and Tip of the Red Giant Branch distances. The authors used their MV method from~\cite{2009MNRAS.392..743W} and employed standardized Gaussian windows of different radii, ranging between $20/h$~Mpc and $50/h$~Mpc to allow direct comparison with the theoretical predictions. The amplitude of the reported bulk flow was between 300 and 400~km/sec extending out to scales of $200/h$~Mpc. A puzzling feature of the detected bulk flows, also reported in~\cite{2009MNRAS.392..743W}, is that they are quite consistent with the $\Lambda$CDM expectations on scales less than $40/h$~Mpc (corresponding to a bulk-flow radius of $20/h$~Mpc), but then the bulk velocity increases as the radius grows. The authors found the probability of observing such a bulk peculiar flow in a $\Lambda$CDM universe to be less than $5\%$ (at $2\sigma$ confidence level).

\begin{figure*}
\begin{center}
\includegraphics[height=2.75in,width=4.5in,angle=0]{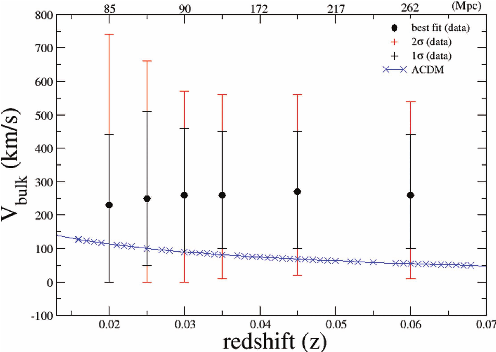}\quad
\end{center}
\caption{The redshift profile of the bulk flow from the likelihood analysis of~\cite{2011MNRAS.414..264C}. The peculiar velocities (black dots) systematically exceed the blue line of the $\Lambda$CDM expectations, but their values also show signs of decrease at lower redshifts (see Fig.~8 in~\cite{2011MNRAS.414..264C}).}  \label{fig:Colin}
\end{figure*}

Using SNIa as standardizable candles, Colin~et~al~\cite{2011MNRAS.414..264C} presented a compelling investigation of cosmological anisotropy. The authors pursued a redshift-binning analysis to investigate potential anisotropy effects across different redshift ranges. They used a sample of more than 500 SNIa from the \textit{Union2} catalog~\cite{2010ApJ...716..712A} with a redshift range of $0.015<z<1.5$. Specifically, the sample was divided into five bins for $z<0.1$ and seven bins for $z>0.1$, ensuring a sufficient number of data points in each bin. In each case the real data were compared with the results from 1000 Monte Carlo realizations assuming a flat $\Lambda$CDM universe. Colin~et~al found that the distant SNIa (with $z>0.045$) were clearly inconsistent with the $\Lambda$CDM scenario. However, at lower redshift the discrepancy dropped down to roughly $1\sigma$ (see Fig.~\ref{fig:Colin} here, or Fig.~8 in~\cite{2011MNRAS.414..264C}). The methodology used in~\cite{2011MNRAS.414..264C} stands out for its rigorous treatment of the observational biases and its careful distance calibration. This was achieved by employing a combination of peculiar velocity modeling and multipole expansion analysis to identify potential directional preferences in the Hubble flow. The results indicated the presence a bulk flow of radius $z\sim0.05$ moving with 260 km/sec towards the CMB dipole (see Fig.~\ref{fig:Colin}). This appears to align with previously identified bulk-flow directions, particularly those reported in~\cite{2009MNRAS.392..743W} by means of galactic peculiar velocities. On scales up to approximately $100/h$~Mpc, the results of~\cite{2011MNRAS.414..264C} are consistent within $1\sigma$ with the current concordance cosmological model (see Fig.~\ref{fig:Colin}). Then, the large-scale inconsistency with the standard $\Lambda$CDM could suggest the presence of a coherent motion at redshifts considerably deeper than expected. Alternatively, the large-scale disagreement seen in Fig.~\ref{fig:Colin} may indicate a different evolution rate for the peculiar-velocity field at higher redshifts, perhaps along the lines discussed in \S~\ref{ssRA} and \S~\ref{ssPVI} earlier.

The growing body of evidence challenging the standard model's predictions on large-scale peculiar motions, was extended by Peery~et~al~\cite{2018MNRAS.481.1368P}. There, building upon methodological frameworks similar to those employed by Nusser and Davis~\cite{2011ApJ...736...93N}, the authors presented a refined analysis of galaxy peculiar velocities, using the \textit{Cosmic Flows~3} (CF3) catalog~\cite{2016AJ....152...50T} to extract the coherent bulk-flow signal across multiple distance scales. The bulk-flow velocity, which was estimated by means of the Minimum Variance (MV) statistical method of~\cite{2009MNRAS.392..743W}, had amplitude close to $280$~km/sec within a sphere of radius $R=150/h$~Mpc (for an ideal survey with a weighting scheme $r^{-2}$). What distinguishes this work is the authors' detailed scale-dependent analysis, demonstrating that the bulk flow amplitude shows little diminution with increasing distance, contrary to the $\Lambda$CDM expectation of rapid convergence at scales beyond the $100/h$~Mpc threshold.

It is important to note that the tension between the survey of Peery~et~al~\cite{2018MNRAS.481.1368P} and the standard cosmological model emerges primarily on the largest probed scales, namely on scales larger than approximately $150/h$~Mpc. On smaller lengths, the measured bulk-flow amplitudes remain broadly consistent with the $\Lambda$CDM predictions. Whether this reflects an underlying difference in the bulk-flow evolution on different scales, or differences in the methodologies employed to estimate them, remains to be seen.

Different analyses apply distinct statistical techniques, weighting schemes and selection criteria that can lead to divergent conclusions, even when based on the same peculiar-velocity datasets. This methodological dependence of the results has fueled the ongoing debate within the cosmology community. The recent study of Watkins~et~al~\cite{2023MNRAS.524.1885W} builds upon this debate and presents several significant methodological innovations that substantially advance the field of bulk-flow measurements. Rather than using a single bulk-flow measurement, the authors performed a multi-scale analysis that employed a series of Gaussian filters on scales varying at $20/h$, $40/h$, $60/h$, $100/h$, $150/h$, and $200/h$~Mpc, in an attempt  to systematically probe how the bulk-flow properties evolve with scale. The study also computes the theoretical expectations for the bulk-flow amplitudes that account for both the specific window functions and the cosmic variance appropriate to each scale.  Then, the new data added from the \textit{Cosmicflows4} (CF4) catalog~\cite{2023ApJ...944...94T} led to an increased bulk-flow velocity relative to its \textit{Cosmicflows-3} (CF3) counterpart (see~\cite{2018MNRAS.481.1368P} and also above).

\begin{figure*}
\begin{center}
\includegraphics[height=2.75in,width=4.5in,angle=0]{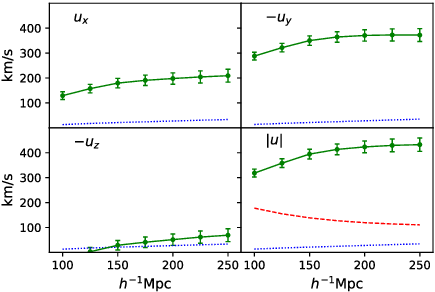}\quad
\end{center}
\caption{The scale dependence of the bulk-flow velocity along the three coordinate axes and of its mean value (green lines), estimated from the \textit{CosmicFlows4} catalog~\cite{2023ApJ...944...94T}. The error bars are due to noise. The dotted blue lines mark the theoretical standard deviation between the bulk-flow estimates and those of an ideal survey based on the standard cosmological model. The red dashed line is the theoretical expectation for the mean bulk velocity according to the standard cosmological model. Note the profound disagreement between the predicted and the measured bulk-flow velocities, as well as the decline in the magnitude of the latter at lower redshifts (see Fig.~7 in~\cite{2023MNRAS.524.1885W}).}  \label{fig:Watkins}
\end{figure*}

The findings of Watkins~et~al~\cite{2023MNRAS.524.1885W} present a significant challenge to the $\Lambda$CDM model. More specifically, the authors reported a bulk-flow faster than 390~km/sec with a radius of $150/h$~Mpc and a directional offset of approximately $20^{\circ}–30^{\circ}$ from the Shapley Concentration. Most importantly, the amplitude of the measured bulk velocity did not seem to drop with increasing redshift, approaching 420~km/sec near the $200/h$~Mpc threshold (see Fig.~\ref{fig:Watkins} here or Fig.~7 in~\cite{2023MNRAS.524.1885W}). On these scales, the probability of observing bulk flows of comparable magnitude within the standard model is notably smaller than that associated with the corresponding CF3 measurements~\cite{2018MNRAS.481.1368P}.

Having said that, we remind the reader once again that the $\Lambda$CDM expectations are based entirely on Newtonian studies, which predict the moderate growth-rate of $\tilde{v}\propto t^{1/3}$ for the linear peculiar-velocity field (e.g.~see~\cite{1976ApJ...205..318P,1980lssu.book.....P} and also \S~\ref{ssNA} here). The relativistic analysis, on the other hand, allows for considerably faster growth, with $\tilde{v}\propto t$ being the minimum rate (see~\S~\ref{ssRA} previously). Moreover, the same studies also suggested that once the accelerated expansion of the universe starts, the aforementioned linear growth should slow down (if not decay) at lower redshifts (see~\cite{2022PhRvD.106h3505M,2026ApJ...997...25T} and also \S~\ref{ssPVI} here). This in turn should lead to a peculiar-velocity profile much like the one reported in~\cite{2023MNRAS.524.1885W} (see Fig.~7 there and also Fig.~\ref{fig:Watkins} here). Note that a similar, though less pronounced, drop in the low-redshift value of the $\tilde{v}$-field was also reported in~\cite{2011MNRAS.414..264C} (see Fig.~8 there or Fig.~\ref{fig:Colin} here)

The advantage of the recent CF4 survey lies in its increased depth, allowing for peculiar-velocity measurements at greater cosmological distances. Using this extended dataset, the probability of observing a bulk flow with radius as large as $200/h$~Mpc (or larger) approaches the $5\sigma$ threshold. This in turn imposes some of the most stringent constraints to date on the consistency of the large-scale velocity fields with the predictions of the $\Lambda$CDM model. While the latter predicts that the bulk-flow amplitudes should decline inversely proportional to their scale, the observations show a much weaker scale dependence. This trend suggests the presence of coherent structures, or motions, extending well beyond the typically expected homogeneity threshold of $100/h$~Mpc. The results of~\cite{2023MNRAS.524.1885W} confirm and substantially strengthen the earlier findings of~\cite{2009MNRAS.392..743W,2010MNRAS.407.2328F},  while addressing many of the methodological criticisms those works received, thus further intensifying the debate on the bulk-flow question.

A complementary investigation into the bulk-flow excess reported in the CF4 dataset has been given by Whitford~et~al~\cite{2023MNRAS.526.3051W}, who focus on the statistical robustness and methodological implications of the different bulk-flow estimators. Specifically, the study compares the Maximum Likelihood Estimation (MLE) and the Minimum Variance (MV) methods, highlighting how the choice of the estimator can influence both the magnitude and the associated uncertainties of the inferred bulk flow. By carefully analyzing the estimator performance, using realistic mock catalogs and applying both techniques to the CF4 data, the authors have quantified systematic biases that may arise from survey geometry, sampling inhomogeneities, and velocity noise. Using the MLE approach, the authors reported a bulk flow as fast as 430~km/sec at a depth of $175/h$~Mpc. This result shows strong consistency with the bulk-flow estimates of~\cite{2023MNRAS.524.1885W}, thus supporting the presence of a coherent motion on scales up to $200/h$~Mpc, or perhaps even more. It should BE noted, however, that the error bars in~\cite{2023MNRAS.526.3051W} are somewhat larger, which reduces the statistical tension with the standard cosmological model. Nevertheless, the probability of such a bulk flow occurring within $\Lambda$CDM limits was estimated to be less the $0.11\%$. All these findings seem to suggest that the severity of any claimed inconsistency with the standard model is sensitive not only to the data, but also to the methodology used to extract cosmological information. Hence, advances in estimator design and validation are not mere technical details, but rather crucial components in developing a reliable understanding of the large-scale velocity field and its implications for cosmic structure formation.

Before closing this section we should also mention the analysis of Salehi~et~al~\cite{2021MNRAS.504.1304S}, using supernovae data from the \textit{Union2} and the \textit{Pantheon} compilations~\cite{2010ApJ...716..712A,2018ApJ...859..101S}. Similar to other related syrveys, the authors also reported bulk peculiar motions consistent with $\Lambda$CDM on small scales, but well in excess of the concordance model on large scales. More specifically, at low redshifts ($0.015<z<0.1$) the \textit{Pantheon} and the \textit{Union2} SNIa data indicated a bulk velocity around 225~km/sec and close to 250~km/sec respectively. In both cases the direction of the bulk motion was in good agreement with that of the CMB dipole. At high redshifts ($0.1<z<0.2$), on the other hand, the respective bulk velocities were 700~km/sec and 1000~km/sec. Intriguingly, these numbers are very close to those in the dark-flow reports of~Kashlinsky~et~al (see~\cite{2008ApJ...686L..49K,2009ApJ...691.1479K,%
2010ApJ...712L..81K,2011ApJ...732....1K,2012arXiv1202.0717K} and \S~\ref{ssDFQ} next). Salehi~et~al suggested a number of possible explanations for the large-scale discrepancy in their results, among which the use of the QCDM (Q for ``quintessense'') cosmological model instead of $\Lambda$CDM~\cite{2021MNRAS.504.1304S}.

\subsection{The dark-flow question}\label{ssDFQ}
%%%%%%%%%%%%%%%%%%%%%%%%%%%%%%%%%%%%%%%%%%%%%%%%
In the literature there are also reports of extreme bulk flows. with sizes and speeds in excess of those of the typical bulk peculiar flows and well beyond the standard $\Lambda$CDM limits. These are the so-called \textit{dark flows}, originally reported in~\cite{2008ApJ...686L..49K,2009ApJ...691.1479K,2010ApJ...712L..81K,%
2011ApJ...732....1K,2012arXiv1202.0717K} and subsequently in~\cite{2012MNRAS.419.3482A}. Dark flows are coherent bulk motions of clusters on very large scales, measured using X-ray galaxy clusters as tracers. The analysis utilised the kinematic Sunyaev-Zel'dovich (kSZ) effect, by measuring small anisotropies in the CMB temperature, triggered by the motion of galaxy clusters relative to the CMB frame~\cite{2000ApJ...536L..67K}. This involves measuring the temperature fluctuations in all-sky CMB maps along the direction of these clusters and fitting for a dipole anisotropy. More specifically, this requires extracting the kinematic component of Sunyaev-Zel'dovich effect, which arises from the Compton scattering of CMB photons off the hot intracluster gas (see \S~\ref{sec:kinematicSN} for further discussion). The method has since been applied to different releases of the WMAP satellite employing various filtering techniques to avoid contamination from the cosmological CMB signal, instrument noise and foreground emissions.

\begin{figure*}
\begin{center}
\includegraphics[height=3in,width=5in,angle=0]{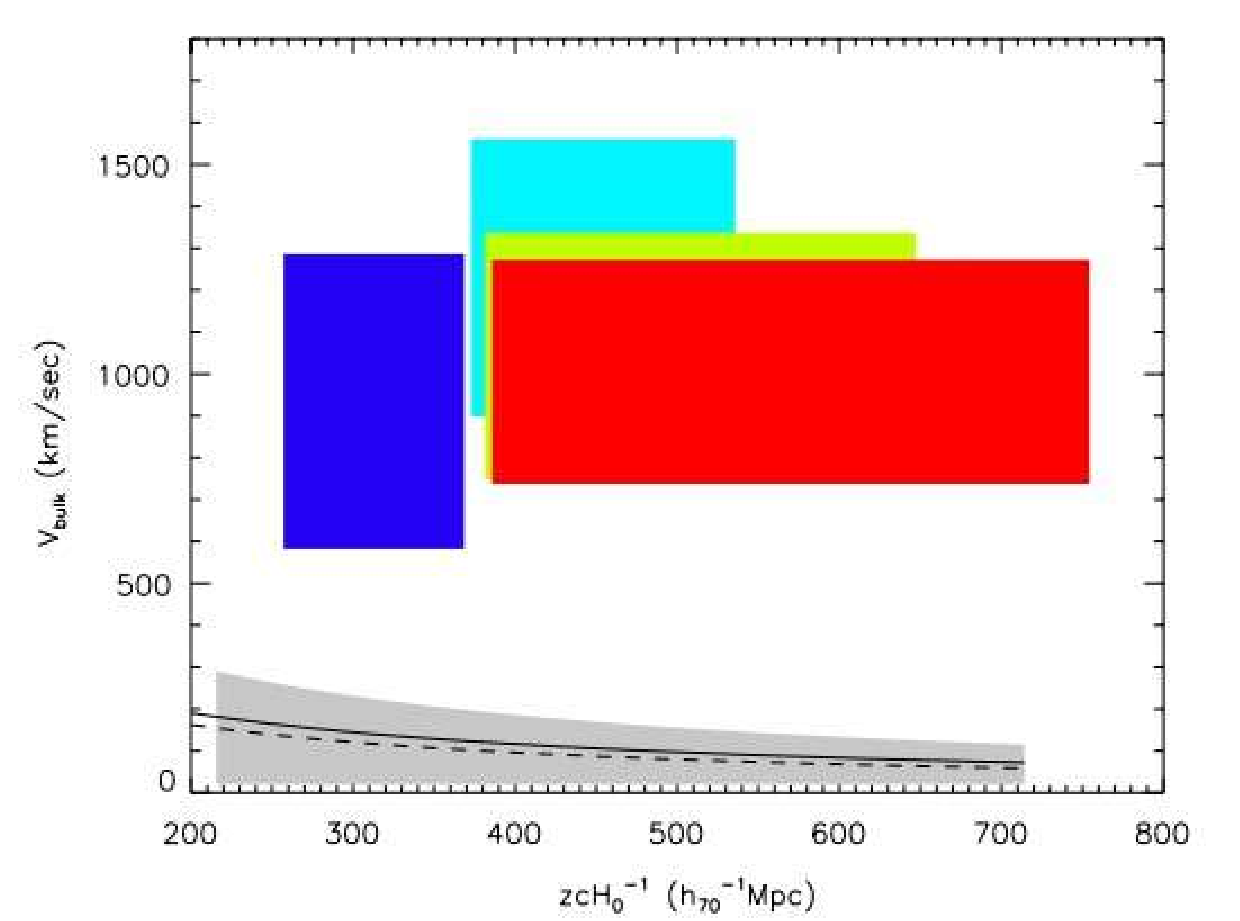}\quad
\end{center}
\caption{The dark-flow velocity plotted against redshift depth. The blue/cyan/green/red regions correspond to measurements at $z\leq0.12/0.16/0.2/0.25$ respectively. The solid and dashed lines correspond to the rms bulk velocity for the concordance
$\Lambda$CDM model for top-hat Gaussian windows, while the lack-shaded regions show the 95\% confidence level of the model. (see also Fig.~2 in~\cite{2010ApJ...712L..81K}).}  \label{fig:darkflows}
\end{figure*}

What makes dark flows particularly intriguing is their amplitude and coherence scale, both of which significantly exceed the predictions of the $\Lambda$CDM model. In particular, by analyzing a catalog covering the entire sky and consisting of 782 X-ray-selected and X-ray-flux-limited galaxy clusters, a statistically significant bulk motion was reported in~\cite{2009ApJ...691.1479K,2010ApJ...712L..81K}. The flow was moving with a velocity in the range of 600-1000~km/sec and extending out to scales of roughly $300/h$~Mpc, which was also the limiting distance of the survey. On the other hand, there was nothing special about the direction of the peculiar motion, which was largely aligned with that of our Local Group relative to the CMB. Using the 5-year and 7-year WMAP data, the authors increased their sample to more than 1000 clusters and came to the same conclusion, namely of a dark flow which appears to move with a nearly constant velocity out to at least 800~Mpc \cite{2010ApJ...712L..81K,2011ApJ...732....1K} (see Fig.~\ref{fig:darkflows} here). It goes without saying that, if confirmed, dark flows could prove an untenable problem for the current cosmological model.

The above results were further discussed and reviewed in~\cite{2012arXiv1202.0717K}, where it was suggested that the reported bulk flow may extend across the entire observable universe. If so, this could potentially point to pre-inflationary structures, or to deviations from standard cosmology. However, with the possible exception of~\cite{2012MNRAS.419.3482A} and perhaps that of~\cite{2021MNRAS.504.1304S}, these findings were not reproduced by other independent studies, which found no statistically significant bulk flow. In~\cite{2009ApJ...707L..42K}, for example, the correlations within the CMB data from the 8-WMAP channels reduced the significance of the dark flow detection down to $0.7\sigma$. Taken at face value, this could suggest an overestimated statistical confidence of the bulk flow and/or that the reported signal may have been an artifact of the measurement techniques.

Building upon the earlier dark-flow studies and analyzing the WMAP 7-year data, in combination with the ROSAT X-ray selected cluster catalogs, Osborne~et~al~ \cite{2011ApJ...737...98O} provided a detailed investigation of the dark-flow dipole. Their methodology involved the use of multiple filtered maps, constructed from different frequency bands, to isolate the kSZ signal at the cluster locations of galaxies. Their findings contradicted those of Kashlinsky~et~al. In particular, the authors detected no significant evidence of a cluster dipole in any of the redshift shells examined, a result consistent with the predictions of the $\Lambda$CDM model. The discrepancy between their findings and those of Kashinsky~et~al. may be attributed to differences in the filtering techniques used for suppressing the CMB component, to systematic error treatments, and to the selection criteria for the galaxy-cluster samples.

The setback for dark flows came from the Planck~13 mission~\cite{2014A&A...561A..97A}, which could not reproduce the kSZ dipole signal reported in the studies of Kashlinsky~et~al. A comparative analysis, using the WMAP 9-year data, also employed in~\cite{2008ApJ...686L..49K}, suggested that the observed signal was unlikely to originate from the peculiar motion of galaxy clusters. Instead, it was attributed to residual contamination - primarily of CMB origin - in the filtered maps. Furthermore, residual foreground contamination, especially from the Milky Way, can induce north-south asymmetries. When projected onto a galaxy cluster catalog with non-uniform sky coverage, these asymmetries may generate spurious dipole signals. Planck adopted alternative filtering techniques that were likely to be more effective in mitigating such dipole-like contamination. By analyzing galaxy clusters within spheres of varying radii centered on the Local Group, Planck placed constraints on the local bulk flow across different cosmic volumes. The kSZ dipole was calculated for cluster sub-samples enclosed within successively larger radii. Then, for each sub-sample, a dipole fit was performed across all directions in the sky. Although the Planck constraints allowed for peculiar velocities as fast as 800~km/sec at redshifts of $z\simeq0.15$, which exceed the typical $\Lambda$CDM limits, Planck concluded that the average cluster velocity was consistent with zero. As a result, no significant bulk-flow detection was reported in support of the standard cosmological model.

In response to the Planck announcements, Atrio-Barandela~et~al~\cite{2013A&A...557A.116A,2015ApJ...810..143A} examined whether the kSZ-signal persists in both the WMAP 9-year data and the then newly released Planck CMB maps. Their analysis utilized high-resolution temperature anisotropy maps from WMAP and Planck, which were pre-processed to minimize foreground contamination and systematic effects that could potentially obscure a kSZ-induced signal. They reported the detection of a dipole signal that correlates with the X-ray properties of galaxy clusters, suggesting a physical origin linked to the cluster gas rather than to the primary CMB foreground residuals, or to instrumental noise. Notably, the direction of the dipole aligned with the all-sky CMB dipole.

The authors argued that the dipole amplitude extracted from their full cluster sample is consistent with that reported by the Planck Collaboration. However, they contended that Planck's error bars were overestimated, which diminished the statistical significance of the Planck results~\cite{2013A&A...557A.116A}. To evaluate the robustness of their findings, they subjected the extracted signals to detailed statistical analyses, including comparisons with simulations based on the $\Lambda$CDM model. These tests aimed to determine whether the observed anisotropies could plausibly arise from random fluctuations, or point to a genuine large-scale flow.

Assuming a kSZ origin for the statistically significant signal, the inferred bulk velocity lies in the range of 600-1000~km/sec, in agreement with the systematic and statistical uncertainties discussed in earlier work by Kashlinsky~et~al~\cite{2011ApJ...732....1K}. The authors emphasized that the dark flow hypothesis cannot yet be conclusively ruled out, as the uncertainties (both statistical and systematic) associated with the peculiar-velocity measurements remained substantial. Moreover, as the velocity-field constraints on scales larger than 100/h~Mpc are sensitive to the assumed value of the Hubble constant, the ongoing discrepancies between the local determinations of $H_0$ and its Planck-inferred value, introduce additional uncertainty into these large-scale measurements.

\subsection{Dipolar asymmetries and peculiar motions}\label{ssDAPMs}
%%%%%%%%%%%%%%%%%%%%%%%%%%%%%%%%%%%%%%%%%%%%%%%%%%%%%%%%%%%%%%%%%%%%
Perhaps the most characteristic imprint of relative motion, their ``trademark signature'' so to speak, is a Doppler-like dipolar anisotropy. The CMB dipole, for example, has been largely treated as an apparent (Doppler-like) effect triggered by the motion of our Local Group relative to the CMB frame. In an analogous way, the same peculiar motion should also induce similar dipolar anisotropies in the sky-distribution of a host of cosmological parameters. Among them, to the Hubble parameter, to the deceleration parameter, as well as to the number counts of distant astrophysical sources.

\subsubsection{The $q$-dipole}\label{sssq-D}
\begin{figure*}
\begin{center}
\includegraphics[height=3.5in,width=3.5in,angle=0]{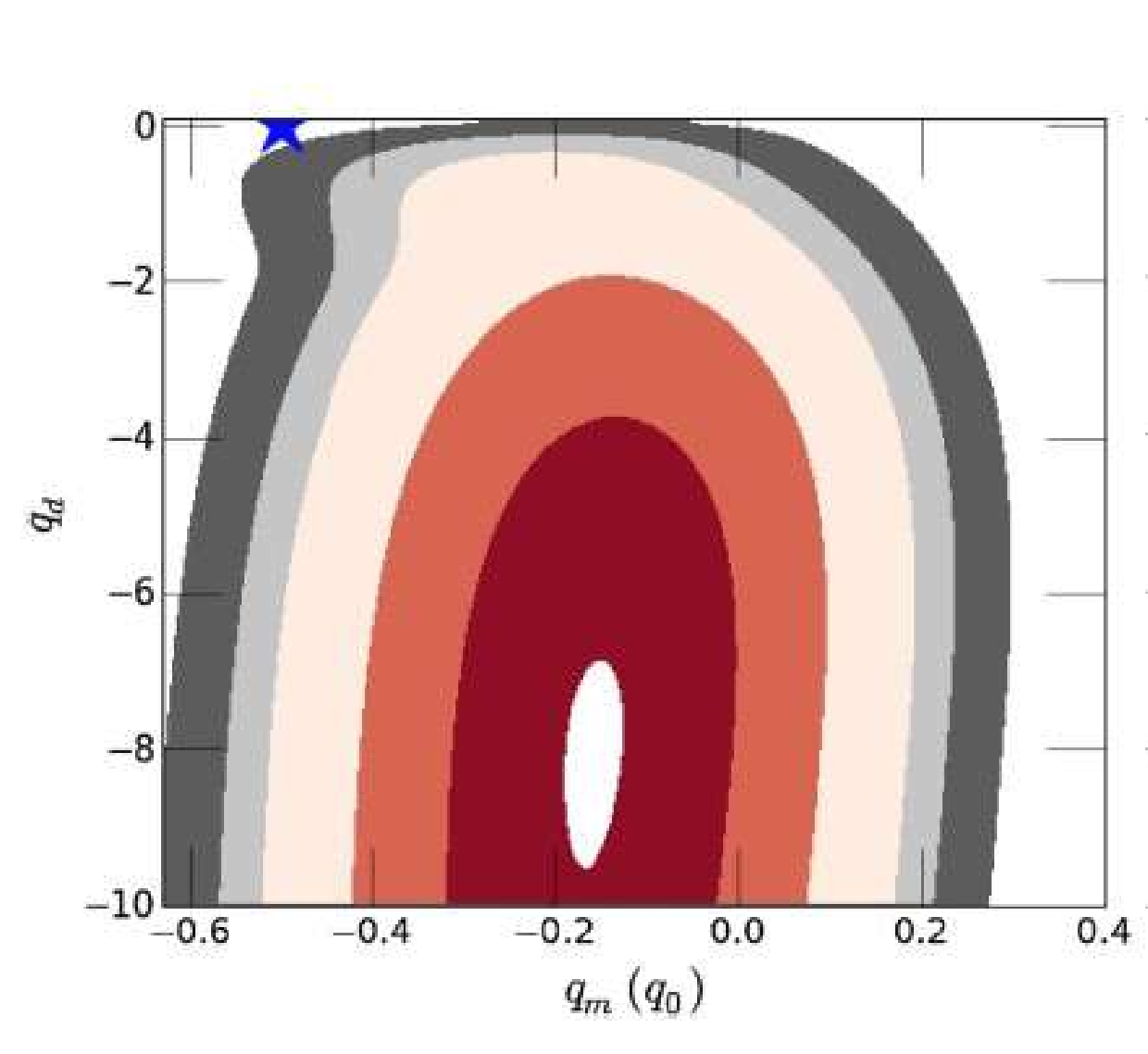}\quad
\end{center}
\caption{Monopole and dipole components of the deceleration parameter (inferred from the JLA catalog). The dipole $q_d=-8.03$ lies near the centre of the white ellipse. The value of $q_0$ for the standard $\Lambda$CDM model is shown as a blue star. Note that the vertical scale of the dipole magnitude is compressed ten-fold relative to the horizontal scale of the monopole (see also Fig.~3 in~\cite{2019A&A...631L..13C}).}  \label{fig:JLA}
\end{figure*}

%%%%%%%%%%%%%%%%%%%%%%%%%%%%%%%%%%%%%%%%%%%%
Over the last fifteen years or so, there have been a number reports claiming the presence of dipolar asymmetries in the sky-distribution of the deceleration parameter, with the first (to the best of our knowledge) coming from the work of Cooke and Lynden-Bell~\cite{2010MNRAS.401.1409C}. The universe appeared to accelerate faster along a certain direction on the celestial sphere and (more or less) equally slower in the antipodal. Nevertheless, the authors attributed the dipole to a statistical coincidence. So, it was not until the work of  of Colin~et~al~\cite{2019A&A...631L..13C} that the dipole in the deceleration parameter was attributed to our peculiar motion relative to the CMB frame. Motivated by the ``tilted universe'' scenario~\cite{2010MNRAS.405..503T,2011PhRvD..84f3503T,%
2015PhRvD..92d3515T}, the authors analyzed the \textit{Joint Light-curve Analysis} (JLA) dataset, which comprised 740 spectroscopically confirmed SNIa, spanning the redshift range $0.01<z<1.3$, to look for anisotropies in the cosmic acceleration. The focus was on the low-redshift domain (with $z\leq0.1$), where the imprint of the local peculiar velocities is expected to be more pronounced. By applying a maximum-likelihood approach to a dipole-modulated deceleration parameter ($q_d$ - see Eq.~(\ref{qd}) below), the analysis identified a statistically significant ($3.9\sigma$) dipolar component with an amplitude of $q_{d}=-8.03$, aligned close to the CMB dipole (see Fig.~3 in~\cite{2019A&A...631L..13C}, or Fig.~\ref{fig:JLA} here). At the same time the monopole component ($q_0$) was found to be consistent with zero (i.e.~no universal acceleration) at $1.4\sigma$ confidence. Notably, the $q$-dipole was found to diminish exponentially with redshift with a decay scale of approximately $100$~Mpc, which implied that the anisotropic signal was confined to relatively local cosmological volumes.

The results of~\cite{2019A&A...631L..13C} were to be expected if we happen to reside in a locally contracting bulk flow, as predicted by the ``tilted'' model (e.g.~see~\cite{2015PhRvD..92d3515T,2021EPJC...81..753T,%
2022MNRAS.513.2394A,2022EPJC...82..521T}). Then, the observed cosmic acceleration may not be a global phenomenon, but a local artifact of our motion relative to the CMB frame, along the lines described in~\S~\ref{ssRMEDP} and \S~\ref{ssD-LDUE} earlier. In other words, we may have been the ``unsuspecting'' observers, who have mistaken the local deceleration of their own bulk flow for global acceleration of the surrounding universe.

A dipolar anisotropy in the $q$-distribution is not the only imprint peculiar motions leave on the observed deceleration parameter. The magnitude of the induced (Doppler-like) dipole should also decrease with increasing scale/redshift. This is another key prediction of the ``tilted universe'' paradigm, as described in \S~\ref{ssD-LDUE} before. Such a redshift-decaying dipole was recently identified after the systematic analysis of the \textit{Pantheon}+ data by~\cite{2025EPJC...85..596S}, with a brief review of that work given in~\cite{2025RSPTA.38340032R}.

\begin{figure*}
\begin{center}
\includegraphics[height=2.5in,width=3in,angle=0]{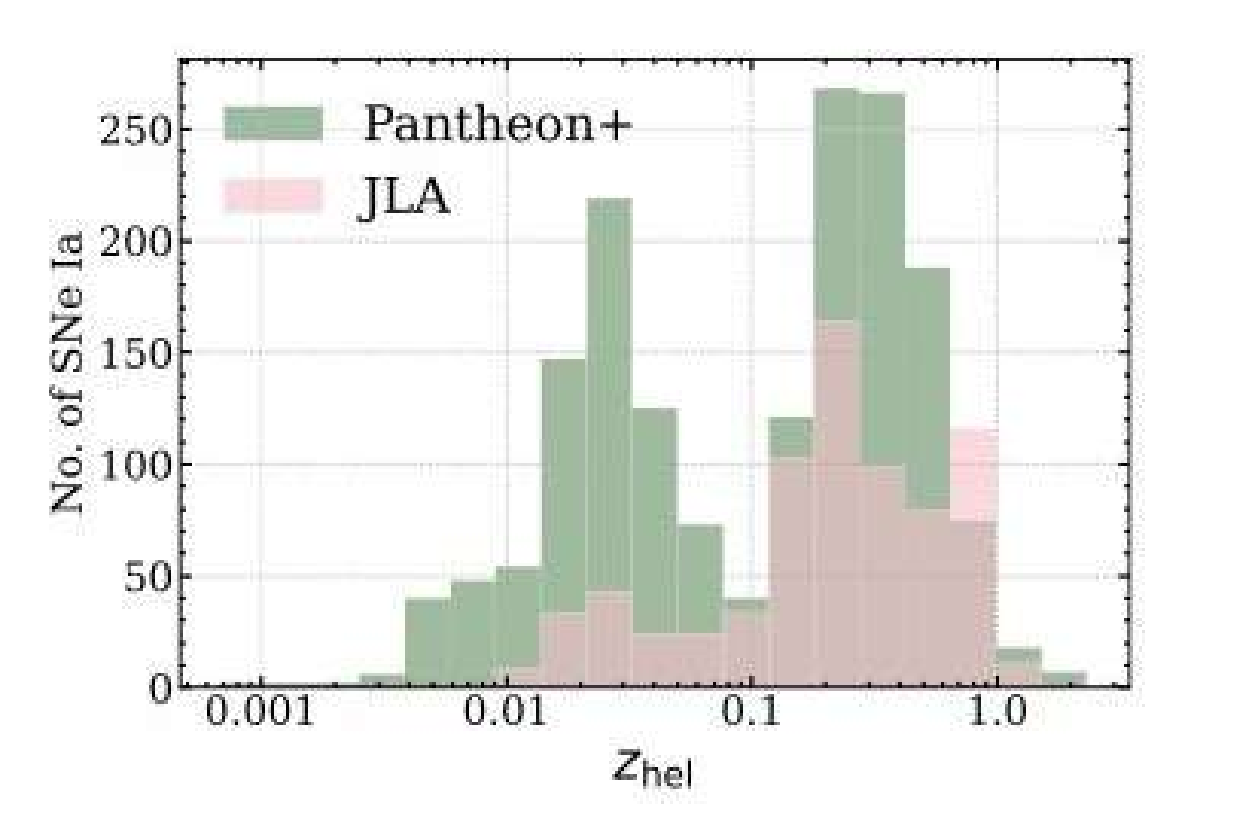}\quad
\end{center}
\caption{Distribution of heliocentric redshifts in the \textit{Pantheon}+ SNIa catalog~\cite{2022ApJ...938..113S}, compared to its
\textit{JLA} counterpart~\cite{2014A&A...568A..22B} (see also Fig.~2 in \cite{2025EPJC...85..596S}).}  \label{fig:SNIa}
\end{figure*}

The \textit{Pantheon}+ catalog contains 1701 spectroscopically confirmed SNIa from several different surveys, including JLA~\cite{2022ApJ...938..113S}. Similarly to the JLA, the \textit{Pantheon}+ data have already implemented peculiar-velocity corrections, both for our motion and that of the SNIa host galaxy~\cite{2022PASA...39...46C}. In addition, unlike JLA, the \textit{Pantheon}+ corrections ensure a smooth bulk-flow decay (in line with the $\Lambda$CDM expectations) and do not introduce an unphysical discontinuity at some maximum distance, as noted in~\cite{2019A&A...631L..13C}. This allowed to extend the peculiar-velocity corrections to all the supernovae in the catalog.

\begin{figure*}
\begin{center}
\includegraphics[height=2.5in,width=6.5in,angle=0]{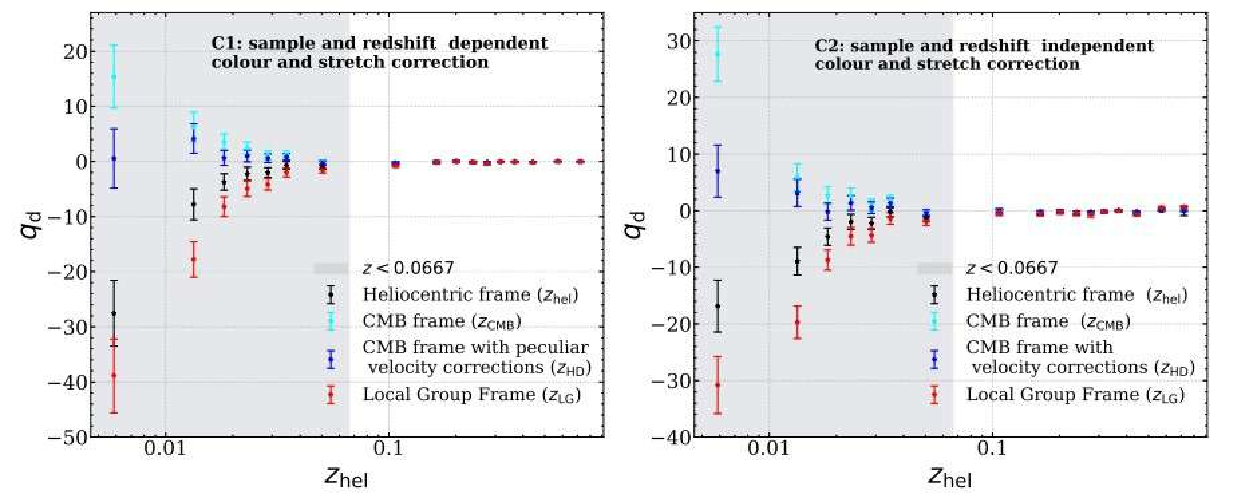}\quad
\end{center}
\caption{The scale-independence of $q_d$ evaluated in 17 shells each containing around 100 supernovae, plotted against the median redshift of the shells (with all the other parameters kept fixed). The analysis is done in the heliocentric, the Local Group and the CMB frame, with the direction fixed to the CMB dipole. The gray shaded region corresponds to $z<0.0667$
(i.e. to distances smaller than 200/h~Mpc). The error bars are at the $1\sigma$ level and the parameterisation is scale-independent within each shell. The left panel corresponds to the C1 analysis and the right panel to C2 (i.e.~with and without sample and redshift-dependence in the light-curve standardisation). The observed decay of the dipole in the deceleration parameter with redshift is a key prediction of the tilted universe scenario, originally discussed in~\cite{2011PhRvD..84f3503T} and later refined in~\cite{2025EPJC...85..596S,2025glc..conf....104T} (see also Fig.~9 in~\cite{2025EPJC...85..596S}).}  \label{fig:sah}
\end{figure*}

In the earlier analysis of the JLA catalog~\cite{2019A&A...631L..13C}, the dipole component in the sky-distribution of the deceleration parameter dominated its monopole counterpart out to redshift $z\simeq0.1$. The tension between this result and the $\Lambda$CDM model, which demands a highly isotropic $q$, was at the $3.9\sigma$ level. In the more recent study of~\cite{2025EPJC...85..596S}, the authors retained 1533 unique events (i.e.~not multiple reports of the same SNIa) from the \textit{Pantheon}+ database, of which 584 were also in the JLA sample. The remaining 949 supernovae contained 446 new events with $z\lesssim0.1$, of which 66 were below the lowest JLA redshift (see Fig.~\ref{fig:SNIa} above). The analysis of the data progressively removed the lower redshift SNIa in incremental steps of around 50 objects per step. This allowed to check the dependence of the deceleration parameter on the redshift range of each supernovae sample.
The anisotropy in the sky-distribution of the deceleration parameter was investigated by introducing the parametrisation
\begin{equation}
q= q_m+ \mathbf{q}_d\cdot\mathbf{n}\,,  \label{qd}
\end{equation}
where $q_m$ is the monopole, $\mathbf{q}_d$ is the dipole and $\mathbf{n}$ is a unitary vector pointing in the direction of the CMB dipole. The results showed a redshift-dependent dipole in the deceleration parameter consistent with the previous findings in the JLA dataset~\cite{2019A&A...631L..13C}, though this time at more than $5\sigma$ significance. Recall that the  $q$-dipole found in the JLA catalog was at the $3.9\sigma$ level. An additional new and potentially most important finding of~\cite{2025EPJC...85..596S} is that the magnitude of the dipolar modulation in the deceleration parameter decayed with scale/redshift, approaching zero for $z>0.1$ (see Fig.~\ref{fig:sah} above). This quantitative result agrees with the qualitative prediction of the ``tilted universe'' scenario (see \S~\ref{ssD-LDUE} earlier).

\begin{figure*}
\begin{center}
\includegraphics[height=2.75in,width=4.75in,angle=0]{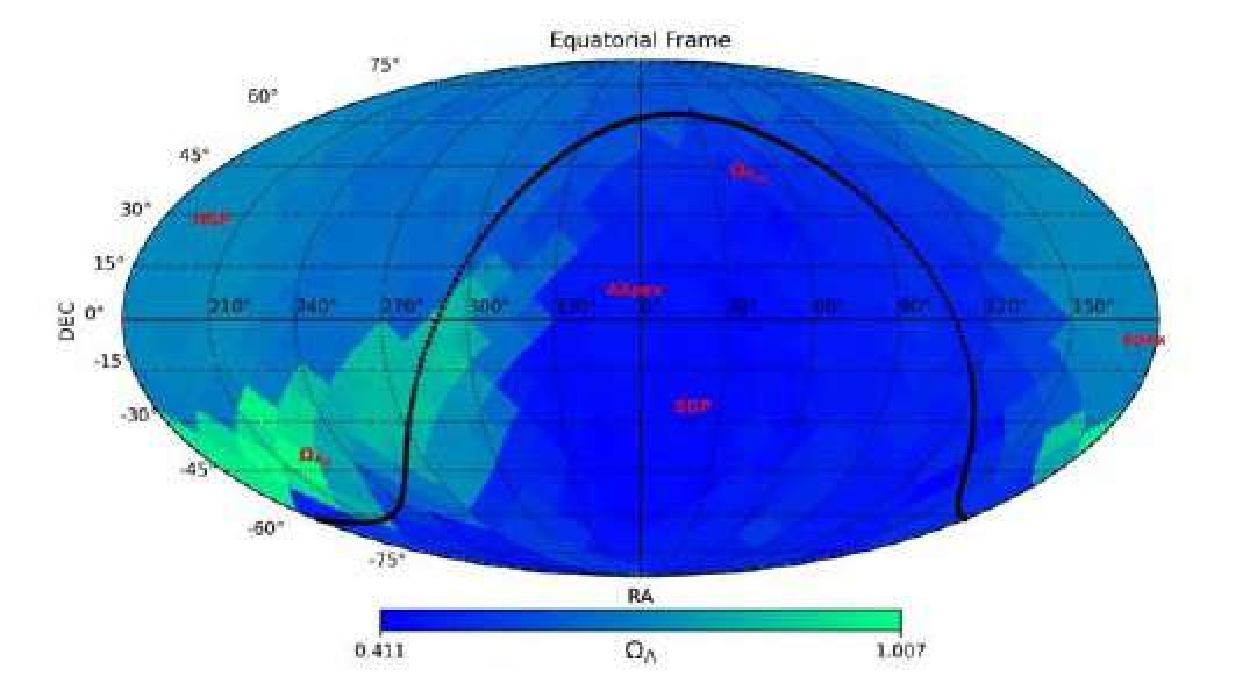}\quad
\end{center}
\caption{The variation of $\Omega_{\Lambda}$ over the whole sky, shown in celestial equatorial coordinates. The solid black line is the Galactic equator, while “NGP” and “SGP” respectively mark the positions of the North and of the South Galactic Poles. The labels Apex and AApex indicate the positions of the CMB Apex and Antiapex, whereas $\Omega_{\Lambda A}$ and $\Omega_{\Lambda AA}$ respectively mark the positions of the Apex and of the Antiapex of the fitted values of $\Omega_{\Lambda}$ (see also Fig.~2 in~\cite{2024ApJ...971...19C}).}  \label{fig:clocchiatti}
\end{figure*}

A complementary data analysis, also looking for dipolar anisotropies in the universal acceleration and also motivated by the ``tilted universe'' model, was pursued by Clocchiatti~et~al~\cite{2024ApJ...971...19C}. The study used the \textit{Pantheon}+ SNIa sample to look for dipolar asymmetries in the distribution of $\Omega_{\Lambda}$. In their report, the authors identified two large-scale asymmetries, the first of which they attributed to systematics related to the distribution of the \textit{Pantheon}+ supernovae and to their anisotropic coverage of the sky. The second dipole, which points approximately $50^{\circ}$ away from the apex of its CMB counterpart, has a statistical significance of $2.8\sigma$ (see Fig.~2 in~\cite{2024ApJ...971...19C} or Fig.~\ref{fig:clocchiatti} here). Intriguingly, the authors claim that the latter dipolar asymmetry is consistent with the predictions of the ``tilted universe'' model - see~\cite{2011PhRvD..84f3503T,2021EPJC...81..753T} for example. Put another way, the aforementioned anisotropy could be the Doppler-like dipole ``seen'' by a tilted observer residing in a bulk flow and moving relative to the universal expansion, as described in~\S~\ref{ssD-LDUE} previously.

Another report of a dipolar asymmetry in the universal acceleration, which was also potentially attributed to our peculiar motion, came from the work of Wang~\&~Wang~\cite{2014MNRAS.443.1680W}. Combining Type~Ia supernovae from the \textit{Union2.1} database with more than hundred Gamma-Ray-Bursts (GRB), the authors found a dipole asymmetry in the rate of cosmic  acceleration, with confidence at the 97.3\% level (i.e.~more than $2\sigma$). In addition, the reported anisotropy was more prominent at lower redshifts, in analogy the one communicated in~\cite{2025EPJC...85..596S} and in (qualitative) agreement with the ``tilted universe'' paradigm. Although an anisotropic dark-energy distribution was not excluded by the authors, their main explanation was a bulk flow with velocity around 270~km/sec on scales of roughly $100/h$~Mpc~\cite{2014MNRAS.443.1680W}.

Before closing this section, we should remind the reader that dipolar asymmetries in the universal acceleration, albeit not explicitly attributed to our peculiar motion, have been reported by other works as well (e.g.~see~\cite{2010JCAP...12..012A,%
2012JCAP...02..004C,2012PhRvD..86h3517M,2013PhRvD..87d3511M,%
2014MNRAS.437.1840Y,2023MNRAS.526.1482C}). These studies, as well as those cited earlier above, use different datasets and employ different techniques in their analyses.

\subsubsection{The H-dipole}\label{sssH-D}
%%%%%%%%%%%%%%%%%%%%%%%%%%%%%%%%%%%%%%%%%%
The isotropy of the universal (Hubble) expansion has been extensively tested by means of type Ia Supernovae (SNIa), with several studies reporting no significant deviations (e.g.~see~\cite{2016MNRAS.456.1881L,2018ApJ...865..119A,%
2018MNRAS.478.5153S,2018MNRAS.474.3516W}) and others claiming anisotropies of relatively mild significance (e.g.~see~\cite{2007A&A...474..717S,2010JCAP...12..012A,%
2013A&A...553A..56K,2015ApJ...801...76A,2015ApJ...808...39B,%
2015ApJ...810...47J}). However, the robustness of the SNIa-based tests has been challenged, primarily because of their highly inhomogeneous spatial distribution~\cite{2017MNRAS.471.1045C,%
2015PhLB..741..168B,2019arXiv190500221R}. There are additional issues as well, such as the sensitivity of the SNIa heliocentric redshifts to the adopted kinematic model and the assumptions involved in calibrating their light-curves. Other probes have used the X-ray background, the distribution of optical and infrared galaxies, radio sources, gamma-ray bursts, etc, albeit without reaching an unambiguous conclusion.

An alternative approach is to use galaxy clusters, since they are more uniformly distributed than the SNIa and extend out to larger scales. This method, which appeals to the directional behaviour of their X-ray luminosity-temperature scaling relation ($L_X-T$), was proposed in~\cite{2018A&A...611A..50M} and employed in~\cite{2020A&A...636A..15M}. In brief, the authors exploit the fact that the cosmological parameters do not directly affect the measurements of the temperature, the flux and the redshift of a galaxy cluster. The cosmological model comes into play when one uses the cluster's luminosity distance, together with its flux and redshift, to measure its luminosity. The latter can be estimated independently from the temperature of the cluster gas as well. Then, demanding that the two luminosity measurements must coincide, one can adjust the data and estimate the cosmological parameters. By repeatedly applying this process to different patches in the sky, one can look for directional dependence in the values of the cosmological parameters. In so doing, the authors also used Monte Carlo simulations to estimate the statistical significance of any detected
anisotropies~\cite{2018A&A...611A..50M}.

\begin{figure*}
\begin{center}
\includegraphics[height=2.5in,width=6.7in,angle=0]{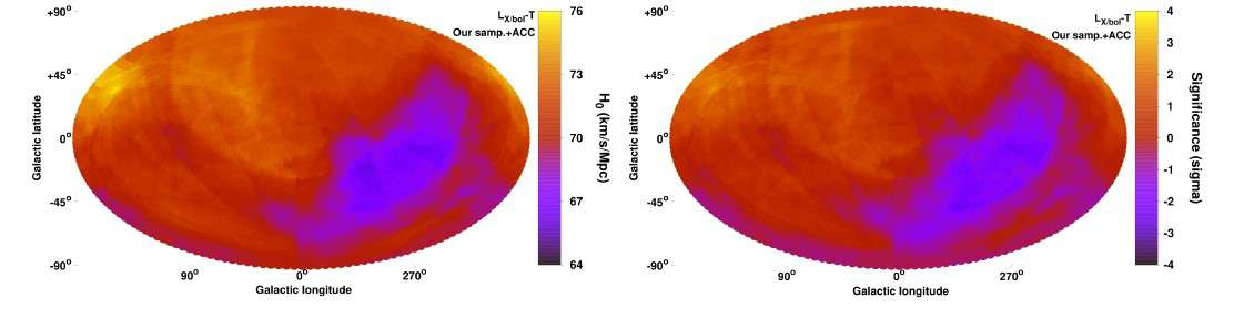}\quad
\end{center}
\caption{Anisotropy in the $H_0$ value based on the $L_X-T$ scaling relation (see also Fig.~4 in~\cite{2021A&A...649A.151M}).}  \label{fig:Migkas}
\end{figure*}

The aforementioned technique was applied to a sample of galaxy clusters (with 313 objects) in~\cite{2020A&A...636A..15M}, looking for regions with significantly different $L_X-T$ relations. The aim was to identify anisotropies in the Hubble expansion, by testing the consistency of the $L_X-T$ relation along different directions. This was achieved after scanning the full sky using cones of different size. The results show a consistent, as well as strong, directional dependence in the value ($H_0$) of the Hubble constant. More specifically, dividing the sky into hemispheres and using Monte Carlo simulations to estimate the statistical significance of their results, the authors found that $H_0$ took systematically lower values towards $(\ell,b)\simeq(277^{\circ},-11^{\circ}$), compared to $(\ell,b)\simeq(32^{\circ},15^{\circ}$). The confidence level level was measured to range between $3.6\sigma$ and $5\sigma$.

In a follow up study~\cite{2021A&A...649A.151M}, the authors analysed an increased sample of up to 570 clusters with measured X-ray, infrared and microwave properties. This significantly improved the sky-coverage and the statistical strength of the study. The new data were used to construct ten different scaling relations, in addition to $L_X-T$, in order to test the isotropy of the Hubble expansion. Employing the same methodology with~\cite{2020A&A...636A..15M}, the results suggested a 9\% variation in the value of the Hubble constant between $(\ell,b)\simeq(280^{\circ},-35^{\circ}$) and the rest of the sky (see Fig.~\ref{fig:Migkas} above). The reported anisotropy had a nearly dipole form and statistical significance greater than $5\sigma$~\cite{2021A&A...649A.151M}. Although the authors could not entirely exclude interference from as yet unknown systematics, they felt confident to claim that their results do not suffer from any known biases. After all, it would be quite unlikely for the systematics to bias all the scaling relations simultaneously and lead to the same results across different observables and samples. Without excluding the possibility the reported $H_0$-dipole to reflect a generic large-scale (cosmic) anisotropy of the universe (a possibility raised in~\cite{2022PhRvD.105f3514K,%
2022PhRvD.105j3510L}), the authors suggested that the observed asymmetry could be due to a bulk flow extending out to more than 500~Mpc and moving with 800~km/sec (much like the dark flows reported by Kashlinsky~et~al - see \S~\ref{ssDFQ} previously). Both alternatives are in serious tension with the standard $\Lambda$CDM model.

\begin{figure*}
\begin{center}
\includegraphics[height=2.5in,width=6.5in,angle=0]{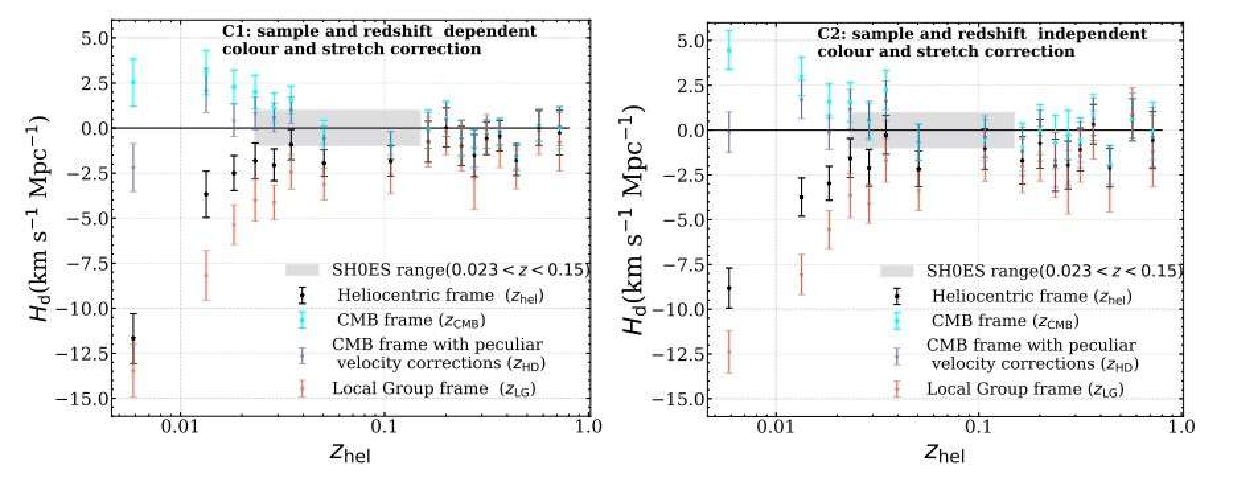}\quad
\end{center}
\caption{The scale dependence of $H_d$, evaluated in 17 shells containing approximately 100 supernovae each and plotted against the median shell redshift, with all other parameters held fixed. The analysis is carried out in the heliocentric, the Local Group, the CMB and Hubble Diagram frames. The parametric form of the fitted dipole (see Eq.~(\ref{Hd}) is scale-independent and its direction is aligned with that of the CMB dipole. The error bars denote $\pm1\sigma$ uncertainties and the gray shaded region corresponds to the redshift range $z=0.023 - 0.15$, while its vertical width indicates the $\pm1$~km/secMpc precision on $H_0$ claimed by the \textit{SH0ES} team~\cite{2022ApJ...934L...7R}. The left and right panels correspond to C1 and the C2 analysis respectively, namely those with and without sample and redshift-dependence in the light curve standardisation. The observed redshift decay of the $H$-dipole is an additional key prediction of the ``tilted universe'' paradigm (see \S~\ref{ssDDHP} here and also Fig.~3 in~\cite{2025EPJC...85..596S}).}  \label{fig:Hdipole}
\end{figure*}

The Hubble and the deceleration parameters are closely related, since the latter is essentially the time derivative of the former. Given this, it makes physical sense to argue that a dipolar anisotropy in the deceleration parameter, should immediately imply the same for the Hubble parameter and vice versa. Looking for verification in the data, Sah~et~al~\cite{2025EPJC...85..596S} employed an analysis closely analogous to the one they used for the $q$-dipole (see \S~\ref{sssq-D} before). More specifically, the anisotropy was investigated by parametrising the Hubble constant as
\begin{equation}
H= H_m+ \mathbf{H}_d\cdot\mathbf{n}\,,  \label{Hd}
\end{equation}
where $H_m$ is the monopole, $\mathbf{H}_d$ is the dipole and $\mathbf{n}$ is the unitary spatial vector along the CMB dipole (compare to expression (\ref{qd}) in \S~\ref{sssq-D} previously). The above parametrisation was fit to the data within the redshift range $0.023<z<0.1$, with 487 SNIa used in the C1 analysis and 480 in the C2. The results revealed a statistically significant $H$-dipole in excess of 1.5~km/secMpc and with a redshift decreasing magnitude, as expected by the ``tilted universe'' model (see \S~\ref{ssDDHP} earlier and consult Fig.~\ref{fig:Hdipole} here).

In the literature one can find more studies that also report dipole-like asymmetries in the sky-distribution of the Hubble parameter, but without relating it to the peculiar motion of our Milky Way. For instance, the analyses of \cite{2022PhRvD.105f3514K,2023PhRvD.108l3533M,2023ChPhC..47l5101T,%
2024A&A...681A..88H,2024A&A...689A.215H} were primarily designed to test the Cosmological Principle (CP).

\subsubsection{Number-count dipoles}\label{sssNCDs}
%%%%%%%%%%%%%%%%%%%%%%%%%%%%%%%%%%%%%%%%%%%%%%%%%%%%
The Cosmological Principle (CP), namely the belief that we are all typical observers in a nearly homogeneous and isotropic (i.e.~almost-FLRW) universe has been the cornerstone of modern cosmology for the last hundred years. The strongest observational support for the CP so far, comes form the CMB spectrum, which exhibits a remarkably high degree of isotropy. Combined with the Copernican Principle, advocating that we are not privileged observers in the cosmos, the isotropy of the CMB leads to the conclusion that we live in a maximally (i.e.~spatially homogeneous and isotropic) universe. There is also theoretical support for CP, primarily coming from the inflationary paradigm, which is supposed to dilute essentially any degree of primordial anisotropy and/or inhomogeneity. Although structure formation has destroyed (to a lesser or larger degree) the overall uniformity of the universe on relatively small scales, the established belief maintains that beyond a certain length our cosmos remains largely homogeneous and isotropic. The uniformity threshold, however, which is typically set close to the 100~Mpc mark, is uncertain and under debate.

Over the last few years, the CP-related discussion has intensified. This follows an increasing number of surveys reporting dipolar asymmetries in the number counts of distant astrophysical sources, like supernovae and quasars for example. The nature of these dipoles raises as yet unanswered questions, regarding their origin and the mechanism responsible for them. For instance, Singal~\cite{2011ApJ...742L..23S} identified a pronounced dipole in both the number counts and the integrated sky-brightness of radio galaxies from the \textit{NRAO VLA Sky Survey} (NVSS). Although the direction of the reported dipole was aligned with that of its CMB counterpart, its amplitude  corresponds to a velocity estimate of the order of $10^{3}$~km/sec. The latter is higher than the velocity estimates based on the amplitude of the CMB dipole at a statistically significant ($\simeq3\sigma$) level. In addition, the dipole amplitude did not seem to decline with increasing survey depth, as it would had been expected if it was solely a peculiar-velocity effect. This result was largely corroborated by the study of Rubart and Schwarz~\cite{2013A&A...555A.117R}, which used data from the (NVSS) and the \textit{Westerbork Northern Sky Survey} (WENSS) to report a radio-dipole as well. Similarly to the study of~\cite{2011ApJ...742L..23S}, the dipole was in the direction of its CMB counterpart and its  amplitude exceeded the expectations by a factor of about 4, which made it inconsistent with a pure kinetic origin at 99.6\% confidence level.

In an analogous study, Rameez~et~al~\cite{2018MNRAS.477.1772R} reported a dipolar anisotropy in the spatial distribution of galaxies listed in the \textit{AllWISE} catalog. The latter comprises close to 1.5 million extragalactic sources across approximately 95\% of the sky and reaching redshifts as far as $z\simeq0.2$, that is considerably deeper than previous infrared surveys. The authors employed an improved hemispheric technique of count-comparison, which revealed a statistically significant dipole in the \textit{AllWISE} galaxy distribution approximately 4-5 times larger than what it would have been expected in the $\Lambda$CDM model. Structures within the redshift range $0.03<z<0.3$ appear to contribute to an unexpectedly large clustering dipole, but the precise sources of this excess remains unidentified. Based on their findings, the authors suggested that we may occupy an atypical location in the universe, which (if true) would have profound implications for the interpretation of essentially all the cosmological data.

More recently, Secrest~et~al~\cite{2021ApJ...908L..51S} identified a prominent large-scale dipole in the all-sky distribution of more than a million quasars, using the \textit{CatWISE} database. Applying the observational test proposed by Ellis and Baldwin~\cite{1984MNRAS.206..377E} to evaluate the isotropy/anisotropy of the universe, they detected an anomalously large dipolar asymmetry aligned closely with the CMB dipole direction. However, the amplitude of the reported dipole is twice larger than it would have been if the responsible agent was our peculiar motion within the $\Lambda$CDM framework. Building upon their earlier work, Secrest~et~al refined and extended their cosmic-dipole investigation in~\cite{2022ApJ...937L..31S}. This study employs similar hemisphere-comparison techniques, but applies them to a vastly improved dataset of quasars identified through \textit{Gaia EDR3}. The new findings are consistent with the previous results, but the amplitude discrepancy becomes even more pronounced. In particular, the quasar-dipole reported in~\cite{2022ApJ...937L..31S} exceeds the expectation by factors of 5-10, depending on sample selection and analysis methods. Most significantly, the amplitude excess appears to persist unchanged with increasing sample depth and out to redshifts of $z>1$. For a recent review and more discussion on the ``Cosmic Dipole Anomaly'', as well as for additional references, the reader is referred to~\cite{2025RvMP...97d1001S}.

So far, the literature contains an appreciable number of surveys reporting anomalous large-scale dipolar anisotropies in the number counts of distant astrophysical sources. In most of the studies (see above), the discrepancy is the amplitude of the dipole and not in the direction, which largely agrees with that of its CMB counterpart. Nevertheless, there have been claims of the opposite as well. More specifically, a dipole with magnitude close to that of the CMB, but significantly different direction, was found in the \textit{Pantheon}+ sample in~\cite{2023JCAP...11..054S}. All this work could suggest that the rest-frame of the matter may not coincide with that of the CMB. It is then conceivable that the aforementioned reports may go far enough to challenge the foundation stones of modern cosmology, like the Cosmological Principle for example, and in so doing change the route of modern cosmology dramatically.

Be that as it may, one needs to keep in mind that dipolar anisotropies are the trademark signature of relative motions. The fact that, in most of the surveys (to the best of our knowledge), the direction of the reported dipole seems to agree with that seen in the CMB spectrum, supports the possibility these number-count dipoles to be artifacts of our galaxy's peculiar flow as well. The discrepancies in the dipole amplitudes may point the opposite way, but they could also reflect the fact that there are relatively few theoretical studies of the issue~\cite{2025arXiv250708931Y}. For instance, as one can easily show (see \S~\ref{ssD-LDNCs} earlier) peculiar motions can trigger dipolar asymmetries in the number-counts of astrophysical sources, the amplitude of which is proportional to the number density of the individual sources (see Eq.~(\ref{nlrels2}) in \S~\ref{ssD-LDNCs}). This simple theoretical dependence may be fairly straightforward to test against the observations.

\section{Impact on astronomical observations}\label{sIAOs}
%%%%%%%%%%%%%%%%%%%%%%%%%%%%%%%%%%%%%%%%%%%%%%%%%%%%%%%%%%
The results and the discussion of the previous sections underscore the necessity of rigorously accounting for the effects of the local kinematics, when interpreting cosmological observables, as these may significantly bias the inferred expansion history of the universe.

\subsection{How peculiar motions can bias astronomical observations}
%%%%%%%%%%%%%%%%%%%%%%%%%%%%%%%%%%%%%%%%%%%%%%%%%%%%%%%%%%%%%%%%%%%%
Peculiar velocities introduce systematic biases across multiple domains of observational cosmology, affecting our measurements of fundamental cosmological parameters and potentially leading to misinterpretations of the underlying physics.

\subsubsection{Redshift and luminosity distance
%%%%%%%%%%%%%%%%%%%%%%%%%%%%%%%%%%%%%%%%%%%%%%%
measurements}\label{sssRLDMs}
%%%%%%%%%%%%%%%%%%%%%%%%%%%%%
\underline{\it Effects on the redshift:} By definition, the redshift of a luminous source is given by the ratio $z=(\lambda_2-\lambda_1)/\lambda_1$, where $\lambda_1$ and $\lambda_2$ are the wavelengths at emission and reception respectively. Then, $1+z=\lambda_2/\lambda_1=\nu_1/\nu_2$, with $\nu_1$ and $\nu_2$ representing the associated frequencies. Given that the latter are proportional to the photon energy (i.e.~$E\propto\nu$), we obtain
\begin{equation}
1+z= {E_1\over E_2}= {\left(k_au^a\right)_1\over\left(k_au^a\right)_2}\,,  \label{z1}
\end{equation}
where $k_a$ is the null vector tangent to the geodesic light-ray (with $k_ak^a=0=k^b\nabla_bk_a$ -- see \S~\ref{sssNC} earlier). Also, recall that $E=-k_au^a$ is the photon energy measured by an observer with 4-velocity $u_a$ -- see \S~\ref{sssNC} earlier as well).

Let us now consider a pair of idealised (CMB) and realistic (tilted) observers, with respective 4-velocities $u_a$ and $\tilde{u}_a$, located at the reception point. Relative to the tilted frame Eq.~(\ref{z1}) reads $1+\tilde{z}= \left(k_au^a\right)_1/\left(k_a\tilde{u}^a\right)_2$, since the photon energy at the emission point ($E_1=(k_au^a)_1$) is unchanged. The latter combines with (\ref{z1}) to give
\begin{equation}
{1+\tilde{z}\over1+z}= {\left(k_au^a\right)_2\over\left(k_a\tilde{u}^a\right)_2}\,.  \label{z2}
\end{equation}
Note that, as stated in \S~\ref{sssNC} and \S~\ref{ssD-LDNCs} before, the involved sources are assumed to have sufficiently high redshifts to ignore their individual peculiar velocities at the point of emission and treat them as comoving with the CMB frame and the universal expansion.

To proceed further, one projects (\ref{4vels4vels}) along the null direction to obtain $k_a\tilde{u}^a=k_au^a+k_a\tilde{v}^a$, which immediately translates into $\tilde{E}=E-k_a\tilde{v}^a$. Here, $E$ and $\tilde{E}$ represent the photon energies measured in the reference (CMB) frame and in the tilted (matter) frames respectively. Using (\ref{k}) and keeping up to $\tilde{v}$-order terms, the relation between the two photon energies reads
\begin{equation}
\tilde{E}= E\left(1+e_a\tilde{v}^a\right)\,,  \label{tE}
\end{equation}
with $e_a$ being the unit spacelike vector in the direction of the source (see \S~\ref{sssNC} previously). Therefore, the scalar $k_a\tilde{v}^a=-Ee_a\tilde{v}^a$ measures the change in the photon energy triggered by the 4-velocity boost (\ref{4vels4vels}), that is by the tilted observer's peculiar flow. Finally, going back to relation $k_a\tilde{u}^a=k_au^a+k_a\tilde{v}^a$, we immediately obtain
\begin{equation}
{k_a\tilde{u}^a\over k_au^a}= 1+ {k_a\tilde{v}^a\over k_au^a}\,,  \label{ks}
\end{equation}
which substituted into the right-hand side of (\ref{z2}) gives
\begin{equation}
1+ \tilde{z}= (1+z)\left[1-\left(e_a\tilde{v}^a\right)_2\right]\,,  \label{z3}
\end{equation}
to first approximation (recall that the index~2 marks the time of reception). In the above $z$ and $\tilde{z}$ are the redshifts of the distant luminous source, as measured by the CMB and the tilted observer at the same spacetime event. Therefore, expression (\ref{z3}) provides the familiar redshift correction due to the observer's peculiar motion relative to the CMB (e.g.~see Eqs.~(5), (6) in~\cite{2011ApJ...741...67D} for comparison). Note that, in the literature, $\tilde{z}$ is also referred to as the total, or the observed, or the heliocentric redshift, while $z$ is also known as the cosmological, or the true redshift. Finally, the scalar $(e_a\tilde{v}^a)_2$ is occasionally referred to as the peculiar, or the Doppler, redshift.

Note that here we have isolated and focused upon the effect of the tilted observer's peculiar motion. The redshift of a distant source is also distorted by additional effects, that occur during the propagation of the radiation signal, like the gravitational redshift and the integrated Sachs-Wolfe effects for example (e.g.~see~\cite{2009PhRvD..80h3514Y}).

\underline{\it Effects on the luminosity distance:} An additional familiar impact of peculiar velocities is the one on distance measurements. For instance, following~\cite{2018PhRvD..98f3505I}, in an FLRW cosmology with no peculiar velocities the luminosity distance of a source at redshift $z$ is given by
\begin{equation}
d_L(z)= {a_0\over\mathcal{H}_0}(1+z)P(z)\,,  \label{FLRWld1}
\end{equation}
where $\mathcal{H}=a^{\prime}/a$ and
\begin{equation}
P(z)= z- {1\over2}\left(2+\mathcal{Q}_0\right)z^2+ {1\over6}\left(3\mathcal{Q}_0^2+6\mathcal{Q}_0+6
-\mathcal{J}_0\right)z^3+ \mathcal{O}(z^4)\,,  \label{FLRWPz}
\end{equation}
with $\mathcal{Q}=-a^{\prime\prime}/a\mathcal{H}^2$ and $\mathcal{J}=a^{\prime\prime\prime}/a\mathcal{H}^3$. In all of the above, the zero suffix indicates the present and the primes denote conformal-time derivatives. Consequently, $\mathcal{H}$, $\mathcal{Q}$ and $\mathcal{J}$ are respectively, the Hubble, the deceleration and the ``jerk'' parameters expressed in terms of conformal time ($\eta$ with $\dot{\eta}=1/a$).

Allowing for (weak) peculiar velocities, but keeping the FLRW nature of the host spacetime unchanged, recasts (\ref{FLRWld1}) into~\cite{2018PhRvD..98f3505I}
\begin{equation}
d_L(z)= {a_0\over\mathcal{H}_0}(1+z)P(z_c)\,.  \label{FLRWld2}
\end{equation}
Here, $z_c=a_0/a$ represents the contribution of the cosmological expansion to the total redshift ($z$) of the source, so that
\begin{equation}
1+z= (1+z_c)(1+z_p)\,,  \label{tz}
\end{equation}
with $z_p$ giving the peculiar-velocity input to the total redshift. Note that after comparing the above result to expression (\ref{z3}), we may write $z_p=-\left(e_a\tilde{v}^a\right)_2$. Therefore, the contribution of the peculiar motion to the total redshift ($z$) of the source vanishes, namely $z_p=0$ and $z=z_c$, when $\left(e_a\tilde{v}^a\right)_2=0$ and vice-versa.

Starting from Eq.~(\ref{tz}), assuming that $z_p\ll1$ and keeping up to first-order terms in $z_p$, we obtain
\begin{equation}
z_c= z- (1+z)z_p+ \mathcal{O}(z_p^2)\,.  \label{lzc}
\end{equation}
On using the above linear expression, a simple Taylor expansion of $P(z_c)$ around $z$, leads to
\begin{equation}
P(z_c)= P(z)- (1+z)z_p{{\rm d}P\over{\rm d}z}+ \mathcal{O}(z_p^2)\,,  \label{FLRWPzc}
\end{equation}
Finally, differentiating (\ref{FLRWPz}) with respect to redshift, substituting its derivative into the right-hand side of the above and then inserting the resulting expression into Eq.~(\ref{FLRWld2}), we arrive at~\cite{2018PhRvD..98f3505I}
\begin{equation}
d_L(z)\simeq -{a_0\over\mathcal{H}_0}\left\{z_p-\left(1+\mathcal{Q}_0z_p\right)z+ {1\over2}\left[\mathcal{Q}_0+\left(3\mathcal{Q}_0^2+2\mathcal{Q}_0
-\mathcal{J}_0\right)z_p\right]z^2\right\}\,,  \label{FLRWld3}
\end{equation}
after dropping terms of order higher than $\mathcal{O}(z_p)$ and $\mathcal{O}(z^2)$. The above expression provides the luminosity distance of a source located at redshift $z$ in an FLRW universe endowed with a weak peculiar-velocity field. Following () and (\ref{FLRWPzc}) and (\ref{FLRWld3}), the effect of teh peculiar motion on the luminosity distance vanishes when $z_p=0$, or equivalently when $\left(e_a\tilde{v}^a\right)_2=0$ (see related discussion above). Alternatively, one may say that there is no relative-motion effect on the luminosity distance when the peculiar velocity is normal to the line-of-sight (see also Eq.~(\ref{ld5}) below).

An additional useful relation follows when the average contribution of the peculiar velocities to the total redshift of the source is assumed to vanish, namely when $\langle z_p\rangle=0$. Then, keeping terms of order up to $\mathcal{O}(z_p)$ and $\mathcal{O}(z^2)$, expression (\ref{FLRWld3}) leads to
\begin{equation}
d_L- \langle d_L\rangle\simeq -{a_0\over\mathcal{H}_0}\left[1-\mathcal{Q}_0z+{1\over2}
\left(3\mathcal{Q}_0^2
+2\mathcal{Q}_0-\mathcal{J}_0\right)z^2\right]z_p\,.  \label{FLRWld4}
\end{equation}
The latter may be seen as the difference between the luminosity distance of a given source and the average luminosity distance of all the sources~\cite{2018PhRvD..98f3505I}.

Adopting an alternative approach, one may assume that $\vec{v}_p$ represents the peculiar velocity of the source and $\vec{v}_o$ is that of the observer. Then, the (linearly perturbed) luminosity distance of the source is~\cite{2011ApJ...741...67D}

\begin{equation}
d_L(z) = d_L^{(0)}(z)- \frac{(1+z)^2}{H(z)}\,\vec{e}\cdot(\vec{v}_o-\vec{v}_p)\,,  \label{ld5}
\end{equation}
where $d_L^{(0)}(z)$ is the unperturbed background value. The second term on the right-hand side of the above represents the perturbation induced by the peculiar motions of the source and the observer. Given that $\vec{e}$ is the unit vector along the line-of-sight, we deduce that the relative-motion effect on $d_L$ vanishes when $\vec{e}$ and $\vec{v}_0-\vec{v}_p$ are normal to each other. Recall that exactly analogous results were obtained via Eqs.~(\ref{tz}), (\ref{FLRWPzc}) and (\ref{FLRWld3}) earlier.

Among others, the aforementioned changes to the luminosity distance directly affect the measurements of type~Ia supernova (SNIa), which serve as critical distance indicators for cosmological parameter inference~\cite{2006PhRvD..73l3526H,2013PhRvL.110b1302B}. Note that, for nearby SNIa (with $z<0.1$), peculiar-velocity effects can introduce errors of 5-10\% in the distance measurements~\cite{2022ApJ...938..112P}. These errors contaminate the determination of the Hubble constant, with the analyses suggesting that peculiar velocities could contribute up to 1-2~km/secMpc to the observed $H_0$-tension~\cite{2022ApJ...934L...7R,2021ApJ...919...16F}.

\subsubsection{Redshift-space distortions}\label{ssR-SDs}
%%%%%%%%%%%%%%%%%%%%%%%%%%%%%%%%%%%%%%%%%%%%%%%%%%%%%%%%%
Galaxy redshift surveys map the cosmic web by using observed redshifts to infer distances through Hubble's law. However, the peculiar velocities of the galaxies induce additional Doppler shifts that contaminate the cosmological redshift signal, creating systematic distortions in the reconstructed three-dimensional galaxy distribution. These redshift-space distortions (RSDs) manifest as anisotropies in what should otherwise be an isotropic clustering pattern, particularly affecting the line-of-sight component of galaxy positions~\cite{1987MNRAS.227....1K,1998ASSL..231..185H,%
2004PhRvD..70h3007S}.

Several distinct redshift-space distortion effects have been identified:

\begin{itemize}
\item \textbf{Fingers of God:} Prominent anisotropic structures appear as radially elongated features in redshift-space galaxy distributions, pointing toward the observer with characteristic needle-like morphology. The physical origin lies primarily in the random, quasi-virialized motions of galaxies within clusters, where one-dimensional velocity dispersions typically reach $\sigma_v\approx1000$~km/sec~\cite{1972MNRAS.156P...1J}. Mathematically, these distortions can be expressed through the mapping between real-space position ($\mathbf{r}$) and redshift-space position ($\mathbf{s}$):
\begin{equation}
\mathbf{s} = \mathbf{r} + \frac{v_\parallel(\mathbf{r})}{aH}\hat{\mathbf{z}}\,,
\end{equation}
with $v_\parallel$ representing the line-of-sight peculiar velocity component, $a$ the cosmological scale factor, and $H=H(a)$ the Hubble parameter~\cite{1998ASSL..231..185H}. The effect manifests itself in the redshift-space power spectrum as a suppression factor, often modeled as:
\begin{equation}
P_s(k,\mu) = P_r(k)(1+\beta\mu^2)^2 F(\sigma_v k \mu)\,,
\end{equation}
where $F(\sigma_v k \mu)$ is a damping function (typically Lorentzian or Gaussian) that encapsulates the Finger of God (FoG) effect, with $\mu$ being the cosine of the angle between the wavevector $\mathbf{k}$ and the line of sight~\cite{2004PhRvD..70h3007S,2015MNRAS.447..234W}.

These redshift-space distortions are especially pronounced in high-density environments and produce a characteristic suppression of clustering power on small scales (i.e.~those with $k\gtrsim 0.1$~h/Mpc)~\cite{1994MNRAS.267.1020P,2005ApJ...630....1Z}. Recent high-precision studies have revealed that the simple dispersion models are often insufficient, requiring more sophisticated treatments that account for the kurtosis of the velocity distribution function and non-linear velocity coupling~\cite{2012JCAP...11..014O,2016IAUS..308..340B}. Advanced techniques, including the distribution function approach~\cite{2011JCAP...11..039S} and effective field theory frameworks~\cite{2019JCAP...03..007V}, have been developed to model such distortions more accurately across a wider range of scales. The FoG effect remains a significant systematic challenge in extracting cosmological information from galaxy clustering measurements, particularly for surveys targeting higher redshifts, where distinguishing between different dark energy models requires percent-level precision~\cite{2016arXiv161100036D,%
2014MNRAS.444..476R}.\\

\item \textbf{Kaiser effect:} While the Fingers of God dominate small-scale distortions, large-scale coherent infall toward overdense regions produces a complementary anisotropic signature. This phenomenon, known as the Kaiser effect~\cite{1987MNRAS.227....1K}, manifests itself as a compression of structures along the line of sight, enhancing the apparent clustering strength in redshift space. Unlike the random motions discussed previously, these distortions arise from coherent flows governed by linear perturbation theory, where peculiar velocities are proportional to the (Newtonian) gravitational acceleration: $\mathbf{v} \propto \mathbf{g} \propto \nabla \Phi \propto \nabla^{-1} \delta$. In the plane-parallel approximation, this leads to the familiar enhancement of the redshift-space power spectrum relative to the real-space spectrum:
\begin{equation}
P_s(k,\mu) = P_r(k)(1+\beta\mu^2)^2\,,
\end{equation}
where $\beta = f/b$ quantifies the relative strength of the distortion, with $f$ representing the (logarithmic) growth factor of structure formation~\cite{1994MNRAS.267.1020P,1994MNRAS.267..785C,2005PhRvD..72d3529L}. The resulting quadrupole-to-monopole ratio of the redshift-space power spectrum provides a direct probe of $\beta$~\cite{1992ApJ...385L...5H}.

Recent theoretical developments have extended this formalism beyond the plane-parallel approximation to account for wide-angle effects~\cite{1998ApJ...498L...1S,2008MNRAS.389..292P}, relativistic corrections~\cite{2009PhRvD..80h3514Y,2011PhRvD..84d3516C} and modifications to gravity~\cite{2008Natur.451..541G,2012MNRAS.420.1079J}. These advancements are particularly relevant for next-generation surveys covering substantial fractions of the sky. The measurement precision has improved dramatically, with current constraints reaching $\sim$1-2\% uncertainty on $f\sigma_8$~\cite{2017MNRAS.470.2617A,2021PhRvD.103h3533A,%
2011MNRAS.418.1707B,2012MNRAS.423.3430B}, facilitating critical tests of gravitational theories on cosmological scales. Furthermore, the combination of redshift-space distortions with weak lensing measurements breaks the degeneracy between bias and growth, enabling model-independent constraints on modified-gravity parameters~\cite{2007PhRvL..99n1302Z,2016MNRAS.456.2806B}. This synergy between different cosmological probes exemplifies how the analysis of large-scale velocity fields contributes to our fundamental understanding of gravity and cosmic acceleration.\\

\item \textbf{Rocket effect:} Beyond the Finger of God (FoG) and the Kaiser effects, the observer's own peculiar motion relative to the CMB rest frame induces another systematic distortion in redshift surveys known as the rocket effect~\cite{1990ApJ...354...18B,2014ApJ...788..157N}. This phenomenon, first formalized by Kaiser~\cite{1987MNRAS.227....1K}, creates a dipolar modulation in the observed galaxy distribution by altering the observed redshifts according to:
\begin{equation}
(1 + z_{\text{obs}}) = (1 + z_{\text{true}})(1 + \mathbf{v}_{\text{obs}} \cdot \hat{\mathbf{n}}/c)
\end{equation}
where $\mathbf{v}_{\text{obs}}$ is the observer's peculiar velocity vector and $\hat{\mathbf{n}}$ is the unit vector pointing toward the observed galaxy. This effect shifts galaxies toward the direction opposite to the observer's motion, creating an apparent dipolar anisotropy in the galaxy distribution. Additionally, the observer's motion induces a modulation in the observed flux and angular positions, further contributing to the overall dipolar pattern~\cite{2011MNRAS.413.2906D,2018JCAP...01..013M}. The resulting contribution to the galaxy power spectrum can be expressed as $P_{\text{rocket}}(k) \propto (\mathbf{v}_{\text{obs}}/c)^2 P_m(k)/k^2$ at linear scales, where $P_m(k)$ is the matter power spectrum~\cite{2014MNRAS.443.1900B}.

While typically smaller than the Kaiser and Fingers of God effects, with the CMB dipole indicating $v_{\text{obs}} \approx 370$ km/s~\cite{2020A&A...641A...6P}, the rocket effect becomes increasingly significant in wide-angle surveys covering large fractions of the sky and at higher redshifts. This effect can contaminate measurements of primordial non-Gaussianity, large-scale clustering, and cosmic dipoles if not properly accounted for~\cite{2012MNRAS.424..472B,2018JCAP...01..013M}. Recent studies have developed sophisticated methods to isolate and extract this signal, potentially using it as a probe of cosmic velocity fields and a consistency check of the cosmological principle. In fact, comparisons between the kinematic dipole measured from galaxy surveys and the CMB dipole have revealed intriguing tensions~\cite{2021ApJ...908L..51S,2021A&A...653A...9S}, highlighting the importance of properly modeling all velocity-induced effects in our interpretation of large-scale structure data. The rocket effect thus complements the previously discussed redshift-space distortions by providing additional insight into peculiar velocity fields on the largest observable scales.\\

\item \textbf{Wide-angle effects:} While the previously described Kaiser and Rocket effects are typically formulated in the plane-parallel (or distant-observer) approximation, this simplification becomes inadequate for modern wide-angle surveys covering large fractions of the sky~\cite{1998ApJ...498L...1S,2000ApJ...535....1M,%
    2012MNRAS.420.2102S}. In the plane-parallel limit, all lines of sight are assumed to be parallel, such that the angle between any pair of galaxies as viewed from the observer is negligible. However, when this angle becomes significant, additional geometric terms arise in the redshift-space correlation function. Mathematically, the correlation function becomes dependent on three variables, namely on the separation ($s$) between galaxies, on the mean distance ($d$) to the pair, and on the angle ($\phi$) between the line connecting the galaxies and the line of sight, introducing a more complex multipole structure~\cite{2008MNRAS.389..292P,%
    2010MNRAS.409.1525R}. These wide-angle contributions can be computed perturbatively as:
\begin{equation}
\xi(s,d,\phi)= \xi^{\text{plane-parallel}}(s,\phi)+ \frac{s}{d}\,\xi^{\text{wide-angle}}(s,\phi) + \mathcal{O}\left(\frac{s^2}{d^2}\right)\,,
\end{equation}
where the correction terms scale with the ratio of separation to distance and introduce mixing between different multipoles of the correlation function~\cite{2018MNRAS.476.4403C,%
2016JCAP...01..048R}.

The importance of these wide-angle effects increases for larger separations, higher redshifts, and surveys with broader angular coverage, acquiring particular significance for measurements of primordial non-Gaussianity through scale-dependent bias~\cite{2019JCAP...03..040B}. Several approaches have been developed to account for these effects, including the use of tripolar spherical harmonics~\cite{2004ApJ...614...51S,%
2007MNRAS.378..119D}, the Fourier-Bessel formalism~\cite{2013PhRvD..88b3502Y}, and direct integration of the wide-angle correlation function~\cite{2014PhRvD..89h3535B,%
2018JCAP...10..032T}. Notably, the interplay between wide-angle effects and other observational systematics, such as magnification bias and evolution effects, further complicates the precise extraction of cosmological information~\cite{2012JCAP...10..025B,%
2012arXiv1206.5809Y}. Modern analyses increasingly incorporate these wide-angle corrections as part of a comprehensive theoretical framework that accounts for all leading observational effects in galaxy clustering. This systematic approach is essential for leveraging the full statistical power of upcoming surveys (like Euclid, DESI, and the Vera C. Rubin Observatory), where sub-percent precision is required to distinguish between competing cosmological models~\cite{2016arXiv161100036D,2018LRR....21....2A}.\\

\item \textbf{Alcock-Paczynski effect:} In addition to the dynamical redshift-space distortions and wide-angle effects, studies of galaxy clustering must contend with a purely geometric distortion known as the Alcock-Paczynski (AP) effect~\cite{1979Natur.281..358A,1996MNRAS.282..877B,%
    2016ApJ...832..103L,2015MNRAS.449..835R}. Unlike the Kaiser, the FoG and Rocket effects, which arise from peculiar velocities, the AP~effect stems from the cosmology-dependent conversion between observed coordinates (angles and redshifts) and physical, or comoving, distances. When analyzing clustering data, one typically assumes a fiducial cosmology to perform this conversion. However, if the latter differs from the true cosmology, it introduces artificial anisotropies in the clustering pattern. Mathematically, the Alcock-Paczynski effect can be parameterized through scaling factors along and perpendicular to the line of sight. These are respectively given by
\begin{equation}
\alpha_{\parallel}= \frac{H^{\text{fid}}}{H^{\text{true}}} \hspace{15mm} {\rm and} \hspace{15mm} \alpha_{\perp} = \frac{D_A^{\text{true}}}{D_A^{\text{fid}}}\,,
\end{equation}
with $H=H(z)$ representing the Hubble parameter and $D_A=D_A(z)$ the angular diameter distance. These scaling factors distort the apparent shape of cosmic structures, transforming a spherical object into an ellipsoid with axis ratio $\alpha_{\parallel}/\alpha_{\perp}$~\cite{2008PhRvD..77l3540P,%
2013MNRAS.431.2834X}.

The AP~effect is particularly challenging because it induces anisotropies that can mimic or interfere with those produced by redshift-space distortions, creating significant degeneracies in parameter estimation~\cite{1996MNRAS.282..877B,%
1996ApJ...470L...1M,2008ApJ...676..889O}. However, this entanglement also presents an opportunity, because the joint analysis of the AP~effect and Redshift-Space Distortions (RSD) can break degeneracies and yield tighter cosmological constraints than either method alone~\cite{2011MNRAS.418.1707B,%
2012MNRAS.426.2719R,2017MNRAS.470.2617A}. Modern analyses typically employ the multipole expansion of the galaxy two-point correlation function, or power spectrum, where different multipoles respond differently to the AP and RSD effects, enabling their separation~\cite{2008PhRvD..77l3540P,2014MNRAS.443.1065B,%
2017MNRAS.464.1640S}. The Alcock-Paczynski test has become increasingly powerful with the growth of large-scale structure surveys, providing constraints on dark energy and modified gravity that complement and reinforce those from other cosmological probes~\cite{2010ApJ...715L.185P,2017ApJ...835..161M,%
2019A&A...621A..69R}. This synergy highlights the importance of developing comprehensive theoretical frameworks that simultaneously account for all anisotropic signals in galaxy clustering, including both the velocity-induced distortions and the geometric AP~effect, particularly for next-generation surveys that will probe ever-larger cosmic volumes with unprecedented precision~\cite{2016ApJ...817...26B,2023AJ....165..253H}.
\end{itemize}

The accurate modeling of redshift-space distortions — including the Kaiser effect, FoG, Rocket effect, Wide-Angle effects, and Alcock-Paczynski distortions - has evolved substantially over the past several decades. This progression reflects both our deepening theoretical understanding and the increasing precision requirements of modern galaxy surveys.

Early approaches relied on the linear theory formalism introduced by Kaiser~\cite{1987MNRAS.227....1K}, which provided an elegant relationship between the real-space power spectrum $P_r(k)$ and its redshift-space counterpart $P_s(k,\mu)$, namely
\begin{equation}
P_s(k,\mu)= (1+\beta\mu^2)^2 P_r(k)\,.
\end{equation}
While remarkably simple, this expression neglects nonlinear effects and breaks down on scales where $k\sigma_v\gtrsim1$. Subsequent refinements incorporated phenomenological treatments of the FoG effect, typically using exponential or Lorentzian damping functions, such as~\cite{1994MNRAS.267.1020P,1994MNRAS.267..785C,1994ApJ...431..569P}:
\begin{equation}
P_s(k,\mu) = (1 + \beta\mu^2)^2 P_r(k) \times \exp[-(k\mu\sigma_v)^2]\,,
\end{equation}
or alternatively
\begin{equation}
P_s(k,\mu) = \frac{(1 + \beta\mu^2)^2 P_r(k)}{1 + (k\mu\sigma_v)^2/2}\,.
\end{equation}

These models, however, proved inadequate in the quasi-linear regime ($k\approx0.1$-$0.5\,h\text{Mpc}^{-1}$), where both the Kaiser effect and nonlinear clustering become important~\cite{2012JCAP...11..014O,2010PhRvD..82f3522T,%
2012ApJ...748...78K}. As a result, more sophisticated approaches have since been developed to address these limitations. These include:

\begin{itemize}
    \item \textbf{Perturbation theory expansions:} These extend the Kaiser formalism into the quasi-linear regime by incorporating higher-order terms in the density and in the velocity fields (see~\cite{2004PhRvD..70h3007S,2010PhRvD..82f3522T,%
        2012JCAP...11..009V,2013MNRAS.429.1674C}). The Taruya-Nishimichi-Saito (TNS) model~\cite{2010PhRvD..82f3522T}, for instance, supplements the Kaiser formula with correction terms $A(k,\mu)$ and $B(k,\mu)$ that account for higher-order coupling between density and velocity fields:
    \begin{equation}
    P_s(k,\mu) = D_{\text{FoG}}(k\mu\sigma_v)[P_{\delta\delta}(k) + 2\mu^2P_{\delta\theta}(k) + \mu^4P_{\theta\theta}(k) + A(k,\mu) + B(k,\mu)]\,,
    \end{equation}
    where $P_{\delta\delta}$, $P_{\delta\theta}$, and $P_{\theta\theta}$ are the auto and cross power spectra of the density ($\delta$) and the velocity divergence ($\theta$). This approach has been further refined by means of Eulerian~\cite{2008PhRvD..77f3530M,2012JCAP...11..029G} and Lagrangian~\cite{2008PhRvD..78h3519M,2013MNRAS.429.1674C,%
    2019JCAP...03..007V} perturbation theory.\\

    \item \textbf{Streaming models:} These provide a more physical description of redshift-space clustering across a wider range of scales, by modeling the pairwise velocity Probability Distribution Function (PDF)~\cite{2011MNRAS.417.1913R,2015MNRAS.447..234W,%
        2015PhRvD..92f3004U}. The general form relates the redshift-space correlation function ($\xi_s$) to its real-space counterpart ($\xi_r$) via the expression
    \begin{equation}
    1 + \xi_s(s_\parallel, s_\perp) = \int_{-\infty}^{+\infty} [1 + \xi_r(r)] \, \mathcal{P}(v_\parallel | \mathbf{r}) \, \delta^D(s_\parallel - r_\parallel - v_\parallel/aH) \, dv_\parallel\,,
    \end{equation}
    with $\mathcal{P}(v_\parallel | \mathbf{r})$ being the PDF of the line-of-sight relative velocity ($v_\parallel$) at separation $\mathbf{r}$. Note that recent implementations incorporate scale-dependent velocity moments and non-Gaussian features in the velocity distribution~\cite{2018MNRAS.479.2256K,2020MNRAS.498.1175C,%
    2016JCAP...12..007V}.\\

    \item \textbf{Distribution function approach:} This framework, developed by Seljak and McDonald~\cite{2011JCAP...11..039S}, describes the galaxy phase-space distribution directly through a Boltzmann-like equation. It expands the redshift-space density in terms of velocity moments, offering a rigorous treatment even in the nonlinear regime, in line with
    \begin{equation}
    \delta_s(\mathbf{k}) = \delta(\mathbf{k}) + \sum_{n=1}^{\infty} \frac{(i k_\parallel)^n}{n!} \int \frac{d^3\mathbf{q}}{(2\pi)^3} T_\delta^{v_\parallel^n}(\mathbf{k-q},\mathbf{q})\,,
    \end{equation}
    where $T_\delta^{v_\parallel^n}$ are velocity-moment kernels. This method has been extended to biased tracers~\cite{2012JCAP...11..014O,2012JCAP...11..009V} and has been combined with Lagrangian perturbation theory~\cite{2019JCAP...03..007V,2020JCAP...07..062C}.\\

    \item \textbf{Effective field theory (EFT):} The approach incorporates unknown small-scale physics through effective counter-terms, extending the validity of perturbative treatments~\cite{2020JCAP...07..062C,2015JCAP...02..013S,2016arXiv161009321P}. The EFT of large-scale structure systematically accounts for the effects of unresolved nonlinear modes on large-scale observables through the controlled expansion
    \begin{equation}
    P_s(k,\mu) = P_s^{\text{SPT}}(k,\mu) + c_s^2 k^2 P_{\text{lin}}(k) + \text{other counter terms}\,.
    \end{equation}
    Here, $P_s^{\text{SPT}}$ is the standard perturbation theory prediction and $c_s^2$ is an effective sound-speed parameter determined from data or simulations. Recent developments have incorporated redshift-space distortions~\cite{2015JCAP...02..013S,2015JCAP...05..019L,%
    2016arXiv161009321P} and extended the formalism to biased tracers~\cite{2018PhR...733....1D,2020JCAP...05..042I}.\\

    \item \textbf{Simulation-calibrated models:} These empirical or semi-analytical models are calibrated against N-body simulations to capture the full nonlinear behavior of RSDs~\cite{2014MNRAS.444..476R,2007MNRAS.374..477T,2019ApJ...874...95Z}. The halo-model approach~\cite{2003MNRAS.341.1311S,2007MNRAS.374..477T} and its extensions~\cite{2019ApJ...884...29N} have been particularly successful at describing galaxy clustering across all scales, with the emulator-based methods~\cite{2019ApJ...874...95Z,2020PhRvD.102f3504K} offering computational efficiency for likelihood analyses.
\end{itemize}

Complementing the theoretical advances, sophisticated statistical techniques have been developed to extract cosmological information from redshift-space distortions. The multipole decomposition methods~\cite{1992ApJ...385L...5H,1994MNRAS.267..785C} expand the anisotropic correlation function or power spectrum in terms of Legendre polynomials:
\begin{equation}
\xi_\ell(s) = \frac{2\ell+1}{2}\int_{-1}^{1} \xi(s,\mu) \mathcal{P}_\ell(\mu) d\mu\,.
\end{equation}
where the monopole ($\ell=0$), quadrupole ($\ell=2$), and hexadecapole ($\ell=4$) contain most of the cosmological information. Alternative approaches include wedge statistics~\cite{2012MNRAS.419.3223K,2017MNRAS.464.1640S} and clustering wedges~\cite{2017MNRAS.467.2085G}, which average the correlation function over wide angular bins, namely
\begin{equation}
\xi_{\mu_1,\mu_2}(s) = \frac{1}{\mu_2-\mu_1}\int_{\mu_1}^{\mu_2} \xi(s,\mu) d\mu\,.
\end{equation}
These complementary techniques help isolate different physical effects and provide valuable cross-checks. Recent analyses have combined multiple statistical measures with sophisticated theoretical models to achieve precision measurements of the growth-rate parameter $f$ and $\sigma_8$ with uncertainties of just a few percent~\cite{2017MNRAS.470.2617A,2021PhRvD.103h3533A,%
2018A&A...619A..17M,2019PhRvR...1c3209B,2016MNRAS.461.3781C}. Such precise constraints provide critical tests of general relativity and alternative gravitational theories on cosmological scales, complementing the geometric probes discussed in previous sections.

Nevertheless, several challenges remain in RSD analyses. These include the proper treatment of nonlinear bias~\cite{2018PhR...733....1D}, the impact of baryonic physics~\cite{2014MNRAS.440.2997V,2016MNRAS.461L..11H}, assembly bias effects~\cite{2019MNRAS.485.1196Z,2017MNRAS.465.2936M} and the influence of massive neutrinos~\cite{2016PhRvD..93f3515U,2019MNRAS.483..734R}. Future surveys, like DESI, Euclid, and the Roman Space Telescope, will require even more sophisticated modeling to fully exploit their statistical power~\cite{2016arXiv161100036D,2018LRR....21....2A,%
2021MNRAS.507.1514E}.

When properly modeled, Redshift-Space Distortions provide valuable cosmological information, allowing constraints on the growth-rate of structure independent of galaxy bias. This makes RSD a powerful probe for testing theories of gravity on cosmological scales~\cite{2008Natur.451..541G,2009JCAP...10..004S,%
2014MNRAS.443.1065B} and for breaking degeneracies in dark-energy constraints~\cite{2008JCAP...05..021W,2008APh....29..336L,%
2009MNRAS.397.1348W}. The combination of RSD measurements with other probes, such as weak lensing, BAO, and peculiar-velocity surveys, leads to particularly powerful constraints on cosmological models~\cite{2007PhRvL..99n1302Z,2013MNRAS.429.2249S,%
2008PhRvD..78f3503J,2009JCAP...10..017K}.

\subsubsection{Contamination of CMB studies}\label{sssCCMBSs}
%%%%%%%%%%%%%%%%%%%%%%%%%%%%%%%%%%%%%%%%%%%%%%%%%%%%%%%%%%%%%
While the previous sections focused on the effects of peculiar velocities in galaxy surveys, these motions also significantly impact the CMB observations. The high precision of modern Cosmic Microwave experiments requires careful accounting for velocity-induced effects, which manifest as both primary and secondary anisotropies~\cite{2002PhRvD..65j3001C}. These effects act both as a contaminant that needs to be removed and as a valuable cosmological signal that is to be extracted.

\begin{enumerate}
\item \textbf{Dipole anisotropy:} The dominant $\sim$3.4 mK dipole pattern in the CMB temperature ($\Delta T/T\simeq10^{-3}$) is predominantly attributed to our motion relative to the rest-frame of the microwave photons at $v_{\text{obs}}\approx370$~km/sec toward the constellation Leo~\cite{2020A&A...641A...6P}. The observed temperature dipole follows the relativistic Doppler formula
\begin{equation}
\frac{\Delta T}{T}\left(\hat{\mathbf{n}}\right) = \frac{1}{\gamma(1-\beta\cos\theta)} - 1 \approx \beta\cos\theta + \mathcal{O}(\beta^2)\,,
\end{equation}
with $\beta=v_{\text{obs}}/c$. Also, $\gamma=(1-\beta^2)^{-1/2}$ and $\theta$ is the angle between the direction of the observation ($\hat{\mathbf{n}}$) and our velocity vector. The precise measurement of this dipole serves as a fundamental calibration reference for CMB experiments~\cite{2018arXiv180401318S}. However, recent studies have questioned whether the CMB dipole is entirely kinematic, suggesting possible contributions from large-scale structure, or even from intrinsic primordial dipoles~\cite{2023CQGra..40i4001A}. We should note that alternative approaches to the dipole effect also exist in the literature (e.g.~see~\cite{2011PhRvD..84d3516B,%
2021EPJST.230.2067M}). However, the covariant tilted-frame analysis offers a unified geometric explanation for the coincidence of dipoles across different cosmological tracers.\\

\item \textbf{Aberration and modulation:} Beyond the dipole, our motion induces two additional effects on the observed CMB: aberration, which causes an apparent deflection in the direction of the arriving photons, and Doppler modulation, producing a direction-dependent rescaling of temperature fluctuations~\cite{2014PhRvD..89b3003J,2014A&A...571A..27P}. For small velocities, these effects couple multipoles $\ell$ and $\ell \pm 1$ in the spherical harmonic decomposition of the temperature field, with coupling coefficients proportional to $\beta$~\cite{2002PhRvD..65j3001C,2011PhRvL.106s1301K}:
\begin{equation}
a_{\ell m}^{\text{obs}} = a_{\ell m} + \beta \left[ \sqrt{\frac{(\ell+1)^2-m^2}{(2\ell+1)(2\ell+3)}} a_{\ell+1,m} - \sqrt{\frac{\ell^2-m^2}{(2\ell-1)(2\ell+1)}} a_{\ell-1,m} \right] + \mathcal{O}(\beta^2)\,,
\end{equation}
These effects have been detected in Planck data at high statistical significance ($\sim 5\sigma$), confirming the expected correlation with the CMB dipole~\cite{2014A&A...571A..27P,%
2020A&A...641A...7P}. However, they can potentially bias measurements of primordial non-Gaussianity and large-scale anomalies if not properly accounted for~\cite{2013PhRvD..87l3005D,2015JCAP...06..047N}.\\

\item \textbf{Kinetic Sunyaev-Zel'dovich (kSZ) effect:} When the CMB photons scatter off free electrons, moving with the peculiar velocity of their host galaxies or clusters, they experience Doppler shifts that induce secondary temperature anisotropies~\cite{1980MNRAS.190..413S,1987ApJ...322..597V}. This kinetic Sunyaev-Zel'dovich effect produces temperature fluctuations of the form
\begin{equation}
\frac{\Delta T}{T}(\hat{\mathbf{n}}) = -\sigma_T \int n_e(\mathbf{r}) \frac{\mathbf{v}_e(\mathbf{r}) \cdot \hat{\mathbf{n}}}{c} e^{-\tau(\mathbf{r})} d\chi\,,
\end{equation}
where $\sigma_T$ is the Thomson cross-section, $n_e$ is the electron number density, $\mathbf{v}_e$ is the electron peculiar velocity, $\tau$ is the optical depth and $\chi$ is the comoving distance. Unlike the thermal Sunyaev-Zel'dovich effect, the kSZ signal is independent of the electron temperature and preserves the blackbody spectrum of the CMB~\cite{2019SSRv..215...17M}. The kSZ effect has been detected both from individual clusters~\cite{2012PhRvL.109d1101H,2019ApJ...880...45S} and statistically through cross-correlations with large-scale structure~\cite{2016PhRvL.117e1301H,2016PhRvD..93h2002S,%
2017JCAP...03..008D}, providing unique information about cosmic velocity fields and missing baryons \cite{2015PhRvL.115s1301H,2016PhRvD..94l3526F}.\\

\item \textbf{Moving lens effects:} Gravitational lensing of the CMB by moving masses produces additional anisotropies through the coupling between lensing and aberration~\cite{1983Natur.302..315B,2019PhRvL.123f1301H}. This ``moving lens effect'' induces a dipolar modulation pattern around massive objects with amplitude proportional to their transverse velocity, according to
\begin{equation}
\frac{\Delta T}{T}(\hat{\mathbf{n}}) \approx -\frac{4GM}{c^2}\frac{\mathbf{v}_{\perp}}{c} \cdot \frac{\hat{\mathbf{n}} - \hat{\mathbf{n}}_0}{|\hat{\mathbf{n}} - \hat{\mathbf{n}}_0|^2}\,.
\end{equation}
Here, $M$ is the mass of the lens, $\mathbf{v}_{\perp}$ is its transverse velocity, and $\hat{\mathbf{n}}_0$ is the direction to the lens center. While challenging to detect for individual objects, the statistical signature of the moving lens effect could be accessible with next-generation CMB experiments and galaxy surveys~\cite{2016PhRvD..94b3513Y,2021PhRvD.104h3529H}. Building on this concept, Cai et al~(2025)~\cite{2025MNRAS.541.2093C} recently presented a unified framework for analyzing such velocity-related dipoles and reported the first detection of a similar effect on much larger scales. The authors distinguish their finding from the small-scale, non-linear moving-lens effect, showing that in the linear regime both the ISW effect and gravitational lensing are expected to exhibit a large-scale dipole aligned with the transverse peculiar velocity. Using a rotational stacking technique on CMB temperature and lensing maps from Planck, centered on galaxies from the SDSS-III BOSS survey, they successfully detected the predicted dipoles in galaxy density, CMB lensing convergence, and the ISW signal~\cite{2025MNRAS.541.2093C}. Crucially, this work demonstrated that, while the standard cross-correlation (monopole) signal depends on galaxy bias, the dipole signal is independent of it, thus offering a new and more robust cosmological probe~\cite{2025MNRAS.541.2093C}. This provides a novel observational method for measuring the effects of transverse peculiar velocities, which have historically been difficult to probe directly.
\end{enumerate}

These velocity-induced effects must be carefully accounted for when analysing the CMB data, particularly in studies of statistical isotropy, large-scale anomalies, and tests of the Cosmological Principle~\cite{2020A&A...641A...7P,2016CQGra..33r4001S}. They offer new opportunities to probe peculiar-velocity fields at different epochs and scales, complementing the Redshift-Space Distortion (RSD) measurements discussed earlier.  For instance, the kSZ effect provides a direct measure of peculiar velocities, unaffected by galaxy bias, while aberration and modulation effects can be used to check the consistency of the observed CMB dipole with its presumed kinematic origin~\cite{2014A&A...571A..27P,2023CQGra..40i4001A}. As CMB experiments and large-scale structure surveys continue to improve, the synergy between these probes will become increasingly important for understanding the dynamics of our Universe and testing fundamental cosmological assumptions.

\subsubsection{Impact on cosmological probes}\label{sssICPs}
%%%%%%%%%%%%%%%%%%%%%%%%%%%%%%%%%%%%%%%%%%%%%%%%%%%%%%%%%%%%
Beyond their direct effects on distance measurements and CMB observations discussed in previous sections, peculiar velocities significantly influence numerous cosmological probes used to constrain the expansion history and growth of structure. Understanding and modeling these velocity-induced systematics is crucial for achieving the precision goals of current and future cosmological surveys.

\begin{itemize}
\item \textbf{Baryon Acoustic Oscillations (BAO):} The BAO feature serves as a standard ruler for measuring cosmic distances and constraining dark energy. However, peculiar velocities can shift the apparent position of the BAO peak in correlation functions by up to 0.5\%~\cite{2012PhRvD..85j3523S,%
    2007ApJ...664..675E}. The shift stems from the nonlinear coupling between the density and the velocity fields, which then generates a mode-coupling that displaces the BAO peak from its linear-theory position~\cite{2008PhRvD..77b3533C,2008PhRvD..77d3525S}. Mathematically, the shift in the BAO scale can be approximated by
\begin{equation}
\Delta \alpha_{\text{BAO}} \approx \frac{\Sigma^2}{2} \frac{d\ln(P_{\text{lin}}(k)/P_{\text{nw}}(k))}{d\ln k}|_{k=k_{\text{BAO}}}\,.
\end{equation}
In the above, $\Sigma$ is the damping scale due to the nonlinear evolution, $P_{\text{lin}}$ is the linear power spectrum, $P_{\text{nw}}$ is the no-wiggle power spectrum, and $k_{\text{BAO}}$ is the BAO scale~\cite{2016MNRAS.460.2453S,%
2013PhRvD..88h3507B}. Reconstruction techniques aim to reverse these nonlinear effects by estimating and removing the displacement field, thereby reducing the shift and sharpening the BAO feature~\cite{2009PhRvD..79f3523P,2015PhRvD..92l3522S}.\\

\item \textbf{Weak lensing:} Cosmic shear measurements are increasingly important for constraining both the growth of structure and geometric distances. Source peculiar velocities introduce additional correlations in weak lensing observables through Doppler lensing effects~\cite{2006PhRvD..73b3523B,2014MNRAS.443.1900B}. The resulting corrections to the lensing convergence power spectrum can be expressed as
\begin{equation}
C_{\ell}^{\kappa\kappa} = C_{\ell}^{\kappa_G\kappa_G} + 2C_{\ell}^{\kappa_G\kappa_v} + C_{\ell}^{\kappa_v\kappa_v}\,,
\end{equation}
where $C_{\ell}^{\kappa_G\kappa_G}$ is the standard gravitational lensing contribution, $C_{\ell}^{\kappa_v\kappa_v}$ is the Doppler lensing auto-correlation, and $C_{\ell}^{\kappa_G\kappa_v}$ is the cross-correlation term~\cite{2006PhRvD..73b3523B}. These velocity-induced terms become particularly important for wide-angle surveys and at low redshifts ($z \lesssim 0.5$), potentially biasing dark-energy constraints if not properly modeled~\cite{2007PhRvD..76j3502H,2017MNRAS.464.4747C}. Conversely, the Doppler lensing signal itself contains valuable cosmological information that can be extracted through cross-correlations with galaxy positions and peculiar velocities \cite{2014MNRAS.443.1900B,2017MNRAS.472.3936B}.\\

\item \textbf{Galaxy clustering:} As extensively discussed in earlier sections, the observed galaxy power spectrum is modulated by peculiar velocity effects through Redshift-Space Distortions (RSD). Beyond these well-known effects, velocities also influence galaxy bias on large scales through relativistic corrections~\cite{2012arXiv1206.5809Y,2019JCAP...03..007V}. These corrections modify the observed galaxy overdensity by terms proportional to the velocity potential and its derivatives, according to
\begin{equation}
\delta_g^{\text{obs}} = b\delta + f\nabla_{\parallel}^2\Phi_v + \mathcal{H}^{-1}\partial_{\parallel}^2\Phi_v + \text{other terms}\,,
\end{equation}
with $\Phi_v$ being the velocity potential, $b$ the galaxy bias and $f$ the growth rate~\cite{2012arXiv1206.5809Y,%
2011PhRvD..84d3516C}. These velocity-dependent terms become increasingly important on the largest observable scales and at high redshifts, affecting searches for primordial non-Gaussianity and tests of the cosmological principle~\cite{2013PhRvL.111q1302C,2015MNRAS.447..400A}. Also, velocity-induced selection effects such as Malmquist bias can significantly impact clustering measurements for flux-limited samples~\cite{2011ApJ...741...67D,2015JCAP...12..033H}.\\

\item \textbf{Void studies:} Cosmic voids are emerging as powerful probes of both the expansion history and modified gravity theories. Peculiar velocities affect void identification and characterization through Redshift-Space Distortions (RSD), which can elongate voids along the line of sight, thus altering their apparent shapes and profiles~\cite{2019PhRvD.100b3504N,2014PhRvL.112y1302H}. These velocity effects must be carefully modeled when using voids for Alcock-Paczynski tests~\cite{2014MNRAS.442..462S,%
    2017ApJ...835..161M}, or when studying void dynamics to test modified-gravity cosmologies~\cite{2017PhRvD..95b4018V,%
    2015JCAP...07..049F}. Interestingly, the velocity field around voids exhibits a coherent outflow pattern that can be used to constrain the growth rate of structure through void-galaxy cross-correlations~\cite{2016PhRvL.117i1302H,%
    2019PhRvD.100l3513A}.\\

\item \textbf{Supernovae cosmology:} Type Ia supernovae (SNe Ia) serve as standardizable candles for measuring the cosmic expansion history. Peculiar velocities introduce additional scatter in the Hubble diagram through Doppler shifts, particularly at low redshifts where $v_{\text{pec}}/cz$ is non-negligible~\cite{2006PhRvD..73l3526H,2011ApJ...741...67D}. This effect can be mitigated by modeling the large-scale flow field using galaxy surveys~\cite{2017MNRAS.464.2517H,%
    2020MNRAS.498.2703B} or through statistical methods that account for the correlated velocity component~\cite{2015JCAP...12..033H}. Moreover, correlations between SNe~Ia and the large-scale structure can potentially bias cosmological parameters if selection effects are correlated with the density field~\cite{2015JCAP...07..025W,%
    2020A&A...644A.176R}.\\

\item \textbf{21cm intensity mapping:} Upcoming radio surveys will map the large-scale distribution of neutral hydrogen through 21cm intensity mapping. Peculiar velocities affect these observations not only through RSD, but also through modulations of the observed brightness temperature~\cite{2012MNRAS.422..926M,2013PhRvD..87f4026H}. In the high-redshift universe, particularly during the epoch of reionization, these velocity effects can create distinctive signatures in the 21cm power spectrum that could potentially offer new cosmological probes~\cite{2006ApJ...653..815M,%
    2012MNRAS.422..926M}. The interplay between peculiar velocities, ionization fronts, and density fluctuations creates rich structures in the observed signal that encode information about early universe physics~\cite{2005ApJ...624L..65B,2013MNRAS.432.2909F}.
\end{itemize}

The ubiquitous nature of peculiar-velocity effects across cosmological probes underscores the need for integrated modeling that consistently accounts for velocity-induced systematics. With the advent of Stage-IV surveys like DESI, Euclid, the Vera Rubin Observatory, and the Square Kilometer Array, the precision frontier of cosmology increasingly demands sophisticated treatments of peculiar velocities, treating them both as a contaminant that needs to be removed, as well as a valuable signal containing complementary cosmological information~\cite{2018LRR....21....2A}. Cross-correlations between different probes offer particularly promising avenues for breaking degeneracies and isolating velocity effects~\cite{2017MNRAS.472.3936B,2019PhRvD..99j3530F}. As we have seen throughout this review, understanding the physics of large-scale peculiar velocities — from their generation through gravitational instability to their diverse observational signatures — remains essential for advancing precision cosmology.

\subsubsection{Large-scale structure mapping
%%%%%%%%%%%%%%%%%%%%%%%%%%%%%%%%%%%%%%%%%%%%
biases}\label{sssLSSMBs}
%%%%%%%%%%%%%%%%%%%%%%%%
While the previous sections examined how peculiar velocities affect various cosmological probes, here we focus specifically on their impact on our attempts to reconstruct the true 3-D distribution of matter in the universe. The mapping between redshift space and real space is non-trivial and introduces systematic biases to our understanding of the cosmic structure.

Large-scale coherent flows can create apparent structures, or mask real ones, in redshift surveys~\cite{1999fsu..conf..250D}. This effect is particularly pronounced in the local universe, where peculiar velocities constitute a significant fraction of the observed redshift. For instance, the ``Great Attractor'' region initially appeared as a massive overdensity in redshift surveys, but subsequent peculiar velocity measurements revealed a more complex structure with coherent infall patterns modifying the apparent distribution~\cite{1988ApJ...326...19L,1987ApJ...313L..37D}. These redshift-space distortions alter our perception of cosmic structures across scales, from the apparent elongation of clusters along the line of sight (the ``Fingers of God'' effect discussed earlier), to the squashing of superclusters in the transverse direction due to coherent infall (the Kaiser effect). Mathematically, the mapping between the real-space position $\mathbf{r}$ and the redshift-space position $\mathbf{s}$ can be expressed as
\begin{equation}
\mathbf{s} = \mathbf{r} + \frac{\mathbf{v}(\mathbf{r}) \cdot \hat{\mathbf{r}}}{aH}\hat{\mathbf{r}}\,,
\end{equation}
where $\mathbf{v}(\mathbf{r})$ is the peculiar velocity field, $a$ is the scale factor and $H=H(a)$ is the Hubble parameter~\cite{1987MNRAS.227....1K,1998ASSL..231..185H}. This seemingly simple transformation has profound implications for cosmic cartography, particularly when attempting to identify the sources of observed bulk flows, or when studying the connectivity and morphology of the cosmic web~\cite{2018MNRAS.473.1195L,2014MNRAS.441.2923C}.

The inverse problem — reconstructing the density field from peculiar - velocity measurements requires sophisticated statistical techniques to handle observational errors and the modeling of the velocity-density relationship~\cite{2017NatAs...1E..36H,%
1995ApJ...449..446Z}. In the linear regime, the peculiar velocity field is related to the density field through
\begin{equation}
\mathbf{v}(\mathbf{r}) = \frac{Hf}{4\pi} \int d^3\mathbf{r}' \delta(\mathbf{r}') \frac{(\mathbf{r}' - \mathbf{r})}{|\mathbf{r}' - \mathbf{r}|^3}\,,
\end{equation}
with $f$ representing the growth rate and $\delta$ the density contrast of the matter~\cite{1980lssu.book.....P,1995PhR...261..271S}. This integral relation can be inverted using Wiener filtering, or Bayesian inference methods, to estimate the underlying density field~\cite{1995ApJ...449..446Z,2010ApJ...709..483L}. However, these reconstructions face several challenges, such as sparse and inhomogeneous sampling of the velocity field, large measurement uncertainties that increase with distance and nonlinear effects that violate the simple linear velocity-density relationship~\cite{1995PhR...261..271S,2006MNRAS.368.1515E}. For instance, the POTENT method~\cite{1999ApJ...522....1D} uses smoothing techniques to reconstruct the potential flow field from radial peculiar velocities, but requires careful error propagation and bias correction, particularly at the survey boundaries~\cite{2011JCAP...10..016E,2012ApJ...744...43C}.

More recent approaches leverage advanced statistical techniques and multiple data sets to improve reconstruction fidelity. The BORG algorithm (Bayesian Origin Reconstruction from Galaxies) implements a full Bayesian analysis framework that jointly infers the initial conditions and their evolution while accounting for observational biases~\cite{2013MNRAS.432..894J,2016MNRAS.455.3169L}. Similarly, the Constrained Local UniversE Simulations (CLUES) project uses constrained realizations of the local density and velocity fields to produce N-body simulations that reproduce observed structures~\cite{2010arXiv1005.2687G,2016MNRAS.455.2078S}. These methods help disentangle true cosmic structures from artifacts induced by peculiar velocities.

Without proper accounting for the complex relationship between peculiar velocities and the underlying matter distribution, we risk misidentifying the true drivers of cosmic flows and misinterpreting the nature of cosmic voids and filaments~\cite{2011JCAP...10..016E,2013AJ....146...86T}. For example, analyses of the local bulk flow have yielded apparently conflicting results regarding its amplitude and scale, with some studies finding flows consistent with $\Lambda$CDM expectations (e.g.~see~\cite{2011ApJ...736...93N,2015PhRvL.115a1301F} and also \S~\ref{ssBFWLCDMLs} here) while others claim evidence for anomalously large flows (e.g.~see~\cite{2009MNRAS.392..743W,%
2010MNRAS.407.2328F} and also \S~\ref{ssBFELCDMLs} here). These discrepancies may partly stem from the way the peculiar-velocity effects are treated in the analysis. Similarly, the apparent alignment of the velocity and the  density fields with the cosmic microwave background dipole direction has been interpreted as both supporting and challenging the standard cosmological model, depending on the methodological details (e.g.~see~\cite{2011MNRAS.414..264C,2013A&A...555A.117R,%
2016JCAP...06..035B}).

The challenge of accurately mapping large-scale structure in the presence of peculiar velocities will become increasingly important as future surveys probe larger volumes with greater precision. Next-generation peculiar-velocity surveys using Type Ia supernovae, the Fundamental Plane, and the Tully-Fisher relation promise to dramatically improve our understanding of the local velocity field~\cite{2017MNRAS.464.2517H,2020MNRAS.497.1275S}. At the same time, advances in the 21cm intensity mapping and in the kinetic Sunyaev-Zel'dovich measurements will provide complementary probes of the velocity field at different redshifts and scales~\cite{2015aska.confE..24B,2018arXiv181013423S}. Combining these diverse datasets through advanced statistical frameworks offers the best hope for overcoming the biases introduced by peculiar velocities and revealing the true structure of our cosmic neighborhood.

\subsection{Historical misinterpretations due to relative-motion
%%%%%%%%%%%%%%%%%%%%%%%%%%%%%%%%%%%%%%%%%%%%%%%%%%%%%%%%%%%%%%%%
effects}\label{ssHMRMEs}
%%%%%%%%%%%%%%%%%%%%%%%%
Throughout the history of astronomy, relative-motion effects have led to significant misinterpretations of observed phenomena. The previous sections have outlined how peculiar velocities influence modern cosmological observations across multiple probes and methodologies. However, the challenge of properly accounting for relative motion has a much longer history, predating modern cosmology by centuries, if not millennia. These historical examples provide valuable cautionary lessons, illustrating how kinematic effects can mask the true nature of astronomical systems and how advances in our understanding often involve recognizing previously overlooked motion-induced biases.

\subsubsection{The geocentric model and planetary retrograde
%%%%%%%%%%%%%%%%%%%%%%%%%%%%%%%%%%%%%%%%%%%%%%%%%%%%%%%%%%%%
motion}\label{sssGMPRM}
%%%%%%%%%%%%%%%%%%%%%%%
Perhaps the most profound example of relative-motion misinterpretation is the persistence of the geocentric model for nearly two millennia. The apparent retrograde motion of planets — their periodic ``backward'' movement across the night sky — was explained through an elaborate system of epicycles in the Ptolemaic model~\cite{1974hama.book.....P,1975hama.book.....N}. The mathematical description of this apparent motion required increasingly complex geometric constructions, with planets moving on small circles (epicycles) whose centers moved along larger circles (deferents) around Earth. The angular position of a planet $\theta_p$ as seen from the Earth could be approximated by
\begin{equation}
\theta_p \approx \omega_d t + e\sin(\omega_d t) + \frac{r_e}{r_d}\sin[(\omega_e - \omega_d)t]\,,
\end{equation}
where $\omega_d$ and $\omega_e$ are the angular frequencies of the deferent and epicycle respectively, $r_e/r_d$ is the ratio of epicycle to deferent radii, and $e$ represents the eccentricity~\cite{1997hama.book.....G}. This system could be extended with additional epicycles to improve accuracy, leading to the famous criticism that the model had become unwieldy with ``circles upon circles''~\cite{1957hama.book.....K}.

These epicycles are not real of course, but merely the result of the Earth's motion relative to other planets, as they orbit the Sun at different rates. In the heliocentric model, the apparent position of a planet results from the combination of its true orbital motion and Earth's orbital motion, which can be expressed as
\begin{equation}
\tan\theta_p = \frac{r_p\sin(\theta_p - \theta_e) - r_e\sin\theta_e}{r_p\cos(\theta_p - \theta_e) - r_e\cos\theta_e}\,.
\end{equation}
Here, $r_p$ and $r_e$ are the orbital radii of the planet and of the Earth, while $\theta_p$ and $\theta_e$ are their true anomalies~\cite{1998hama.book.....E}. The Copernican revolution fundamentally changed our understanding by recognizing that these complex apparent motions could be elegantly explained by relative motion in a heliocentric system~\cite{1957hama.book.....K,%
1997hama.book.....G}. The subsequent refinements by Kepler, replacing circular orbits with ellipses, and Newton's theory of gravitation fully clarified the picture.

This historical example demonstrates how the assumption of a privileged reference frame led to increasingly complex models, trying to explain what - at the end of the day - were simple kinematic effects. It serves as a reminder that observational phenomena may have simpler and more elegant explanations when viewed from an appropriate reference frame. Deceptions caused by relative-motion effects in the analysis and the interpretation of the observations, have a long history in astronomy and they may still occur in modern cosmology (e.g.~see related discussion in \S~\ref{ssRMEDP} earlier).

\subsubsection{The Solar System's motion and the CMB
%%%%%%%%%%%%%%%%%%%%%%%%%%%%%%%%%%%%%%%%%%%%%%%%%%%%
dipole}\label{SSMCMBD}
%%%%%%%%%%%%%%%%%%%%%%
The discovery of the cosmic microwave background radiation (CMB) in 1964 by Penzias and Wilson~\cite{1965ApJ...142..419P} revealed a nearly isotropic radiation field permeating the universe. However, subsequent measurements detected a significant dipolar asymmetry of approximately 3.4~mK~\cite{1976BAAS....8Q.351C,1977PhRvL..39..898S,%
1993ApJ...419....1K}, which was initially thought to be a sign of a large-scale cosmic anisotropy. It was later understood, however, that the Cosmic Microwave dipole is a Doppler-like effect, due to our motion relative to the CMB frame. Since the Solar System moves at approximately 370~km/sec in the direction of the Leo constellation, the CMB photons appear slightly blueshifted along our direction of motion and redshifted along the antipodal.\footnote{Modern analyses have decomposed the total motion into contributions from the Earth's orbit around the Sun (30~km/sec), the Sun's motion around the Galactic center (220~km/sec), and the Galaxy's motion relative to the Local Group and larger-scale structures~\cite{1993ApJ...419....1K,%
1999AJ....118..337C}. This decomposition reveals a hierarchy of relative motions that must be disentangled to isolate the cosmic signal.} The temperature variation as a function of direction takes the form
\begin{equation}
\frac{\Delta T}{T}(\hat{\mathbf{n}})= \frac{T(\hat{\mathbf{n}}) - T_0}{T_0} = \beta\cos\theta + \mathcal{O}(\beta^2)\,,
\end{equation}
where $\beta = v/c$ is the observer's velocity relative to the CMB frame, $\theta$ is the angle between the direction of observation ($\hat{\mathbf{n}}$) and the velocity and $T_0\simeq2.72$~K is the monopole temperature.

This realization transformed our understanding of cosmic isotropy, confirming the predictions of the Cosmological Principle while simultaneously providing a unique standard of absolute rest in the universe. The initial misinterpretation illustrates how easily relative-motion effects can be mistaken for intrinsic properties of the observed system.

Interestingly, recent high-precision measurements of large-scale structure have revealed potential tensions with the standard interpretation of the CMB dipole as purely kinematic. Surveys of radio galaxies and quasars for example (see \S~\ref{sssNCDs} for a discussion and references), have reported dipole amplitudes and directions that deviate from CMB-based predictions, raising questions about whether the observed CMB dipole might contain a non-kinematic component.

\subsubsection{Spiral nebulae and the expanding
%%%%%%%%%%%%%%%%%%%%%%%%%%%%%%%%%%%%%%%%%%%%%%%
universe}\label{sssSNEU}
%%%%%%%%%%%%%%%%%%%%%%%%
Another instructive historical example involves the debate over the nature of the spiral nebulae in the early 20th century. Before Hubble's landmark observations, astronomers were divided on whether these objects were relatively small structures within our Galaxy, or distant ``island universes'' comparable to the Milky Way. The detection of apparent rotational motion in some nebulae initially seemed to favor the former interpretation, as the implied velocities for extragalactic objects would be implausibly high~\cite{1928asco.book.....J}.

The resolution came with improved distance measurements and the recognition that Doppler shifts in spectral lines revealed both rotational and recessional velocities~\cite{1929PNAS...15..168H}. Hubble's discovery of the distance-redshift relation reframed these observations in the context of an expanding universe~\cite{1929PNAS...15..168H,1931ApJ....74...43H}. For small velocities, the observed redshift ($z$) relates to the recessional velocity ($v$) through
\begin{equation}
1 + z = \frac{\lambda_{\text{obs}}}{\lambda_{\text{emt}}}\approx 1 + \frac{v}{c} + \mathcal{O}(v^2/c^2)\,,
\end{equation}
with $\lambda$ representing the wavelength. This relation, combined with the linear distance-redshift relation $v=H_0d$, established the paradigm of cosmic expansion that continues to frame our understanding today.

These historical episodes highlight a recurring pattern, namely that significant advances in astronomical understanding often come from correctly identifying and accounting for relative-motion effects. This remind us that, as we interpret the complex phenomena discussed throughout this review - redshift-space distortions, peculiar velocity surveys, bulk flows, and cosmic web mapping - we should remain vigilant about potential kinematic effects that may still be overlooked or misinterpreted. The progression from geocentrism to heliocentrism, from static universe to an expanding cosmos and from anisotropic to isotropic CMB, exemplifies the fact that distinguishing between apparent and intrinsic properties through proper-motion analysis has repeatedly transformed our cosmic perspective.

\subsubsection{Local Group dynamics and the ``Great
%%%%%%%%%%%%%%%%%%%%%%%%%%%%%%%%%%%%%%%%%%%%%%%%%%%
Attractor''}\label{sssLGDGA}
%%%%%%%%%%%%%%%%%%%%%%%%%%%%
Following the historical examples of geocentrism and the CMB dipole, another instructive case of relative-motion misinterpretation emerged in the study of local cosmic flows in the 1980s. Observations of peculiar velocities of galaxies in our cosmic neighborhood revealed a systematic motion toward a region in the constellation Centaurus, with a remarkably coherent flow pattern~\cite{1988ApJ...326...19L}. This discovery stemmed from analyses of the Fundamental Plane and the Tully-Fisher relations for elliptical and spiral galaxies, respectively, which revealed peculiar velocities of several hundred km/sec relative to the Hubble flow. The inferred peculiar velocity field $\mathbf{v}_p(\mathbf{r})$ appeared to converge toward a localized region, suggesting a gravitational attraction from a massive concentration beyond the visible distribution of galaxies - dubbed the ``Great Attractor'' - with an estimated mass of approximately $5\times10^{16}\,{\rm M}_{\odot}$.

The initial modeling employed a simple point-mass approximation for the attractor, where the peculiar velocity field followed from the relation
\begin{equation}
\mathbf{v}_p(\mathbf{r}) \approx \frac{GfM}{r^2}\, \hat{\mathbf{r}}_{GA}\,.
\end{equation}
Here, $M$ is the attractor mass, $r$ is the distance from the attractor, $\hat{\mathbf{r}}_{GA}$ is the unit vector pointing toward the Great Attractor and $f=f(\Omega_m)$ is the growth-rate of structure~\cite{1987ApJ...313L..37D,1990ApJ...354...18B}. However, subsequent studies revealed a more nuanced picture. The apparent convergence of peculiar velocities toward this region resulted from a complex superposition of multiple gravitational influences, including effects from the Shapley Supercluster and other structures~\cite{1987ApJ...313L..37D,2006ApJ...645.1043K}. The initial interpretation overestimated both the mass concentration and its gravitational influence, due to insufficient accounting for the complex web of relative motions in the local universe \cite{2017NatAs...1E..36H}.

Modern analyses employing more sophisticated modeling techniques, including the POTENT method~\cite{1999fsu..conf..250D} and its successors. The related studies have revealed that the local velocity field is better described by multiple attractors at different distances. This can be represented by the following more complex velocity-density relation
\begin{equation}
\mathbf{v}_p(\mathbf{r}) = \frac{H_0 f}{4\pi} \int d^3\mathbf{r}' \delta(\mathbf{r}') \frac{(\mathbf{r}' - \mathbf{r})}{|\mathbf{r}' - \mathbf{r}|^3}\,,
\end{equation}
where $\delta(\mathbf{r})$ is the density contrast field~\cite{2017NatAs...1E..36H,2019MNRAS.488.5438G}. Today, we understand the ``Great Attractor'' as part of a more extensive network of structures, with the Shapley Concentration playing a dominant role in shaping the local velocity field~\cite{2013AJ....146...69C,2014Natur.513...71T}. The ``Laniakea Supercluster,'' a basin of attraction containing the Milky Way and approximately 100,000 other galaxies, provides a more accurate framework for understanding our local cosmic neighborhood~\cite{2014Natur.513...71T}. This case demonstrates how the interpretation of peculiar-velocity patterns requires careful consideration of multiple reference frames and gravitational sources, as well as the dangers of oversimplified models when analyzing complex dynamical systems.

\subsubsection{Misinterpreted velocity dispersions in galaxy
%%%%%%%%%%%%%%%%%%%%%%%%%%%%%%%%%%%%%%%%%%%%%%%%%%%%%%%%%%%%
clusters}\label{sssMVDGCs}
%%%%%%%%%%%%%%%%%%%%%%%%%%
Early measurements of velocity dispersions in galaxy clusters yielded surprisingly high values, implying much larger masses than could be accounted for by visible matter~\cite{1933AcHPh...6..110Z}. While this ultimately led to the correct inference of dark matter's existence, initial interpretations sometimes overestimated cluster masses due to contamination from large-scale peculiar-velocity fields and projection effects.

The observed line-of-sight velocity dispersion ($\sigma_v$) in clusters can be artificially enhanced by coherent flows that add to the genuine dynamical dispersion~\cite{1987MNRAS.227....1K}. In the standard virial mass estimation, the cluster mass ($M$) and the velocity dispersion are related by
\begin{equation}
M\approx \frac{3\pi\sigma_v^2 R}{2G}\,,
\end{equation}
with $R$ being the characteristic radius of the cluster~\cite{1988ARA&A..26..631B}. When large-scale bulk motions were not properly accounted for, they led to systematic overestimation of dynamical masses by as much as 15-20\% in some cases~\cite{2000MNRAS.313..229C,2007A&A...466..437W}. Additional complications arise from interlopers, that is from galaxies that appear to be cluster members due to projection effects, but they actually lie in the foreground or the in background. These can further inflate the measured velocity dispersion~\cite{2007A&A...466..437W,2010MNRAS.408.1818W}.

Another source of misinterpretation stems from the assumption of dynamical equilibrium. The measured velocity dispersion represents a snapshot of a potentially non-equilibrium system, where mergers and ongoing accretion can produce temporary enhancements in the velocity dispersion that do not reflect the true gravitational potential~\cite{2008ApJ...672..122E,2013ApJ...772...47S}. N-body simulations have shown that clusters can exhibit significant departures from virial equilibrium, with the virial parameter $\eta=2T/|W|$ - where $T$ is kinetic energy and $W$ is potential energy - deviating from the equilibrium value $\eta=1$ by up to 40\% during major mergers~\cite{2008ApJ...672..122E}.

Modern analyses carefully separate the various components of the measured velocity dispersion, distinguishing between internal cluster dynamics and superimposed large-scale flows~\cite{2011Natur.477..567W,2014MNRAS.442.1887F}. Advanced techniques include phase-space analysis to identify and remove interlopers~\cite{2007A&A...466..437W}, the caustic method to measure mass profiles~\cite{1999MNRAS.309..610D} and joint analysis of velocity dispersion with other mass proxies, such as X-ray temperature and weak lensing~\cite{2013ApJ...767...15R,%
2015MNRAS.454.3938S}. This systematic approach has refined our understanding of cluster masses, while providing valuable information about the large-scale velocity field, all of which have ultimately strengthen rather than weaken the case for dark matter.

\subsubsection{Supernova distance measurements and the Hubble
%%%%%%%%%%%%%%%%%%%%%%%%%%%%%%%%%%%%%%%%%%%%%%%%%%%%%%%%%%%%%
tension}\label{sssSDMHT}
%%%%%%%%%%%%%%%%%%%%%%%%
The apparent acceleration of cosmic expansion, discovered through Type Ia supernova observations in the late 1990s~\cite{1998AJ....116.1009R,1999ApJ...517..565P}, revolutionized cosmology. However, subsequent refinements in the cosmic distance ladder have revealed tensions in measurements of the Hubble constant, with local measurements yielding consistently higher values ($H_0\simeq73$~km/secMpc) than those inferred from CMB observations ($H_0\simeq67$~km/secMpc)~\cite{2019NatAs...3..891V,%
2019ApJ...876...85R}.

While this ``Hubble tension'' may point to new physics beyond the standard cosmological model, peculiar velocity effects cannot be ruled out as contributing factors. Local bulk flows could systematically bias distance measurements if not properly accounted for in the calibration of standard candles~\cite{2015JCAP...07..025W,2018PhRvD..97j3529B}. In the presence of a bulk-flow velocity ($\mathbf{v}_{\text{bulk}}$), the observed redshift ($z_{\text{obs}}$) of a source at cosmological redshift $z_{\text{cos}}$ is modified according to the relation
\begin{equation}
1 + z_{\text{obs}} = (1 + z_{\text{cos}})(1 + z_{\text{pec}}) \approx (1 + z_{\text{cos}})\left(1 + \frac{\mathbf{v}_{\text{bulk}} \cdot \hat{\mathbf{n}}}{c}\right)\,,
\end{equation}
where $\hat{\mathbf{n}}$ is the unit vector along the line of sight~\cite{2011ApJ...741...67D}. If this effect is not properly accounted for, it can introduce a directional bias in the inferred Hubble constant:
\begin{equation}
H_0^{\text{obs}}(\hat{\mathbf{n}}) \approx H_0^{\text{true}}\left(1 + \frac{\mathbf{v}_{\text{bulk}} \cdot \hat{\mathbf{n}}}{cz}\right)\,,
\end{equation}
which becomes more significant at lower redshifts~\cite{2017JCAP...03..022O,2017MNRAS.471.4946W}.

Recent analyses by Perivolaropoulos and Skara~\cite{2022NewAR..9501659P} suggest that a dipolar anisotropy in the Hubble expansion rate could partially explain the observed tension. This dipole can be parameterized as
\begin{equation}
H_0(\hat{\mathbf{n}}) = H_0^{\text{iso}}(1 + A \hat{\mathbf{d}} \cdot \hat{\mathbf{n}})\,,
\end{equation}
where $H_0^{\text{iso}}$ is the isotropic component, $A$ is the dipole amplitude, and $\hat{\mathbf{d}}$ is the dipole direction. Several studies have found evidence for such a dipole with amplitude $A \approx 0.1$~\cite{2020A&A...636A..15M,2021CQGra..38r4001K}, though its statistical significance and interpretation remain debated. This anisotropy might reflect the influence of local large-scale structures and the associated peculiar velocity field, rather than a genuine departure from the standard cosmological model.

Alternative explanations for the Hubble tension include local void models~\cite{2019ApJ...875..145K}, where the Milky Way resides in an underdense region that expands faster than the cosmic average, creating an apparent discrepancy between local and global expansion rates. While simple void models struggle to fully account for the observed tension~\cite{2019ApJ...875..145K,2020MNRAS.495.2630C}, more complex scenarios involving multiple structures or non-standard void profiles remain under investigation. The potential role of peculiar velocities in the Hubble tension highlights the continued relevance of reference-frame considerations in modern precision cosmology.

\subsubsection{Misinterpreted void dynamics}\label{sssMVDs}
%%%%%%%%%%%%%%%%%%%%%%%%%%%%%%%%%%%%%%%%%%%%%%%%%%%%%%%%%%%
Cosmic voids, namely large underdense regions in the universe, have emerged as important probes in cosmology and gravity. Early observations suggested anomalously high outflow velocities from voids, which were initially interpreted as potential evidence for modified gravity theories~\cite{2006MNRAS.367.1629S,%
2001ApJ...557..495P}. In the linear regime, the radial velocity profile of a spherical void is
\begin{equation}
v_r(r) = -\frac{1}{3}Hfr\delta(r)\,,
\end{equation}
with $\delta=\delta(r)$ representing the profile of the density contrast~\cite{2014PhRvL.112y1302H}. For a top-hat void with density contrast $\delta_v$ and radius $R_v$, this yields the velocity
\begin{equation}
v_r(R_v) = -\frac{1}{3}HfR_v\delta_v\,,
\end{equation}
at the void boundary. Deviations from this relationship could potentially indicate modifications to gravity, or to dark energy physics~\cite{2006MNRAS.367.1629S,2015MNRAS.451.1036C}.

Subsequent analyses revealed that these apparent high velocities resulted from selection effects and reference frame ambiguities~\cite{2019PhRvD.100b3504N,2016MNRAS.455L..99L}. The choice of reference frame significantly affects the interpretation of void dynamics, with peculiar velocities relative to the CMB frame sometimes creating the appearance of enhanced expansion rates~\cite{2016MNRAS.455L..99L}. Furthermore, the identification of voids themselves is sensitive to the reference frame, with different void-finding algorithms yielding different results when applied in redshift space versus real space~\cite{2019PhRvD.100b3504N,%
2014MNRAS.443.3238P}.

Modern void studies carefully account for relative-motion effects, using reconstructed velocity fields to separate genuine dynamical features from apparent kinematic effects \cite{2016PhRvL.117i1302H,%
2017PhRvD..96h3506A}. The observed redshift-space void-galaxy cross-correlation function $\xi^s_{vg}$ can be modeled as
\begin{equation}
\xi^s_{vg}(s_\perp, s_\parallel) = \int_{-\infty}^{+\infty} [1 + \xi^r_{vg}(r)] \mathcal{P}(v_\parallel | \mathbf{r}) \delta^D(s_\parallel - r_\parallel - v_\parallel/aH) dv_\parallel\,,
\end{equation}
where $\mathcal{P}(v_\parallel | \mathbf{r})$ is the probability distribution of line-of-sight velocities at real-space separation $\mathbf{r}$~\cite{2016PhRvL.117i1302H,2017PhRvD..96h3506A}. This approach has placed stronger constraints on modified gravity theories while highlighting the importance of reference frame considerations in void dynamics.

Advanced statistical techniques have further refined our understanding of void kinematics. The void-galaxy cross-correlation function exhibits a distinctive quadrupole moment in redshift space, which can constrain the growth rate of structure~\cite{2016PhRvL.117i1302H,2016MNRAS.462.2465C}. Similarly, the velocity profile around voids can be measured directly by the kinematic Sunyaev-Zel'dovich effect~\cite{2015PhRvL.115s1301H}. According to these studies, void dynamics are largely consistent with the $\Lambda$CDM predictions, once the proper reference-frame corrections are applied~\cite{2019PhRvD.100b3504N,%
2016PhRvL.117i1302H}, thus illustrating how apparent anomalies can be often resolved by careful consideration of the relative-motion effects.

\subsubsection{Lessons for contemporary cosmology}\label{sssLCC}
%%%%%%%%%%%%%%%%%%%%%%%%%%%%%%%%%%%%%%%%%%%%%%%%%%%%%%%%%%%%%%%%
These historical misinterpretations are also valuable cautionary lessons for modern cosmological analyses. In particular, the lessons highlight the crucial importance of:

\begin{itemize}
    \item Carefully identifying the relevant reference frames in any analysis of cosmic motions.\vspace{5pt}
    \item Distinguishing between intrinsic properties and relative-motion effects.\vspace{5pt}
    \item Accounting for the hierarchical nature of cosmic flows, which operate across multiple scales.\vspace{5pt}
    \item Recognizing that apparent anomalies may reflect incomplete modeling of relative-motion effects rather than new physics.\vspace{5pt}
    \item Employing multiple, complementary observational probes to break degeneracies between genuine physical effects and kinematic artifacts.\vspace{5pt}
\end{itemize}

As cosmological observations achieve unprecedented precision, the subtle effects of peculiar velocities become increasingly significant. Contemporary challenges that may involve similar considerations include the hemispherical power asymmetry in the CMB~\cite{2020A&A...641A...7P}, the alignment of low multipoles~\cite{2016CQGra..33r4001S}, and various tensions between early and late-universe probes~\cite{2020JCAP...01..013D}. While some of these anomalies may ultimately point to new physics, history teaches us to thoroughly exhaust conventional explanations based on reference frame effects and observational systematics before invoking more exotic solutions.

The evolution of our understanding of the ``Great Attractor'', cluster velocity dispersions, the Hubble tension, and void dynamics demonstrates both the persistent challenges and the analytical progress in disentangling genuine cosmic phenomena from relative-motion effects. As we have seen throughout this review, peculiar velocities represent both a challenge to be overcome and a valuable cosmological probe in their own right. By learning from historical misinterpretations and applying sophisticated modeling techniques, modern cosmology continues to refine our understanding of cosmic motions across all scales, from the local neighborhood to the largest observable structures in the universe.

\subsection{Methods to account for, or to mitigate, the
%%%%%%%%%%%%%%%%%%%%%%%%%%%%%%%%%%%%%%%%%%%%%%%%%%%%%%%
biases}\label{MAMBs}
%%%%%%%%%%%%%%%%%%%%
Given the significant impact of peculiar velocities on cosmological observations, researchers have developed various sophisticated techniques to account for, or/and mitigate these biases. These approaches range from statistical corrections to advanced modeling methods that incorporate the complex physics of large-scale flows.

\subsubsection{Peculiar velocity field reconstruction}\label{PVFR}
%%%%%%%%%%%%%%%%%%%%%%%%%%%%%%%%%%%%%%%%%%%%%%%%%%%%%%%%%%%%%%%%%%
\begin{figure*}
\begin{center}
\includegraphics[height=2.5in,width=3in,angle=0]{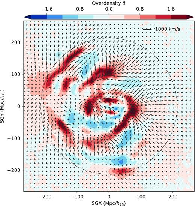}
\end{center}
\caption{The central part of the Cosmic Flows~3 velocity field reconstruction in the $SGZ=0$~Mpc$\,h^{-1}\,_{75}$ slice, within the Supergalactic Plane (SGX-SGY) coordinates. The background gradient represents the matter density field ($\delta$), with the red regions corresponding to overdensities ($\delta >0$) and the blue to underdensities ($\delta <0$). These respectively indicate areas of gravitational attraction and repulsion. The dotted black circle illustrates the edge of the data at $z=0.054$, which corresponds to $\sim200$~Mpc (see also Fig.~6 in~\cite{2019MNRAS.488.5438G}.}
\label{fig:graziani}
\end{figure*}

Peculiar velocity analyses are well known for their susceptibility to significant errors, as they are highly sensitive to a variety of biases and systematic uncertainties-including Malmquist bias, calibration uncertainties on distance indicators (like Tully-Fisher, Fundamental Plane, SNIa), incomplete sky coverage, or cosmic variance. To mitigate these biases, one of the most robust methodologies involves reconstructing the underlying velocity field using data from galaxy redshift surveys. Such reconstructions typically exploit the theoretical relationship between the mass density and velocity fields as established in the framework of gravitational instability theory~\cite{1994ARA&A..32..371D,%
1995MNRAS.272..885F}
\begin{equation}
\vec{v}(\vec{r}) \approx \frac{f H_0}{4\pi} \int d^3r' \frac{\vec{r}'-\vec{r}}{|\vec{r}'-\vec{r}|^3} \delta(\vec{r}')\,,
\label{eq:pecden}
\end{equation}
with $f\simeq\Omega_m^{0.55}$ representing the growth rate of structure, $\delta(\vec{r})$ the density contrast and $H_0$ the Hubble constant. Next, we will briefly summarize some of the most widely used approaches.

\begin{enumerate}
\item The VELMOD method~\cite{1997ApJ...486..629W,1998ApJ...507...64W} is a robust maximum-likelihood framework designed to estimate parameters of the peculiar-velocity field from combined measurements of galaxy distances and redshifts. The technique is particularly effective for analyzing the local universe (within $cz\sim3000$~km/sec) and provides a statistically rigorous approach to account for observational uncertainties and selection biases. In~\cite{2001MNRAS.326.1191B}, VELMOD was applied to compare the velocity field predicted from the spatial distribution of galaxies in the IRAS PSCz redshift survey with peculiar velocity measurements from TF spiral galaxies in the SFI catalog. More recently, Carrick et al~\cite{2015MNRAS.450..317C} employed VELMOD in conjunction with SFI++ spiral galaxy data to constrain both the amplitude of a residual bulk flow and the best-fitting value of the parameter $\beta^*$. The latter is defined as $\beta^{*}=f/b^*$, where $f=f(\Omega_{m})$ is the linear growth rate of cosmic structure and $b^*$ is the linear bias parameter of the galaxy sample. By tightly linking the observed galaxy velocity and density fields within the framework of gravitational instability theory, VELMOD yields more precise constraints on cosmological parameters and the structure of the nearby universe.\\

\item The POTENT reconstruction, originally proposed in~\cite{1990ApJ...364..349D}, has been a cornerstone in the reconstruction of large-scale structure in the local universe, with applications to datasets like the Mark III catalog~\cite{1999ApJ...522....1D} and the IRAS galaxy surveys~\cite{1998ApJ...495..516S}. The POTENT procedure recovers the underlying mass-density fluctuation field from a whole-sky sample of observed radial peculiar velocities. The main goal of the POTENT analysis is to accurately reconstruct the velocity and density fields with minimal systematic errors~\cite{1997ApJ...486..629W}. The process involves the following steps: (a) preparing the data for POTENT analysis, which includes grouping galaxies and applying corrections for Malmquist bias; (b) smoothing the peculiar-velocity data to produce a uniformly smoothed radial velocity field with minimal bias; (c) utilizing the gravitational potential flow assumption to reconstruct the gravitational potential and the three-dimensional velocity field; (d) estimating the underlying density field using an approximation based on the nonlinear gravitational-instability theory. The fixed smoothing scale, a distinctive feature of the POTENT method, ensures spatial statistical uniformity. This uniformity is valuable both for cosmographic studies and for straightforward, direct comparisons of the reconstructed fields with theoretical models and other observational data.\\

\item The VIRBIUS framework (VelocIty Reconstruction using Bayesian Inference Software - see~\cite{2016MNRAS.457..172L}) offers a comprehensive Bayesian approach for inferring the full three-dimensional velocity field from galaxy distance measurements and spectroscopic catalogs. Unlike the traditional methods, which are typically limited by their reliance on linear theory and their vulnerability to systematic biases, VIRBIUS employs a hierarchical Bayesian model that jointly reconstructs the density and velocity fields, cosmological parameters, and the observational characteristics of distance indicators. By integrating these components within a unified inference scheme, VIRBIUS aims to systematically account for and mitigate sources of bias and uncertainty, thereby providing a more robust and accurate reconstruction of the 3-D peculiar-velocity field.\\

\item The Wiener Filter (WF) methodology \cite{1995ApJ...449..446Z,1995MNRAS.272..885F,%
    2012ApJ...744...43C,2015MNRAS.449.4494H} has emerged as a powerful tool for estimating the velocity field, particularly in cases where observational data are sparse, noisy, or exhibit incomplete sky coverage - conditions that are frequently encountered in peculiar velocity surveys. The WF method is a Bayesian statistical approach that reconstructs cosmic fields by optimally combining observational measurements with prior knowledge of the statistical properties of large-scale structure. In this framework, the underlying density field is modeled as a Gaussian random field, characterized by a known power spectrum $P=P(k)$, while observational noise is typically assumed to be Gaussian and uncorrelated. The WF utilizes the observed galaxy distribution to infer the underlying density field, by incorporating the galaxy bias parameter ($b$ - such that $\delta_{g} = b\delta$). From the reconstructed density field, the peculiar-velocity field is then derived via the relation $\nabla\cdot\mathbf{v}=-f\delta$, where $f$ is the linear growth rate of structure. This method offers a robust means of mapping the large-scale structure of the Universe, by facilitating direct comparisons between galaxy and mass distributions and by enabling estimates of key cosmological parameters, such as $\Omega_m$ and the bias factor $b$. However, it is important to note that the results are inherently sensitive to the choice of cosmological parameters, which introduces a degree of model-dependence into the analysis.\\

\item The state-of-the-art forward modeling techniques~\cite{2012MNRAS.425.2443K,2013ApJ...772...63W,%
    2019MNRAS.488.5438G,2022MNRAS.513.5148V} improve upon VIRBIUS and reconstruct the late-time velocity and density-contrast fields from late-time observations of galaxies with distance indicators. An example of the forward modeling method is given in~\cite{2019MNRAS.488.5438G}.The linear peculiar velocity reconstruction method is based on the CF3 galaxy catalog amd it is used to map the velocity field of the local universe, up to a redshift $\sim0.05$, from the peculiar velocities of galaxies. The methodology allows for a more accurate reconstruction of the velocity field by incorporating the effects of large-scale structure and the distribution of matter in the universe. The study assumes a $\Lambda$CDM model and a Gaussian error model for the observations, for the distance moduli and for the redshift data to derive the peculiar velocities. These are essential for understanding the dynamics of the local universe and for testing cosmological models.

The methodology builds upon Bayesian forward-modeling frameworks, similar in spirit to the methods developed for large-scale structure reconstruction, but it is tailored to the specific characteristics of the peculiar-velocity data. The peculiar velocity ($v=v(r)$) of a galaxy is related to the surrounding matter density field ($\delta=\delta(r)$) through Eq.~(\ref{eq:pecden}) and forms the basis of their reconstruction. Graziani et al.~\cite{2019MNRAS.488.5438G} construct the density field of the local universe from observed galaxy positions and luminosities. This process involves correcting for galaxy bias (the difference between the galaxy and matter distributions) and smoothing the density field to reduce noise. Instead of directly inverting the peculiar-velocity data, the authors predict velocities from the modeled density field and compare these predictions to the observations. This forward-modelling approach minimizes systematic biases that arise in traditional inversion methods.

To validate the uncertainty of their results, the atudies create mock catalogs with simulated galaxy distributions and velocity fields, applying the same selection function and observational uncertainties as in the real CF3 dataset. They then use a Markov Chain Monte Carlo (MCMC) exploration to infer the most likely parameters (such as the amplitude and coherence scale of the velocity field) that reproduce the observed distribution of distances and redshifts. This approach is robust against the Malmquist bias and accommodates non-Gaussian error distributions. The reconstructed peculiar-velocity field reveals coherent bulk flows on scales of tens of Mpc, consistent with the gravitational influence of large-scale structures like the Shapley Supercluster, the Great Attractor and the Perseus-Pisces Supercluster. The velocity field is dominated by flows toward these massive structures, as expected from the $\Lambda$CDM model and and shown in Fig.~\ref{fig:graziani}. Prominent overdense regions, such as clusters or superclusters (in red), are associated with converging peculiar velocities, while underdense voids (in blue) show diverging velocities. This work is a significant step in leveraging the increasing quality of peculiar-velocity catalogs, thus providing a more accurate and systematic method to map the local velocity field and to test cosmological models against detailed observations.

The 2M++ velocity field reconstruction~\cite{2015MNRAS.450..317C} represents one of the most comprehensive applications of these techniques, providing a detailed map of peculiar velocities throughout the local universe out to depths of approximately 200~Mpc.
\end{enumerate}

\subsubsection{Mitigating redshift-space distortions}\label{sssMRSDs}
%%%%%%%%%%%%%%%%%%%%%%%%%%%%%%%%%%%%%%%%%%%%%%%%%%%%%%%%%%%%%%%%%%%%%
Peculiar velocities distort the mapping between real and redshift space, affecting galaxy clustering measurements. Several techniques have been developed to address these effects.

\begin{itemize}
\item \textbf{Multipole expansion:} By decomposing correlation functions into Legendre polynomials, researchers can isolate and model the anisotropic components (dipole, quadrupole, etc) introduced by peculiar velocities~\cite{1992ApJ...385L...5H,2004PhRvD..70h3007S}. Additionally, these multipole moments can be compared against theoretical models, offering critical insights into both the linear and nonlinear regimes of structure formation. This analysis helps disentangle the effects of peculiar velocities from those of cosmic expansion.\\

\item \textbf{Wedges analysis:} This approach divides the two-dimensional correlation function into angular wedges with respect to the line of sight, enabling cleaner separation of radial and transverse components~\cite{2012MNRAS.419.3223K}. Each wedge captures distinct facets of clustering information. More specifically, the radial wedge is particularly sensitive to distortions along the line of sight and is closely linked to the Hubble parameter ($H$), whereas the transverse wedge, being less influenced by peculiar velocities, is more directly associated with geometric distances and is thus sensitive to the angular diameter distance. In~\cite{2014MNRAS.439...83A}, the authors utilize this technique to conduct an in-depth study of the large-scale clustering of galaxies from the \textit{SDSS-III BOSS Data Release 9}. By measuring the BAO features in the two-point correlation function, they derive robust constraints on $H$ and the angular diameter distance, offering an independent validation of the standard $\Lambda$CDM cosmological model.\\

\item \textbf{Density-field reconstruction:} Prior to BAO analysis, peculiar velocity effects can be mitigated by applying a reconstruction procedure that estimates and removes nonlinear displacements \cite{2007ApJ...664..675E,2012MNRAS.427.2132P}. This method begins with the observed galaxy density field and (under the assumption that large-scale structures have primarily evolved through gravitational instability) applies linear perturbation theory to infer the displacement field responsible for moving galaxies away from their primordial (nearly uniform) distribution. To mitigate the influence of small-scale nonlinearities, a smoothing filter is generally applied to the density field. By effectively reversing the effects of gravitational clustering, the reconstruction process sharpens the BAO feature in the density field relative to the unreconstructed data. This approach both sharpens the BAO feature and restores the isotropy of the correlation function on large scales.\\

\item \textbf{Phase-space distribution function modeling:} The method models redshift-space distortions using the full phase-space distribution of tracers, by simultaneously incorporating both density and velocity fields~\cite{2011JCAP...11..039S,2017JCAP...10..009H}. The redshift-space density field is expressed as a series of moments of the distribution function - such as density, momentum and higher-order velocity moments - enabling a systematic expansion of the redshift-space power spectrum in terms of their correlations. This formalism offers a clean separation of physical effects, like coherent flows and small-scale velocity dispersions (Fingers~of~God), while allowing for the controlled inclusion of non-linearities. Crucially, it does not rely on specific assumptions about velocity fields or bias models, making it broadly applicable and well-suited to quasi-linear scales.
\end{itemize}

Modern BAO analyses typically employ reconstruction techniques that reduce the peculiar velocity smearing of the acoustic peak, thus improving distance constraints by around 50\%~\cite{2017MNRAS.470.2617A}.

\subsubsection{Correcting distance ladder
%%%%%%%%%%%%%%%%%%%%%%%%%%%%%%%%%%%%%%%%%
measurements}\label{sssCDLMs}
%%%%%%%%%%%%%%%%%%%%%%%%%%%%%
While the reconstruction techniques discussed in the previous section primarily address large-scale velocity fields, accurate measurements of cosmic distances (particularly those used in the cosmic distance ladder) require additional specialized methods to mitigate the peculiar-velocity effects. Distance indicators such as Type~Ia supernovae, Tully-Fisher relation, Fundamental Plane and surface brightness fluctuations are all affected by peculiar motions, which introduce both systematic biases and additional scatter in the derived distances.

For a galaxy with a peculiar velocity $\vec{v}_{\text{pec}}$, the observed redshift $z_{\text{obs}}$ differs from the cosmological redshift $z_{\text{cos}}$ according to:
\begin{equation}
(1 + z_{\text{obs}}) = (1 + z_{\text{cos}})(1 + z_{\text{pec}}) \approx (1 + z_{\text{cos}})\left(1 + \frac{\vec{v}_{\text{pec}} \cdot \hat{r}}{c}\right)\,,
\end{equation}
where $\hat{r}$ is the unit vector along the line of sight. This relation becomes increasingly important at low redshifts, where the peculiar velocity contribution can be a significant fraction of the observed redshift. At $z\simeq0.01$, a typical peculiar velocity of 300 km/s induces a redshift perturbation of $\Delta z\simeq0.001$, corresponding to a distance error of approximately 10\%~\cite{2011ApJ...741...67D,2006PhRvD..73l3526H}. Several specialized techniques have been developed to address these biases.

\begin{itemize}
\item \textbf{Flow model corrections:} This approach incorporates reconstructed peculiar velocity fields to correct individual distance measurements. The methodology typically follows the steps\vspace{5pt}
    \begin{enumerate}
        \item Construct a density field from a galaxy redshift survey, using methods like those described in Section 4.1.\vspace{5pt}
        \item Derive the peculiar velocity field using linear theory, or more advanced reconstruction techniques.\vspace{5pt}
        \item For each distance indicator (e.g.~SNIa), predict the line-of-sight peculiar velocity $v_{\text{pec}}=\vec{v}_{\text{pec}}\cdot\hat{r}$ at its position.\vspace{5pt}
        \item Correct the observed redshift to obtain the cosmological redshift: $z_{\text{cos}}\simeq z_{\text{obs}}-v_{\text{pec}}/c$.\vspace{5pt}
    \end{enumerate}

    Neill et al~\cite{2007ApJ...661L.123N} applied this method to low-redshift SNIa, using the IRAS PSCz density field to predict peculiar velocities. The authors found that flow-model corrections reduced the scatter in the Hubble diagram by up to 20\% for SNIa within $z<0.028$. More recent work by Davis et al~\cite{2011ApJ...741...67D} used the 2M++ redshift compilation to derive more accurate flow models, finding that peculiar-velocity corrections can shift individual distance measurements by up to $\pm8\%$ at $z\simeq0.01$. The flow-model approach has been refined by Peterson et al~\cite{2022ApJ...938..112P}, who demonstrated that incorporating directional anisotropies in the local flow field is crucial for accurate corrections.

    Incorporating the model for velocities of Hudson et al~\cite{2004MNRAS.352...61H}, we can express the expected peculiar velocity at position $\vec{r}$ as
    \begin{equation}
    v_{\text{model}}(\vec{r}) = \beta V_{\text{ext}}(\vec{r}) + V_{\text{bulk}} \cdot \hat{r} + \sum_{i=1}^{N_{\text{att}}} \frac{V_i}{r_i^2 + R_i^2} \frac{\vec{r}_i \cdot \hat{r}}{r_i}\,,
    \end{equation}
    where $\beta=f/b$ is the redshift distortion parameter, $V_{\text{ext}}$ is the velocity predicted from the external density field, $V_{\text{bulk}}$ is a residual bulk flow and the sum represents the contributions from $N_{\text{att}}$ individual attractors with characteristic velocities $V_i$ and core radii $R_i$.\\

\item \textbf{Thermal energy correction:} Beyond coherent flows, random motions in the peculiar-velocity field introduce additional scatter in distance measurements. This ``thermal'' component effectively adds an uncertainty term to distance measurements, particularly for low-redshift objects. Peterson et al~\cite{2022ApJ...938..112P} formalized this approach, showing that the effective variance in the measured distance modulus ($\mu$) due to peculiar velocities is:
    \begin{equation}
    \sigma_{\mu,v}^2 = \left(\frac{5}{\ln 10}\right)^2 \left(\frac{\sigma_v^2}{c z_{\text{cos}}}\right)^2\,,
    \end{equation}
    with $\sigma_v$ being the one-dimensional peculiar-velocity dispersion. For typical values of $\sigma_v\simeq250$-$300$~km/sec, this introduces an additional uncertainty of $\sigma_{\mu,v}\simeq0.46$~mag at $z\simeq0.01$, decreasing to $\sigma_{\mu,v}\simeq0.09$~mag at $z\simeq0.05$~\cite{2022ApJ...938..112P}.

    The total uncertainty in the distance modulus then becomes
    \begin{equation}
    \sigma_{\mu,\text{total}}^2 = \sigma_{\mu,\text{int}}^2 + \sigma_{\mu,\text{obs}}^2 + \sigma_{\mu,v}^2\,,
    \end{equation}
    where $\sigma_{\mu,\text{int}}$ is the intrinsic scatter of the distance indicator and $\sigma_{\mu,\text{obs}}$ is the observational uncertainty. Accounting for this additional uncertainty term is crucial for proper weighting of low-redshift objects in cosmological analyses and for accurate error estimation in the Hubble constant measurements.\\

\item \textbf{Malmquist bias correction:} Peculiar velocities interact with selection effects in flux-limited samples, creating complex biases in distance measurements.\vspace{5pt}
    \begin{enumerate}
        \item \textit{Homogeneous Malmquist bias}: Arises from the fact that the volume element increases with distance, thus causing more objects from greater distances to scatter into a sample than from lesser distances.\vspace{5pt}
        \item \textit{Inhomogeneous Malmquist bias}: Occurs due to the coupling between peculiar velocities and the underlying density field, as galaxies in overdense regions tend to have infall velocities that make them appear closer than they actually are \cite{1995PhR...261..271S,%
            1994MNRAS.266..468H}.\vspace{5pt}
    \end{enumerate}

    For a flux-limited sample with selection function $S(r)$, the corrected distance $r_{\text{true}}$ relates to the observed distance $r_{\text{obs}}$ according to the formula
    \begin{equation}
    \langle r_{\text{true}} | r_{\text{obs}} \rangle r_{\text{obs}} \exp(3.5\sigma^2)\,,
    \end{equation}
    with $\sigma$ representing the fractional distance error~\cite{1994MNRAS.266..468H}. More sophisticated corrections account for the inhomogeneous distribution of galaxies and the coupling between density and velocity fields, as implemented in the VELMOD method~\cite{1997ApJ...486..629W}.

    Recent work by Kourkchi et al~\cite{2020ApJ...896....3K} demonstrated that careful treatment of Malmquist bias is essential for accurate distance measurements using the Tully-Fisher relation, particularly when combining multiple surveys with different selection functions. They implemented a comprehensive bias correction methodology that accounts for both homogeneous and inhomogeneous components, reducing the systematic errors in the derived distances below 3\%.\\

\item \textbf{Bayesian hierarchical modeling:} This framework represents a more comprehensive approach to peculiar-velocity corrections, simultaneously constraining cosmological parameters, distance-redshift relations and peculiar velocity fields within a unified probabilistic model. The hierarchical structure typically includes\vspace{5pt}
    \begin{enumerate}
        \item A cosmological model that predicts the true distance-redshift relation.\vspace{5pt}
        \item A peculiar velocity model that accounts for deviations from the Hubble flow\vspace{5pt}.
        \item Distance indicator models that relate observable quantities to true distances.\vspace{5pt}
        \item Selection effects and measurement uncertainties.\vspace{5pt}
    \end{enumerate}

    Mandel et al~\cite{2017ApJ...842...93M} pioneered this approach for SNIa, developing the BAYESN framework that simultaneously models the intrinsic SNIa population, measurement uncertainties and peculiar velocities. The method naturally accounts for correlated peculiar velocities through the spatial covariance function
    \begin{equation}
    \mathbf{C}_v = \sigma_v^2 \exp\left(-\frac{|\mathbf{r}_i - \mathbf{r}_j|^2}{2\lambda_v^2}\right)\,,
    \end{equation}
    where $\sigma_v$ is the peculiar-velocity dispersion and $\lambda_v$ is the correlation length.

    Howlett et al~\cite{2017ApJ...847..128H} extended this approach to peculiar-velocity surveys, developing a Bayesian framework that jointly analyzes redshift and peculiar velocity data to constrain the growth rate of structure, while accounting for systematic biases. Their method has been applied to the 6dFGS velocity survey, yielding constraints on the growth rate parameter ($f\sigma_8$) with approximately 15\% precision.

    More recently, Boruah et al~\cite{2020MNRAS.498.2703B} implemented a sophisticated Bayesian framework for modeling the local peculiar-velocity field, by using both direct measurements from distance indicators and constraints from the galaxy density field. The proposed approach reconstructs the full 3-D velocity field, while properly accounting for uncertainties and selection effects, thus enabling more accurate corrections for distance ladder measurements.
\end{itemize}

The SH0ES measurements of the Hubble constant~\cite{2022ApJ...934L...7R} now incorporate several of these correction techniques, including flow-model corrections based on the 2M++ density field and a careful treatment of the additional uncertainty from random peculiar velocities. The latest analysis uses simulations to verify that these corrections effectively mitigate peculiar-velocity biases, reducing this source of systematic uncertainty in the value of $H_0$ to approximately 0.5\%. This level of precision is essential for meaningful comparisons with CMB-derived values in the context of the Hubble tension.

Looking forward, new approaches continue to emerge for addressing peculiar-velocity effects in distance-ladder measurements. Pesce et al.~\cite{2020ApJ...891L...1P} demonstrated that megamaser-based distance measurements provide a promising alternative for measuring $H_0$ with reduced sensitivity to peculiar-velocity effects, by targeting galaxies at intermediate redshifts ($0.03 \lesssim z \lesssim 0.05$) where such effects are naturally suppressed. In parallel, Bayesian forward-modelling frameworks that reconstruct the initial density field and its subsequent velocity field evolution~\cite{2019A&A...625A..64J} offer a route to incorporate physically consistent peculiar-velocity predictions directly into distance-ladder analyses, potentially tightening constraints on $H_0$.

As distance ladder measurements continue to play a crucial role in the Hubble tension debate, the development of increasingly sophisticated methods to address peculiar-velocity biases remains an active and important area of research, sitting at the intersection of observational techniques and theoretical modeling of large-scale structure.

\subsubsection{Advanced statistical methods}\label{sssASMs}
%%%%%%%%%%%%%%%%%%%%%%%%%%%%%%%%%%%%%%%%%%%%%%%%%%%%%%%%%%%
Beyond specific correction techniques, several advanced statistical approaches help mitigate biases in cosmological analyses.

\begin{itemize}
\item \textbf{Minimum Variance estimators:} These optimally weighted statistics minimize the impact of peculiar velocity noise on bulk flow measurements~\cite{2009MNRAS.392..743W,%
    2010MNRAS.407.2328F}.\vspace{5pt}

\item \textbf{Gaussian process regression:} Provides a flexible framework for modeling spatially correlated peculiar-velocity fields, without assuming a specific parametric form~\cite{2014MNRAS.444.3926J}.\vspace{5pt}

\item \textbf{Machine Learning approaches:} Neural networks and other ML methods can learn complex patterns in peculiar-velocity fields, thus enabling more accurate corrections \cite{2020MNRAS.498.2703B}.\vspace{5pt}

\item \textbf{MCMC sampling with velocity likelihood:} This approach marginalizes over peculiar-velocity uncertainties when constraining cosmological parameters~\cite{2019A&A...625A..64J}.
\end{itemize}

\subsubsection{Multi-probe analysis}
%%%%%%%%%%%%%%%%%%%%%%%%%%%%%%%%%%%%
Combining multiple observational probes can help break degeneracies and reduce the impact of peculiar-velocity biases.

\begin{itemize}
\item \textbf{Joint analysis of redshift-space distortions and direct peculiar velocity measurements:} This approach provides complementary constraints on the growth rate of structure and helps mitigate systematic uncertainties~\cite{2017MNRAS.471..839A,%
    2020MNRAS.497.1275S}.\vspace{5pt}

\item \textbf{Cross-correlation of kSZ and galaxy surveys:} The kinematic Sunyaev-Zel'dovich effect provides a redshift-independent probe of peculiar velocities that can be combined with galaxy surveys to constrain velocity fields at higher redshifts~\cite{2012PhRvL.109d1101H,%
    2017JCAP...03..008D}.\vspace{5pt}

\item \textbf{Combining lensing and peculiar velocity data:} Gravitational lensing provides a complementary probe of the matter distribution that is not affected by peculiar velocities, helping to break degeneracies in velocity field models~\cite{2010Natur.464..256R,2017JCAP...05..015H}.
\end{itemize}

\subsubsection{Specialized survey design}\label{sssSSD}
%%%%%%%%%%%%%%%%%%%%%%%%%%%%%%%%%%%%%%%%%%%%%%%%%%%%%%%
Future surveys are being designed with peculiar velocity mitigation strategies in mind.\vspace{5pt}

\begin{itemize}
\item \textbf{Wide sky coverage:} Surveys with nearly complete sky coverage, such as WALLABY~\cite{2020Ap&SS.365..118K} and TAIPAN~\cite{2017PASA...34...47D}, help reduce cosmic variance in peculiar-velocity measurements.\vspace{5pt}

\item \textbf{Dense sampling:} Higher galaxy density reduces shot noise in velocity field reconstructions, as demonstrated by the 6dFGS peculiar-velocity survey~\cite{2014MNRAS.445.2677S}.\vspace{5pt}

\item \textbf{Multi-wavelength observations:} Combining optical, infrared, and radio measurements helps constrain systematic uncertainties in distance indicators used for peculiar velocity measurements~\cite{2016AJ....152...50T}.\vspace{5pt}

\item \textbf{Deep spectroscopic coverage:} Deeper, higher-redshift spectroscopy reduces the fractional impact of local peculiar velocities on cosmological parameters; upcoming facilities like DESI and Taipan’s full-depth program exemplify this strategy~\cite{2022AJ....164..207D,2016arXiv161100036D}.
\end{itemize}

\subsubsection{Forward modeling approaches}\label{sssFMAs}
%%%%%%%%%%%%%%%%%%%%%%%%%%%%%%%%%%%%%%%%%%%%%%%%%%%%%%%%%%
Rather than trying to correct for peculiar velocities after measurement, forward modeling approaches incorporate these effects directly into the analysis pipeline.

\begin{itemize}
\item \textbf{Constrained simulations:} These are numerical simulations designed to reproduce the observed local universe, including its peculiar velocity field~\cite{2017MNRAS.465.4886C,2018NatAs...2..680H}.\vspace{5pt}

\item \textbf{Bayesian origin reconstruction from galaxies:} The BORG algorithm reconstructs the initial conditions that would evolve into the observed galaxy distribution, naturally accounting for peculiar velocities~\cite{2013MNRAS.432..894J,%
    2016MNRAS.455.3169L}.\vspace{5pt}

\item \textbf{Simulation-based inference:} Rather than applying analytical corrections, these methods use numerical simulations to create mock observations that include peculiar-velocity effects and then compare directly to the data~\cite{2020MNRAS.495.4227K}.
\end{itemize}

The COSMIC FLOWS program exemplifies this approach, integrating observational data with constrained simulations to understand the complex three-dimensional motions in the local universe (e.g.~see~\cite{2013AJ....146...86T,2016AJ....152...50T,%
2023ApJ...944...94T}).

\subsubsection{Future prospects}\label{sssFPs}
%%%%%%%%%%%%%%%%%%%%%%%%%%%%%%%%%%%%%%%%%%%%%%
Several promising developments are poised to significantly improve our ability to account for the peculiar-velocity biases.

\begin{itemize}
\item The combination of LSST optical observations with SKA radio measurements will provide unprecedented constraints on peculiar-velocity fields across multiple scales~\cite{2020PASA...37....7S}.\vspace{5pt}

\item Next-generation CMB experiments will enable more precise measurements of the kSZ effect, offering a complementary probe of peculiar velocities at higher redshifts~\cite{2019JCAP...02..056A}.\vspace{5pt}

\item Standard sirens from gravitational wave sources with electromagnetic counterparts will provide distance measurements independent of the cosmic distance ladder and less susceptible to peculiar-velocity biases~\cite{2018Natur.562..545C}.\vspace{5pt}

\item Advances in computational techniques, particularly in Bayesian hierarchical modeling and simulation-based inference, will enable more sophisticated treatment of peculiar-velocity effects in cosmological analyses~\cite{2017arXiv170101467T}.
\end{itemize}

These developments promise to transform peculiar velocities from a source of systematic uncertainty into a powerful cosmological probe in their own right, providing unique information on the growth of structure and on the underlying cosmological model.

\section{Future directions and emerging technologies}\label{sFDETs}
%%%%%%%%%%%%%%%%%%%%%%%%%%%%%%%%%%%%%%%%%%%%%%%%%%%%%%%%%%%%%%%%%%%
The previous section detailed the sophisticated methods developed to account for peculiar-velocity biases in cosmological measurements. Building upon these techniques, the field is poised for significant advancements through upcoming surveys, technological innovations and novel theoretical approaches.

\subsection{Overview of upcoming peculiar velocity
%%%%%%%%%%%%%%%%%%%%%%%%%%%%%%%%%%%%%%%%%%%%%%%%%%
surveys}\label{ssOUPVSs}
%%%%%%%%%%%%%%%%%%%%%%%%
This section explores the future directions and their potential to transform the science of peculiar velocities from a source of systematic uncertainty into a precision cosmological probe.

\subsubsection{Ground-based spectroscopic surveys}
%%%%%%%%%%%%%%%%%%%%%%%%%%%%%%%%%%%%%%%%%%%%%%%%%%
The Dark Energy Spectroscopic Instrument (DESI)~\cite{2023MNRAS.525.1106S} is an ongoing peculiar-velocity survey, commenced its five-year observational campaign in May 2021. Hosted on the 4-meter Mayall Telescope at Kitt Peak National Observatory, DESI represents a significant step forward in the precision mapping of the Universe's expansion and the growth of cosmic structure~\cite{2019BAAS...51g..57L}. With its highly multiplexed fiber-optic spectrographs, which are capable of obtaining spectra for 5000 objects simultaneously over a wavelength range from 360 to 980 nm, DESI is considered one of the world's best facilities for wide-field spectroscopy throughout the decade.

A central objective of DESI is to construct the most comprehensive catalog of galaxy peculiar velocities and spectroscopic redshifts to date. By the end of its campaign, DESI is expected to collect high-precision redshift measurements for approximately 34 million galaxies and quasars over a total sky coverage of 14,000 square degrees, though the peculiar-velocity measurements are confined to a lower-redshift subset of roughly 4,000 deg$^2$. The data will enable a 3-D reconstruction of the Universe's large-scale structure and impose stringent constraints on cosmological parameters such as the Hubble constant, the growth rate of structure and the dark-energy equation of state.

The DESI survey uses a sequence of target classes to map the large-scale structure of the Universe from redshift 0 up to 3.5. These targets include:
\begin{itemize}
    \item Red luminous galaxies, covering a redshift range of $0.3< z<1$.\vspace{5pt}
    \item Emission-line galaxies, covering $0.6< z<1.6$ and being the largest sample.\vspace{5pt}
    \item Quasars and their Lyman-$\alpha$ forest, yielding measurements out to $z>2$.
\end{itemize}
In addition to its primary cosmological objectives, DESI includes a target program aimed at measuring galaxy peculiar velocities through the use of redshift-independent distance indicators. While a variety of well-established methods exist for estimating extragalactic distances - such as Cepheid variable stars~\cite{1912HarCi.173....1L}, the tip of the red giant branch~\cite{1993ApJ...417..553L}, Type~Ia supernovae~\cite{1993ApJ...413L.105P} and surface brightness fluctuations~\cite{1988AJ.....96..807T} - the DESI velocity program focuses specifically on the Fundamental Plane relation for early-type galaxies and on the Tully-Fisher relation for spiral galaxies. By applying these techniques to a statistically significant sample of several hundred thousand galaxies, the survey aims to construct the most comprehensive and precise peculiar-velocity catalog to date. A key innovation of DESI is the simultaneous use of both distance indicators within a unified framework and with consistent target selection criteria. The aim is to enhance the precision and robustness of distance and velocity estimates across a broad range of galaxy types.

\subsubsection{Upcoming radio surveys}\label{sssURSs}
%%%%%%%%%%%%%%%%%%%%%%%%%%%%%%%%%%%%%%%%%%%%%%%%%%%%%
The Square Kilometre Array (SKA) and its precursors will revolutionize the study of peculiar velocities through 21cm surveys of neutral hydrogen in galaxies. WALLABY (Widefield ASKAP L-band Legacy All-sky Blind surveY)~\cite{2020Ap&SS.365..118K}, using the Australian Square Kilometre Array Pathfinder (ASKAP), is expected to detect H~I in over 500,000 galaxies across 75\% of the sky, with approximately 10\% of these galaxies suitable for Tully-Fisher measurements. The full SKA will extend this capability dramatically, potentially measuring peculiar velocities for millions of galaxies across unprecedented volumes.

These radio surveys complement optical programs through their ability to probe gas-rich spiral galaxies, their immunity to dust extinction and their natural provision of spectral-line widths needed for Tully-Fisher analyses. The combination of optical and radio peculiar-velocity surveys, will provide a more complete census of galaxy motions across different types of galaxies and host environments.

\subsubsection{Synergies with photometric surveys}\label{sssSPSs}
%%%%%%%%%%%%%%%%%%%%%%%%%%%%%%%%%%%%%%%%%%%%%%%%%%%%%%%%%%%%%%%%%
The Vera C. Rubin Observatory's Legacy Survey of Space and Time (LSST)~\cite{2009arXiv0912.0201L} will provide deep, multi-band photometry for billions of galaxies. While not directly measuring peculiar velocities, LSST data will enhance their science through improved photometric calibration for distance indicators, identification of transient distance markers (like SNe~Ia) and more precise characterization of galaxy properties relevant to distance indicator relations.

The Nancy Grace Roman Space Telescope (formerly WFIRST)~\cite{2015arXiv150303757S}, will complement the ground-based efforts with space-based infrared imaging and spectroscopy, particularly valuable for measuring distances to galaxies in obscured regions and at higher redshifts than accessible to the current surveys. Similarly, the SPHEREx mission~\cite{2014arXiv1412.4872D} will conduct an all-sky near-infrared spectral survey, creating a three-dimensional map of the universe by measuring redshifts for hundreds of millions of galaxies. This extensive dataset will be invaluable for mapping large-scale structure and cross-correlating with peculiar velocity surveys to reduce cosmic variance, particularly on large scales.

\subsection{Novel observational techniques for peculiar-velocity
%%%%%%%%%%%%%%%%%%%%%%%%%%%%%%%%%%%%%%%%%%%%%%%%%%%%%%%%%%%%%%%%
measurements}\label{ssNOTPVMs}
%%%%%%%%%%%%%%%%%%%%%%%%%%%%%%
While the previous sections focused on the established methods for measuring peculiar velocities mainly through classical distance indicators, recent years have witnessed a remarkable diversification of cosmic rulers that hold significant promise for future peculiar-velocity studies.

\subsubsection{Cosmic rulers beyond traditional distance
%%%%%%%%%%%%%%%%%%%%%%%%%%%%%%%%%%%%%%%%%%%%%%%%%%%%%%%%
indicators}\label{sssCRBTDIs}
%%%%%%%%%%%%%%%%%%%%%%%%%%%%%
The emerging approaches complement traditional methods, by accessing different redshift regimes, galaxy populations and systematic uncertainties, thereby strengthening our ability to reconstruct the three-dimensional velocity field across the cosmic history.

\begin{itemize}
\item \textbf{Surface Brightness Fluctuations (SBF)}: First proposed by Tonry and Schneider~\cite{1988AJ.....96..807T}, the SBF technique exploits the statistical fluctuations in surface brightness arising from the Poisson distribution of unresolved stars in early-type galaxies. The measured SBF magnitude $\bar{m}$ is related to the distance modulus $(m-M)$ through:
\begin{equation}
\bar{m}= \bar{M}+ (m-M)+ P_{\mathrm{SBF}}\,,
\end{equation}
where $\bar{M}$ is the absolute SBF magnitude (calibrated through Cepheid distances, or other primary indicators) and $P_{\mathrm{SBF}}$ represents population corrections based on galaxy color. Advances in imaging capabilities and calibration techniques have substantially enhanced the utility of SBF as a distance indicator~\cite{2021ApJ...911...65B}. Blakeslee et al~\cite{2009ApJ...694..556B} demonstrated that SBF measurements with HST can achieve precision of 5-8\% on distances out to $\sim$100 Mpc.

The SBF method offers several unique advantages for peculiar-velocity studies. Unlike the Fundamental Plane, it does not require spectroscopy, making it observationally more efficient for large samples. Jensen et al~\cite{2021ApJS..255...21J}, have shown that with next-generation facilities like JWST and the Nancy Grace Roman Space Telescope, the SBF method can potentially reach $z \sim 0.1$ with percent-level precision, dramatically extending the volume over which direct peculiar velocity measurements are feasible. Recently, Garnavich et al~\cite{2023ApJ...953...35G} have proposed combining SBF with Type Ia supernovae to constrain peculiar-velocity fields with reduced systematic uncertainties, potentially helping to resolve ongoing tensions in the measurement of the Hubble constant.\\

\item \textbf{Type II Supernovae}: While Type~Ia supernovae have long been the gold standard for cosmological distance measurements, Type~II supernovae (core-collapse explosions of massive stars) offer complementary capabilities for peculiar-velocity science. Though intrinsically more diverse than their Type~Ia counterparts, supernovae of Type~II-P (characterized by a plateau in their light-curves) can be standardized through several techniques.

The Expanding Photosphere Method (EPM)~\cite{1974ApJ...193...27K} uses the photospheric expansion of the supernova to derive a distance via:
\begin{equation}
\theta = \frac{R}{D} = \frac{\sqrt{f_{\lambda}/(\pi B_{\lambda}}}{D \xi_{\lambda}}\,.
\end{equation}
Here, $\theta$ is the angular size of the photosphere, $R$ is the photospheric radius, $D$ is the distance, $f_{\lambda}$ is the observed flux, $B=B_{\lambda}(T)$ is the Planck function at temperature $T$, and $\xi_{\lambda}$ is a dilution factor accounting for the non-blackbody nature of the photosphere.

Alternatively, the Standardized Candle Method (SCM)~\cite{2002ApJ...566L..63H} employs the following empirical correlation between the luminosity and the expansion velocity of the supernova during its plateau phase:
\begin{equation}
M = \alpha \log(v_{\mathrm{Fe\,II}}/5000) + \beta(V-I) + \gamma\,,
\end{equation}
with $M$ being the absolute magnitude, $v_{\mathrm{Fe\,II}}$ the expansion velocity (measured from the iron lines), $(V-I)$ the color, while $\alpha$, $\beta$ and $\gamma$ are calibration parameters.

Recently, de Jaeger et al~\cite{2020MNRAS.496.3402D} have demonstrated that Type~II supernovae can achieve distance precisions of 12-14\%, thus facilitating measurements of the Hubble constant with 4.4\% precision. While this accuracy remains less precise than that of the Type~Ia supernovae, the Type~II offer important advantages. More specifically, Type~II events re approximately five times more common, they occur in a wider range of galaxy types (including low-mass star-forming galaxies that rarely host SNe~Ia) and they are subject to different astrophysical systematics. Rodriguez et al~\cite{2014AJ....148..107R} have shown that combining SNe~II with traditional peculiar-velocity tracers can significantly improve the reconstruction of local flow fields, particularly in low-density environments underrepresented in early-type galaxy samples.\\

\item \textbf{Fast Radio Bursts (FRBs)}: These enigmatic millisecond-duration radio transients offer a novel and potentially revolutionary approach to measuring cosmic distances. FRBs experience a frequency-dependent delay due to dispersion by free electrons along the line of sight. The effect is quantified by the dispersion measure (DM), given by
\begin{equation}
\mathrm{DM} = \int_0^d n_e(l)\,{\rm d}l = \int_0^z \frac{n_e(z)(1+z)}{H}\,{\rm d}z\,,
\end{equation}
where $n_e$ is the electron density and $H=H(z)$ is the Hubble parameter. The observed DM consists of contributions from the Milky Way, from the host galaxy and form the intergalactic medium (IGM), with the latter directly related to the cosmic distance.

Macquart et al~\cite{2020Natur.581..391M} empirically confirmed the predicted relationship between redshift and the DM contribution from the IGM (the ``Macquart relation''), demonstrating that FRBs can indeed serve as cosmological rulers. While individual FRB distances currently have uncertainties of 30-40\% (due to variations in the IGM density and host galaxy contributions), Wu et al~\cite{2020ApJ...895...33W} have shown that statistical samples of hundreds of localized FRBs could constrain the cosmic expansion history to 2-3\% precision.

For peculiar velocity studies, James et al~\cite{2022MNRAS.510L..18J} demonstrated that correlating the residuals from the Macquart relation with the reconstructed velocity field can constrain the growth-rate of structure. Furthermore, Madhavacheril et al~\cite{2019PhRvD.100j3532M} proposed using FRBs to break degeneracies in the kSZ measurements, creating a powerful synergy between these two novel probes. As localization capabilities improve with instruments like CHIME/FRB~\cite{2018ApJ...863...48C} and the Deep Synoptic Array (DSA-2000)~\cite{2019BAAS...51g.255H} - expected to localize up to 10,000 FRBs per year - these cosmic messengers will become increasingly valuable for mapping the large-scale velocity field at intermediate redshifts ($0.1< z<1$), where traditional distance indicators become impractical.\\

\item \textbf{GW standard sirens}: The detection of gravitational waves (GWs) from compact binary mergers has opened a fundamentally new window on the cosmos, providing ``standard sirens'' whose distance can be measured directly from the GW signal alone~\cite{1986Natur.323..310S}. For a binary merger, the GW strain amplitude ($h$) is related to the luminosity distance ($D_L$) by
\begin{equation}
h \propto \frac{\mathcal{M}^{5/3} \omega^{2/3}}{D_L}\,,
\end{equation}
with $\mathcal{M}$ being the chirp mass and $\omega$ the orbital frequency. Unlike electromagnetic distance indicators, this measurement requires no cosmic distance ladder and is independent of assumptions about the source's intrinsic luminosity.

The first gravitational-wave event with an electromagnetic counterpart, namely GW170817 \cite{2017Natur.551...85A}, enabled a measurement of the Hubble constant with around 15\% precision. Nicolaou et al~\cite{2020MNRAS.495...90N} demonstrated that the peculiar velocity of the host galaxy NGC~4993 contributed a systematic uncertainty of about 7\% to this measurement, highlighting both a challenge and an opportunity. Put another way, while peculiar velocities must be accurately modeled to extract cosmological parameters from standard sirens, standard sirens with independently measured redshifts can be conversely used to probe the peculiar velocity field.

Future gravitational wave observatories will dramatically expand these capabilities. Indeed, Mukherjee et al~\cite{2021PhRvD.103d3520M} projected that with 100 binary neutron-star mergers with electromagnetic counterparts, as expected from next-generation detectors like the Einstein Telescope and Cosmic Explorer, the velocity power spectrum could be constrained with precision comparable to contemporary galaxy surveys. These measurements would probe scales of 20-50~Mpc, complementing both larger-scale measurements from redshift-space distortions and smaller-scale measurements from traditional peculiar-velocity surveys. It was additionally shown that by correlating standard siren residuals with the reconstructed density field can provide constraints on the growth-rate parameter ($\beta=f/b$) with 5-10\% precision, offering a novel test of gravity on cosmological scales~\cite{2021MNRAS.508.4512L}.\\

\item \textbf{Extragalactic distance scale from water megamasers}: Water megamasers in the accretion disks of supermassive black holes provide a geometric distance-measurement method of exceptional precision. By measuring the Keplerian rotation curve of the maser spots through VLBI observations, combined with long-term monitoring of the acceleration of spectral features, researchers can determine the angular diameter distance to the host galaxy
\begin{equation}
D_A = \frac{v_r^2 r}{a \sin^2 i}\,,
\end{equation}
where $v_r$ is the rotational velocity, $r$ is the radius of the massing disk, $a$ is the centripetal acceleration, and $i$ is the inclination~\cite{2013ApJ...767..154R}. The Megamaser Cosmology Project has measured distances to several galaxies with precision of 5-10\%~\cite{2020ApJ...891L...1P}.

These measurements are particularly valuable for peculiar-velocity studies, because they provide accurate distances to spiral galaxies on intermediate scales (of 30-200~Mpc), where the traditional methods face significant challenges. Kuo et al~\cite{2019arXiv191014314K} demonstrated that incorporating megamaser distances can significantly improve the constraints on the bulk flow and shear of the local universe. While the sample size of suitable megamaser systems remains limited (only about a dozen systems are currently known), upcoming facilities like the next-generation Very Large Array (ngVLA)~\cite{2018ASPC..517....3M}, will enhance our ability to detect and characterize these systems, thus expanding their role in the science of peculiar velocities.\\

\item \textbf{Baryon Acoustic Oscillations (BAO) as relative distance indicators}: While traditionally used as cosmic rulers for measuring the expansion history, BAO can also provide constraints on the peculiar velocities through their anisotropic clustering pattern. The measured correlation function in redshift space exhibits anisotropies that depend on both the Alcock-Paczynski effect and redshift-space distortions, according to
\begin{equation}
\xi^s(s_\perp, s_\parallel) = \xi^r\sqrt{s_\perp^2 + \frac{s_\parallel^2}{(1+\beta\mu^2)^2}}\,.
\end{equation}
Here, $\xi^s$ and $\xi^r$ are the redshift-space and real-space correlation functions, $s_\perp$ and $s_\parallel$ are the transverse and line-of-sight separations, $\beta = f/b$ is the redshift-space distortion parameter and $\mu$ is the cosine of the angle between the line of sight and the separation vector.

Combining baryon acoustic oscillation (BAO) measurements with traditional peculiar-velocity (PV) surveys helps disentangle cosmic expansion from peculiar motions, improving distance–redshift modeling and reducing systematics that affect $H_0$ in the local Universe~\cite{2015MNRAS.449..848H}.
\end{itemize}

The diversification of cosmic rulers beyond traditional distance indicators, is a key development in the study of peculiar velocities. Each method has different systematic uncertainties, applies to different galaxy populations and its optimal redshift range differs as well. By combining these complementary probes, we can construct a more complete and robust picture of the cosmic velocity field across scales and throughout cosmic history. This in turn enhances our ability to address fundamental questions about the growth of structure, the nature of gravity and the composition of the universe.

\subsubsection{Kinematic Sunyaev-Zel'dovich
%%%%%%%%%%%%%%%%%%%%%%%%%%%%%%%%%%%%%%%%%%%
measurements}\label{sssKS-ZMs}
%%%%%%%%%%%%%%%%%%%%%%%%%%%%%%
While the distance indicator techniques discussed previously offer precise peculiar-velocity measurements in the local universe, they become increasingly challenging at higher redshifts where distance uncertainties grow and sample sizes diminish. The kinematic Sunyaev-Zel'dovich (kSZ) effect offers a complementary approach that can probe peculiar velocities at substantially higher redshifts, providing a unique window into the growth of large-scale structure across cosmic time.

The kSZ effect arises from the Doppler shifting of CMB photons as they scatter off moving free electrons~\cite{1980MNRAS.190..413S}. The resulting temperature fluctuation is
\begin{equation}
\frac{\Delta T}{T_{\text{CMB}}}(\hat{\mathbf{n}}) = -\sigma_T \int n_e(\mathbf{r}) \frac{\mathbf{v}_e(\mathbf{r}) \cdot \hat{\mathbf{n}}}{c} e^{-\tau(\mathbf{r})} {\rm d}\chi\,.
\end{equation}
In the above, $\sigma_T$ is the Thomson cross-section, $n_e$ is the electron number density, $\mathbf{v}_e$ is the peculiar velocity of the electrons, $\tau$ is the optical depth and $\chi$ is the comoving distance along the line of sight $\hat{\mathbf{n}}$. Unlike the thermal SZ effect, the kSZ signal preserves the blackbody spectrum of the CMB, making it challenging to separate from the primary CMB anisotropies. Furthermore, the kSZ effect is sensitive only to the line-of-sight component of the velocity field, and its amplitude is proportional to both the electron density and the velocity, creating a fundamental degeneracy between these quantities.

Despite these challenges, significant progress has been made in detecting and utilizing the kSZ effect for cosmological inference. The first statistical detection was achieved by Hand et al~\cite{2012PhRvL.109d1101H} using a pairwise estimator that measures the relative motion of galaxy pairs through their kSZ signatures. This approach exploits the fact that, on average, pairs of galaxies approach each other due to their mutual gravitational attraction, creating a dipolar pattern in the kSZ signal. The pairwise kSZ signal can be expressed as
\begin{equation}
T_{\text{pkSZ}}(r) = -T_{\text{CMB}} \sigma_T \bar{n}_e \frac{v_{12}(r)}{c}\,,
\end{equation}
where $v_{12}(r)$ is the mean pairwise velocity of galaxies separated by distance $r$ and $\bar{n}_e$ is the average electron density in the galaxies. Subsequent work by De Bernardis et al~\cite{2017JCAP...03..008D} and by Soergel et al~\cite{2016MNRAS.461.3172S} confirmed this detection using improved datasets and methodology.

Recent advances in kSZ methodology have significantly expanded the utility of this probe for peculiar velocity science.

\begin{itemize}
\item \textbf{Velocity reconstruction techniques:} Smith et al~\cite{2018arXiv181013423S} developed a framework for reconstructing the full 3D velocity field from a combination of galaxy positions and kSZ measurements. Their approach uses the fact that, in the linear regime, the velocity field is directly related to the density field through:
\begin{equation}
\mathbf{v}(\mathbf{k}) = i\frac{Hf}{k}\frac{\mathbf{k}}{k}\delta(\mathbf{k})\,,
\end{equation}
with $f$ representing the growth rate, $H$ the Hubble parameter and $\delta(\mathbf{k})$ the density contrast in Fourier space. By using galaxy positions to trace the density field and kSZ measurements to constrain the velocity field, they can break the optical depth degeneracy and reconstruct the full 3D velocity field. This technique has been further refined by Cayuso et al~\cite{2018PhRvD..98f3502C}, who incorporated non-linear corrections to the velocity-density relationship.\\

\item \textbf{Projected field methods:} Hill et al~\cite{2016PhRvL.117e1301H} introduced a novel approach that circumvents the optical depth degeneracy by working with projected fields. Instead of trying to extract the velocity for individual objects, they measure the correlation between the projected kSZ field and the projected momentum field derived from galaxy surveys:
\begin{equation}
\left\langle \left(\frac{\Delta T}{T_{\text{CMB}}}\right)_{\text{kSZ}} \hat{q}_{\parallel} \right\rangle = -\sigma_T \bar{\tau}_e \frac{\langle v_{\parallel}^2 \rangle}{c}\,,
\end{equation}
where $\hat{q}_{\parallel}$ is the normalized, projected galaxy momentum field, and $\bar{\tau}_e$ is the average optical depth. This method effectively isolates the kSZ signal by leveraging the directional information in the galaxy momentum field, significantly enhancing the signal-to-noise ratio. Refinements to this approach by Ferraro et al~\cite{2016PhRvD..94l3526F} have demonstrated that it can provide constraints on the growth rate of structure competitive with traditional redshift-space distortion measurements.\\

\item \textbf{Cross-correlation approaches:} Schaan et al~\cite{2016PhRvD..93h2002S} developed a method to extract the kSZ signal through cross-correlation with reconstructed velocity fields from galaxy redshift surveys. By stacking CMB temperature measurements at galaxy positions and weighting by the reconstructed line-of-sight velocity, they effectively isolate the kSZ signal, namely
\begin{equation}
\left\langle \frac{\Delta T}{T_{\text{CMB}}} \frac{v_{\text{rec}}}{c} \right\rangle = -\sigma_T \bar{\tau}_e \frac{\langle v_{\text{true}} v_{\text{rec}} \rangle}{c^2}\,,
\end{equation}
with $v_{\text{rec}}$ being the reconstructed velocity and $v_{\text{true}}$ the true velocity. This approach was successfully applied to ACT CMB data and BOSS galaxy data, providing evidence for the kSZ effect at $>3\sigma$ significance. Recent work by Tanimura et al~\cite{2021A&A...645A.112T}  has extended this approach to larger datasets and higher redshifts, demonstrating its robustness and versatility.\\

\item \textbf{Moving lens techniques:} Beyond the kSZ effect from galaxy clusters, peculiar velocities also induce a dipolar pattern in the CMB lensing signal around moving objects. This ``moving lens'' effect, first described by Birkinshaw and Gull~\cite{1983Natur.302..315B}, provides a complementary probe of transverse velocities that is independent of the electron density. Hotinli et al~\cite{2019PhRvL.123f1301H} demonstrated that this effect could be detected statistically in future CMB experiments, offering a novel probe of peculiar velocities that complements the kSZ effect.\\

\item \textbf{Component separation methods:} A major challenge in kSZ studies is separating the signal from the primary CMB and other foregrounds. Advanced component separation techniques developed by Remazeilles et al.~\cite{2019MNRAS.483.3459R} leverage the unique spectral and spatial properties of the kSZ signal to isolate it from other components. These methods will be particularly important for future CMB experiments with improved frequency coverage and sensitivity.
\end{itemize}

The scientific potential of kSZ measurements extends well beyond traditional peculiar-velocity science. Deutsch et al.~\cite{2018PhRvD..98l3501D} demonstrated that kSZ observations can constrain the sum of neutrino masses through their effect on the growth of structure, while Battaglia et al~\cite{2015ApJ...812..154B} showed that the kSZ effect can probe the distribution of baryons in the circumgalactic medium, addressing the ``missing baryon problem''. Furthermore, Munchmeyer et al~\cite{2019PhRvD.100h3508M} proposed using kSZ tomography to constrain primordial non-Gaussianity through its impact on the large-scale velocity field.

Next-generation CMB experiments like CMB-S4~\cite{2019arXiv190704473A} and the Simons Observatory~\cite{2019JCAP...02..056A} will dramatically improve kSZ measurements through their combination of high sensitivity, large sky coverage, and improved angular resolution. Simulations by Flender et al~\cite{2016ApJ...823...98F} predict that these experiments, combined with upcoming galaxy surveys like LSST and Euclid, will enable kSZ detections with signal-to-noise ratios exceeding 100, transforming kSZ measurements from a challenging niche technique into a robust and powerful cosmological probe.

Recent work has begun to realize this potential, with preliminary results from Planck and ACT data demonstrating the ability to constrain the growth rate of structur~\cite{2021PhRvD.104d3502C} and to detect the kSZ signal from filamentary structures in the cosmic web~\cite{2020MNRAS.491.2318T}. Looking ahead, Sehgal et al~\cite{2019BAAS...51g...6S} projected that CMB-S4 will constrain the growth rate parameter $f\sigma_8$ to sub-percent precision across a wide range of redshifts ($0.3<z<3$), providing a powerful test of General Relativity on cosmic scales.

These developments will transform kSZ measurements from a promising but challenging technique into a robust probe of cosmic flows across a wide range of scales and redshifts, complementing the local peculiar velocity surveys discussed in previous sections and extending our ability to trace the evolution of large-scale structure throughout cosmic history.

\subsubsection{Redshift-space distortion tomography}\label{sssRSDT}
%%%%%%%%%%%%%%%%%%%%%%%%%%%%%%%%%%%%%%%%%%%%%%%%%%%%%%%%%%%%%%%%%%%
While the previous sections have explored direct measurements of peculiar velocities through distance indicators and novel probes like the kSZ effect, complementary information about the cosmic velocity field can be extracted from the anisotropic clustering of galaxies in redshift surveys. These redshift-space distortions (RSDs), first theoretically formalized by Kaiser~\cite{1987MNRAS.227....1K}, arise from the same peculiar velocities that we have been discussing throughout this review, but they provide an independent statistical probe of these motions rather than direct measurements for individual objects.

In redshift space, the observed galaxy overdensity field $\delta_g^s(\mathbf{s})$ is related to the real-space matter overdensity field $\delta_m(\mathbf{r})$ through the mapping:
\begin{equation}
\delta_g^s(\mathbf{s}) = \int d^3\mathbf{r} \, \delta_D\left(\mathbf{s} - \mathbf{r} - \frac{\mathbf{v}(\mathbf{r}) \cdot \hat{\mathbf{r}}}{aH} \hat{\mathbf{r}}\right) [1 + \delta_g(\mathbf{r})]\,,
\end{equation}
where $\mathbf{v}(\mathbf{r})$ is the peculiar velocity field, $H$ is the Hubble parameter and $a$ is the scale factor. In the linear regime and under the distant observer approximation, this leads to the well-known Kaiser formula for the redshift-space power spectrum:
\begin{equation}
P_g^s(k, \mu) = (b + f\mu^2)^2 P_m(k)\,,
\end{equation}
with $b$ being the galaxy bias, $f$ the linear growth rate, $\mu = \hat{\mathbf{k}} \cdot \hat{\mathbf{r}}$ the cosine of the angle between the wavevector and the line of sight and $P_m(k)$ the real-space matter power spectrum.

Traditional RSD analyses measure the parameter combination $\beta = f/b$ or $f\sigma_8$, where $\sigma_8$ is the amplitude of matter fluctuations on 8/hMpc scales. However, recent methodological innovations have significantly enhanced the information that can be extracted from RSDs, effectively enabling tomographic reconstruction of the velocity field across cosmic time and spatial scales. These advances include:

\begin{itemize}
\item \textbf{Multi-tracer techniques:} Proposed by McDonald and Seljak~\cite{2009JCAP...10..007M}, this approach leverages the fact that different galaxy populations trace the same underlying velocity field but with different bias factors. By simultaneously analyzing multiple tracers, one can partially cancel the cosmic variance that typically limits the precision of RSD measurements. The observed density fields for two tracers, A and B, are respectively written as
\begin{equation}
\delta_A^s(\mathbf{k}) = b_A \delta_m(\mathbf{k}) + f\mu^2 \delta_m(\mathbf{k}) + \epsilon_A(\mathbf{k})
\end{equation}
and
\begin{equation}
\delta_B^s(\mathbf{k}) = b_B \delta_m(\mathbf{k}) + f\mu^2 \delta_m(\mathbf{k}) + \epsilon_B(\mathbf{k})\,,
\end{equation}
where $\epsilon_A$ and $\epsilon_B$ represent stochastic noise terms. The ratio of these fields depends only on the bias difference and is independent of the underlying density field, enabling precise measurements of the relative bias.

Beutler et al~\cite{2019JCAP...03..040B} applied this technique to BOSS data, combining luminous red galaxies and emission line galaxies to improve constraints on $f\sigma_8$ by approximately 25\%.  Alarcon et al~\cite{2020MNRAS.498.2614A} further showed that incorporating photometric surveys as additional tracers can enhance constraints even when redshift precision is limited. The fundamental advantage of the multi-tracer approach is that it becomes limited by shot noise rather than cosmic variance, making it particularly powerful for studying large scales where the velocity field carries unique cosmological information.\\

\item \textbf{Void-galaxy correlation methods:} Cosmic voids—large underdense regions in the universe—offer a complementary environment for studying redshift-space distortions. The velocity field around voids exhibits a coherent outflow pattern, leading to distinctive RSD signatures in the void-galaxy cross-correlation function. Hamaus et al~\cite{2016PhRvL.117i1302H} developed a theoretical framework for analyzing these distortions
\begin{equation}
\xi^s_{vg}(s, \mu) = \xi^r_{vg}(r) + \frac{1}{3}\,f\bar{\xi}^r_{vg}(r) + f\mu^2 \left[ \xi^r_{vg}(r) - \bar{\xi}^r_{vg}(r) \right]\,.
\end{equation}
In the above, $\xi^s_{vg}$ and $\xi^r_{vg}$ are the void-galaxy correlation functions in redshift and real space, respectively, and $\bar{\xi}^r_{vg}$ is the volume-averaged correlation function.

This approach offers several advantages over traditional galaxy clustering analyses. First, the velocity field around voids is predominantly linear even at scales where the overall matter distribution is nonlinear, simplifying theoretical modeling. Second, the void-galaxy correlation function has a well-defined functional form that provides a robust reference for measuring distortions. Nadathur et al~\cite{2019PhRvD.100b3504N} applied this method to BOSS data, obtaining constraints on $f\sigma_8$ competitive with galaxy clustering measurements while subject to different systematic uncertainties.

Recent work by Pisani et al~\cite{2019BAAS...51c..40P} has shown that void RSDs are particularly sensitive to modified gravity theories that incorporate screening mechanisms, as these mechanisms are least effective in low-density environments. Correa et al~\cite{2021MNRAS.500..911C} further demonstrated that combining void-galaxy correlations with traditional galaxy clustering can break degeneracies between gravity modifications and galaxy bias, potentially providing smoking-gun tests of General Relativity on cosmic scales. The upcoming Euclid mission, with its unprecedented void catalog containing millions of voids across a wide redshift range, is expected to improve constraints on gravity theories by an order of magnitude using these techniques~\cite{2022A&A...660A..67N}.\\

\item \textbf{Higher-order statistics:} While the power spectrum captures the Gaussian information in the density field, higher-order statistics like the bispectrum access additional non-Gaussian information that is particularly valuable for breaking degeneracies in RSD analyses. Gil-Marín et al~\cite{2014JCAP...12..029G} developed a comprehensive framework for modeling the redshift-space galaxy bispectrum, so that
\begin{eqnarray}
B^s_g(k_1, k_2, k_3, \mu_1, \mu_2, \mu_3) &=& Z_1(k_1, \mu_1)Z_1(k_2, \mu_2)Z_2(k_1, k_2, \mu_1, \mu_2)P_L(k_1)P_L(k_2)\nonumber\\ &&+~\text{cyc.}\,,
\end{eqnarray}
where $Z_1$ and $Z_2$ are redshift-space kernel functions that depend on the bias parameters and growth rate, and $P_L$ is the linear power spectrum.

The bispectrum is particularly powerful for studying the velocity field because it captures the mode coupling induced by gravitational evolution, which directly reflects the nonlinear relationship between density and velocity. Furthermore, D'Amico et al~\cite{2021JCAP...01..006D} showed that the bispectrum can help distinguish between different theories of modified gravity that predict the same linear growth rate but different nonlinear couplings.

Beyond the bispectrum, innovative summary statistics like marked power spectra~\cite{2021PhRvL.126a1301M}, position-dependent power spectra~\cite{2014JCAP...05..048C}, and wavelet phase harmonics~\cite{2020PhRvD.102j3506A} have been developed to efficiently extract non-Gaussian information relevant to the velocity field. These statistics are particularly sensitive to the environmental dependence of structure formation, providing unique windows into the relationship between density and velocity across different cosmic environments.\\

\item \textbf{Velocity field reconstruction:} Beyond constraining summary statistics such as $f$ and $\sigma_8$, modern redshift-space distortion (RSD) analyses can recover the full three-dimensional peculiar-velocity field jointly with the underlying matter density. Bayesian forward-modelling frameworks cast this as a high-dimensional inference problem on latent fields given the observed redshift-space galaxy distribution. Representative implementations include the BORG family, which infers initial conditions and forward-evolves them to the present epoch to obtain self-consistent density and velocity fields, and the VIRBIuS framework, which is tailored to peculiar-velocity applications and combines redshift-space information with distance-derived velocities~\cite{2019A&A...625A..64J,2016MNRAS.455.3169L}.

\begin{equation}
P(\mathbf{v}, \delta \mid \mathbf{d}) \propto P(\mathbf{d} \mid \mathbf{v}, \delta)\, P(\mathbf{v} \mid \delta)\, P(\delta)\,,
\end{equation}
where $\mathbf{d}$ are the observed galaxy positions in redshift space (and, in extended setups, distance-indicator data). The likelihood $P(\mathbf{d} \mid \mathbf{v}, \delta)$ is a forward model that maps latent fields to observables, accounting for redshift-space mapping (coherent flows and Fingers-of-God), selection functions, survey masks, tracer bias, and Poissonian shot noise. The conditional $P(\mathbf{v} \mid \delta)$ encodes the gravity-based velocity–density relation (e.g., from the continuity/Euler equations in linear or mildly non-linear regimes), enforcing physically consistent large-scale, nearly irrotational flows. The prior $P(\delta)$ typically places a Gaussian prior on the initial conditions that are evolved with perturbative or particle-mesh dynamics to capture non-linearity and mode coupling.

In practice, these frameworks use gradient-based samplers (e.g., Hamiltonian Monte Carlo) or variational/MAP schemes with adjoint gradients, block-sampling nuisance parameters (bias, noise) and marginalizing over survey systematics. The output is a posterior over 3D density and velocity fields with quantified uncertainties, enabling corrections to redshift–distance relations, cross-correlation with direct PV measurements, and tests of gravity via the velocity–density consistency relation.

This approach has been extended by Lilow et al~\cite{2021MNRAS.507.1557L} to incorporate non-Gaussian likelihoods and non-linear bias models, enabling more accurate reconstruction on smaller scales. Modi et al~\cite{2021A&C....3700505M} further demonstrated that neural networks can learn to predict the velocity field directly from the redshift-space galaxy distribution, potentially capturing complex non-linear relationships that are difficult to model analytically.

The reconstructed velocity fields provide rich datasets that can be analyzed well beyond simple summary statistics. For example, the topology and kinematic decomposition of the cosmic flow—via divergence, shear, and vorticity—encode information about structure growth and possible non-Gaussian features~\cite{2012MNRAS.425.2049H}.\\

\item \textbf{Synergies with peculiar velocity surveys:} Perhaps most relevant to this review, Adams and Blake~\cite{2017MNRAS.471..839A} pioneered methods to combine redshift-space distortion measurements with direct peculiar velocity surveys. The two probes provide complementary information: RSD measurements constrain the parameter combination $f\sigma_8$, while peculiar velocity surveys directly measure $f\sigma_8^2$. By combining them, one can break the degeneracy between the growth rate and the amplitude of fluctuations.

Building on this work, Said et al~\cite{2020MNRAS.497.1275S} performed a joint analysis of the 6dFGS peculiar velocity survey and redshift-space distortions from SDSS, obtaining 9\% precision on $\gamma = 0.566 \pm 0.052$, where $\gamma$ is the growth index parameter in the parameterization $f \approx \Omega_m^\gamma$. This provides one of the most stringent tests of General Relativity on cosmological scales to date. Looking ahead, Howlett et al~\cite{2017MNRAS.464.2517H} projected that combining WALLABY's 40,000 peculiar velocities with redshift-space distortions from future surveys could constrain $\gamma$ to 3\% precision, potentially revealing signatures of modified gravity.

More sophisticated joint analyses have been developed by Boruah et al~\cite{2020MNRAS.498.2703B}, who incorporated both datasets within a unified Bayesian framework that accounts for their different noise properties and systematic uncertainties. This approach not only improves parameter constraints but also enables consistency tests between the two probes, potentially revealing systematics in either dataset or genuine new physics.\\

\item \textbf{Tomographic analyses:} Modern galaxy surveys spanning wide redshift ranges enable tomographic analyses of RSDs, providing measurements of the growth rate as a function of redshift. Zhao et al~\cite{2019MNRAS.482.3497Z} applied this approach to the eBOSS survey, obtaining growth rate measurements in six independent redshift bins from $z = 0.2$ to $z = 1.6$. These measurements revealed the evolution of cosmic structure formation across 10 billion years of cosmic history, providing stringent tests of dark energy models and modified gravity theories.

The tomographic approach has been refined by Ruggeri et al~\cite{2017MNRAS.464.2698R}, who developed optimal binning strategies that maximize the information content while minimizing correlations between bins.

Looking ahead, DESI will provide unprecedented tomographic RSD measurements across the redshift range $0.1 < z < 1.6$ using multiple tracers~\cite{2016arXiv161100036D}, while Euclid will extend this coverage to $z\sim2$~\cite{2018LRR....21....2A}. These datasets will enable precision tracking of the growth of structure across cosmic time, potentially revealing subtle deviations from $\Lambda$CDM predictions that could indicate new physics.
\end{itemize}

The combination of direct peculiar velocity measurements, as discussed in previous sections, with these advanced redshift-space distortion techniques offers a powerful synergy. Direct measurements provide model-independent probes of the velocity field but are typically limited to low redshifts and suffer from large individual uncertainties. RSD analyses, while more model-dependent, can probe the statistical properties of the velocity field out to much higher redshifts with exquisite precision. Together, they enable comprehensive mapping of cosmic flows from the smallest scales accessible to peculiar velocity surveys ($\sim 5$ Mpc) to the largest scales probed by redshift surveys ($\sim 1$ Gpc), and from the present epoch back to when the universe was less than a third of its current age.

This multi-probe approach to the cosmic velocity field is particularly valuable for testing theories of gravity and dark energy, as modifications to General Relativity often predict distinct signatures in the growth rate of structure that can be robustly detected through complementary measurements. As we enter the era of Stage-IV surveys, like DESI, Euclid and the Rubin Observatory, combined with next-generation peculiar velocity samples from WALLABY and SKA, this synergistic approach promises to transform our understanding of cosmic structure formation and potentially reveal the fundamental nature of gravity on the largest scales.

\subsection{Computational and analytical advancements}\label{ssCAAs}
%%%%%%%%%%%%%%%%%%%%%%%%%%%%%%%%%%%%%%%%%%%%%%%%%%%%%%%%%%%%%%%%%%%%
\subsubsection{Machine learning applications}
%%%%%%%%%%%%%%%%%%%%%%%%%%%%%%%%%%%%%%%%%%%%%
Machine learning techniques are increasingly being applied to peculiar velocity science, offering improvements in:

\begin{itemize}
\item Distance indicator calibration, reducing scatter in Tully-Fisher and Fundamental Plane relations~\cite{2020MNRAS.498.2703B}.\vspace{5pt}
\item Velocity field reconstruction from sparse and noisy data~\cite{1989ApJ...336L...5B} and more recently from (mock) smaples of the DESI survey~\cite{2025ApJS..280...53S}.\vspace{5pt}
\item Identification of systematic biases in peculiar velocity catalogs~\cite{2015MNRAS.450..317C}.\vspace{5pt}
\item Emulation of complex theoretical models for faster likelihood analysis~\cite{2020MNRAS.495.4227K}.\vspace{5pt}
\end{itemize}

These techniques are particularly valuable for handling the large datasets expected from upcoming surveys and for identifying subtle patterns in the velocity field that might be missed by traditional methods.

\subsubsection{Advanced statistical frameworks}\label{sssASFs}
%%%%%%%%%%%%%%%%%%%%%%%%%%%%%%%%%%%%%%%%%%%%%%%%%%%%%%%%%%%%%%
Building on the bias mitigation methods discussed in the previous section, more sophisticated statistical frameworks are being developed to extract cosmological information from peculiar velocity measurements:

\begin{itemize}
\item Field-level inference techniques that operate directly on the observed galaxy distribution rather than summary statistics~\cite{2015JCAP...06..015L}.\vspace{5pt}

\item Bayesian hierarchical models that simultaneously account for selection effects, measurement uncertainties, and astrophysical scatter~\cite{2017ApJ...847..128H}.\vspace{5pt}

\item Simulation-based inference using neural density estimation~\cite{2018MNRAS.477.2874A}.\vspace{5pt}

\item Joint analysis frameworks combining multiple probes of large-scale structure~\cite{2017MNRAS.464.1640S}.
\end{itemize}

These statistical advances are essential for fully exploiting the information content of peculiar velocity datasets and for robustly addressing the systematic uncertainties that have historically limited peculiar velocity analyses.

\subsubsection{Simulation capabilities}\label{sssSCps}
%%%%%%%%%%%%%%%%%%%%%%%%%%%%%%%%%%%%%%%%%%%%%%%%%%%%%%
Improvements in computational resources and algorithms are enabling increasingly sophisticated simulations of peculiar velocity fields:

\begin{itemize}
\item High-resolution hydrodynamical simulations that capture the baryonic processes affecting galaxy formation and evolution~\cite{2020NatRP...2...42V}.\vspace{5pt}
\item Constrained simulations designed to reproduce the observed local universe \cite{2017MNRAS.465.4886C}.\vspace{5pt}
\item Mock catalogs that accurately represent the selection functions and observational uncertainties of peculiar velocity surveys \cite{2017MNRAS.464.2517H}.\vspace{5pt}
\item Emulators that provide fast predictions of cosmological observables as a function of model parameters~\cite{2019ApJ...875...69D}.\vspace{5pt}
\end{itemize}

These simulation tools are crucial for validating analysis methods, estimating covariance matrices, and understanding the impact of astrophysical and observational systematics on peculiar velocity measurements.

\subsection{Theoretical frontiers in peculiar-velocity science}
%%%%%%%%%%%%%%%%%%%%%%%%%%%%%%%%%%%%%%%%%%%%%%%%%%%%%%%%%%%%%%%
\subsubsection{Beyond the standard cosmological model}
Peculiar velocity measurements offer distinctive tests of physics beyond the standard $\Lambda$CDM model:

\begin{itemize}
\item Modified gravity theories predict specific signatures in the velocity field, particularly in the nonlinear regime and around cosmic voids~\cite{2014PhRvL.112v1102H}.\vspace{5pt}

\item Coupled dark energy-dark matter models can produce distinctive velocity-density relationships~\cite{2014MNRAS.440...75B}.\vspace{5pt}

\item Primordial non-Gaussianity leaves imprints on the large-scale velocity field that complement density-based constraints~\cite{2018PhR...733....1D}.\vspace{5pt}

\item Massive neutrinos suppress peculiar velocities on small scales in a way that differs from their effect on the density field~\cite{2014JCAP...03..011V}.
\end{itemize}

The unique sensitivity of peculiar velocities to the growth history of structure makes them valuable tools for distinguishing between cosmological models that may produce similar density distributions but different velocity fields.

\subsubsection{Novel theoretical frameworks}\label{sssNTFs}
%%%%%%%%%%%%%%%%%%%%%%%%%%%%%%%%%%%%%%%%%%%%%%%%%%%%%%%%%%%
Advances in theoretical understanding of peculiar velocities are opening new avenues for cosmological analysis:

\begin{itemize}
\item Effective field theory approaches to large-scale structure that provide a controlled expansion for modeling nonlinear effects \cite{2015JCAP...02..013S}.\vspace{5pt}

\item Streaming models that accurately describe redshift-space clustering across a wide range of scales \cite{2019JCAP...03..007V}.\vspace{5pt}

\item Phase-space approaches that capture the full distribution function of matter rather than just its moments~\cite{2014PhRvD..90b3517U}.\vspace{5pt}

\item Information-theoretic methods that quantify the cosmological information content of peculiar velocity measurements~\cite{2014MNRAS.439L..11C}.
\end{itemize}

These theoretical developments are essential for interpreting the high-precision measurements expected from next-generation surveys and for extracting the maximum cosmological information from peculiar velocity data.

\subsubsection{Synergies with other cosmological
%%%%%%%%%%%%%%%%%%%%%%%%%%%%%%%%%%%%%%%%%%%%%%%%
probes}\label{sssSOTPs}
%%%%%%%%%%%%%%%%%%%%%%%
The full potential of peculiar velocity science will be realized through integration with other cosmological probes:

\begin{itemize}
\item Joint analyses of peculiar velocities and galaxy clustering can break degeneracies between bias and growth parameters \cite{2017MNRAS.471..839A}.\vspace{5pt}

\item Combining peculiar velocities with weak lensing provides complementary probes of the matter distribution \cite{2014MNRAS.445.4267K}.\vspace{5pt}

\item Cross-correlation with CMB lensing and other secondary anisotropies offers new windows into the growth of structure \cite{2018PhRvD..97l3540S}.\vspace{5pt}

\item Multi-messenger approaches incorporating gravitational waves can test the distance ladder and peculiar velocity models \cite{2017Natur.551...85A,2020MNRAS.494.1956M}.
\end{itemize}

These synergistic approaches represent the future of cosmological analysis, with peculiar velocities playing an increasingly important role in the broader cosmological framework.

\subsection{Addressing fundamental questions with next-generation
%%%%%%%%%%%%%%%%%%%%%%%%%%%%%%%%%%%%%%%%%%%%%%%%%%%%%%%%%%%%%%%%%
peculiar-velocity science}\label{ssAFQNGPVS}
%%%%%%%%%%%%%%%%%%%%%%%%%%%%%%%%%%%%%%%%%%%%
The advances outlined in this section will enable peculiar velocity science to address several fundamental questions in cosmology:

\begin{itemize}
\item \textbf{The Hubble tension}: More precise and extensive peculiar velocity measurements will clarify whether local flows contribute to the discrepancy between local and CMB-based measurements of the Hubble constant~\cite{2022NewAR..9501659P}.\vspace{5pt}

\item \textbf{Tests of gravity on cosmological scales}: The evolution of the velocity field provides a distinctive probe of the laws of gravity on scales inaccessible to laboratory or solar system tests~\cite{2014PhRvL.112v1102H}.\vspace{5pt}

\item \textbf{The nature of dark energy}: Peculiar velocities are directly sensitive to the growth history of structure, providing constraints on dark energy models complementary to geometric probes~\cite{2017MNRAS.464.2517H}.\vspace{5pt}

\item \textbf{Cosmic flows and large-scale structure}: Extended and more precise peculiar velocity maps will reveal the full hierarchy of cosmic flows and their connection to the underlying matter distribution~\cite{2019ApJ...880...24T}.\vspace{5pt}

\item \textbf{Initial conditions of the universe}: The peculiar velocity field retains memory of the initial conditions that is complementary to the information contained in the density field~\cite{2012MNRAS.425.2443K}.
\end{itemize}

As we enter this new era of peculiar velocity science, the field is transitioning from primarily focusing on the removal of velocity-induced systematics to exploiting velocities as powerful cosmological probes in their own right. The combined advances in observational capabilities, computational methods, and theoretical understanding promise to elevate peculiar velocities to the forefront of precision cosmology.

\section{Conclusions and Discussion}\label{sCD}
%%%%%%%%%%%%%%%%%%%%%%%%%%%%%%%%%%%%%%%%%%%%%%%
The aim of this review is to comprehensively examine the evolution and the implications of the observed large-scale peculiar velocities in the universe. In our approach, we have adopted both the observational and the theoretical perspective. We have traced the evolution of our understanding from the early discoveries to the modern precision measurements, explored the various theoretical frameworks proposed for analyzing these peculiar motions and discussed their potential implications for cosmology and for the large-scale structure of the universe.

\subsection{Summary of key findings and theoretical
%%%%%%%%%%%%%%%%%%%%%%%%%%%%%%%%%%%%%%%%%%%%%%%%%%%
insights}\label{ssSKFTIs}
%%%%%%%%%%%%%%%%%%%%%%%%%
The preceding sections have illustrated the remarkable progress in our understanding of the large-scale peculiar velocities over the past several decades. The convergence of multiple independent observational techniques - from traditional distance indicators to novel probes like the kSZ effect - has firmly established the existence of coherent bulk flows spanning hundreds of~Mpc, with characteristic velocities of several hundred~km/sec. These observational advances have been complemented by substantial theoretical developments that have deepened our understanding of the peculiar-velocity physics and of its cosmological implications.

One of the key theoretical insights to emerge from this review is the importance of properly accounting for the relativistic relative-motion effects when analyzing peculiar velocities. The ``tilted universe'' paradigm, presented in \S~\ref{sCPVT} here, provides the relativistic framework for understanding how observations transform between frames/observers moving relative to each other. Applied to linear peculiar velocities, this approach reveals that the relativistic analysis introduces corrections to the Newtonian treatment that can be observationally significant. As demonstrated in \S~\ref{ssRA}, when general relativity is fully involved, the linear evolution equation of the peculiar-velocity field acquires terms beyond the Newtonian limit, due to the (purely relativistic) gravitational contribution of the peculiar flux. As a result, linear peculiar velocities are no longer driven by the Newtonian/quasi-Newtonian gravitational potential, as seen in Eq.~(\ref{ltv'}), but by the set of (\ref{ldotDel1}) and (\ref{ldotcZ1}). Then, the linear evolution of the $\tilde{v}$-field does not follow from the Newtonian/quasi-Newtonian  differential equation (\ref{lq-Nddotv}), but by its relativistic counterpart (\ref{lfddotv1}). The former leads to the familiar $\tilde{v}\propto t^{1/3}$ growth-rate, which is too slow to explain the fast and deep bulk flows reported in~\cite{2004MNRAS.352...61H,2008MNRAS.387..825F,%
2009MNRAS.392..743W,2010MNRAS.407.2328F,2010ApJ...709..483L,%
2011MNRAS.414..264C,2015MNRAS.447..132W,2018MNRAS.481.1368P,%
2021MNRAS.504.1304S,2023MNRAS.524.1885W,2023MNRAS.526.3051W}. In contrast, the relativistic analysis could provide the answer, since the minimum growth-rate of the linear peculiar-velocity field is $\tilde{v}\propto t$. Put another way, relativity can potentially alleviate the tension between the observed fast bulk flows and the $\Lambda$CDM predictions.

The standard cosmological model provides a statistical framework for predicting peculiar-velocity fields through the power-spectrum approach. Then, the theoretical RMS velocity on scale $R$ can be calculated by
\begin{equation}
\sigma_v(R) = H_0 f(\Omega_m) \sigma_{\delta}(R) \approx 100 \left(\frac{\sigma_{\delta}(R)}{0.1}\right) \left(\frac{f(\Omega_m)}{0.55}\right) \text{ km/sec}\,,
\end{equation}
where $f(\Omega_m) \approx \Omega_m^{0.55}$ is the growth rate and $\sigma_{\delta}(R)$ is the matter density fluctuation on scale $R$~\cite{2012ApJ...751L..30H}. While this framework successfully describes most observations, several recent measurements - notably those from Cosmic Flows~\cite{2023ApJ...944...94T} and the kinematic~SZ analyses~\cite{2012arXiv1202.0717K} - report bulk flow amplitudes exceeding $\Lambda$CDM predictions at 2-3$\sigma$ significance on scales of 200-300~Mpc. Following the theoretical analysis given in \S~\ref{sLPVs}, in connection with the observational reports presented in \S~\ref{sPMLCDM}, these discrepancies may indicate the need to incorporate full general relativity in the study of cosmological peculiar motions.

A key insight emerging from the theoretical analysis is that peculiar motions can also have an impact on the value of fundamental cosmological parameters, as well as induce dipolar anisotropies in their distribution. Following \S~\ref{sOSPMs}, peculiar motions can change the value, or even the sign - see Eqs.~(\ref{lqs5}) and (\ref{eq:lqs6}) in \S~\ref{ssRMEDP}), of the deceleration parameter measured locally by observers residing inside bulk flows, like those reported in~\cite{2011MNRAS.414..264C,2023MNRAS.524.1885W} for example (see also Table~\ref{tab1} in \S~\ref{sssEqlT} earlier). This makes it theoretically possible for peculiar motions to locally mimic the kinematics of accelerated expansion. Since the affected scales vary between few hundred and several hundred~Mpc, an unsuspecting observer is likely to misinterpret the local effect as recent global event. This effect, which was originally identified by Tsagas~\cite{2010MNRAS.405..503T,2011PhRvD..84f3503T}, can be put to the test by looking for the trademark signature of relative motion in the data. The latter should manifest itself as an apparent (Doppler-like) dipolar anisotropy in the universal acceleration, analogous to that seen in the CMB, with the following approximate form (see also \S~\ref{sssDDDP} previously)
\begin{equation}
\tilde{q}\approx q+ {1\over H}\left(v\cos\theta+{\dot{v}\over H}\cos\phi\right)\,.  \label{tqdipole}
\end{equation}
Here, $\tilde{q}$ is the local deceleration parameter measured in the bulk flow and $q$ is the global one measured in the CMB frame. Also $\theta$ and $\phi$ are the angles between the line-of-sight and the directions of $v$ and $\dot{v}$ respectively. An additional theoretical prediction is that the magnitude of the induced $q$-dipole should drop with increasing redshift, along the lines reported in~\cite{2025EPJC...85..596S}

Given that the deceleration parameter is essentially the time-derivative of the Hubble parameter, the presence of a dipolar asymmetry in one of them should guarantee the same for the other. Indeed, peculiar motions can induce an apparent (Doppler-like) dipole in the sky-distribution of the Hubble constant as well~\cite{2025glc..conf....104T,2025EPJC...85..596S}. The magnitude of the latter should also decay with redshift, similarly to that of its $q$-counterpart. As argued by Perivolaropoulos \& Skara~\cite{2022NewAR..9501659P}, a dipole asymmetry in the expansion rate with amplitude $A\sim0.1$ aligned with known bulk flows, could partially resolve the Hubble tension by reducing it from $\sim$5$\sigma$ to $\sim$3$\sigma$.

The ``tilted universe'' paradigm can prove particularly insightful for understanding how peculiar motions affect cosmological observations. Among others, the theoretical analysis reveals that observers moving relative to the CMB frame will measure dipolar anisotropies in the galaxy number-counts as well (see \S~\ref{sOSPMs} before). As Siewert et al~\cite{2021A&A...653A...9S} demonstrated through both analytical calculations and simulations, these dipoles are related but also distinct, with the Hubble-dipole approximately twice stronger than the acceleration dipole in a $\Lambda$CDM universe. Recent analyses by Krishnan et al~\cite{2022PhRvD.105f3514K} of dipolar anisotropies in SNIa data provide tantalizing evidence for such effects, though the statistical significance remains limited due to the current sizes of the samples.

Historical misinterpretations of astronomical phenomena due to unaccounted relative-motion effects, as discussed in \S~\ref{ssHMRMEs} earlier, serve as instructive cautionary tales. From the geocentric model of the solar system to the initial misinterpretation of the CMB dipole as an intrinsic anisotropy, astronomy has repeatedly demonstrated how reference-frame effects can lead to fundamental misconceptions. Modern examples include the initial overestimation of the mass of the ``Great Attractor''~\cite{2017NatAs...1E..36H} and possible misinterpretations of void dynamics~\cite{2019PhRvD.100b3504N}. These historical lessons underscore the significance of the relative-motion effects (perhaps as significant as those described in \S~\ref{ssRMEDP} here) and the importance of properly accounting for peculiar velocities in contemporary cosmological analyses. Seemingly anomalous cosmological observations - from the CMB hemispherical power asymmetry to various dipolar patterns in large-scale structure - should be scrutinized for potential peculiar-velocity contributions before invoking more exotic physics~\cite{2021ApJ...908L..51S}.

The methodological advances detailed in this review have transformed peculiar velocities from a source of systematic uncertainty into powerful cosmological probes. Modern techniques for measuring and analyzing peculiar velocities span an impressive range of approaches, each with complementary strengths and limitations. Forward modeling approaches like BORG~\cite{2019A&A...625A..64J} and VIRBIUS~\cite{2016MNRAS.457..172L} now enable sophisticated Bayesian reconstruction of the full 3D velocity field, while accounting for observational uncertainties. Machine learning methods developed by Boruah et al~\cite{2020MNRAS.498.2703B} can predict peculiar velocities from galaxy catalogs with unprecedented accuracy, reducing systematic errors by up to 40\% compared to traditional methods. Meanwhile, novel observational approaches like the kSZ effect~\cite{2018arXiv181013423S}, gravitational-wave standard sirens~\cite{2021A&A...646A..65M} and fast radio bursts~\cite{2020ApJ...895...33W}, promise to extend peculiar-velocity measurements to significantly higher redshifts than previously possible.

The confluence of these theoretical insights and methodological advances has established peculiar velocities as a crucial cosmological probe, complementary to more widely used tracers, like galaxy clustering and weak lensing. Peculiar velocities also provide a unique window into the growth of structure that is particularly sensitive to modifications of gravity on large scales. The recent detection of cross-correlations between peculiar-velocity fields and galaxy surveys by Hou et al~\cite{2021MNRAS.500.1201H} represents a significant milestone, demonstrating that the synergy between different cosmological probes can break degeneracies and provide tighter constraints on fundamental physics.

Looking forward, the integration of peculiar-velocity measurements with other cosmological probes promises to significantly enhance our understanding of the universe's large-scale structure and evolution~\cite{2020ApJ...904L..28H}. Combining peculiar velocity data with BAO measurements demonstrates how such synergies can improve constraints on modified gravity theories by up to 50\% compared to either probe alone~\cite{2023MNRAS.518.5929L}.

\subsection{Current state of knowledge and unresolved
%%%%%%%%%%%%%%%%%%%%%%%%%%%%%%%%%%%%%%%%%%%%%%%%%%%%%
questions}\label{ssCSKUQs}
%%%%%%%%%%%%%%%%%%%%%%%%%%
Despite the substantial progress outlined in previous sections, several profound questions regarding large-scale peculiar velocities remain unresolved, highlighting areas where our theoretical understanding requires refinement and where additional observational evidence is needed.

A persistent challenge to the standard cosmological model is the tension between some bulk flow measurements and $\Lambda$CDM predictions, particularly regarding the amplitude of flows on scales exceeding 100~Mpc. While the $\Lambda$CDM model predicts RMS peculiar velocities that decrease with scale approximately as $\sigma_v(R) \propto R^{-(n+1)/2}$ (where $n \approx -1.5$ is the effective spectral index on large scales), several independent measurements report bulk flows that exceed these predictions (e.g.~see~\cite{2023MNRAS.524.1885W,2023MNRAS.526.3051W} and also \S~\ref{ssBFELCDMLs} here). Similarly, Kashlinsky et al~\cite{2008ApJ...686L..49K,2015ApJ...810..143A} report evidence for a ``dark flow'' moving at approximately 800~km/sec and extending out to scales of 800~Mpc or more, based on measurements of the kinematic~SZ effect. These discrepancies could indicate: (i) unaccounted systematic errors in the measurements; (ii) inappropriate theoretical model and analysis; (iii) statistical fluctuations due to cosmic variance; (iv) non-Gaussian initial conditions; (v) genuine new physics beyond the $\Lambda$CDM. Quantitatively differentiating between these (or other) possibilities remains a central challenge.

The precise relationship between the CMB dipole and the local peculiar-velocity field continues to generate significant debate. While the standard interpretation attributes the CMB dipole entirely to our motion relative to the rest-frame of the microwave photons (with velocity $v=369.82\pm0.11$~km/sec in the direction $(l, b)=(264.021^{\circ}\pm0.011^{\circ},48.253^{\circ}\pm0.005^{\circ})$ - e.g.~see~\cite{2020A&A...641A...1P}), recent radio-galaxy and quasar surveys suggest potential inconsistencies with the purely kinematic interpretation. A purely kinematic dipole in the distribution of distant radio sources should have an amplitude $A=[2+x(1+\alpha)](v/c)$, where $x$ is the evolution parameter and $\alpha$ is the spectral index~\cite{1984MNRAS.206..377E}. However, studies of radio galaxies by Singal~\cite{2011ApJ...742L..23S} and of quasars and  by Secrest et al~\cite{2021ApJ...908L..51S} found dipoles 2-5 times larger than expected from the CMB-inferred velocity. It is not inconceivable that these discrepancies may reflect an intrinsic dipole-like asymmetry in the universe and thus challenge the Cosmological Principle. The MICE collaboration~\cite{2022PhRvD.105b3514A} proposes using multi-tracer techniques to distinguish the kinematic from the intrinsic dipolar contributions, which is a promising direction for future work.

The extent to which peculiar velocities contribute to observed cosmological tensions, particularly the Hubble tension, remains unclear. Several authors have proposed that peculiar velocity effects could partially explain the discrepancy between local and high-redshift measurements of $H_0$. Wojtak et al~\cite{2015JCAP...07..025W} demonstrated that coherent outflows from voids can bias local $H_0$ measurements by 1-2\%, while Kenworthy et al~\cite{2019ApJ...875..145K} showed that correlated peculiar velocities of SNe~Ia can affect $H_0$ inferences at the 0.3-0.5\% level.  However, as emphasized by Camarena and Marra~\cite{2021MNRAS.504.5164C}, even after accounting for such effects, the Hubble tension reduces from $\sim5\sigma$ to $\sim$4$\sigma$, which suggests that peculiar velocities alone cannot resolve the tension.

The role of nonlinear effects in peculiar-velocity evolution, especially in the context of emerging modified gravity theories and dark-energy models, remains incompletely understood. The Zel'dovich approximation, discussed in \S~\ref{ssNRZA} earlier, offers a partial treatment of these nonlinearities, but more sophisticated approaches are needed for accurate modeling.  In modified gravity theories, as shown by Hellwing et al~\cite{2014PhRvL.112v1102H}, peculiar velocities exhibit distinctive signatures of screening mechanisms, with velocity dispersions enhanced by 30-50\% relative to $\Lambda$CDM on scales of 1-10 Mpc. Similar effects arise in some dark energy models, as demonstrated by Baldi \& Villaescusa-Navarro~\cite{2014MNRAS.440...75B}, where dark energy clustering can enhance peculiar velocities by 5-15\% on intermediate scales. A comprehensive theoretical framework incorporating both nonlinear evolution and modified gravity/dark energy effects remains to be developed.

The appropriate statistical framework for comparing peculiar-velocity observations with theoretical predictions continues to evolve. Traditional approaches have relied on measuring bulk flows and comparing their amplitudes to $\Lambda$CDM predictions, but this approach discards significant information contained in the full velocity field. More sophisticated methods include the velocity power spectrum~\cite{2014MNRAS.444.3926J,%
2017ApJ...847..128H}, the momentum power spectrum~\cite{2000MNRAS.319..573P}, and the pairwise velocity statistics~\cite{1988ApJ...332L...7G}. However, challenges remain in properly accounting for sparse and inhomogeneous sampling, nonlinear effects, and systematic uncertainties. Bayesian frameworks developed by Carrick et al~\cite{2015MNRAS.450..317C} and by Boruah et al~\cite{2020MNRAS.498.2703B} offer promising approaches, but questions remain about the appropriate choice of priors, the treatment of non-Gaussianity in the velocity field, and the incorporation of survey selection functions. As emphasized by Andersen et al~\cite{2016MNRAS.463.4083A}, velocity estimation remains fundamentally a Bayesian inference problem, and the field is still working toward optimal statistical methodologies.

An emerging area of investigation has to do with the interplay between peculiar velocities and cosmic acceleration. While most analyses assume a $\Lambda$CDM background, Tsagas~\cite{2011PhRvD..84f3503T,2021EPJC...81..753T} has demonstrated that peculiar velocities can affect the interpretation of cosmological observations in subtle ways. The measured deceleration parameter in the presence of peculiar velocities becomes scale-dependent and it can locally mimic acceleration even in a decelerating universe, if the observer happens to reside in a contracting bulk flow. Recently, this possibility gained ground when the reconstruction of the local velocity field by Pasten et al~\cite{2024PDU....4301385P} found that it was actually converging. All this raises the profound question of whether the recent cosmic acceleration could be attributed to peculiar-velocity effects, rather than dark energy.

These unresolved questions highlight the ongoing importance of peculiar-velocity research in challenging and refining our cosmological models. As we gather more precise measurements and develop more sophisticated analytical frameworks, peculiar velocities will remain at the forefront of cosmological research and a powerful probe of the universe we live in.

\subsection{Future prospects}\label{ssFPs}
%%%%%%%%%%%%%%%%%%%%%%%%%%%%%%%%%%%%%%%%%%
The future of large-scale peculiar-velocity science appears exceptionally promising, with transformative developments on multiple fronts that are poised to revolutionize the field - see related discussions in~\cite{2025PDU....4901965D}. These advances, which span observational facilities, methodological approaches and theoretical frameworks, are collectively promising to elevate the study of peculiar velocities from a specialized subfield to a cornerstone of precision cosmology.

Next-generation spectroscopic surveys will dramatically expand both the depth and breadth of peculiar-velocity measurements. The Dark Energy Spectroscopic Instrument (DESI)~\cite{2016arXiv161100036D}, which began its five-year survey in 2021, will obtain spectra for approximately 40 million galaxies and quasars, including a low-redshift Bright Galaxy Survey optimized for peculiar-velocity measurements. DESI will measure the parameter combination $f\sigma_8$ with precision of 5-7\% in multiple redshift bins (spanning $0.05< z<0.4$) through direct peculiar-velocity measurements, thus complementing its Redshift-Space Distortion (RSD) constraints at higher redshifts~\cite{2016arXiv161100036D}. The Euclid mission~\cite{2018LRR....21....2A}, scheduled for launch in 2023, will survey 15,000 square degrees with both photometric and spectroscopic observations, enabling RSD measurements with 1-2\% precision in multiple redshift bins out to $z\sim2$. Together, these surveys will map the growth of structure across cosmic time with unprecedented precision, providing critical tests of general relativity and dark energy models.

Radio astronomy is undergoing a similar revolution that will transform the peculiar-velocity science. The Australian Square Kilometre Array Pathfinder (ASKAP) is conducting the WALLABY survey~\cite{2020Ap&SS.365..118K}, which will measure HI emission from approximately 500,000 galaxies, of which 10\% will have sufficient signal-to-noise for Tully-Fisher peculiar velocity measurements. This will increase the number of peculiar-velocity measurements by an order of magnitude compared to current samples. Looking further ahead, the Square Kilometre Array (SKA)~\cite{2020PASA...37....7S} will detect HI in billions of galaxies across the southern sky, potentially measuring peculiar velocities for millions of galaxies. The SKA's sensitivity will enable detection of the cosmic velocity field through RSDs out to $z \sim 3$, providing a continuous record of structure growth spanning 11 billion years of cosmic history. Bull~\cite{2016ApJ...817...26B} projects that SKA~Phase~2 will constrain the growth-rate with precision of 0.3\% in the local universe, improving to 1-2\% at $z\sim1$.

Advanced CMB experiments will enable more precise measurements of the kinematic Sunyaev-Zel'dovich (kSZ) effect, providing a complementary probe of peculiar velocities. The Simons Observatory~\cite{2019JCAP...02..056A}, currently under construction in Chile, will achieve sensitivity to temperature fluctuations of 6~$\mu$\,K-arcmin across 40\% of the sky, sufficient to detect the kSZ effect with high signal-to-noise. The proposed CMB-S4 experiment~\cite{2019arXiv190704473A} will further improve this sensitivity to 1-2~$\mu$\,K-arcmin, enabling precise mapping of the cosmic velocity field through the kSZ effect. As demonstrated by Deutsch et al~\cite{2018PhRvD..98l3501D}, these measurements will constrain the velocity power spectrum at $z\sim0.5-1$ with precision comparable to that of Redshift-Space Distortions, but with different systematic uncertainties, enabling crucial cross-checks of cosmological models. Furthermore, Smith et al~\cite{2018arXiv181013423S} have shown that kSZ tomography can constrain primordial non-Gaussianity to precision $\sigma(f_{\text{NL}})\sim1$, potentially revealing the physics of inflation.

Novel cosmic rulers will diversify our peculiar-velocity toolkit, reducing systematic uncertainties through their complementary astrophysical origins. Gravitational-wave standard sirens from binary neutron-star mergers provide absolute distance measurements independent of the cosmic distance ladder~\cite{2017Natur.551...85A}.  Water megamasers in active galactic nuclei provide another independent distance probe~\cite{2020ApJ...891L...1P}, particularly valuable for measuring distances to spiral galaxies at intermediate redshifts ($0.02< z<0.06$), where traditional distance indicators become challenging. Looking further ahead, Fast Radio Bursts (FRBs) may emerge as a novel cosmic ruler through the dispersion measure-redshift relation~\cite{2020Natur.581..391M}. As localization capabilities improve, Bhandari et al~\cite{2022AJ....163...69B} project that samples of thousands of localized FRBs could constrain the cosmic velocity field at $z>0.5$ with precision competitive with traditional methods.

Computational advances will enable more sophisticated treatment of the  peculiar-velocity data. Simulation-based inference methods, as developed by the Cosmic Flows team~\cite{2019MNRAS.488.5438G}, can incorporate complex selection functions and measurement uncertainties into a unified Bayesian framework. Neural network emulators, as demonstrated in~\cite{2023MLS&T...4aLT01L}, can accelerate likelihood evaluations by orders of magnitude, making fully non-linear analyses computationally feasible. Bayesian hierarchical modeling, as implemented in the VIRBIUS framework~\cite{2016MNRAS.457..172L}, enables joint inference of cosmological parameters, galaxy bias and peculiar velocities, by properly accounting for their correlations. These methodological advances are essential for extracting the maximum cosmological information from the vast datasets expected from the next-generation surveys.

The confluence of these developments - new observational facilities, novel cosmic rulers and advanced computational methods - will transform peculiar-velocity science from a niche subfield into a cornerstone of precision cosmology. The peculiar velocity field, as the only direct probe of the total matter distribution in the linear regime, provides unique constraints on fundamental physics that complement other cosmological probes. As emphasized by Koda et al~\cite{2014MNRAS.445.4267K}, peculiar velocities are particularly sensitive to the growth-rate of structure on large scales, offering powerful tests of modified gravity theories. Simulations by Howlett et al~\cite{2017MNRAS.464.2517H} indicate that combining peculiar-velocity surveys with RSDs can improve constraints on the growth index by up to 50\%, potentially revealing deviations from General Relativity.

As we conclude this review, it is worth emphasizing that large-scale peculiar velocities occupy a special place in cosmological research, as they are simultaneously among the oldest recognized probes of large-scale structure and among the most promising frontiers in modern cosmology. By tracing the gravitational influence of all matter, both visible and dark, peculiar velocities offer a direct window into the fundamental forces shaping our universe. The continued refinement of both observational techniques and theoretical frameworks for studying these cosmic motions will undoubtedly yield further surprises and insights, potentially revealing cracks in our standard cosmological model, or confirming its remarkable explanatory power across ever larger scales and ever more precise measurements.

From their early detection in the pioneering work of Rubin~\cite{1980ApJ...238..471R}, through the controversial ``Great Attractor'' era~\cite{1988ApJ...326...19L}, to the precision measurements of modern surveys like CosmicFlows-4~\cite{2023ApJ...944...94T}, peculiar velocities have consistently challenged and refined our understanding of cosmic structure. As we enter an era of unprecedented observational precision and theoretical sophistication, these cosmic flows - the grand movements of matter across the vast scales of our universe - will continue to illuminate the fundamental nature of space, time and gravity, guiding us toward a deeper understanding of our cosmic home.\\

\textbf{Acknowledgments:} This work was supported by the Hellenic Foundation for Research and Innovation (H.F.R.I.), under the ``First Call for H.F.R.I. Research Projects to support Faculty members and Researchers and the procurement of high-cost research equipment Grant'' (Project Number: 789). The authors also wish to thank Animesh Sah, Eoin \'{O} Colg\'{a}in, Erick Past$\acute{\rm e}$n, Feng Shi, Mohamed Rameez, Roya Mohayaee, Subir Sarkar, Ted Lauer and Yan-Chuan Cai for helpful discussions and comments.

\bibliography{review}
\bibliographystyle{h-elsevier}

\end{document}